\documentclass[useAMS,usenatbib,usegraphicx]{mn2e}
\usepackage[usenames,dvipsnames]{color}

\title[Cluster entropy profiles]
  {How AGN feedback and metal cooling shape cluster entropy profiles}

\author[Y. Dubois et al.]
  {Yohan~Dubois$^1$\thanks{E-mail: yohan.dubois@physics.ox.ac.uk}, Julien~Devriendt$^{1,2}$, Romain Teyssier$^{3,4}$ and Adrianne~Slyz$^1$ \\
  $^1$Astrophysics, University of Oxford, Denys Wilkinson Building, Keble Road, Oxford, OX13RH, United Kingdom\\
  $^2$Centre de Recherche Astrophysique de Lyon, Universit\'e de Lyon I, CNRS UMR 5574, ENS-Lyon, 9 Avenue Charles Andr\'e,\\ 69561, St-Genis-Laval Cedex, France\\
  $^3$Universit\"at Z\"urich, Institute f\"ur Theoretische Physik, Winterthurerstrasse 190, CH-8057 Z\"urich, Switzerland\\
  $^4$CEA Saclay, DSM/IRFU/SAP, B\^atiment 709, F-91191 Gif-sur-Yvette, Cedex, France
}
\date{Accepted . Received ; in original form }

\pagerange{\pageref{firstpage}--\pageref{lastpage}} \pubyear{2010}

\def\LaTeX{L\kern-.36em\raise.3ex\hbox{a}\kern-.15em
    T\kern-.1667em\lower.7ex\hbox{E}\kern-.125emX}

\begin{document}

\label{firstpage}

\maketitle

\begin{abstract}
Observed clusters of galaxies essentially come in two flavors: non cool core clusters characterized by an isothermal
temperature profile and a central entropy floor, and cool-core clusters where temperature and entropy in the central region
are increasing with radius. Using cosmological resimulations of a galaxy cluster,  we study
the evolution of its intracluster medium (ICM) gas properties, and through them we assess the effect
of different (sub-grid) modelling of the physical processes at play, namely gas cooling, star formation,
feedback from supernovae and active galactic nuclei (AGN). More specifically we show that AGN feedback
plays a major role in the pre-heating of the proto-cluster 
as it prevents a high concentration of mass from collecting in the center of the future galaxy cluster at
early times. However, AGN activity during the cluster's later
evolution is also required to regulate the mass flow into its
core and prevent runaway star formation in the central galaxy. Whereas the energy deposited by supernovae alone
is insufficient to prevent an overcooling catastrophe, supernovae are responsible for spreading a large
amount of metals at high redshift, enhancing the cooling efficiency of the ICM gas. As the AGN energy
release depends on the accretion rate of gas onto its central black hole engine, the AGN responds to this
supernova enhanced gas accretion by injecting more energy into the surrounding gas, and as a result increases
the amount of early pre-heating. We demonstrate that the interaction between an AGN jet and the ICM gas
that regulates the growth of the AGN's BH, can naturally produce cool core clusters
 if we neglect metals. However, as soon as metals are allowed to contribute to
the radiative cooling, only the non cool core solution is produced.

\end{abstract}


\section{Introduction}

Galaxy clusters are the largest virialised structures in the Universe directly
observed using radio relics, the Sunyaev Zel'dovich effect,
gravitational lensing, and X-ray measurements. These observations
provide support to a cold dark matter like bottom-up 
formation scenario, by showing that a large variety of clusters, spanning an 
increasingly wider mass range, are present at different epochs, from smooth relaxed objects
to disturbed and merging ones. However, such support comes at a cost:
as one goes back further in time, the build-up of clusters becomes
a progressively more complex process that involves accretion of cold material
from the diffuse inter-galactic medium (IGM) and filaments, shock heating 
in the proto-cluster potential well along with multiple mergers potentially stripping 
a significant amount of gas and stars from galaxies.

Although the assembly histories of different clusters are generally 
extremely different, X-ray measurements of unperturbed galaxy clusters
feature gas with similar properties, especially entropy\footnote{We
take the usual definition of the entropy in astrophysics
$K=T/n_e^{2/3}$, where $T$ is the gas temperature and $n_e$ is the
electron number density assuming a mean molecular weight for electrons
of $\mu_e=1.167$ in the hot cluster gas. The logarithm of this entropy
is the standard definition of entropy in thermodynamics $S=\log K$.}
profiles, which are strikingly self-similar down to the central 100 
(or even 10 in some cases) kiloparsecs \citep{lloyd-daviesetal00, piffarettietal05, donahueetal06,
morandi&ettori07, cavagnoloetal09, sandersonetal09, prattetal10}.
This is interpreted as the consequence of the formation of a large-scale
gravitational-shock which heats the cold infalling gas from the IGM
(~$10^4$ K) into an extremely hot and turbulent intra-cluster medium
(ICM) with temperatures around $10^7$-$10^8$ K \citep{tozzi&norman01}.

However in the cores ($< 100$ kpc) of these clusters there seem to exist
a dichotomy in gas properties which has led observers to split cluster 
samples into 'cool core' and 'non-cool core' clusters. Arguably the most 
relevant of these gas properties is the entropy, as non-cool
core clusters exhibit a well defined entropy floor in their centre, whereas
in cool-core clusters the entropy profile decreases with decreasing
radius on all scales \citep{morandi&ettori07, sandersonetal09}.  This
lack of entropy floor in the central regions of some clusters is
believed to be the signature of a cooling core because of the 
decreasing central temperature profile associated to it \citep{sandersonetal06, vikhlininetal06,
prattetal07}.

On the other hand, one of the long-standing problems of cosmological hydrodynamics
simulations of cluster formation is the so-called `cooling
catastrophe' which takes place when gas is allowed to radiatively cool and `standard'
sub-grid physics, such as star formation and feedback from supernovae
(SN), is implemented \citep{suginohara&ostriker98, lewisetal00,
pearceetal00, borganietal02, ettorietal04, kravtsovetal05, borganietal05, nagaietal07}.  Runaway cooling
occurs, with severe consequences on both ICM and galaxy
properties: the formation of an enormous cooling flow of gas ($\sim1000$ M$_\odot$/yr) into the
core triggers an anomalously high emission of X-rays, the presence of
a persistent galactic disc of gas, and tremendous episodes of star formation. 
Central galaxies rapidly become too massive and blue.  Thus, getting realistic 
thermodynamical properties of the ICM by numerical means constitutes a major challenge.

Several processes have been invoked to solve this theoretical puzzle:
pre-heating from early feedback processes in high-redshift galaxies
\citep{babuletal02}, Active Galactic Nuclei (AGN) feedback from
quasars or radio jets \citep{binney&tabor95, rephaeli&silk95}, conduction of thermal energy from the outer shock-heated regions
carried by electrons \citep{voigt&fabian04,rasera&chandran08}, and gas sloshing from
minor and major mergers \citep{fabian&daines91}.
Whilst the verdict is still out for early pre-heating and thermal 
conduction, the ability of AGN feedback to stem cooling flows has already 
been demonstrated in several hydrodynamics simulations \citep{sijacki&springel06, 
sijackietal07, khalatyanetal08, puchweinetal08, fabjanetal10, duboisetal10, mccarthyetal10, 
mccarthyetal11, teyssieretal11}.  It seems therefore natural to consider such a
 feedback mechanism as key to account for the
self-regulation of the cold baryonic content of massive galaxies.  
However it remains to be demonstrated whether or not it can also explain
the wide variety of observed ICM properties.  Especially, it is not obvious how 
such a physical process can reproduce the cool/non-cool core dichotomy observed in the population 
of galaxy clusters.

In a recent work \citep{duboisetal10}, we demonstrated that 
jet-mechanical feedback from a central AGN is able to suppress the cooling
catastrophe in a re-simulated cosmological galaxy cluster.  Using a
different numerical approach for gas dynamics and AGN feedback,
\cite{mccarthyetal11} recently showed that the excess of entropy in
the cores of a few galaxy groups is generated by the selective ejection of
low-entropy material at high redshift $z=2$-4 caused by both SN and AGN
feedback.  However much work is still needed to assess (i) whether the same mechanism works for
more massive structures such as galaxy clusters, where mass loss from large outflows 
can be more easily prevented and (ii) to which extent such a mechanism depends 
on the numerical method and the specific subgrid model implemention of the AGN feedback.  

Therefore the aim of the present paper is to explore how various implementations 
of AGN feedback, which differ either by the nature of the energy injected (kinetic or thermal) 
or by the epoch at which this energy injection occurs, impact the ICM gas properties.  
More specifically, we focus on the time evolution of thermodynamical quantities  
and mass distributions in our simulations, comparing and contrasting our results with observations
whenever possible. The paper is organised as follows.  In section~\ref{numerics},
we describe the basic numerical ingredients of the simulations which
are analysed.  In section~\ref{results}, we present
our results and compare the evolution of entropy, density and temperature profiles
of the simulated galaxy clusters. 
Finally in section \ref{conclusion}, we discuss the consequences of
these numerical experiments.

\section{Numerics}
\label{numerics}

The simulations presented in this paper use variations around the sub-grid physics
and the exact same initial conditions as in
~\cite{duboisetal10}. For completeness' sake we briefly recall in this
section the basics of the numerical models that we employ.

\subsection{Physics of galaxy formation}

Gas radiates energy through atomic collisions assuming a H/He
primordial composition \citep{sutherland&dopita93}. As a result it can
collapse into dark matter (DM) potential wells to form galaxies
\citep{silk77}. To model reionization, a homogeneous
 UV background heating is imposed from $z=8.5$ using the prescription of
\cite{haardt&madau96}.  Star formation occurs in high gas density
regions $\rho> \rho_0$ ($\rho_0=0.1\, \rm H.cm^{-3}$). When the
density threshold is surpassed, a random Poisson process spawns star
cluster particles according to a Schmidt law $\dot \rho_*= \epsilon
\rho/t_{\rm ff}$, where $t_{\rm ff}$ is the gas free-fall time and
$\epsilon$ is the star formation efficiency, taken to be
$\epsilon=0.02$ \citep{krumholz&tan07} in order to reproduce the
observational surface density laws \citep{kennicutt98}. The reader can
consult \cite{rasera&teyssier06} and \cite{dubois&teyssier08winds} for
more information on the star formation method.

Feedback from SN type II is included in a similar fashion as 
\cite{dubois&teyssier08winds}: we model SN explosions  by 
modifying the gas mass, momentum, and energy of surrounding cells
following a Sedov blast wave solution.  We adopt this approach because
since we do not resolve SN remnants, 
thermal energy input in local cells would be quickly radiated away by the
efficient gas cooling in the inter-stellar medium (ISM,
\citealp{navarro&white93}).  Our sub-grid model yields a
physically motivated description of SN bubble expansion, and 
produces large-scale galactic winds in low-mass
halos \citep{dubois&teyssier08winds}.  As we adopt a Salpeter IMF for star formation,
we release an average of $10^{51}$ ergs per 150 $M_{\odot}$ of stars formed in SN explosions.
We return all the gas expelled by the SN explosion after
10 Myr, 10 \% of which is assumed to be metals which we advect as a passive scalar.
These metals also contribute to the cooling function of the gas 
assuming a Solar abundance ratio for the different elements.

We use a modified equation of state (EoS) at high gas density
$\rho> \rho_0$ to prevent numerical instabilities from artificially growing
on the smallest grid scales \citep{trueloveetal97}.  More
specifically, the minimum temperature in dense regions is set to $T_{\rm
min}=T_0 (\rho/\rho_0)^{n-1}$, with $T_0=10^4K$, and $n=4/3$ which
leads to a constant Jeans mass $M_{\rm J}=1.3\, 10^9 \, \rm
M_{\odot}$. Such a value of the polytropic index $n$ is a rough
approximation of the complex functional form of the EoS obtained by
analytical modeling of the multiphase structure of the ISM in
\cite{springel&hernquist03}.

We introduced a novel numerical scheme for AGN feedback in
\cite{duboisetal10} based on bipolar kinetic outflows (or jets). We
assume that a unique black hole (BH) engine located at the center of each galaxy
pumps a fraction of the energy it gains by accreting gas into such a jet.
BHs, are modeled as sink particles \citep{bateetal95, krumholzetal04}, and
can grow both by mergers with other BHs, and by accreting some of the
surrounding gas according to the Bondi accretion rate
\begin{equation} \dot M_{\rm BH}=4\pi \alpha G^2 M_{\rm BH}^2 {\bar
\rho \over (\bar c_s^2 + \bar u_r^2)^{3/2}} \, ,
\label{dMBH}
\end{equation} where $\alpha=(\rho/\rho_0)^2$ is a dimensionless boost
factor that accounts for the unresolved small scale structure of the
ISM \citep{booth&schaye09}, $G$ is the gravitational constant, $M_{\rm
BH}$ is the black hole mass, $\bar \rho$ is the average gas density,
$\bar c_s$ is the average sound speed, and $\bar u_r$ is the average velocity of
the black hole with respective to the surrounding gas.
The accretion rate onto a BH is
Eddington-limited
\begin{equation} \dot M_{\rm Edd}={4\pi G M_{\rm BH}m_{\rm p} \over
\epsilon_{\rm r} \sigma_{\rm T} c}\, ,
\label{dMEdd}
\end{equation} where $\sigma_{\rm T}$ is the Thompson cross-section,
$c$ is the speed of light, $m_P$ is the proton mass, and
$\epsilon_{\rm r}$ is the radiative efficiency, assumed to be equal to
$0.1$ for the \cite{shakura&sunyaev73} accretion onto a Schwarzschild
BH.

The total AGN luminosity is simply proportional to the rest-mass
accreted energy 
\begin{equation} \dot E_{\rm AGN}=\epsilon_f \epsilon_r \dot M_{\rm
BH}c^2\, ,
\label{E_BH}
\end{equation} where $\epsilon_f$ is set to 1 to recover $M_{\rm BH}$--$M_*$,
$M_{\rm BH}$--$\sigma_*$ relations consistent with observational
findings \citep{tremaineetal02, haring&rix04} using AGN jets (Dubois et al., in prep.). 
The gas properties are locally modified around the BH to account for the presence of a jet with $10^4$ km/s velocity (as in the original prescription of \cite{Ommaetal04}, see \citealp{duboisetal10} for more details on the scheme) within a cylinder of radius $\Delta x$ and height $2 \Delta x$.

We have also introduced a thermal implementation of AGN feedback
\citep{teyssieretal11} based on earlier works performed with
smoothed-particle hydrodynamics (SPH) simulations
\citep{sijackietal07, booth&schaye09}.  In this alternative subgrid modelling, we release the AGN energy  in
a thermal form as soon as it reaches a level high enough to reheat the surrounding gas to temperatures above $10^7$ K. 
 The radius of the injection bubble centred on the BH is chosen to be $\Delta x$.
As the energy is more efficiently coupled to the gas, we
have to assume a different efficiency $\epsilon_f$. As in the jet case, its value of $0.15$
to reproduce the $M_{\rm BH}$--$M_*$, $M_{\rm BH}$--$\sigma_*$ relations.  All
other parameters of the model are kept identical to the AGN jet
feedback.

\subsection{Initial conditions and simulation runs}

\begin{figure*}
  \centering{\resizebox*{!}{5.7cm}{\includegraphics{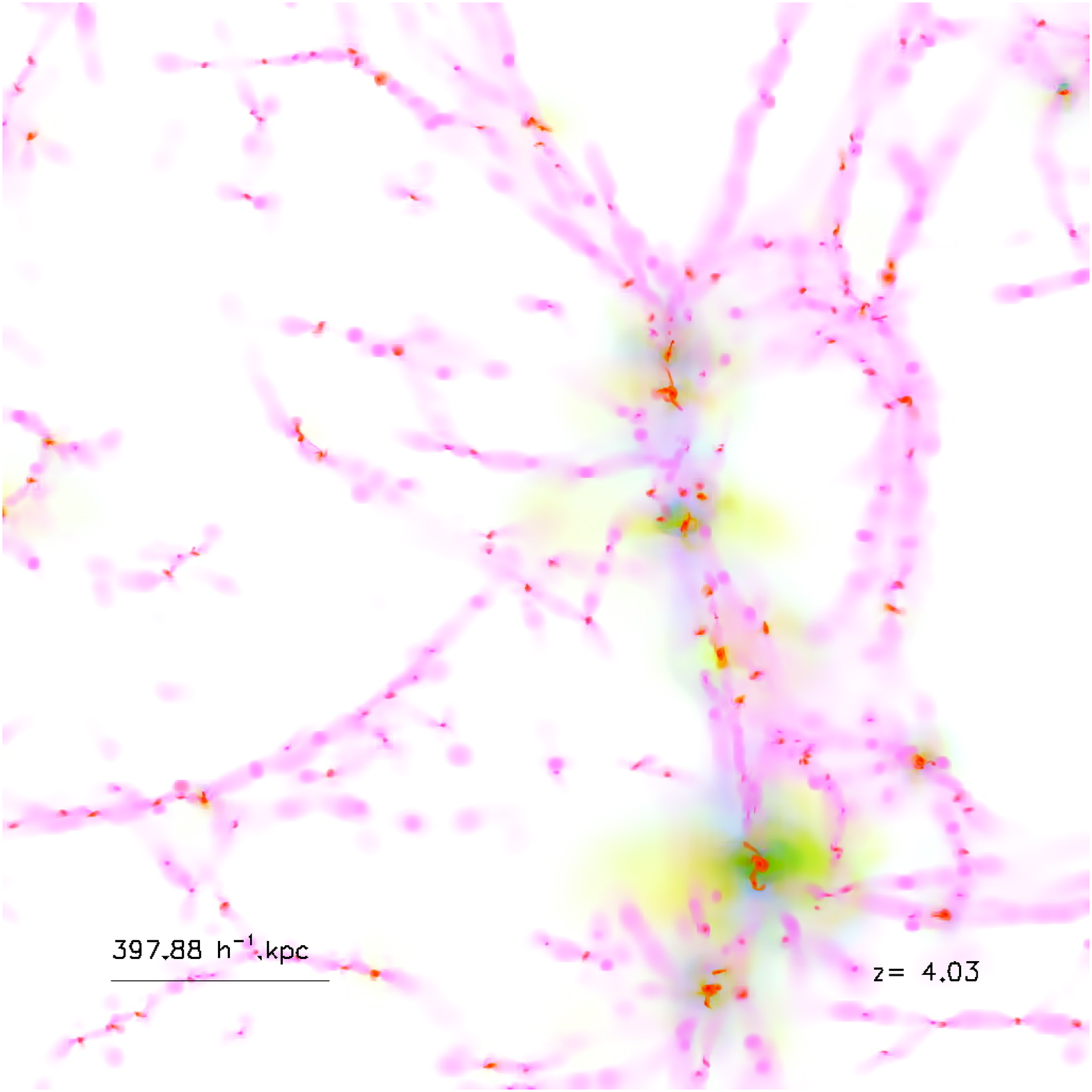}}}
  \centering{\resizebox*{!}{5.7cm}{\includegraphics{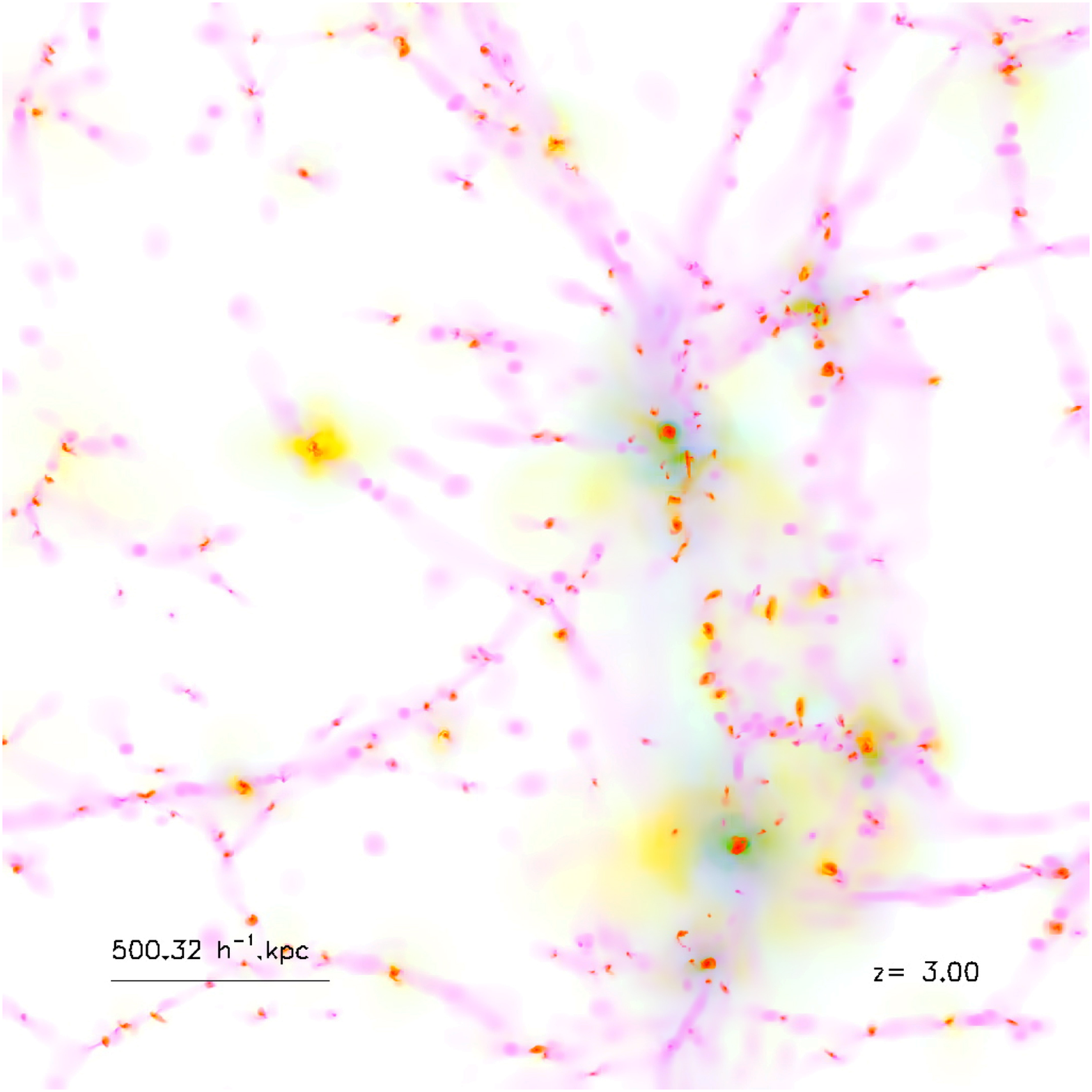}}}
  \centering{\resizebox*{!}{5.7cm}{\includegraphics{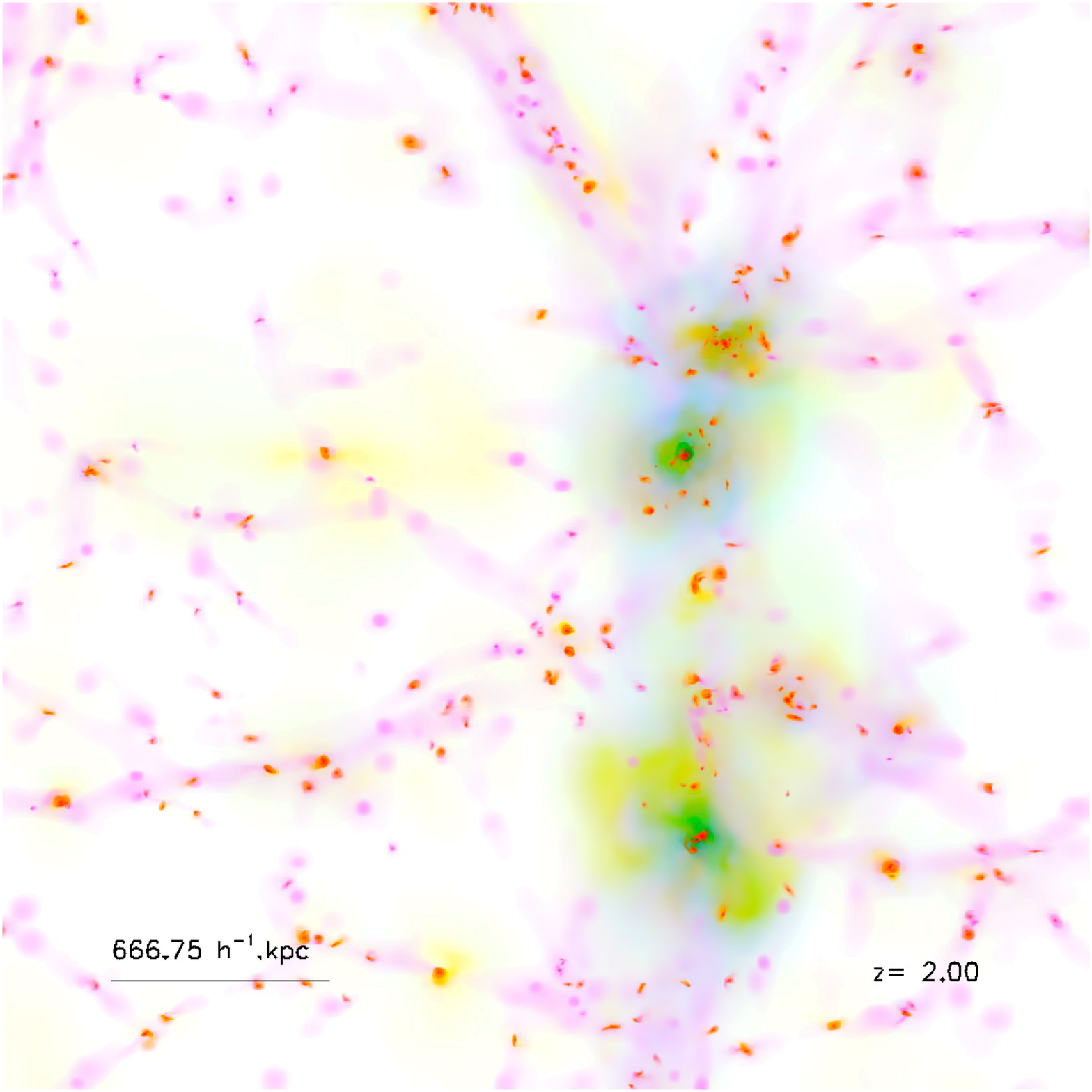}}}\\
  \centering{\resizebox*{!}{5.7cm}{\includegraphics{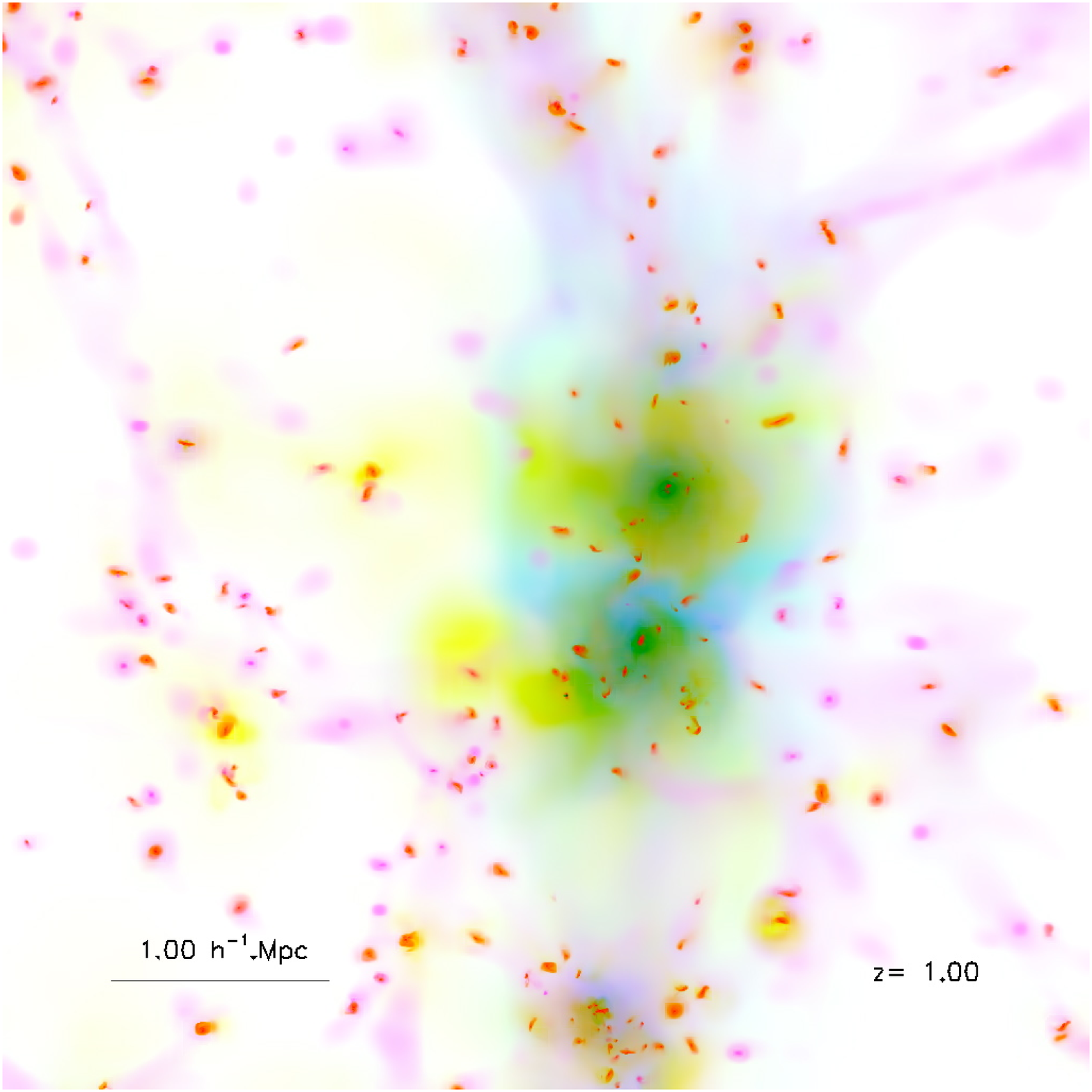}}}
  \centering{\resizebox*{!}{5.7cm}{\includegraphics{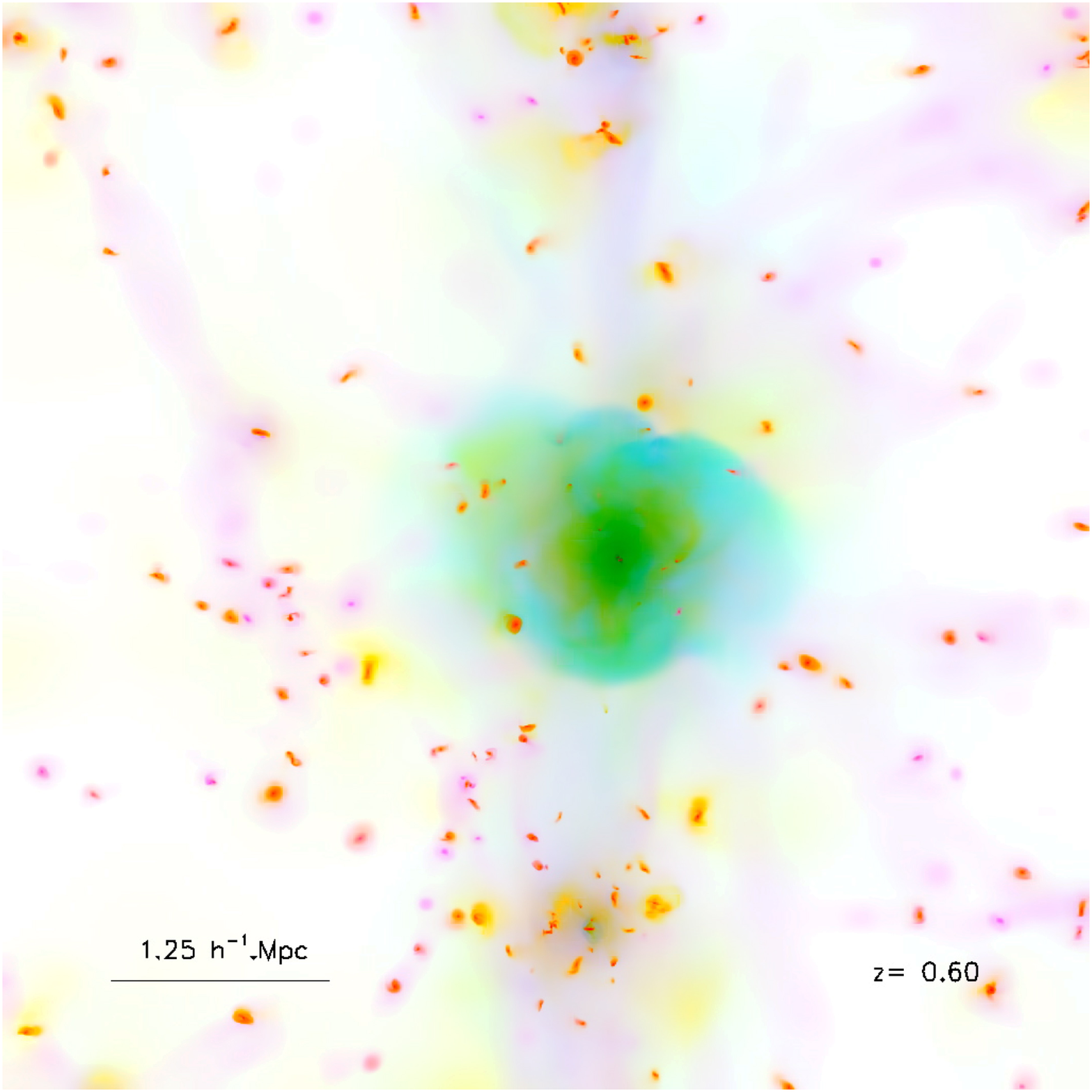}}}
  \centering{\resizebox*{!}{5.7cm}{\includegraphics{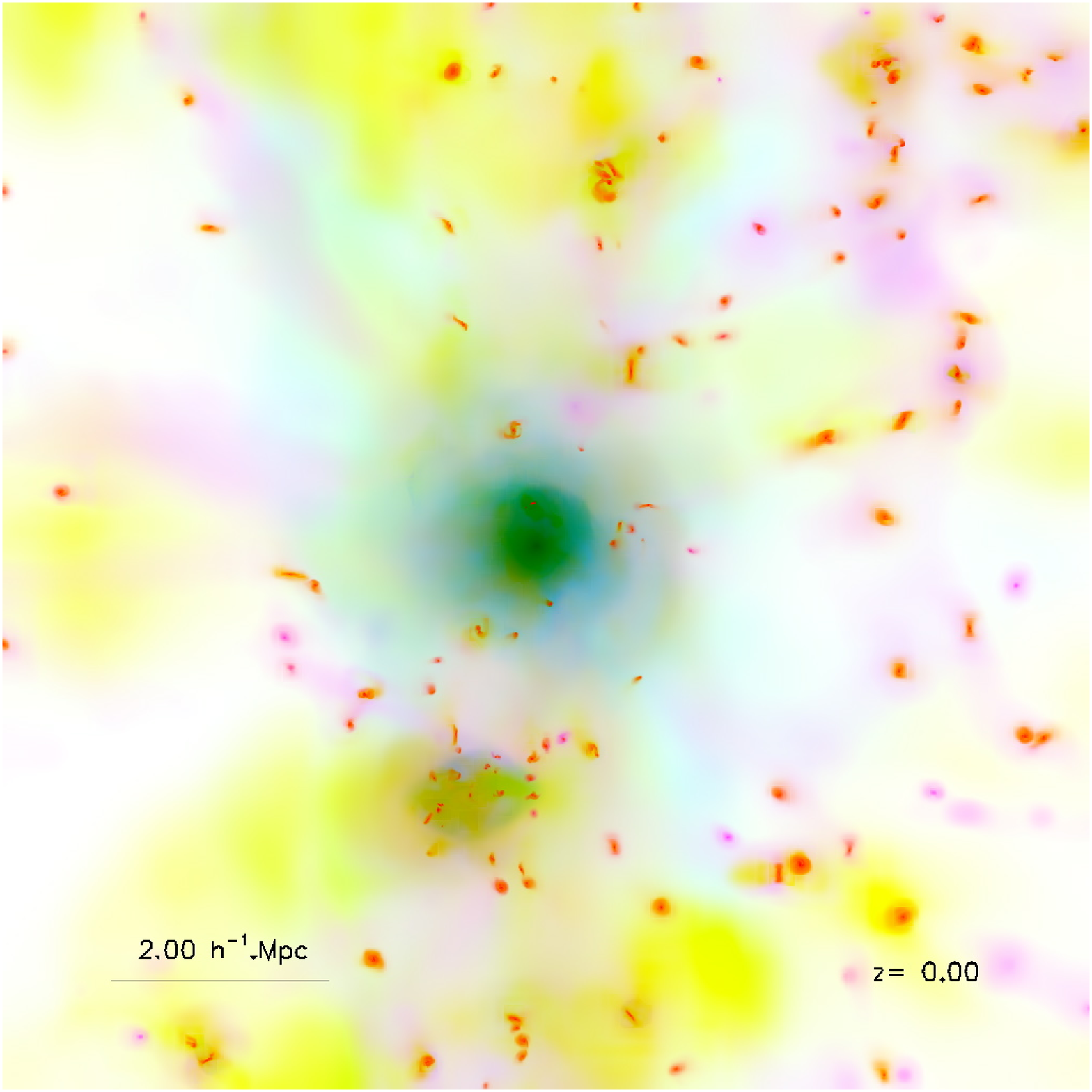}}}
  \caption{Time sequence of the galaxy cluster resimulation ZAGNJETrun at $z=4.03$ (upper left panel), $z=3$ (upper middle panel), $z=2$ (upper right panel), $z=1$ (bottom left panel), $z=0.6$ (bottom middle panel), and $z=0$ (bottom right panel). This simulation includes gas cooling, UV background, star formation, SN feedback, metal enrichment, and jet AGN feedback. The gas density is shown in magenta, gas temperature in cyan, and gas metallicity in yellow. The image is a eight times zoom into the simulation box.}
    \label{nice_zoom}
\end{figure*}

\begin{figure*}
  \centering{\resizebox*{!}{8cm}{\includegraphics{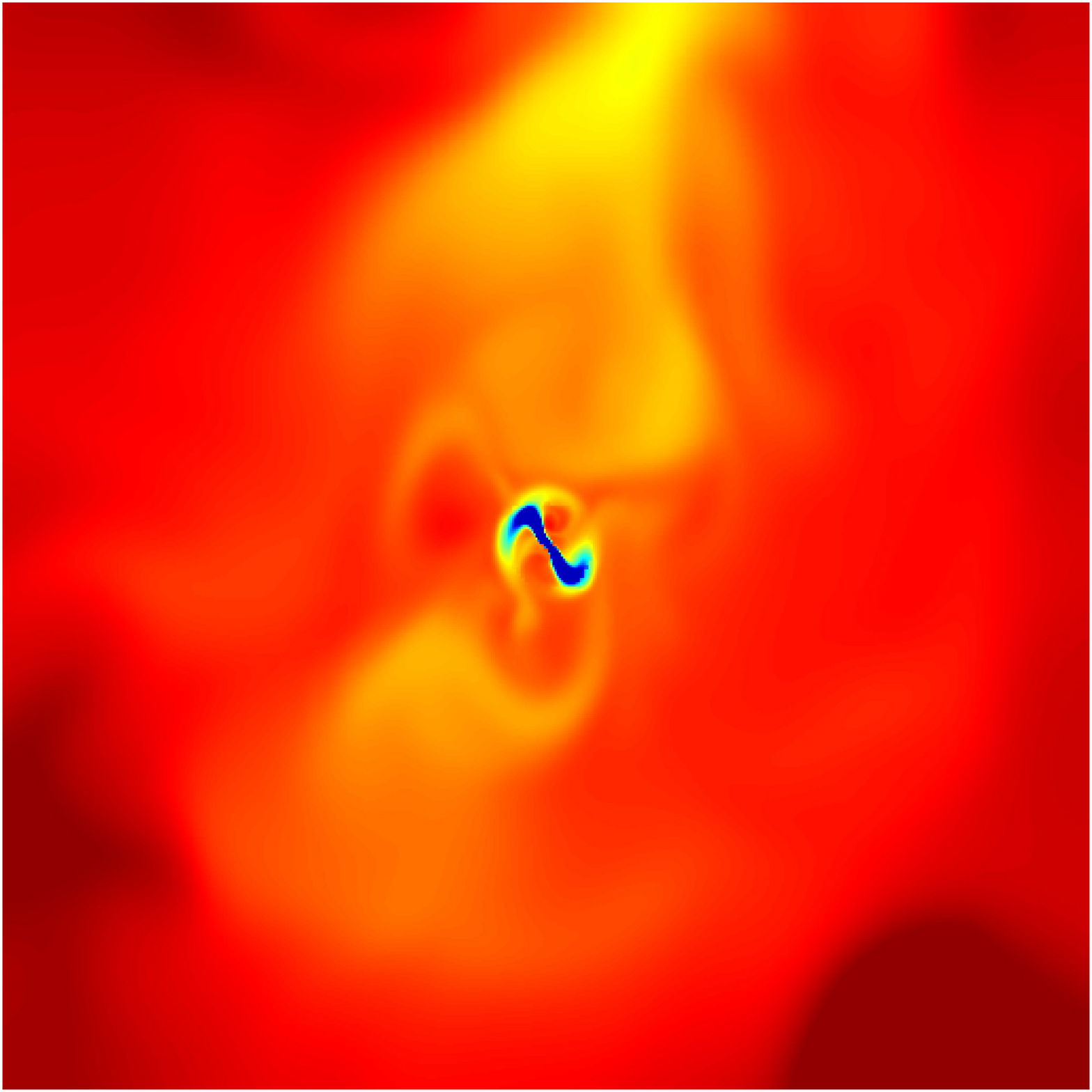}}}
  \centering{\resizebox*{!}{8cm}{\includegraphics{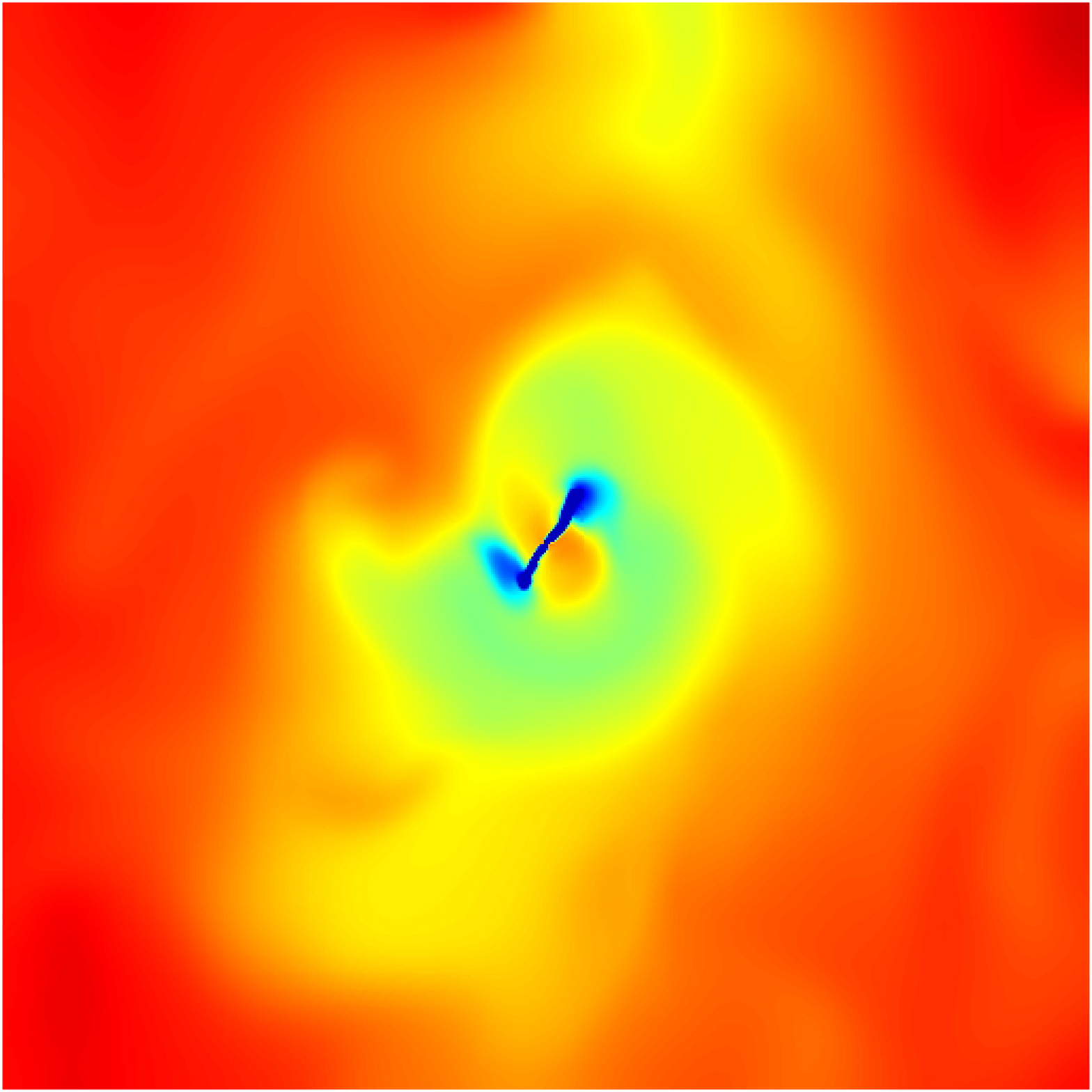}}}
  \centering{\resizebox*{!}{8cm}{\includegraphics{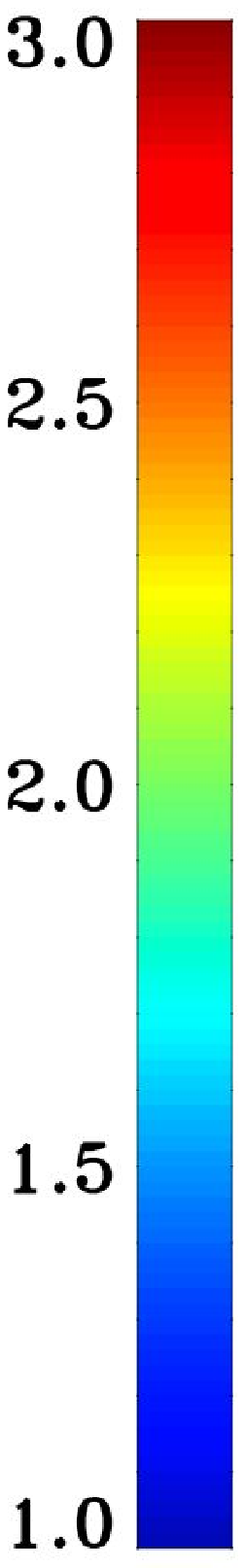}}}\\
  \centering{\resizebox*{!}{8cm}{\includegraphics{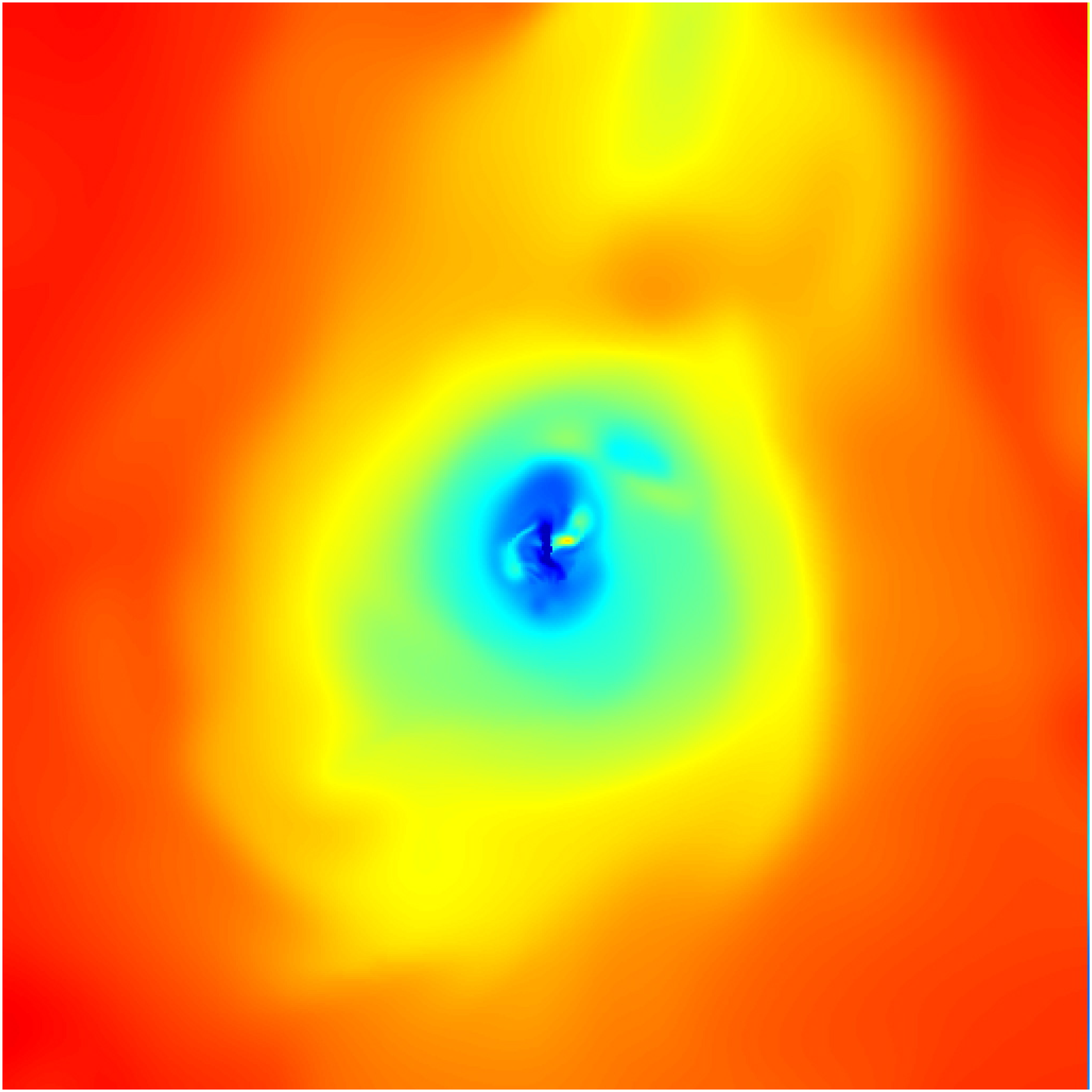}}}
  \centering{\resizebox*{!}{8cm}{\includegraphics{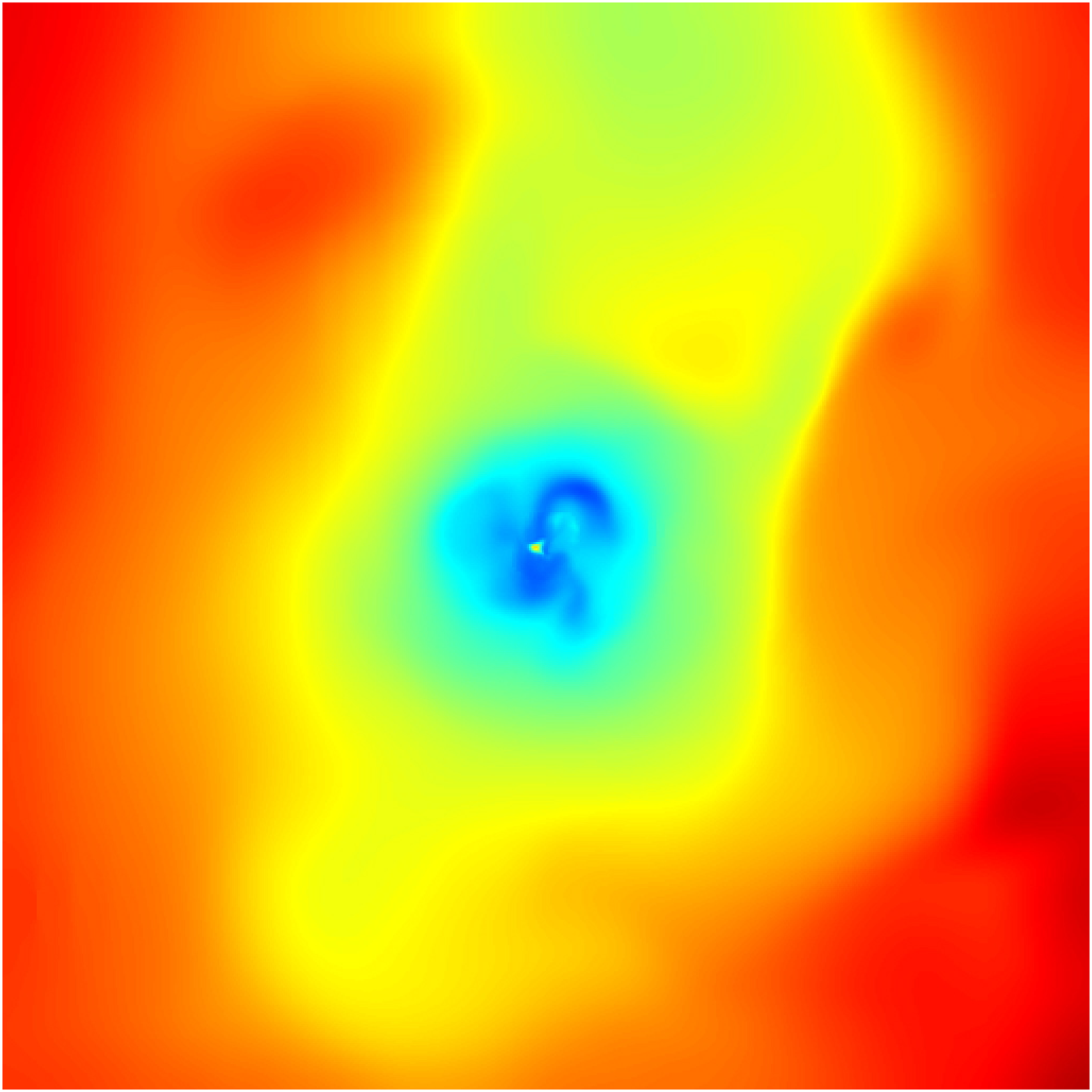}}}
  \centering{\resizebox*{!}{8cm}{\includegraphics{./ct_entro.ps}}}\\
  \caption{Cuts of the gas entropy through the cluster core at $z=0$ for
the NOAGNrun (upper left), the AGNOFFrun (upper right), the
AGNJETrun (bottom left), and the AGNHEATrun (bottom right). Color bar
units are in log(keV.cm$^2$). The picture size is $893$ kpc.}
    \label{entro_map_nice}
\end{figure*}

\begin{figure*}
  \centering{\resizebox*{!}{8cm}{\includegraphics{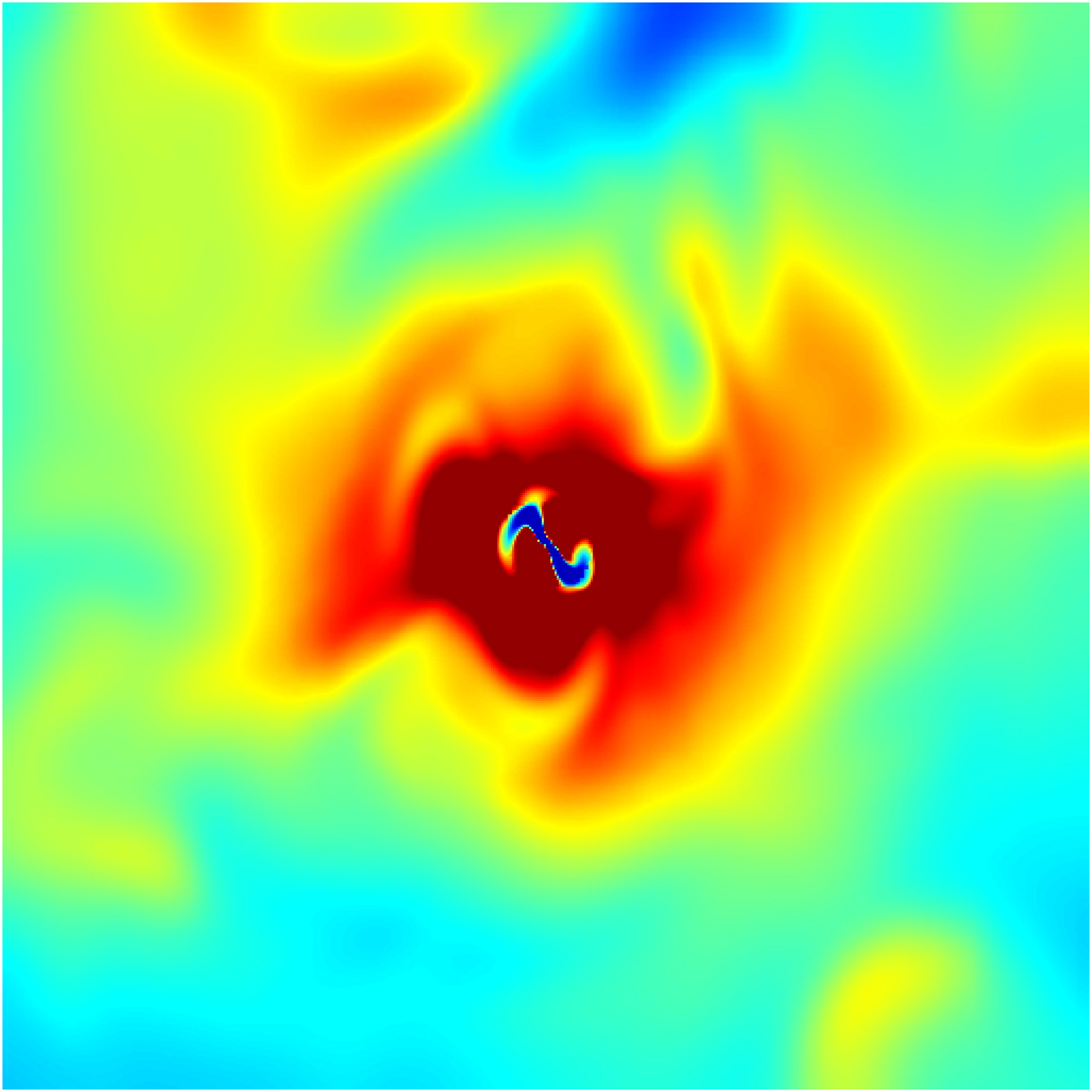}}}
  \centering{\resizebox*{!}{8cm}{\includegraphics{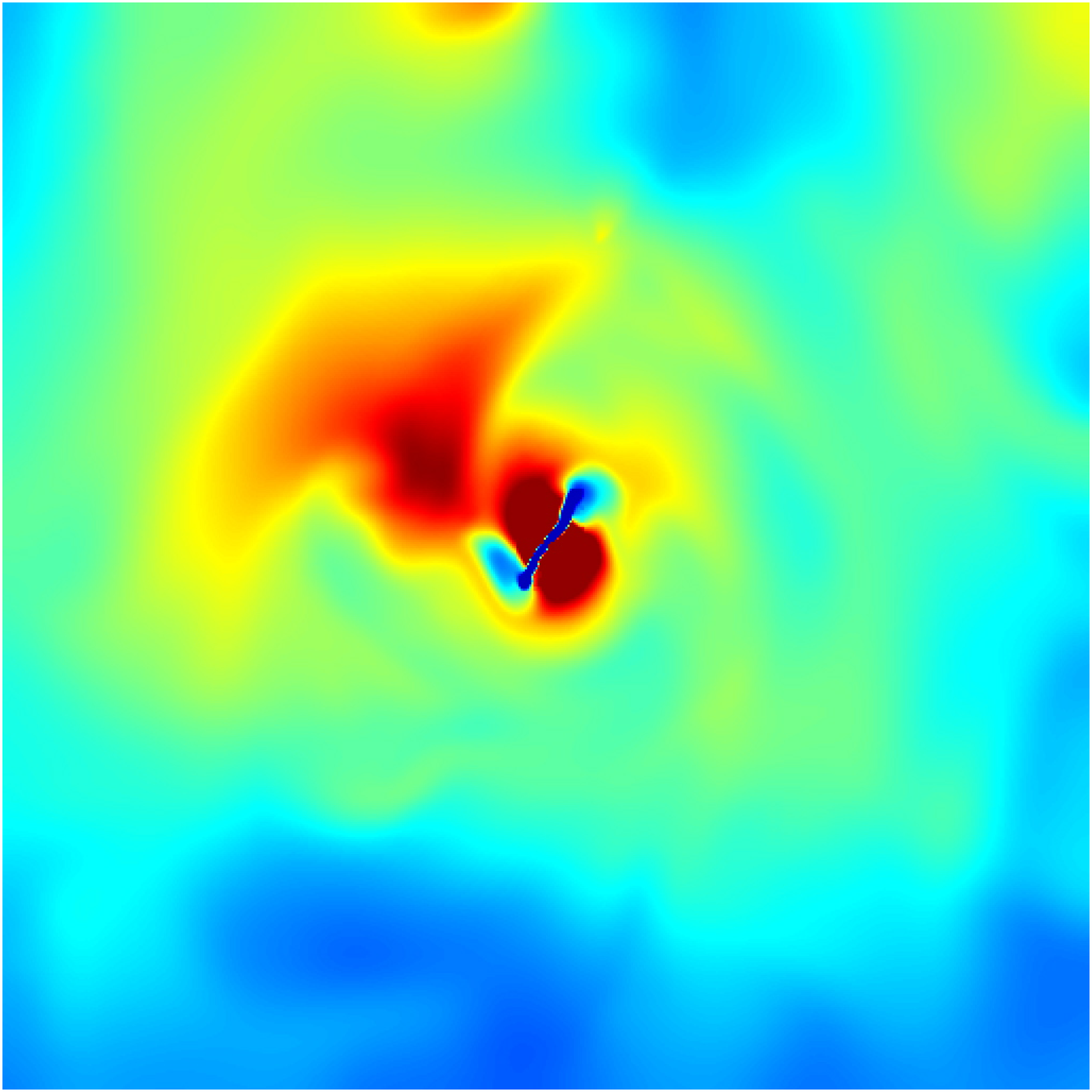}}}
  \centering{\resizebox*{!}{8cm}{\includegraphics{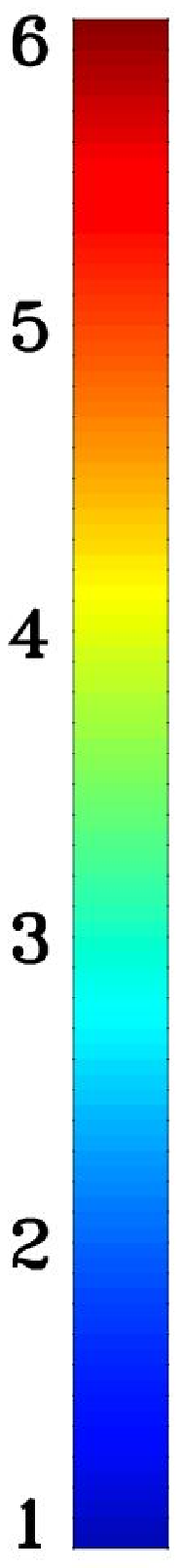}}}\\
  \centering{\resizebox*{!}{8cm}{\includegraphics{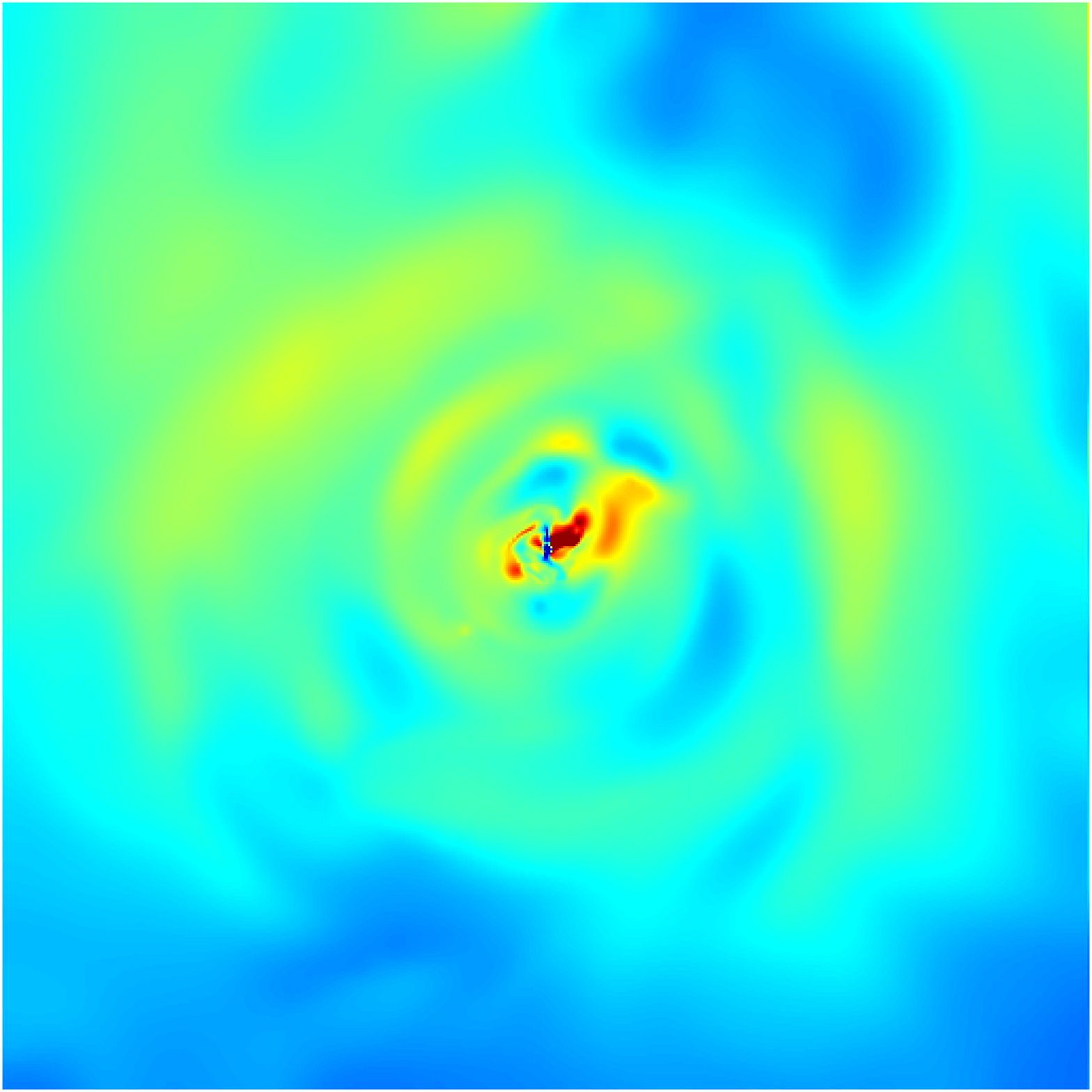}}}
  \centering{\resizebox*{!}{8cm}{\includegraphics{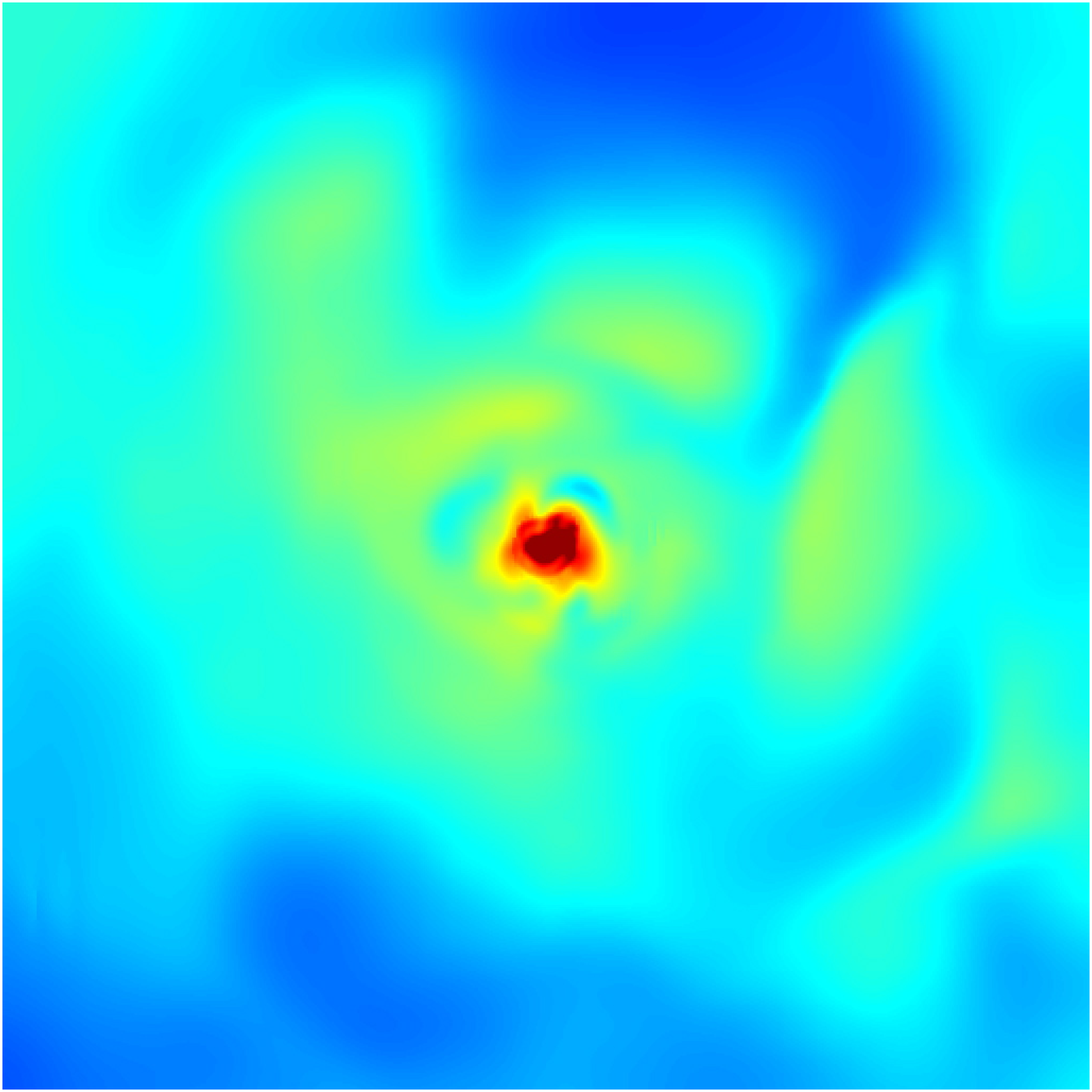}}}
  \centering{\resizebox*{!}{8cm}{\includegraphics{./ct_temp.ps}}}\\
  \caption{Cuts of the gas temperature through the cluster core at $z=0$
for the NOAGNrun (upper left), the AGNOFFrun (upper right), 
the AGNJETrun (bottom left), and the AGNHEATrun (bottom right). Color
bar units are in keV. The picture size is $893$ kpc}
    \label{temp_map_nice}
\end{figure*}

The simulations are run with the Adaptive Mesh Refinement (AMR) code
{\sc ramses} \citep{teyssier02}. The Euler equations are solved
using a second-order unsplit Godunov scheme. More specifically, fluxes at cell
interfaces are computed using an approximate Riemann solver where variables interpolated from
cell-centered values are reconstructed with a first-order MinMod Total Variation Diminishing
scheme. The Poisson equation is solved using a particle-mesh method where collisonless particles (DM, stars and sink
particles) are added to the AMR grid with a Cloud-in-Cell algorithm.

We assume a flat $\Lambda$CDM cosmology with total matter density
$\Omega_{m}=0.3$, baryon density $\Omega_b=0.045$, dark energy density
$\Omega_{\Lambda}=0.7$, rms mass fluctuation amplitude in spheres of $8 \, h^{-1} \rm Mpc$
$\sigma_8=0.90$ and Hubble constant $H_0=70\, \rm km.s^{-1}.Mpc^{-1}$
that corresponds to the Wilkinson Microwave Anisotropies Probe (WMAP)
1 year best-fit cosmological model \citep{spergeletal03}. The simulations
are performed using a resimulation (zoom) technique: the coarse region
is a $128^3$ grid with $M_{\rm DM}=2.9\times10^{10} \, \rm M_{\odot}$
DM resolution in a $80\,\rm h^{-1} Mpc$ simulation box. This region
contains a smaller $256^3$ equivalent grid in a sphere of radius
$20\,\rm h^{-1} Mpc$ with $M_{\rm DM}=3.6\times10^9 \, \rm M_{\odot}$
DM resolution, which in turn encloses the final high resolution sphere
with radius $6\,\rm h^{-1} Mpc$, a $512^3$ equivalent grid and $M_{\rm
DM}=4.5\times10^8 \, \rm M_{\odot}$ DM resolution.

Cells in the highest resolution region may be adaptively
refined up to $\ell_{\rm max}=16$ levels of refinement, maintaining a
constant physical resolution of $1.2 \, \rm h^{-1} kpc$ at all times. Refinement follows a
quasi-Lagrangian criterion: if more than 8 DM particles lie in a cell,
or if the baryon mass exceeds 8 times the initial DM mass resolution,
the cell is refined. A Jeans length criterion is also added to ensure
the numerical stability of the scheme on all levels $\ell < \ell_{\rm
max}$ \citep{trueloveetal97} where $\delta \rho=\rho/\bar \rho >
10^5$: we force the cells fulfilling these conditions to sample the local
Jeans length with more than 4 cells.

With this particular set of initial conditions we follow the formation of a galaxy cluster with mass $M_{500}=2.4\,
10^{14}\,Ê\rm M_{\odot}$, and radius $r_{500}=940$ kpc at $z=0$, where
the $500$ subscript stands for the mass and radius at 500 times the
average density of the Universe (see AGNJETrun simulation data in
table~\ref{tabnames}).  This cluster undergoes
a 1:1 major merger from $z=1$ down to $z=0.6$ (see fig.~2
of \citealp{duboisetal10}), naturally separating early time evolution
where the pre-heating in the proto-cluster structures
takes place, from late-time evolution during which the gas of the
cluster relaxes.

\begin{table*}
\caption{Simulations performed with different gas physics and sub-grid models. Column 1 gives the name of the simulation, column
2 indicates the form of cooling (none, H/He primordial composition, or metal), column 3 specifies whether star formation is activated,
column 4 if supernovae are included, column 5 the form of AGN feedback, columns 6 and 7 list $r_{500}$ and $M_{500}$ at $z=0$ respectively, columns
8 and 9 list $r_{2000}$ and $M_{2000}$ at $z=0$ respectively, column 10 lists the most massive BH mass $M_{\rm BH}$ in the central galaxy at $z=0$.}
\label{tabnames}
\begin{tabular}{@{}|l|c|c|c|c|c|c|c|c|c|}
  \hline
  Name & Cooling & SF & SN & AGN & $r_{500}$ & $M_{500}$ &$r_{2000}$ & $M_{2000}$ & $M_{\rm BH}$ \\
	& & & & & (kpc) & ($\rm M_{\odot}$) & (kpc) & ($\rm M_{\odot}$) & ($\rm M_{\odot}$) \\
  \hline
  \hline
  ADIArun & No & No & No & No & 944 & $2.36 \, 10^{14}$& 486 & $1.29 \, 10^{14}$ & --\\
  \hline
  NOAGNrun & Primordial & Yes & No & No& 956 & $2.46 \, 10^{14}$& 514 & $1.52 \, 10^{14}$ & --\\
  AGNJETrun & Primordial &Yes & No & Jet \citep{duboisetal10} & 944 & $2.36 \, 10^{14}$& 486 & $1.29 \, 10^{14}$& $1.7\, 10^{10}$\\
  AGNHEATrun & Primordial &Yes & No & Heat \citep{teyssieretal11} & 949 & $2.41 \, 10^{14}$& 496 & $1.37 \, 10^{14}$ & $7.8\, 10^{10}$\\
  AGNOFFrun & Primordial & Yes & No & Jet ($z>0.58$) \& No ($z\le0.58$) & 946 & $2.38 \, 10^{14}$& 496 & $1.37 \, 10^{14}$& $1.1\, 10^{10}$\\
  \hline
  ZNOAGNrun & Metal & Yes & Yes & No & 961 & $2.50 \, 10^{14}$& 536 & $1.73 \, 10^{14}$ & --\\
  ZAGNJETrun & Metal &Yes & Yes & Jet \citep{duboisetal10} & 939 & $2.33 \, 10^{14}$& 474 & $1.19 \, 10^{14}$& $3.3\, 10^{10}$\\
  ZAGNHEATrun & Metal &Yes & Yes & Heat \citep{teyssieretal11}& 936 & $2.31 \, 10^{14}$ & 479 & $1.23 \, 10^{14}$& $1.3\, 10^{11}$ \\
  ZAGNOFFrun & Metal &Yes & Yes & Jet ($z>0.58$) \& No ($z\le0.58$) & 946 & $2.38\, 10^{14}$ & 486 & $1.29\, 10^{14}$ & $6.4\, 10^{10}$ \\
  \hline
\end{tabular}
\end{table*}

In \cite{duboisetal10}, we performed two simulations to explore
the role of AGN feedback on the regulation of the cooling
catastrophe: one simulation with AGN feedback, cooling and star
formation (AGNJETrun), and another identical to the first where AGN feedback
was switched off (NOAGNrun).  In this paper, we add several new
simulations to this set to understand the impact of the AGN sub-grid modelling and
the influence of early pre-heating on our results. Namely, we run an adiabatic simulation that 
ignores cooling and star formation (ADIArun), a simulation with
thermal AGN feedback as in \cite{teyssieretal11} (AGNHEATrun), and a
simulation that involves restarting the AGNJETrun simulation at
$z=0.58$ (after the major merger has taken place) and thereafter deactivating the gas accretion onto
the BHs and the related AGN feedback (AGNOFFrun).  The first new
simulation (ADIArun) allow us to study the thermal properties of this
cluster under the simplest physical conditions (gravity and gas
dynamics only).  The second new simulation (AGNHEATrun) allows us to compare the
impact of the type of AGN feedback on the evolution of the
cluster.  Finally with the third new simulation (AGNOFFrun) we assess
whether it is the AGN feedback at early or
late times that shapes the thermodynamics  of the cluster at $z=0$
and prevents a cooling catastrophe.

On top of that, we also add a new set of simulations including metal cooling 
where the metals are produced self-consistently in SN explosions and advected with the flow (ZNOAGNrun, ZAGNJETrun, and
ZAGNHEATrun). These enable us to quantify the importance of metal cooling on the
ICM gas dynamics.

The properties of the whole set of simulations are summarized in table~\ref{tabnames}.

Fig.~\ref{nice_zoom} shows a time sequence of the ZAGNJETrun simulation inside the zoom region (10 $\rm h^{-1}.Mpc$ comoving on a side).
The early evolution of the progenitors of the cluster at high redshift $z>1$ (upper panels) shows the filamentary structure of the early Universe  (gas density in magenta), and the deposit of metals (yellow color) in the IGM through SN feedback and AGN feedback, which correspond to a strong pre-heating phase.
The merger of the two clusters between $z=1$ (pre-merger phase, bottom left panel) and $z=0.6$ (post-merger phase, bottom middle panels) produces a strong shock (temperature coded in cyan) during the encounter and shock waves develop at larger distance during the post-merger.
The $z=0$ final output of the simulation shows the relaxed cluster (bottom right panel).

\section{Results}
\label{results}

We follow the time evolution of the entropy
profiles, as well as other thermodynamical properties of the ICM, for
our simulations without metal cooling (figure~\ref{allin}) and with metal cooling (figure~\ref{allin_Z}) in order to understand
what drives the $z=0$ entropy profiles displayed in fig.~\ref{entro_comp}.  
Each of the following sections systematically evaluates the
consequences of including increasingly more physics into the simulations.
First, in section~\ref{2D-1Dprofiles}, we map out the gas structure of the galaxy cluster at $z=0$, and 
discuss how we separate the ISM of the central galaxy from the ICM gas.
Section~\ref{adiabatic} presents results from
the simplest case of cluster evolution under the influence of pure gravity and hydrodynamics (ADIArun).
Section~\ref{cooling} considers the addition of atomic cooling, star formation, and UV reionization (NOAGNrun).
Section~\ref{AGNfbk} studies the cluster evolution with the additional complexity of AGN feedback, comparing effects of two
different subgrid models: AGN jet feedback (AGNJETrun)
and AGN feedback modeled as a thermal intput (AGNHEATrun). Section~\ref{agnoff_section} assesses the role of pre-heating (AGNOFFrun), and
finally section~\ref{metals} examines how our results change with the inclusion of metal enrichment from supernovae (ZNOAGNrun, ZAGNJETrun, ZAGNHEATrun, ZAGNOFFrun). 

\subsection{Azimuthal structure and radial profiles}
\label{2D-1Dprofiles}

Fig.~\ref{entro_map_nice} shows a two-dimensional entropy cut through the centre of the cluster at $z=0$ for the four 
simulations which include primordial gas cooling (NOAGNrun, AGNOFFrun, AGNJETrun, AGNHEATrun).
It appears that away from the galaxy disk interface, the entropy 
distribution is smooth but non-uniform. 
Lower entropy gas is preferentially located in the central regions.
Due to radiative losses from atomic cooling, a large disc
component appears in two cases (NOAGNrun and AGNOFFrun), signaling a cooling catastrophe (massive disc and high
star-formation).  This cold ISM component is also visible in
fig.~\ref{temp_map_nice} with gas temperatures below 1 keV.  In
the simulations where AGN feedback proceeds down to $z=0$ (AGNJETrun
and AGNHEATrun), the galactic disc is much smaller and the spatial extent of low entropy material is greater than
in the cases where catastrophic cooling proceeds unimpeded (NOAGNrun
and AGNOFFrun).  The AGN jet in the AGNJETrun simulation manifests itself close to the disc
by launching a bipolar outflow of high entropy and high temperature
gas which sends sound waves out to large radii in the ICM (see \citealp{duboisetal10}).

Interestingly, the ICM in the NOAGNrun simulation has the highest level of 
entropy within the core, even though this is the simulation that does not
include any feedback process at anytime.  Compared to the other simulations, 
the central ICM  also has a  much larger temperature 
(fig.~\ref{temp_map_nice}).  The AGNOFFrun simulation shows similar
characteristics: the ICM develops a high-entropy, high-temperature phase on both sides of the disc.

\begin{figure*}
  \centering{\resizebox*{!}{3.4cm}{\includegraphics{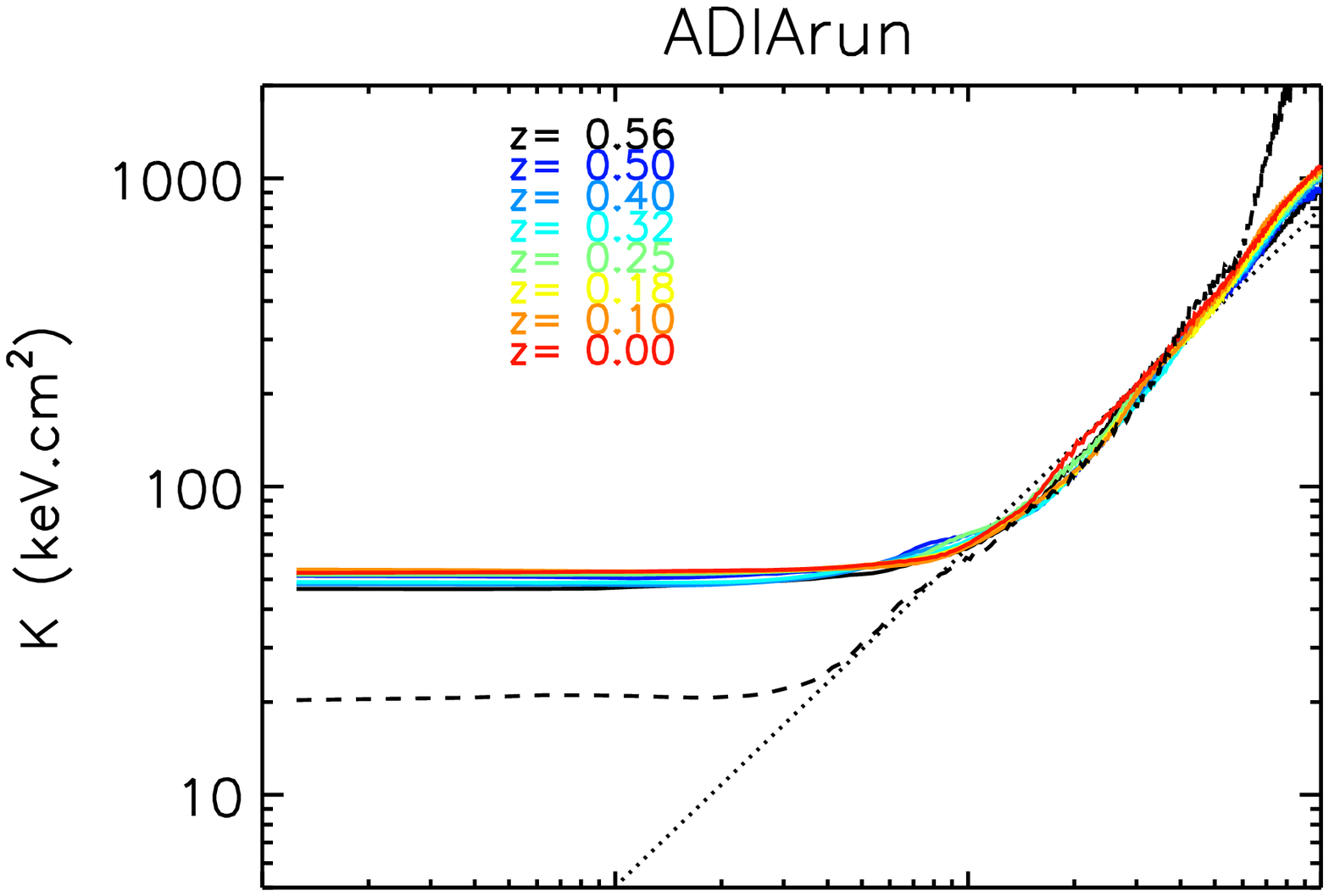}}\hspace{-1.40cm}}
  \centering{\resizebox*{!}{3.4cm}{\includegraphics{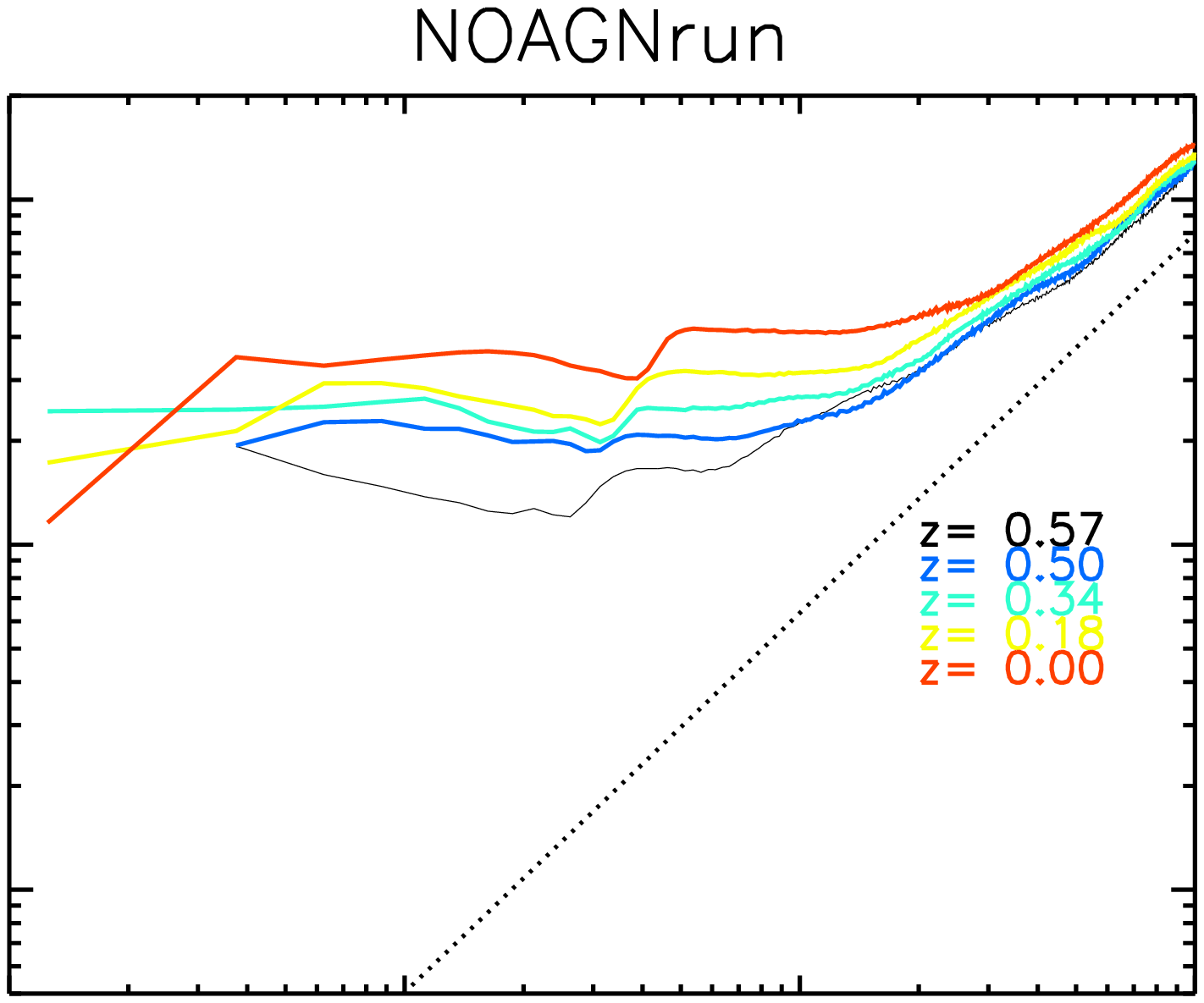}}\hspace{-1.40cm}}
  \centering{\resizebox*{!}{3.4cm}{\includegraphics{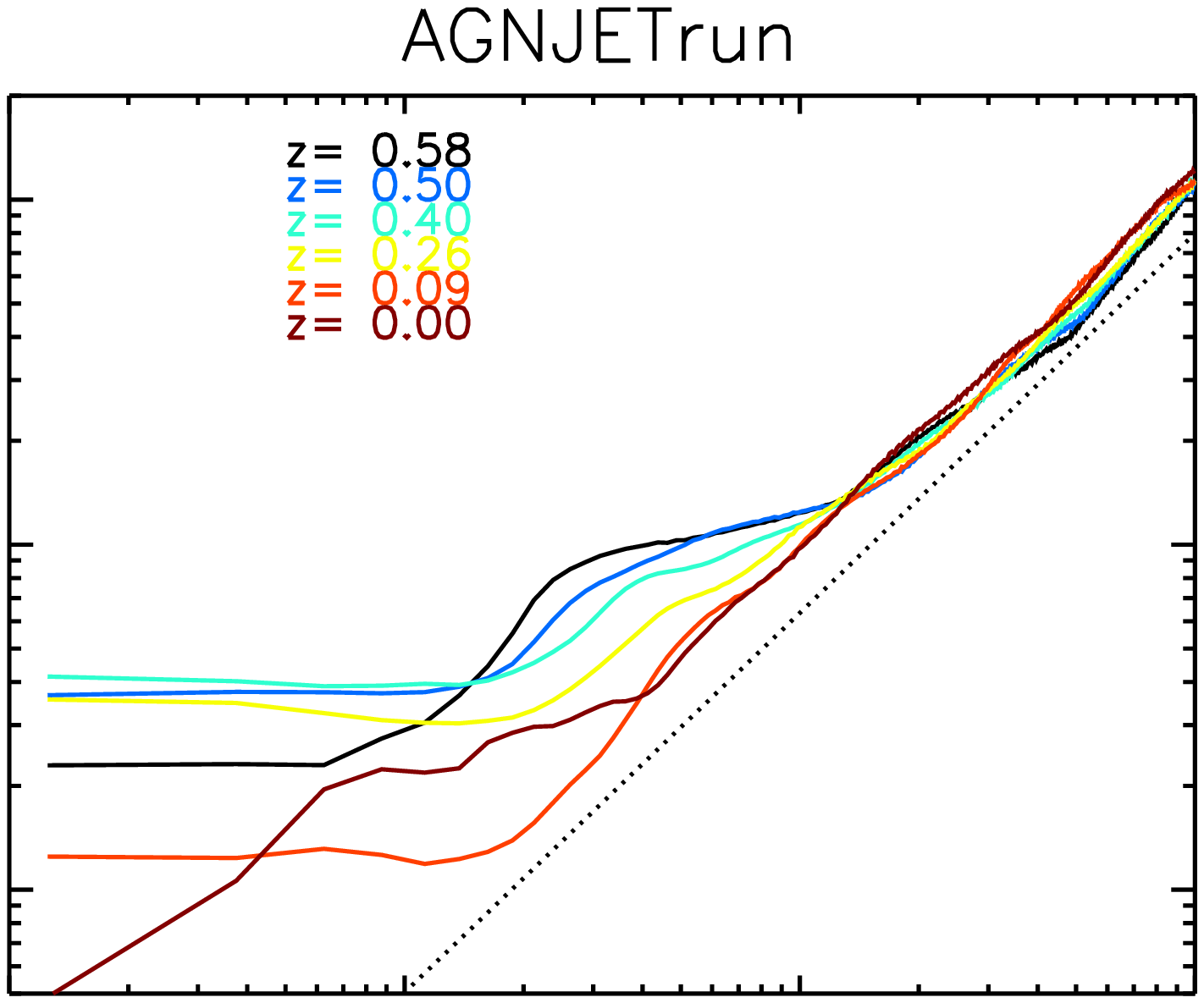}}\hspace{-1.40cm}}
  \centering{\resizebox*{!}{3.4cm}{\includegraphics{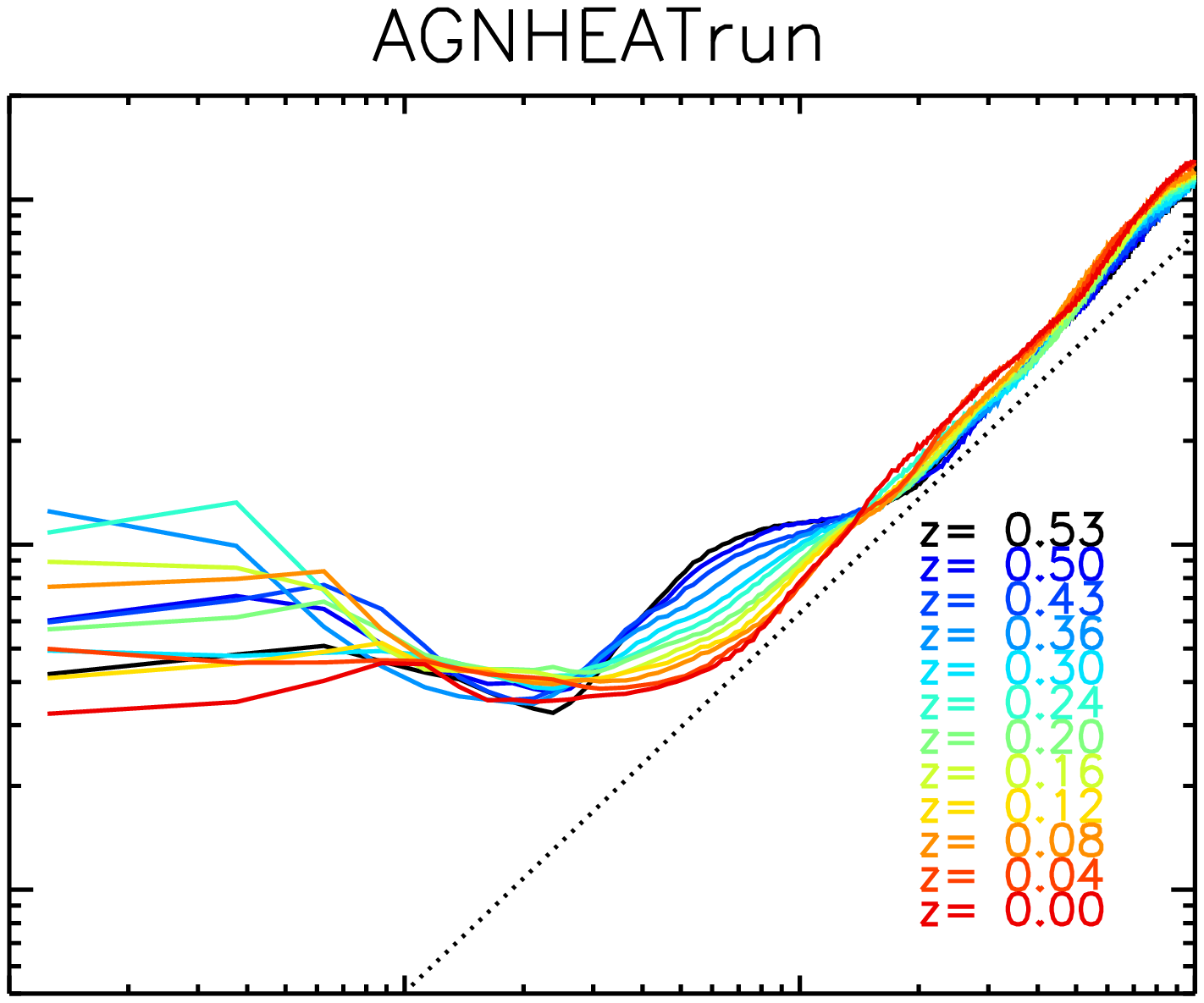}}\hspace{-1.40cm}}
  \centering{\resizebox*{!}{3.4cm}{\includegraphics{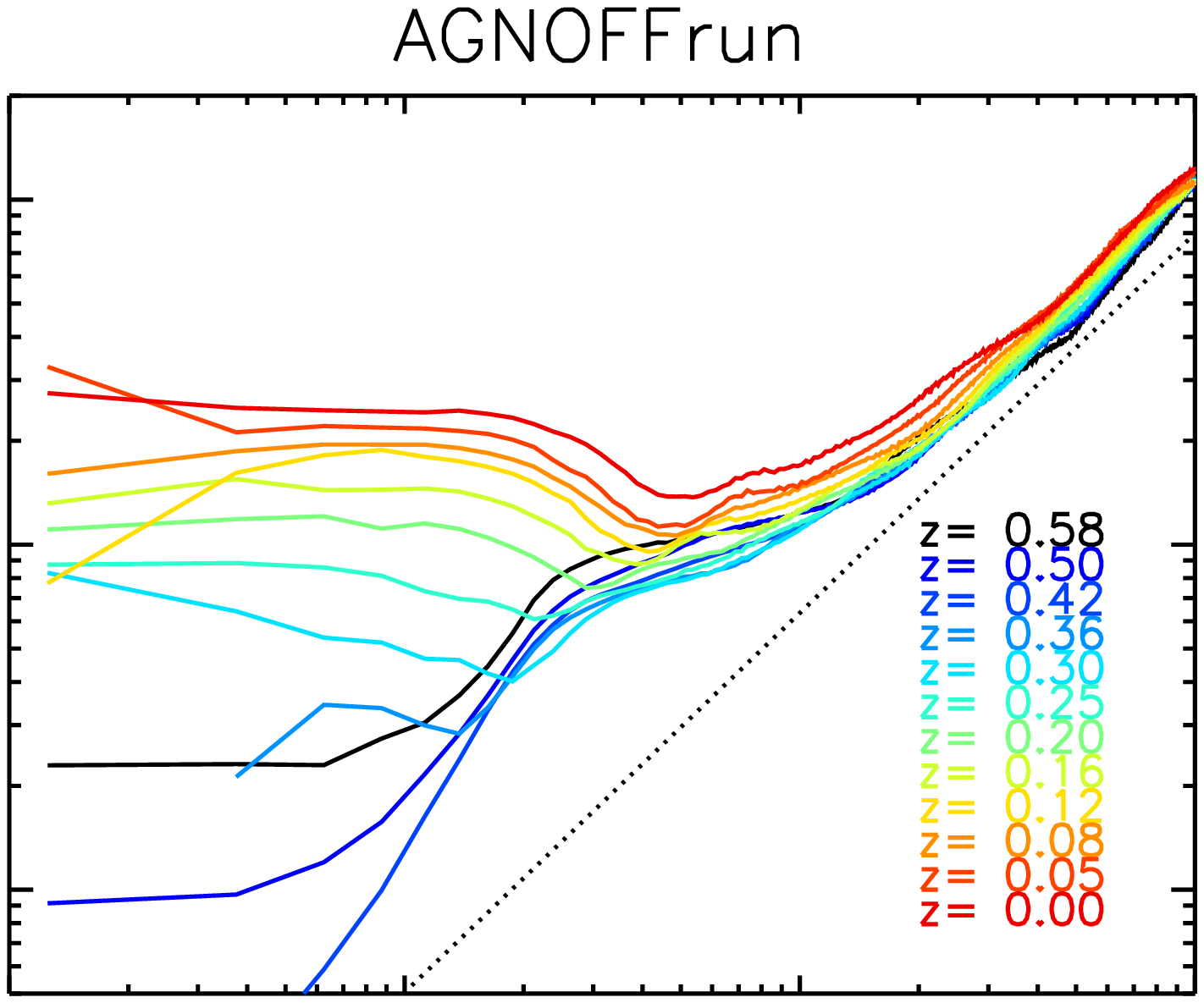}}\vspace{-0.99cm}}\\
  \centering{\resizebox*{!}{3.4cm}{\includegraphics{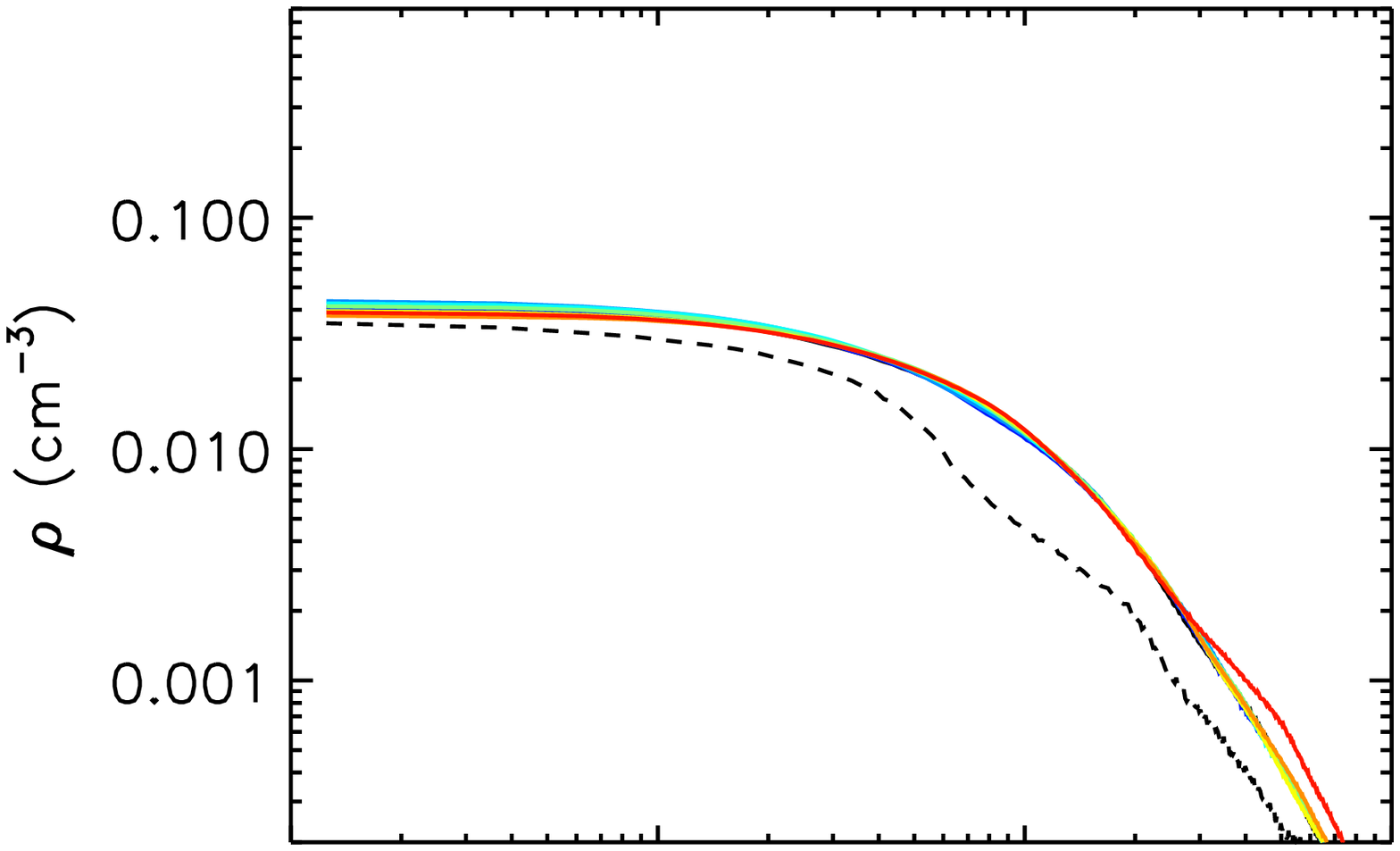}}\hspace{-1.40cm}}
  \centering{\resizebox*{!}{3.4cm}{\includegraphics{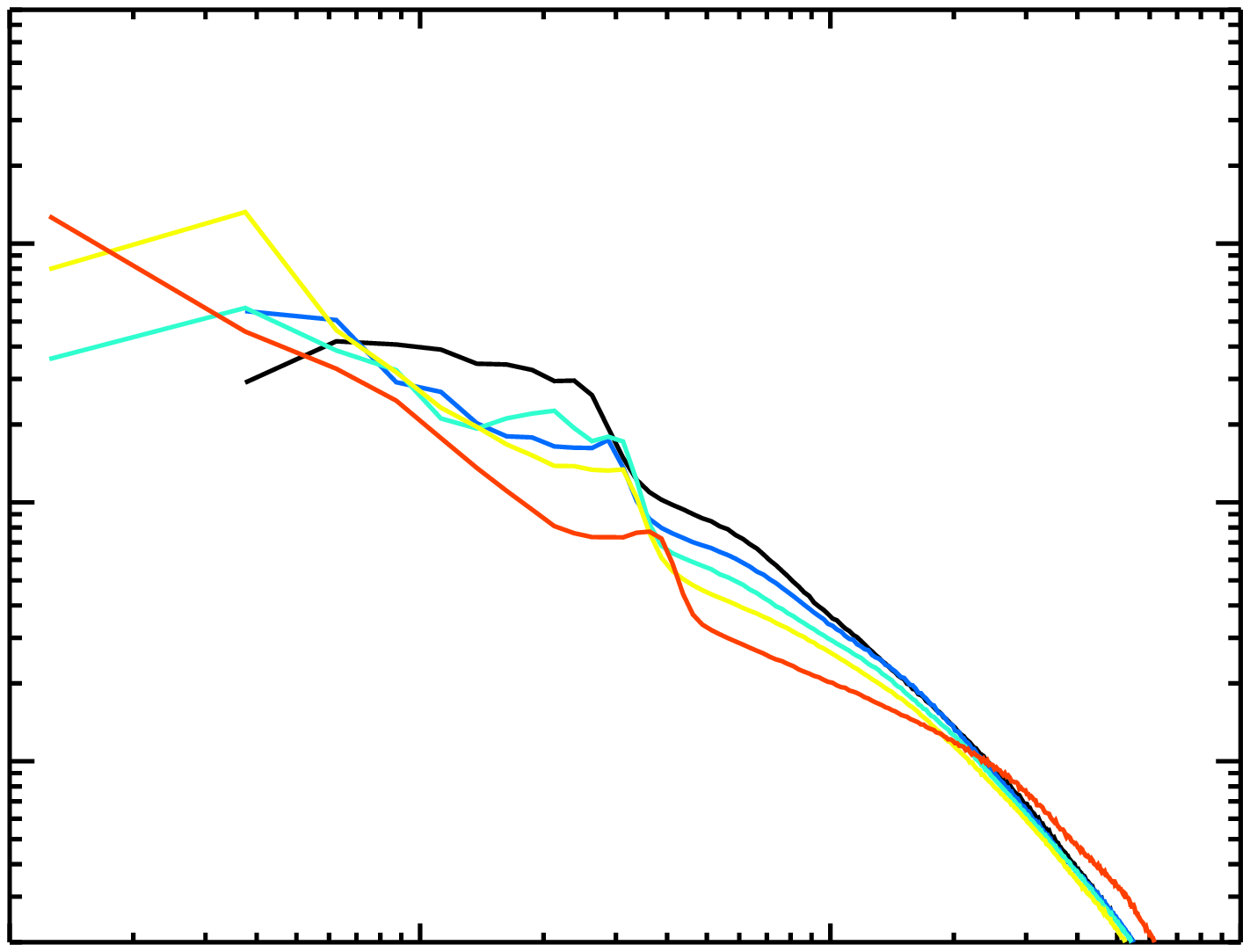}}\hspace{-1.40cm}}
  \centering{\resizebox*{!}{3.4cm}{\includegraphics{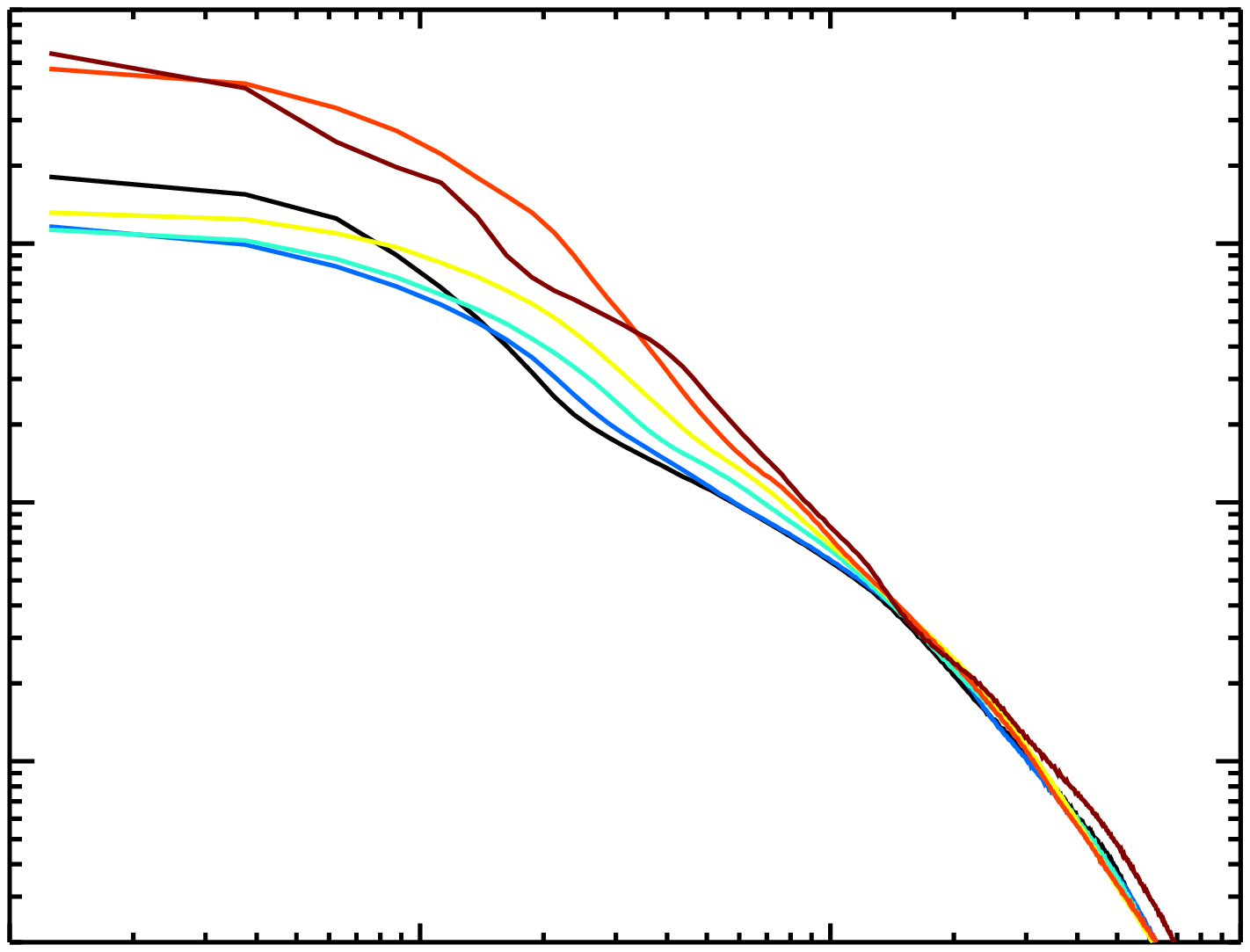}}\hspace{-1.40cm}}
  \centering{\resizebox*{!}{3.4cm}{\includegraphics{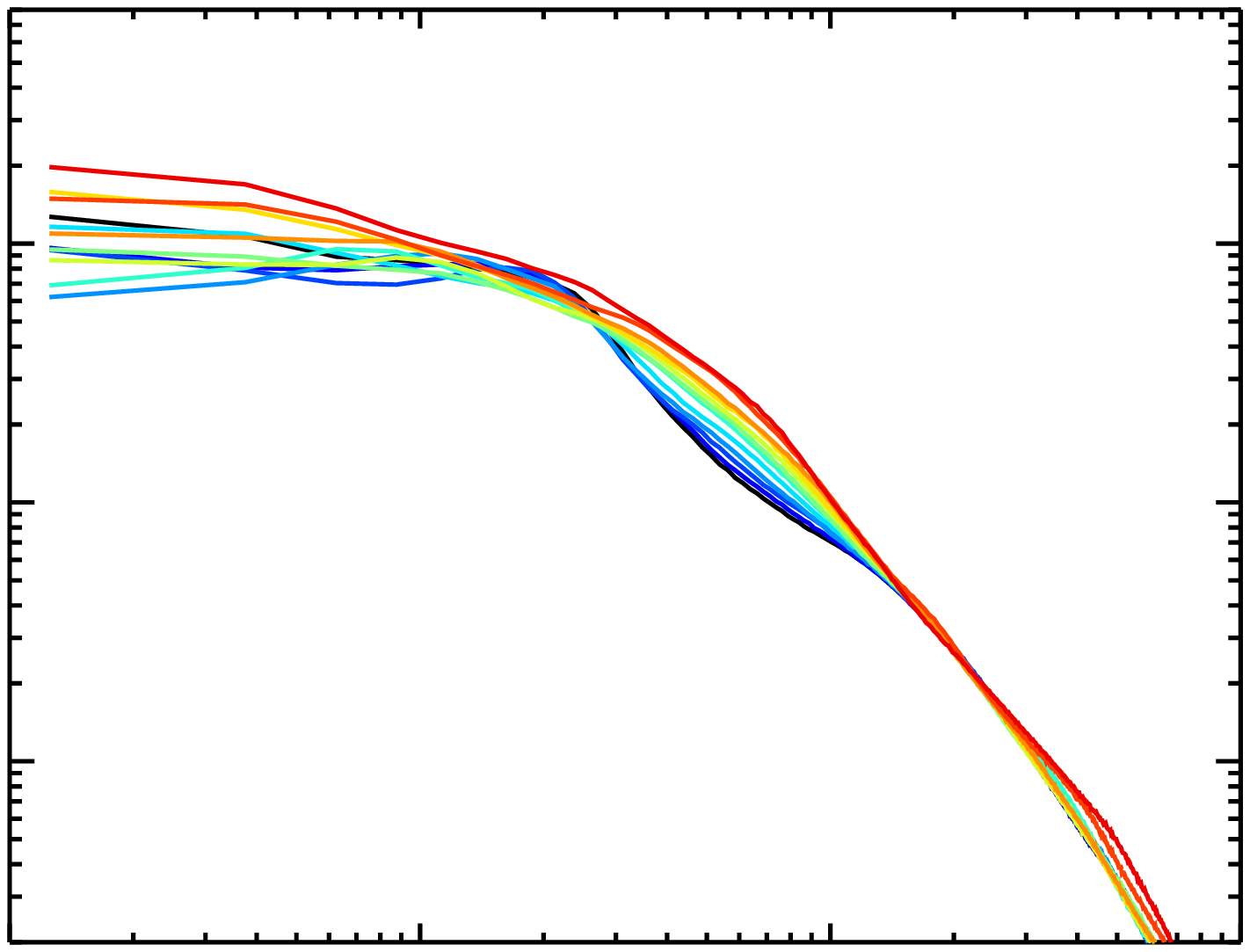}}\hspace{-1.40cm}}
  \centering{\resizebox*{!}{3.4cm}{\includegraphics{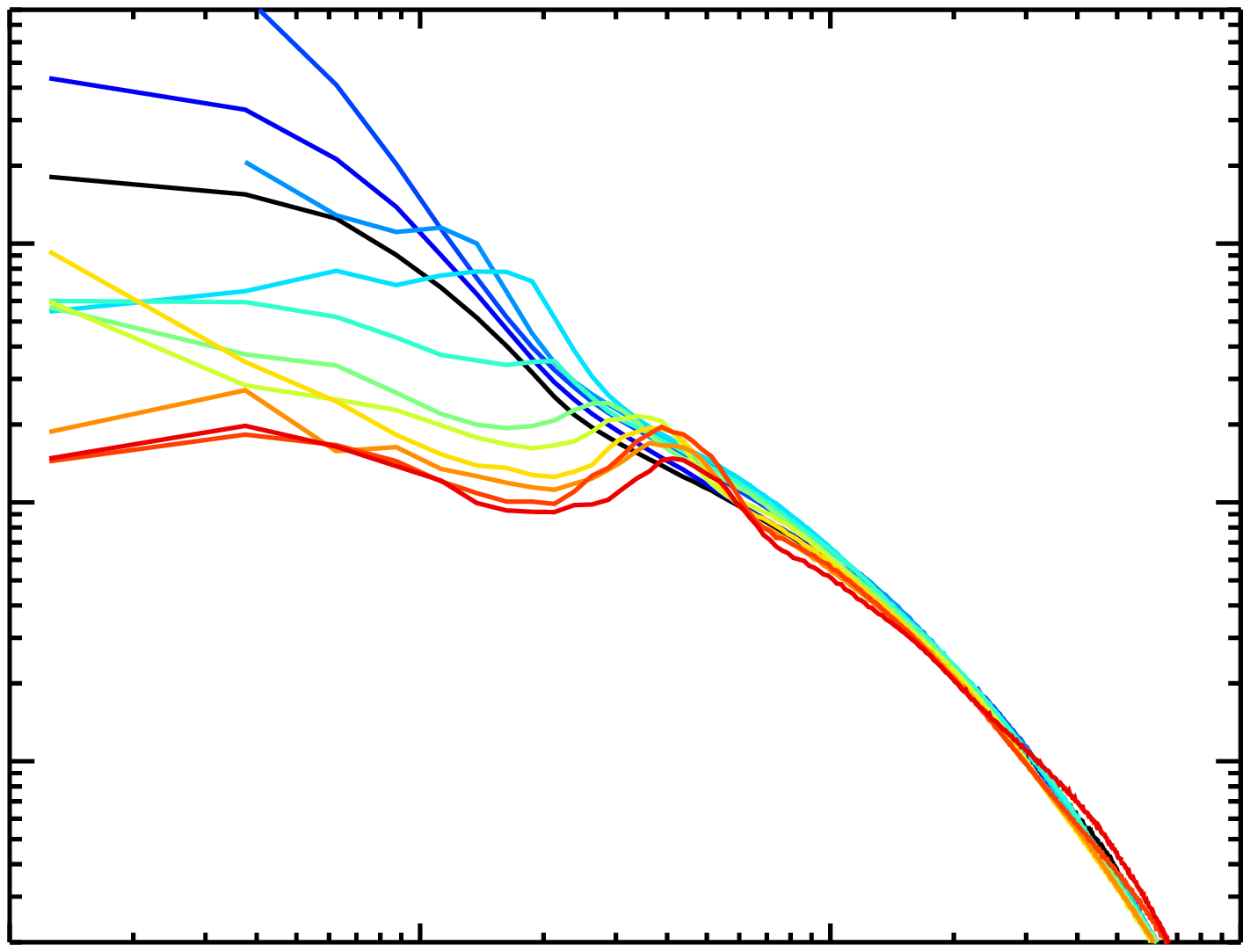}}\vspace{-0.99cm}}\\
  \centering{\resizebox*{!}{3.4cm}{\includegraphics{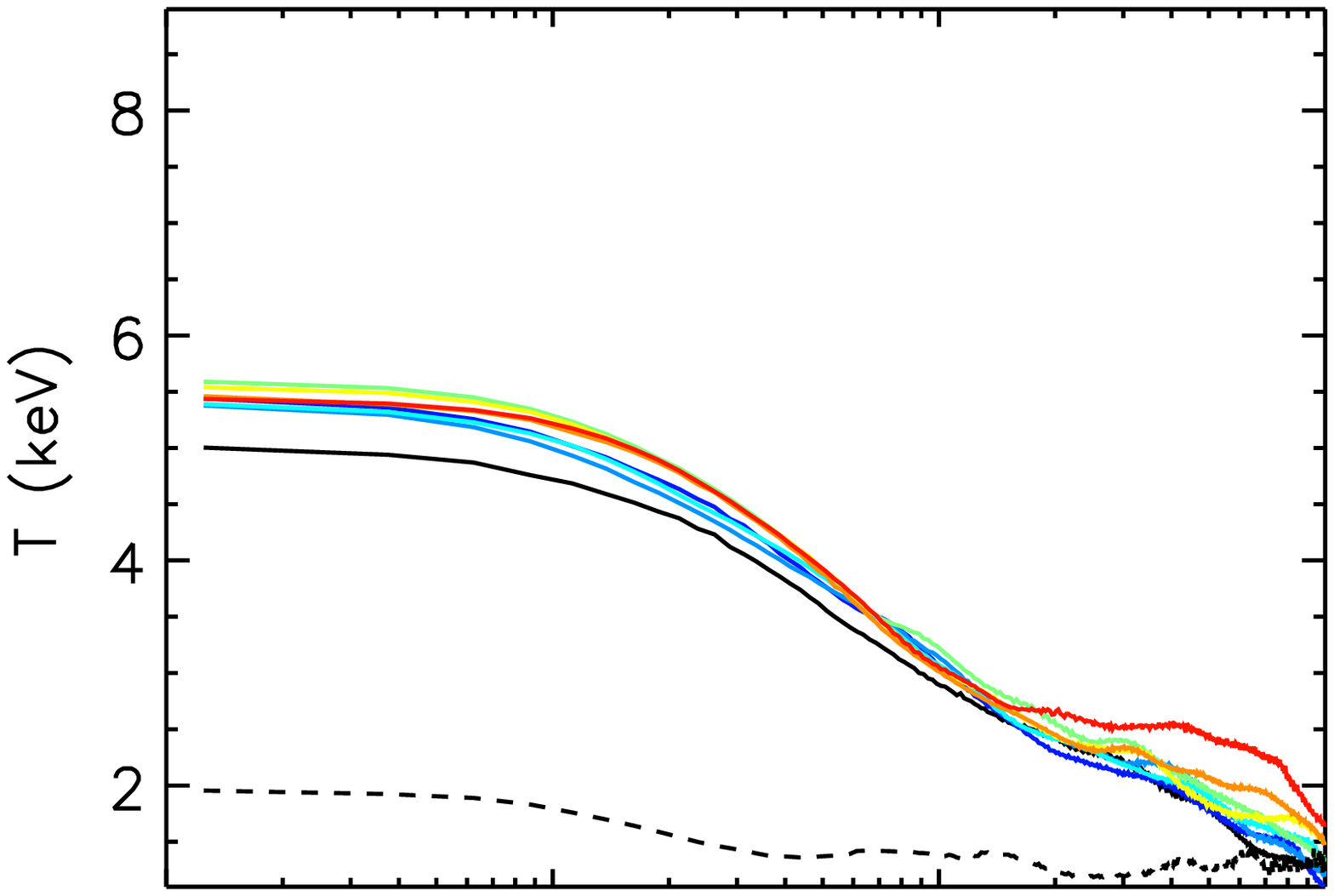}}\hspace{-1.40cm}}
  \centering{\resizebox*{!}{3.4cm}{\includegraphics{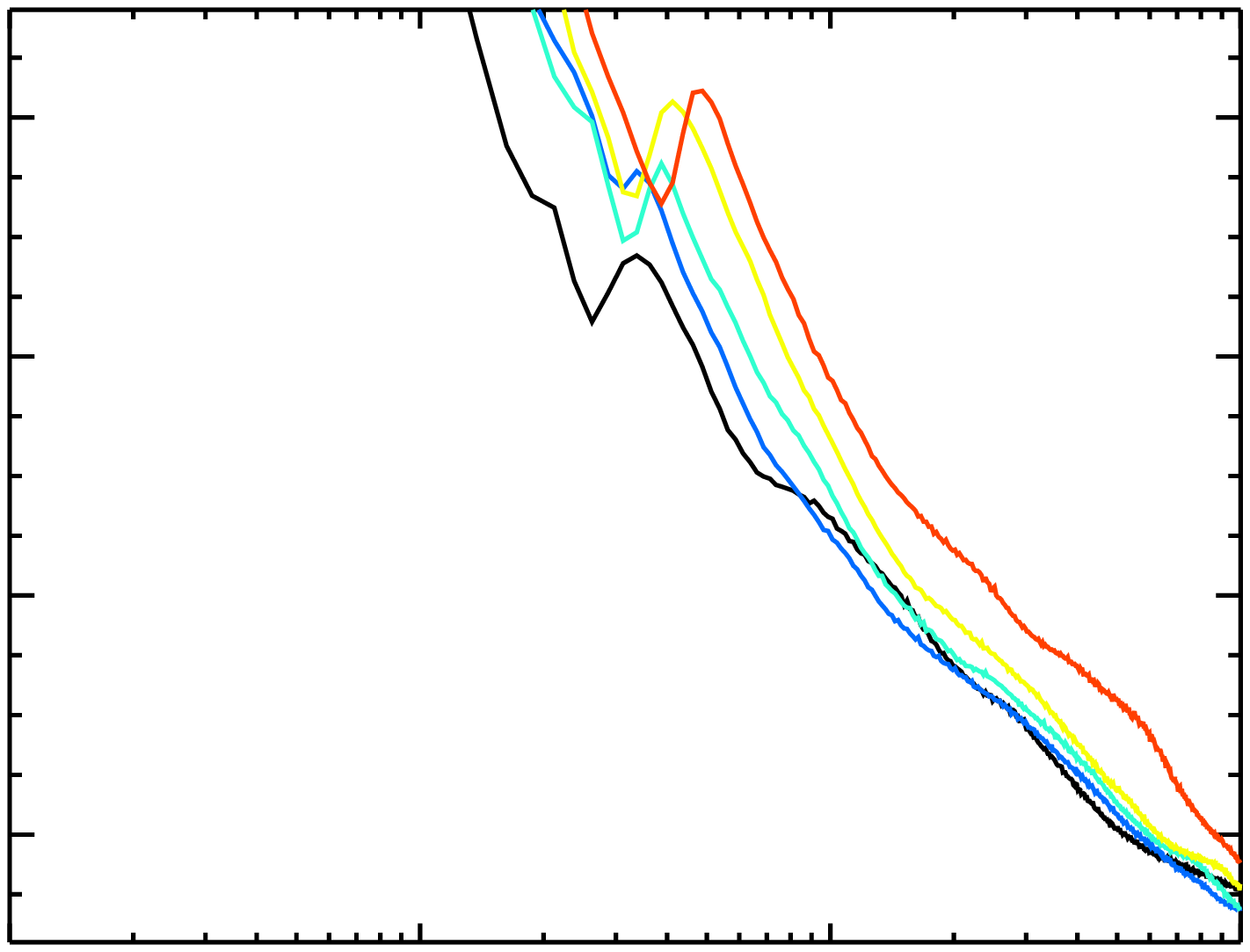}}\hspace{-1.40cm}}
  \centering{\resizebox*{!}{3.4cm}{\includegraphics{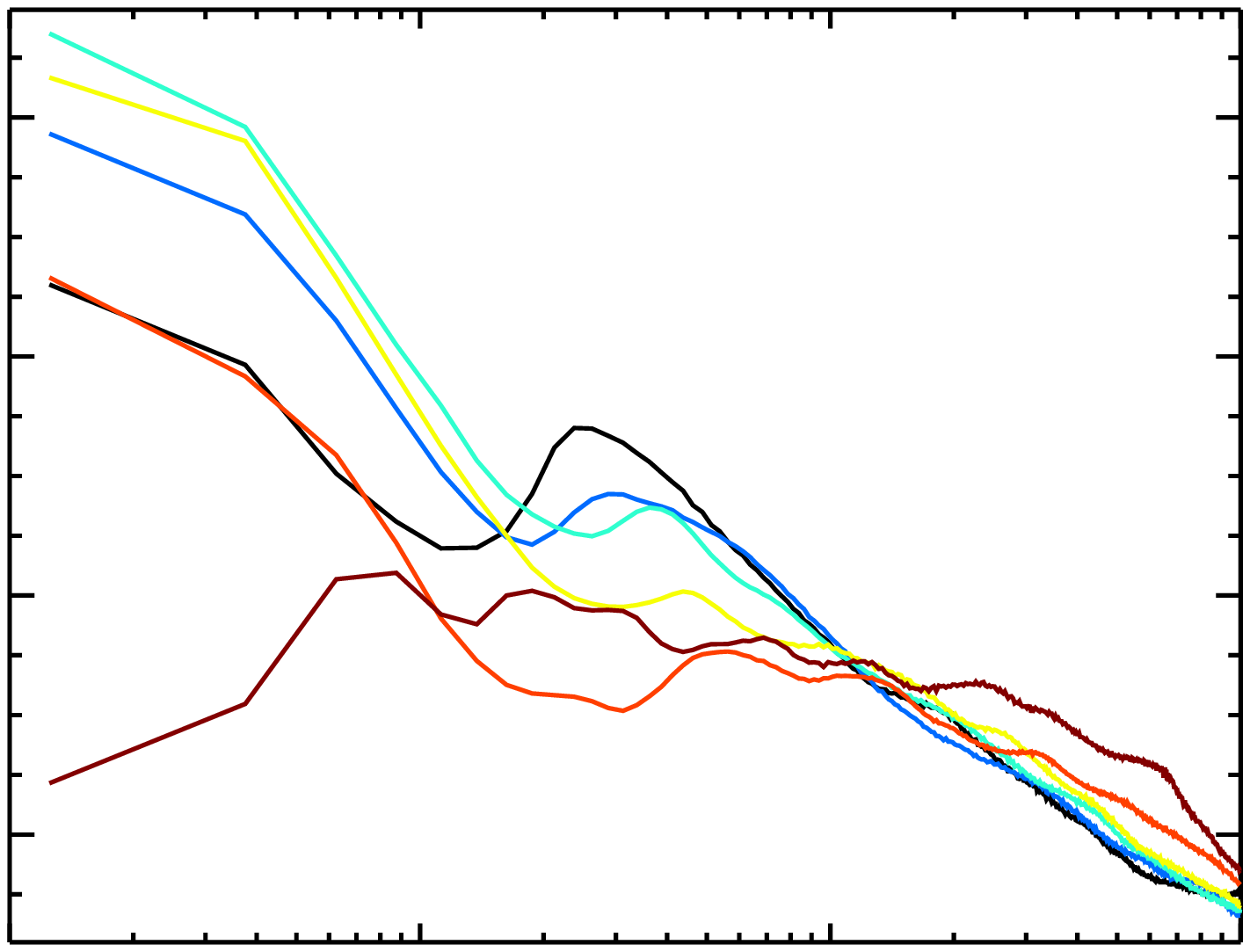}}\hspace{-1.40cm}}
  \centering{\resizebox*{!}{3.4cm}{\includegraphics{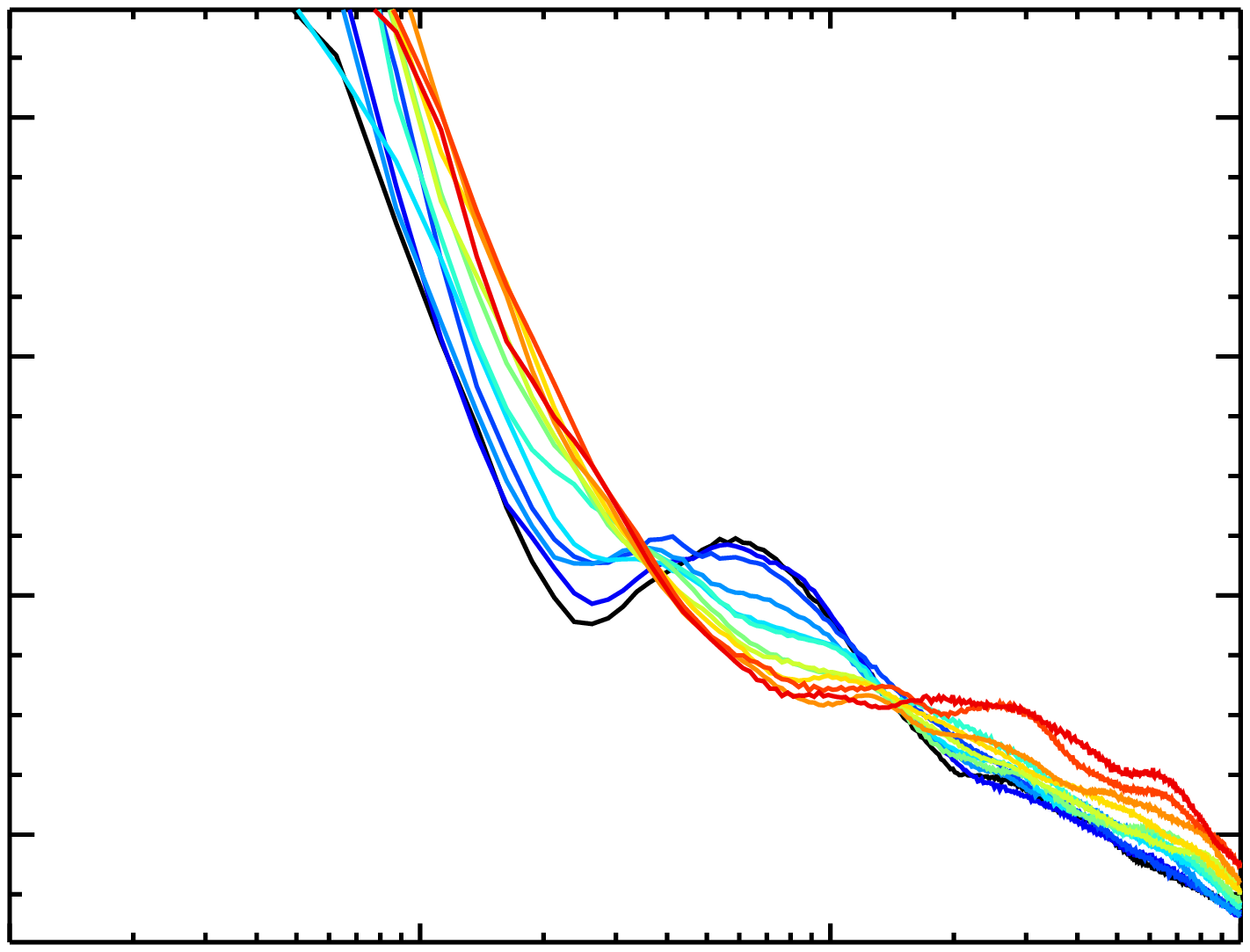}}\hspace{-1.40cm}}
  \centering{\resizebox*{!}{3.4cm}{\includegraphics{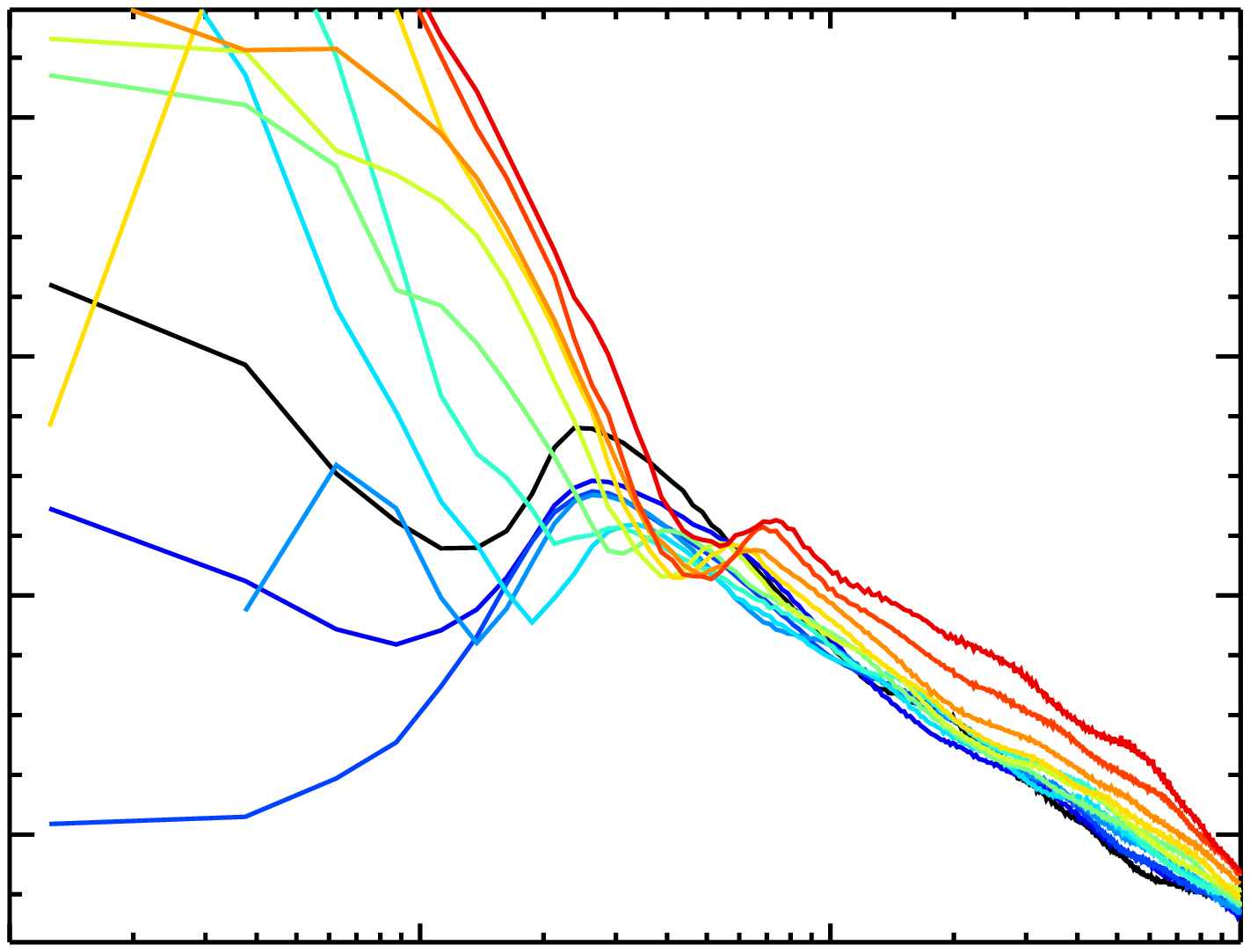}}\vspace{-0.99cm}}\\
  \centering{\resizebox*{!}{3.4cm}{\includegraphics{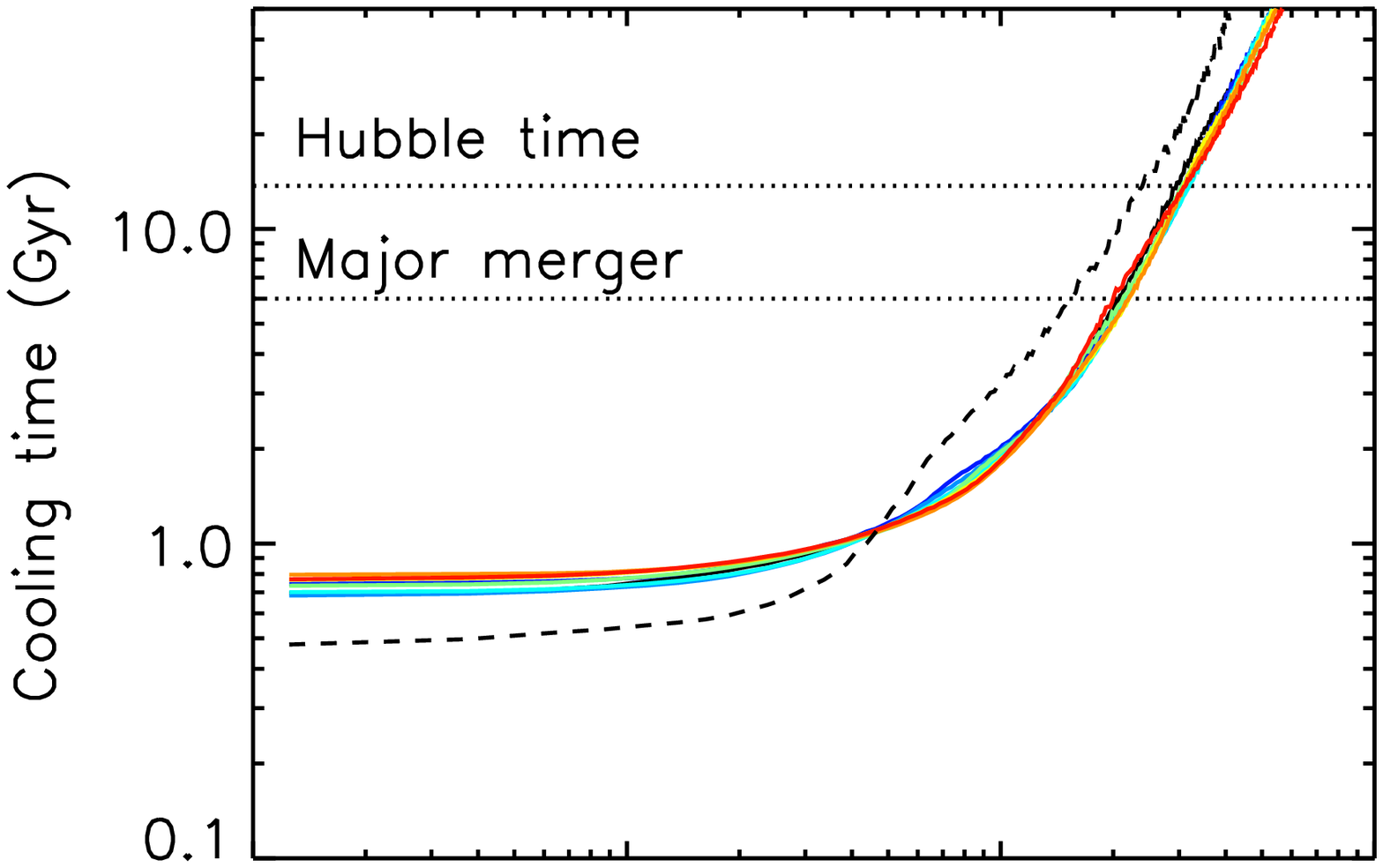}}\hspace{-1.40cm}}
  \centering{\resizebox*{!}{3.4cm}{\includegraphics{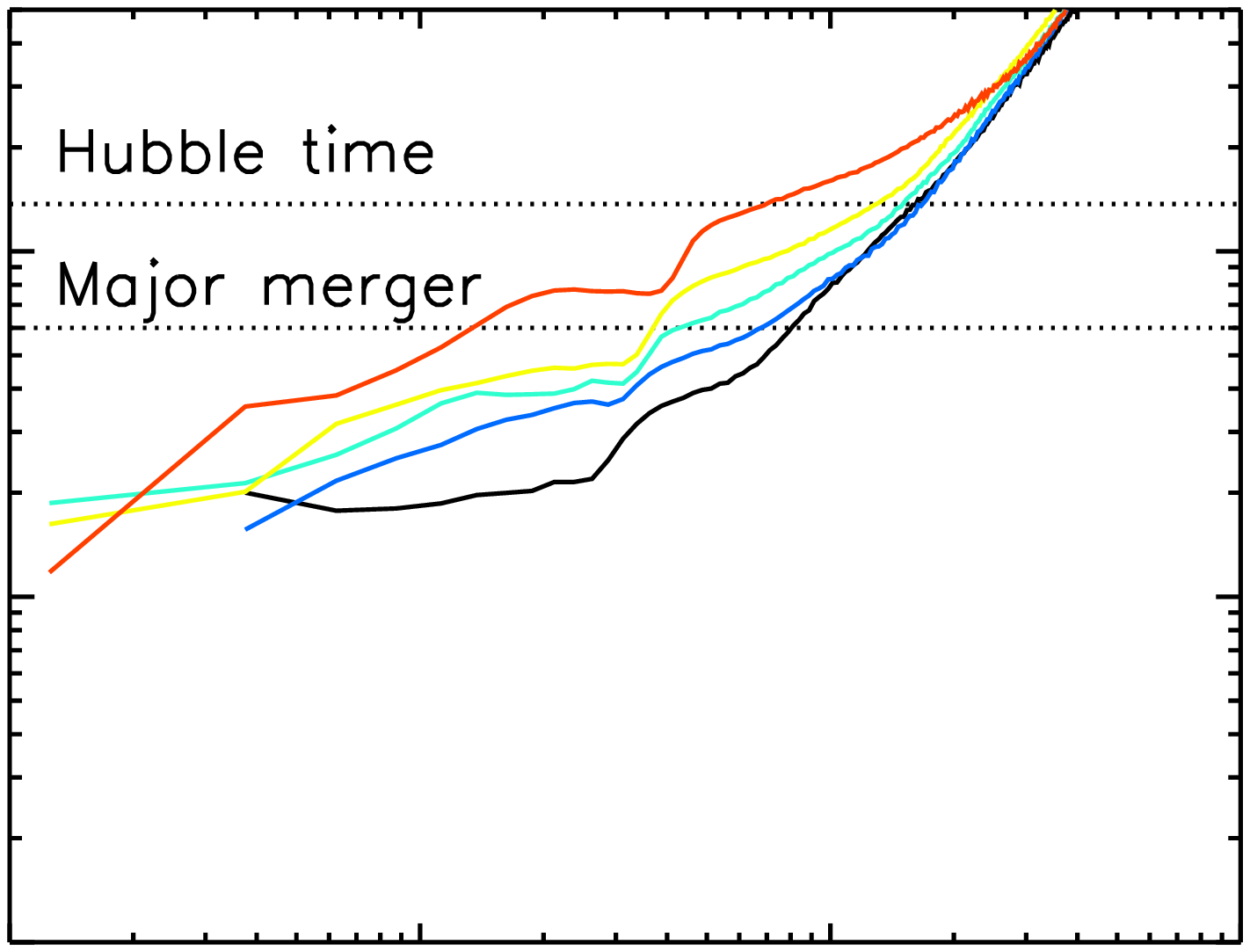}}\hspace{-1.40cm}}
  \centering{\resizebox*{!}{3.4cm}{\includegraphics{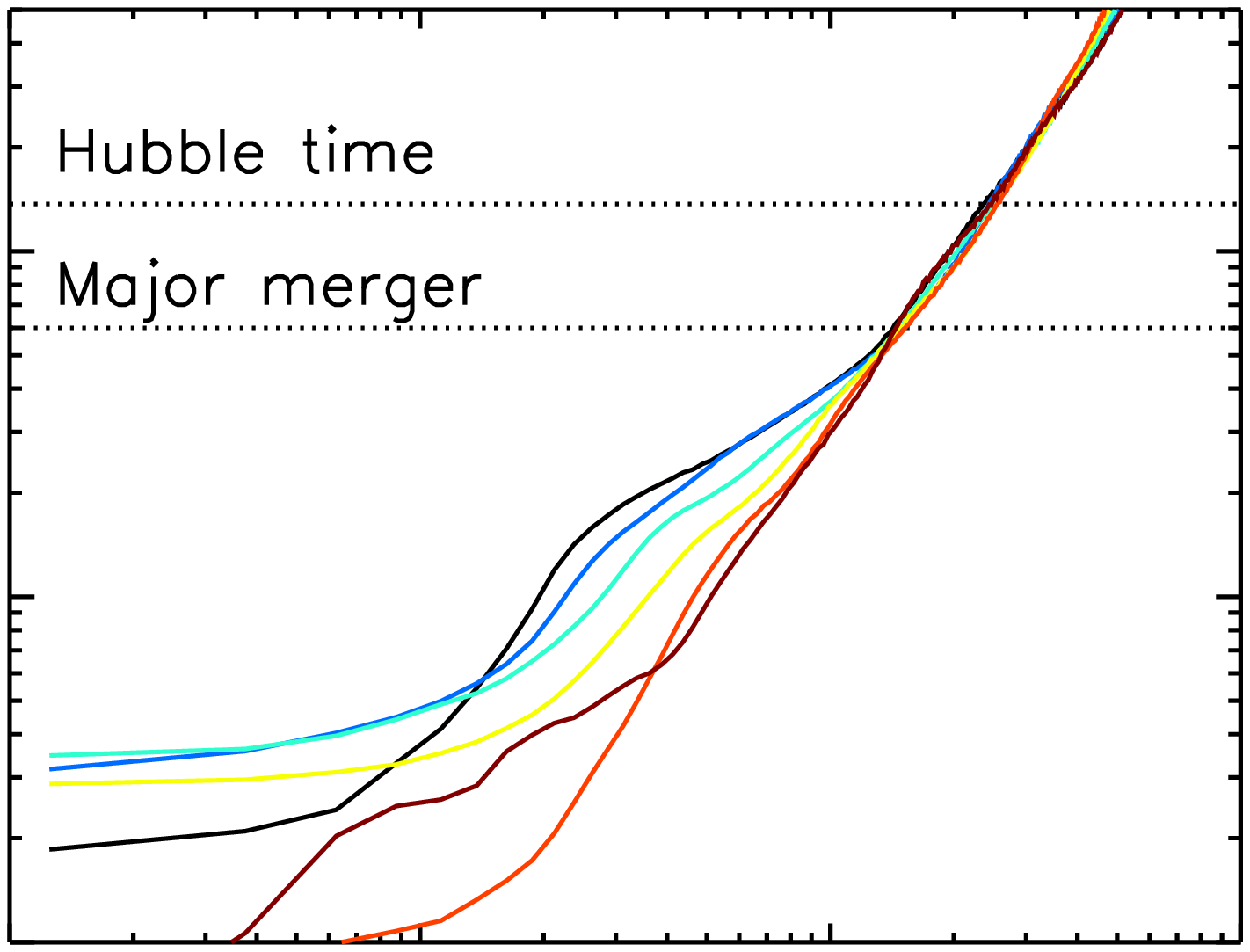}}\hspace{-1.40cm}}
  \centering{\resizebox*{!}{3.4cm}{\includegraphics{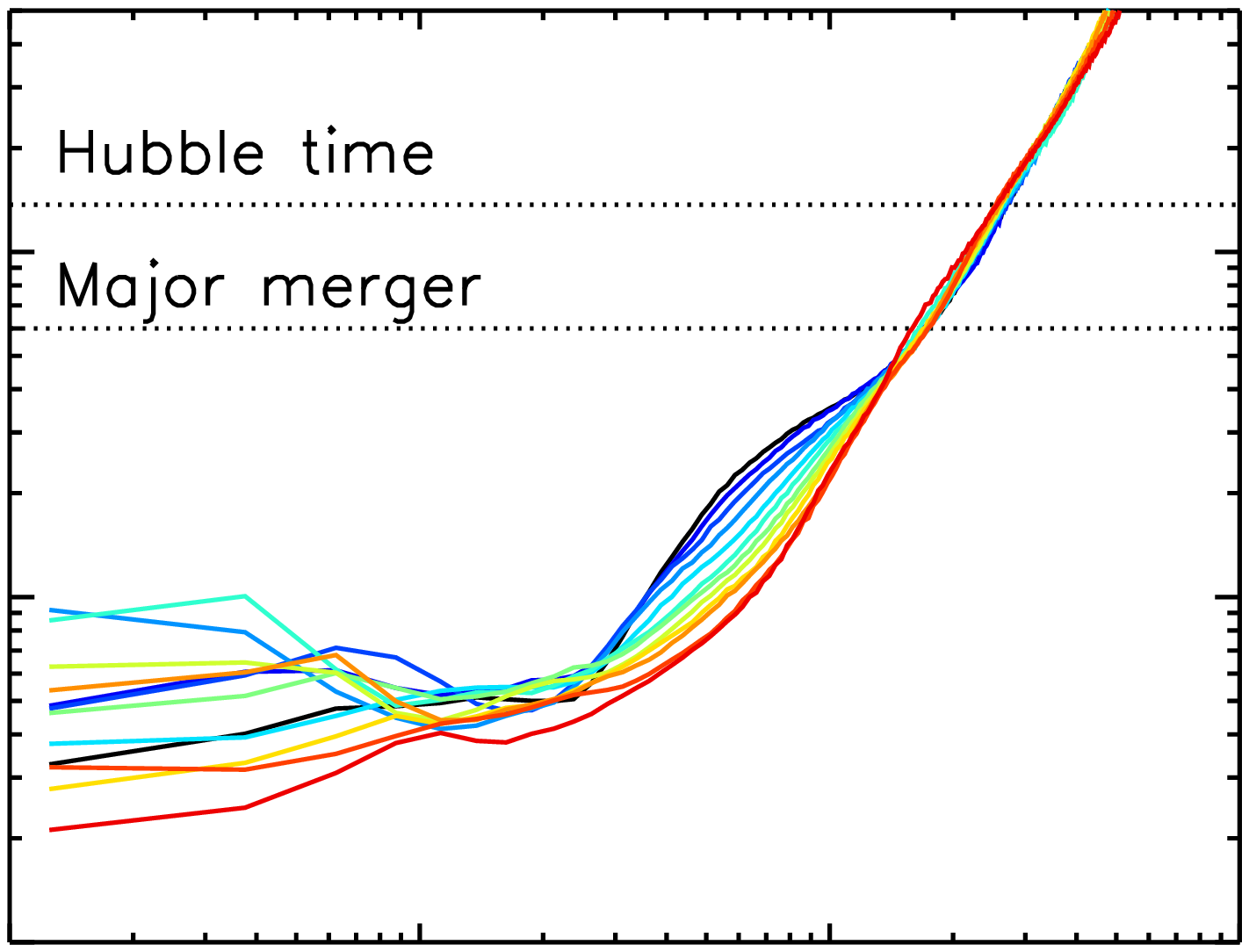}}\hspace{-1.40cm}}
  \centering{\resizebox*{!}{3.4cm}{\includegraphics{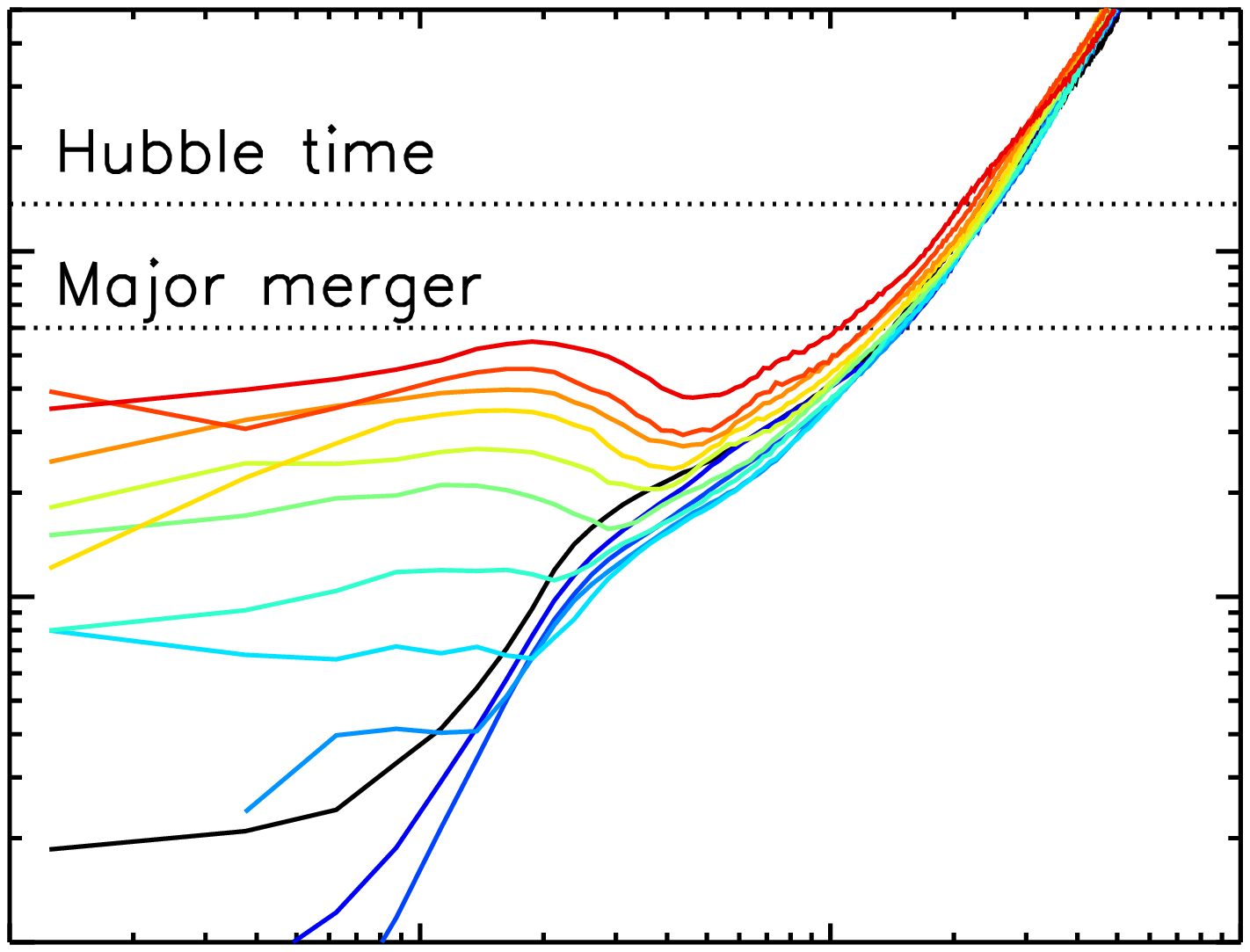}}\vspace{-0.99cm}}\\
  \centering{\resizebox*{!}{3.4cm}{\includegraphics{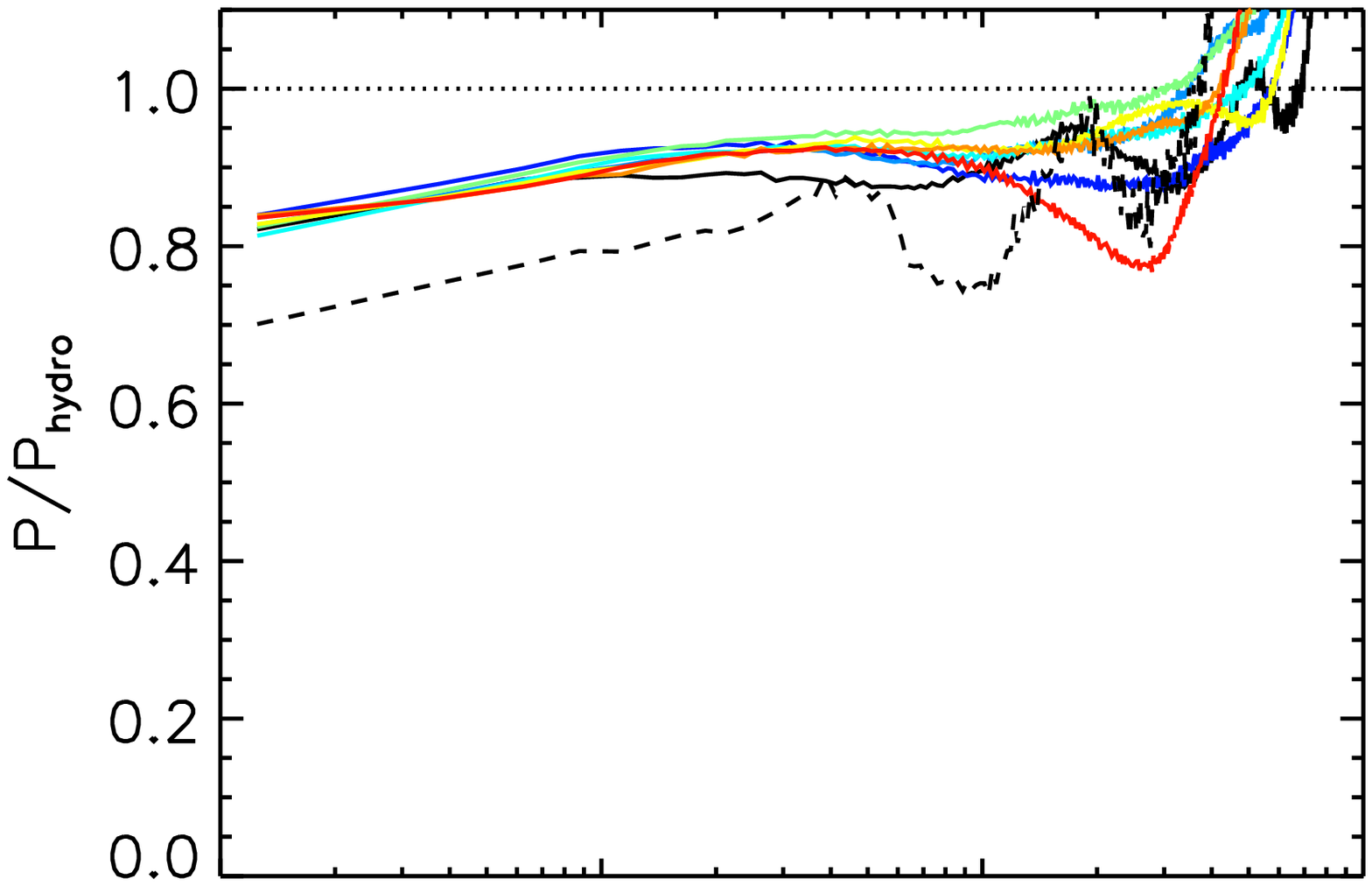}}\hspace{-1.40cm}}
  \centering{\resizebox*{!}{3.4cm}{\includegraphics{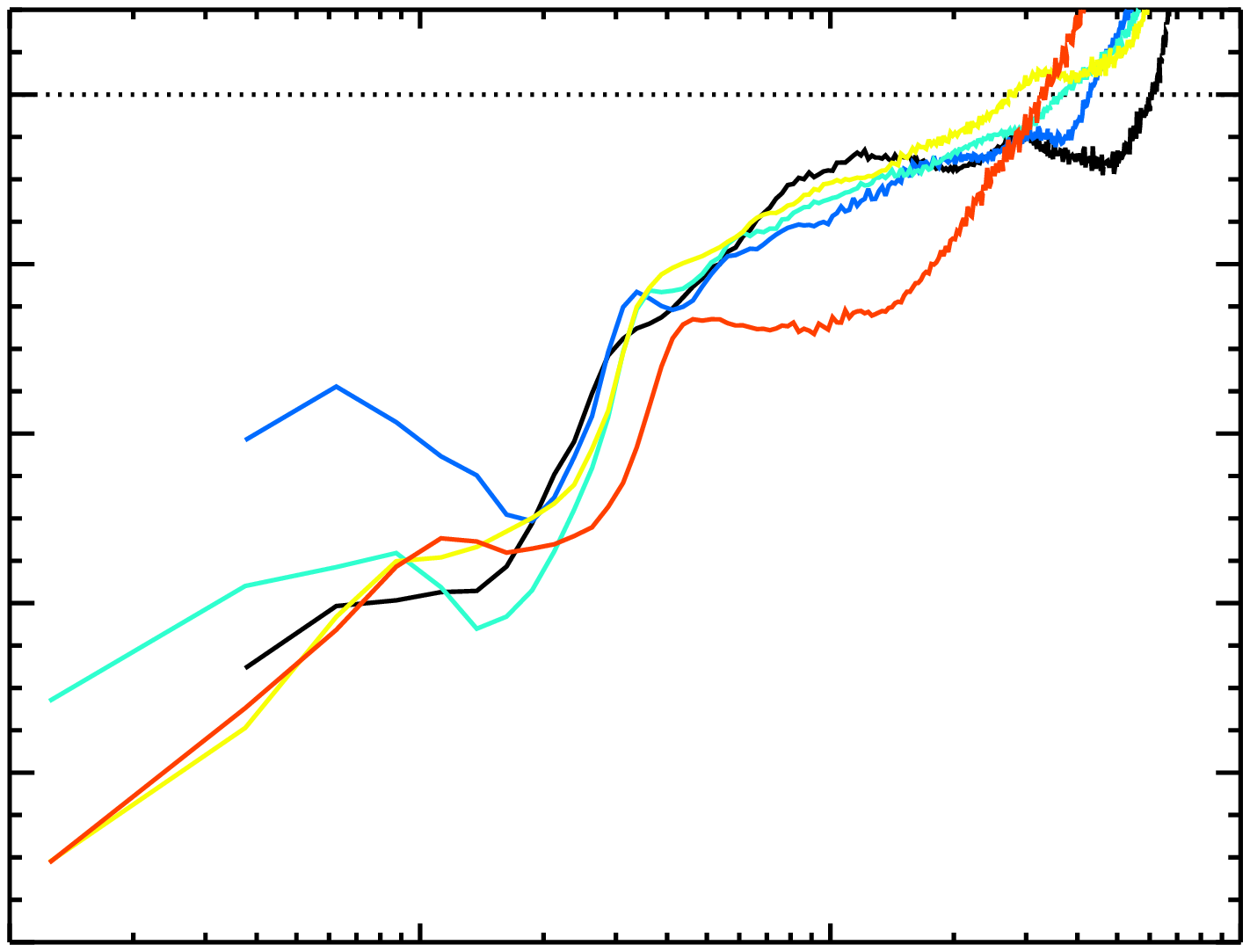}}\hspace{-1.40cm}}
  \centering{\resizebox*{!}{3.4cm}{\includegraphics{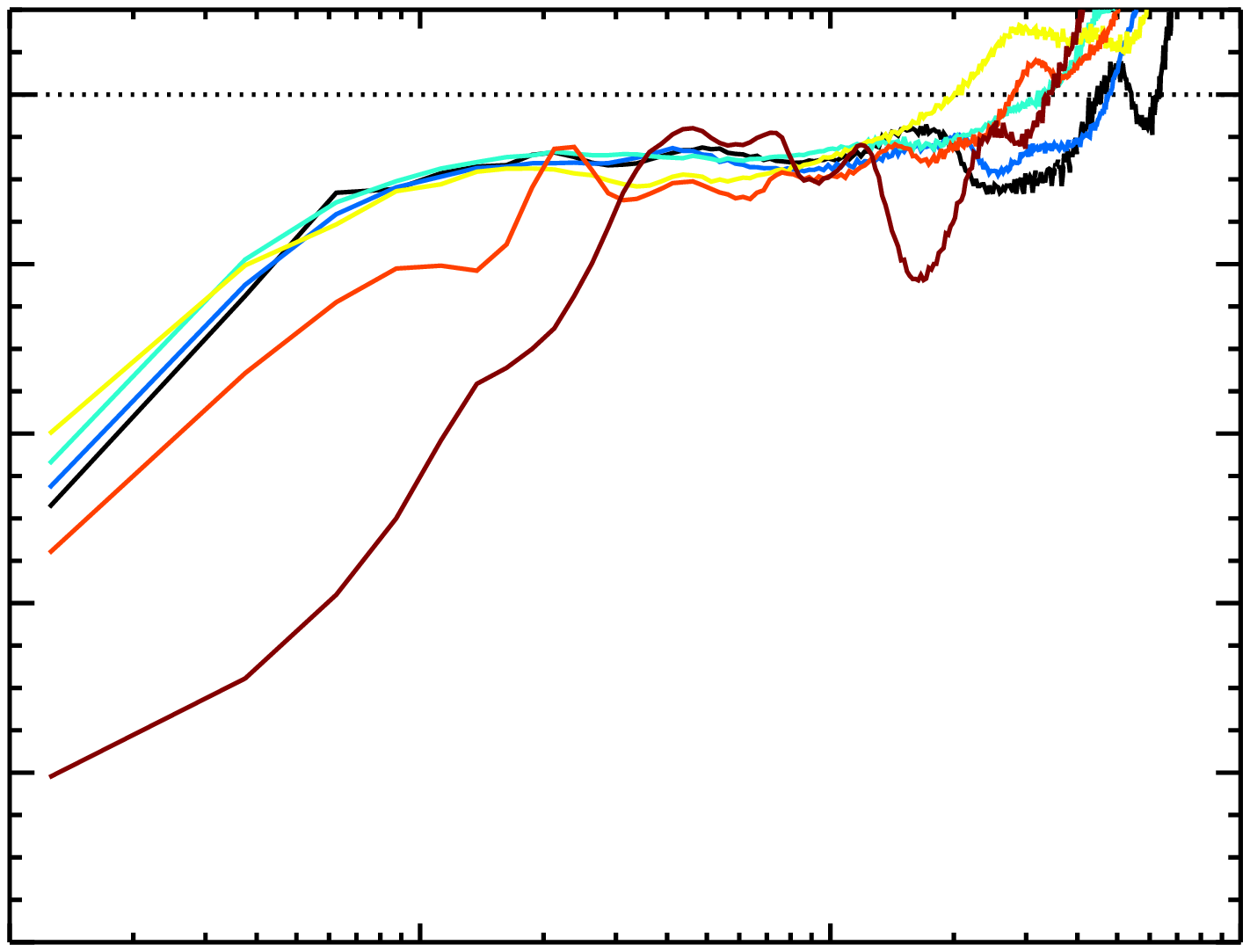}}\hspace{-1.40cm}}
  \centering{\resizebox*{!}{3.4cm}{\includegraphics{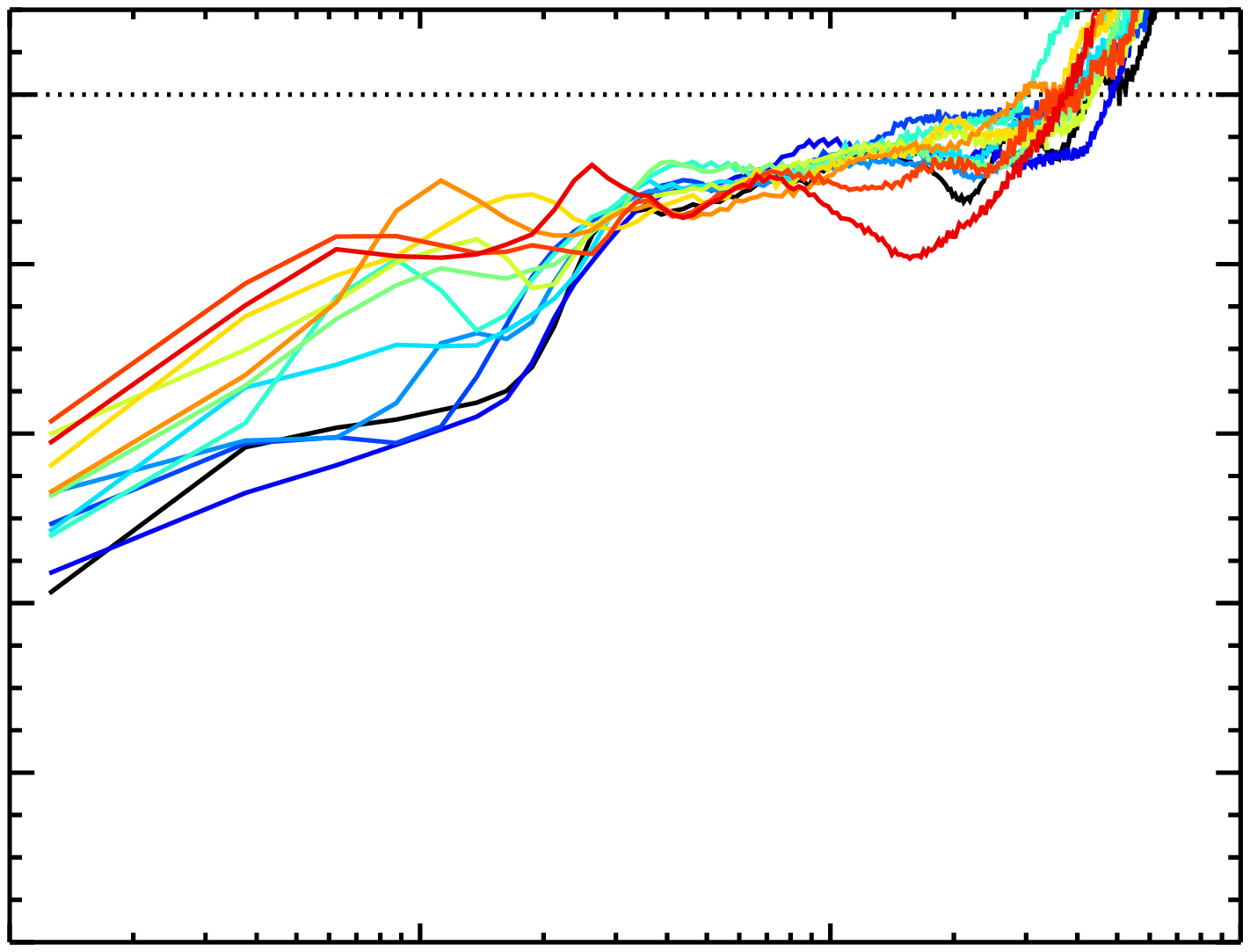}}\hspace{-1.40cm}}
  \centering{\resizebox*{!}{3.4cm}{\includegraphics{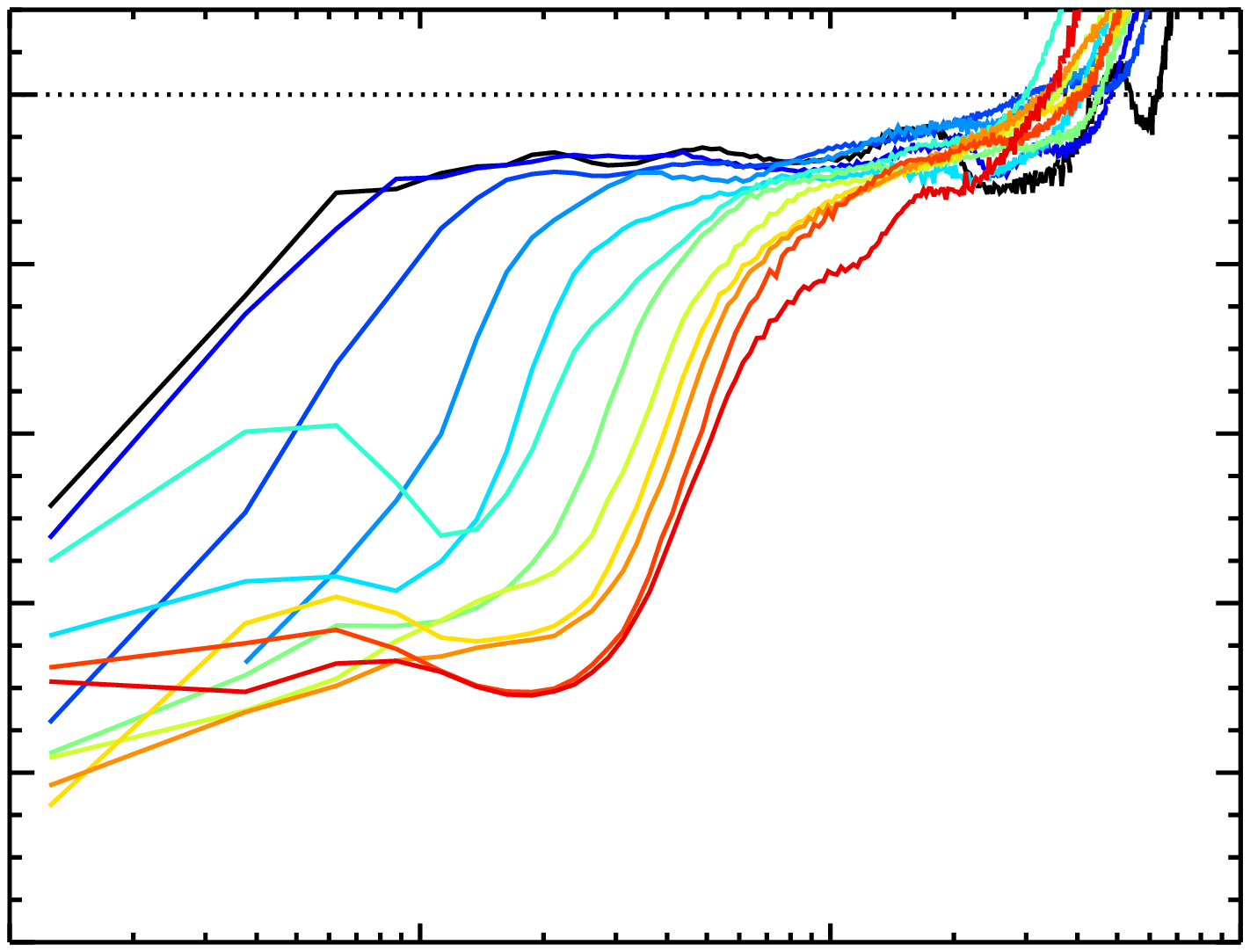}}\vspace{-0.99cm}}\\
  \centering{\resizebox*{!}{3.4cm}{\includegraphics{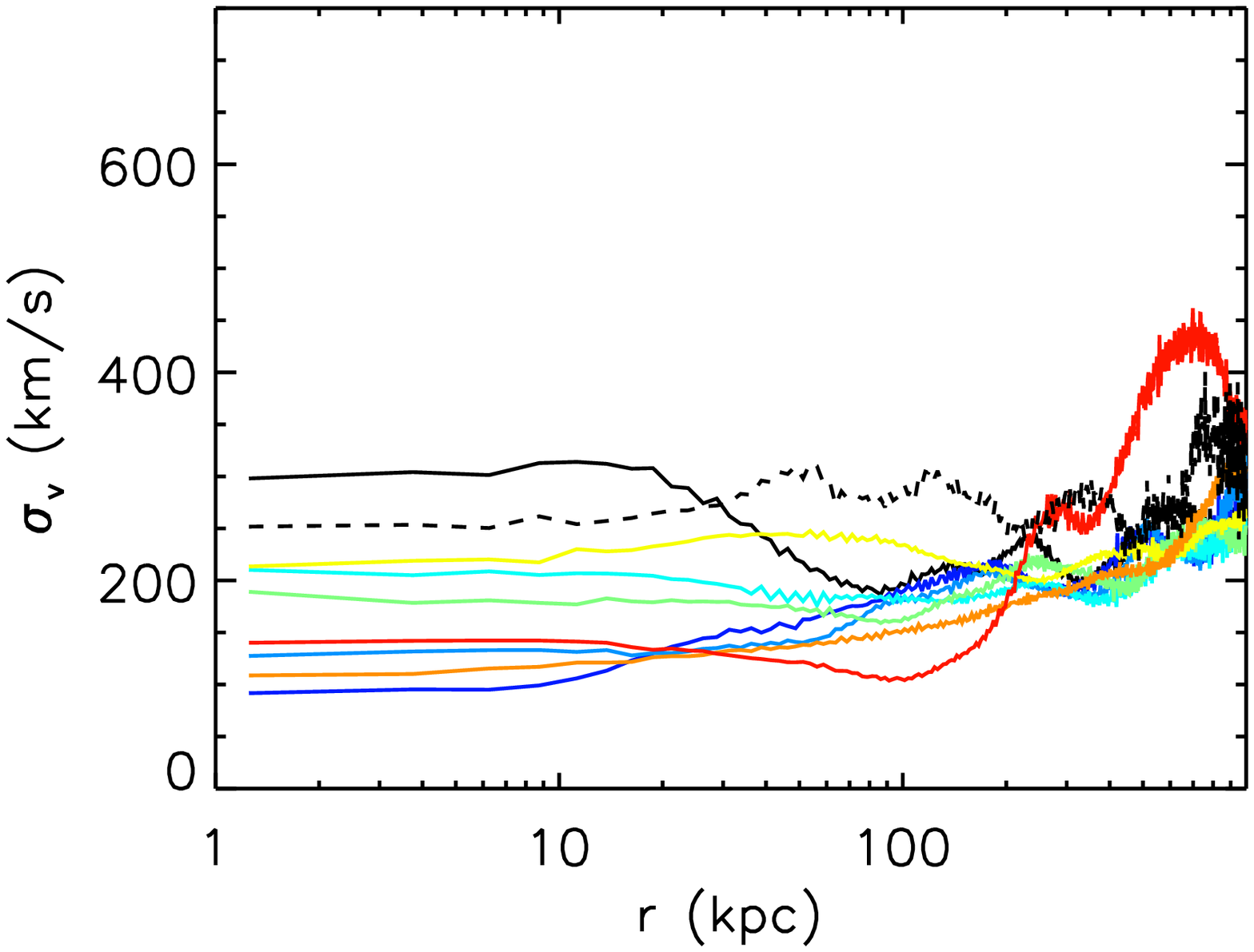}}\hspace{-1.40cm}}
  \centering{\resizebox*{!}{3.4cm}{\includegraphics{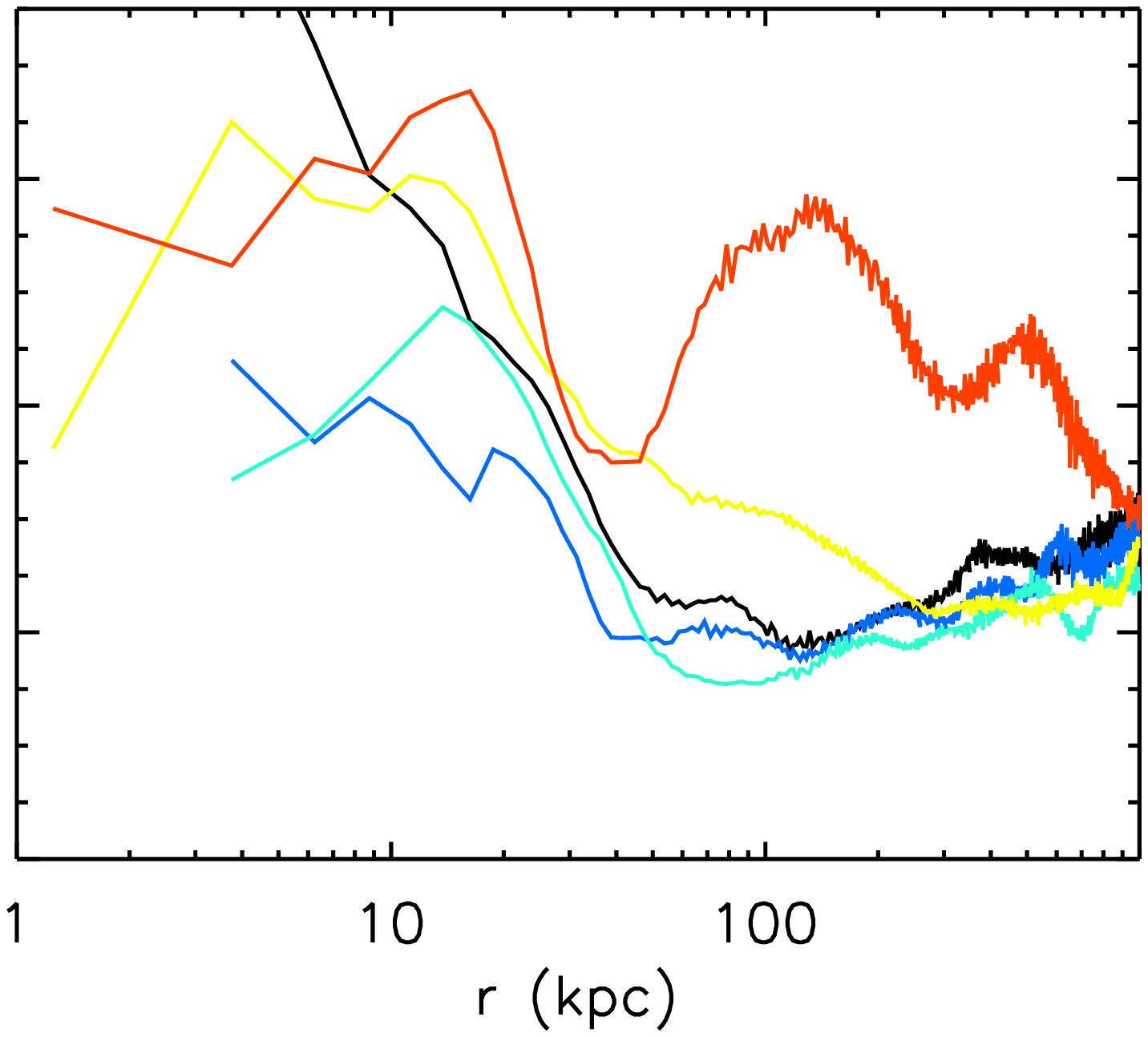}}\hspace{-1.40cm}}
  \centering{\resizebox*{!}{3.4cm}{\includegraphics{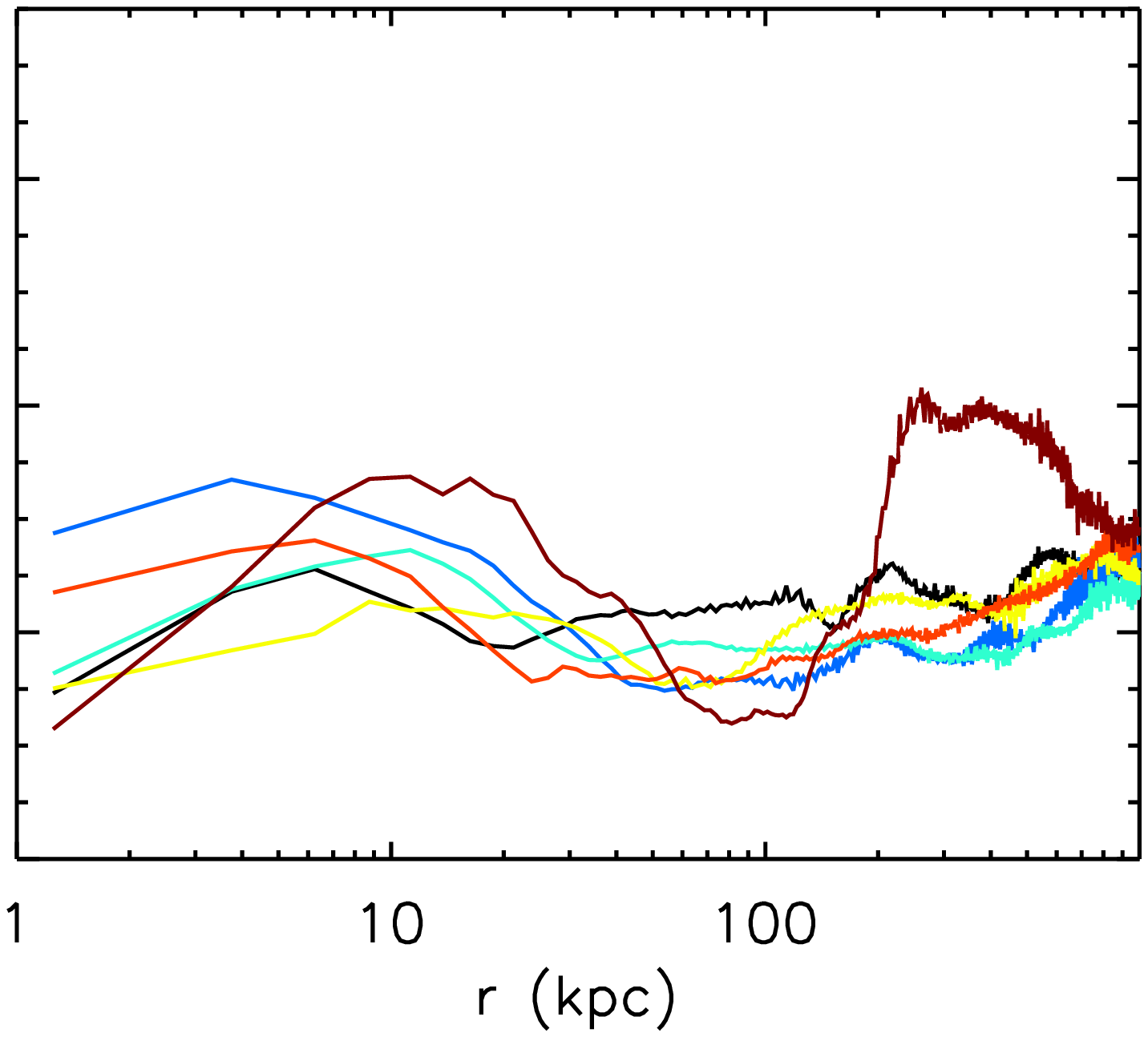}}\hspace{-1.40cm}}
  \centering{\resizebox*{!}{3.4cm}{\includegraphics{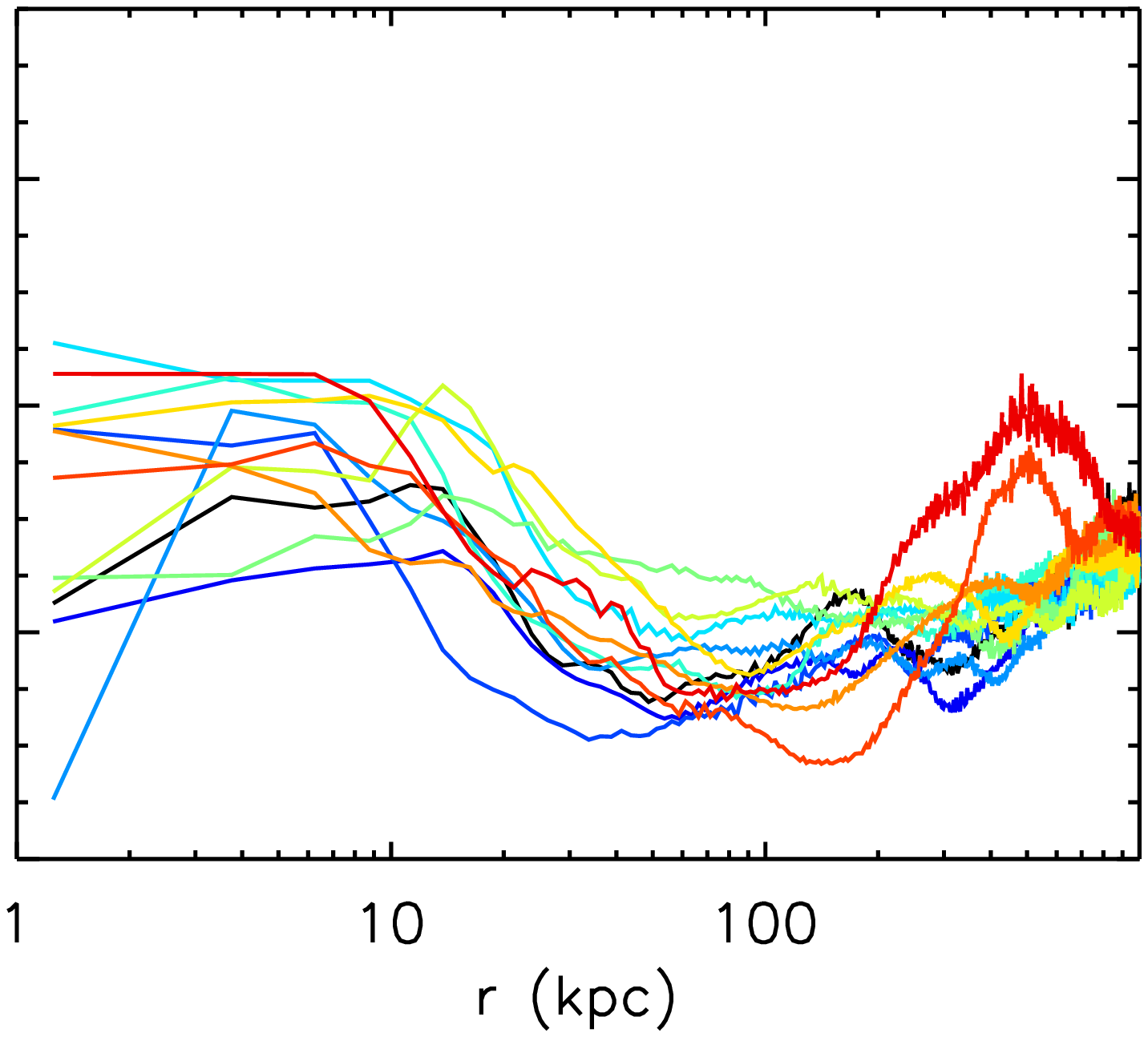}}\hspace{-1.40cm}}
  \centering{\resizebox*{!}{3.4cm}{\includegraphics{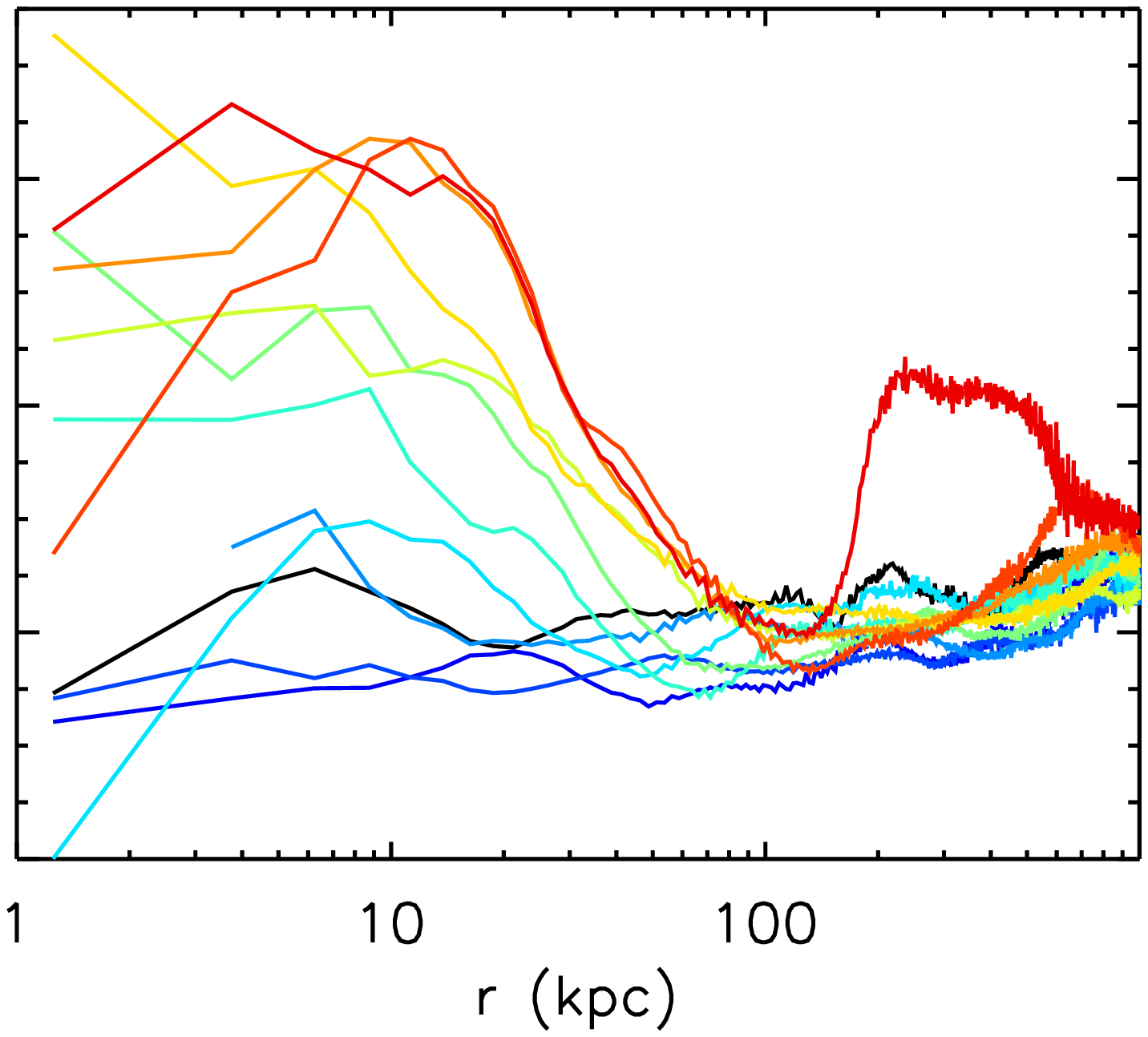}}}
  \caption{A comparison of the time evolution post major merger ($z=0.58$) of the ICM volume-weighted, angular averaged entropy (first row), density (second row), temperature (third row), gas cooling time (fourth row), ratio of gas pressure to hydrostatic pressure (fifth row) and radial gas velocity dispersion (sixth row) profiles for the ADIArun (first column), NOAGNrun (second column), AGNJETrun (third column), AGNHEATrun (fourth column), AGNOFFrun (fifth column). Colors correspond to profiles at different redshifts listed on the entropy plots (first row). The dotted line in the entropy profiles corresponds to the $r^{1.1}$ power-law inferred at large radii from both SPH and AMR simulations of galaxy clusters by Voit et al. (2005). We have also plotted the pre-merger profiles of the cluster for the ADIArun ($z=1.2$) (dashed lines in left column panels). }
    \label{allin}
\end{figure*}

Fig.~\ref{allin}, and subsequent figures including 1-D radial gas profiles
are obtained using volume-weighted angular-averaged profiles centered
on the most massive galaxy at each time step of the simulation. The
galaxies are identified using a halo finder (HaloMaker from
~\citealp{tweedetal09} using the Most-Massive Sub-halo Method) on star
particles or on DM particles when no stars are present.

To fairly compare gas properties from simulations to observations
(mostly from X-ray emission), one should use X-ray-weighted profiles
instead of volume-weighted profiles.  But the dense disc component
in some simulations (AGNJETrun and AGNOFFrun) pollutes the 
profiles even when cold cells
($T< 1$ keV) are removed. For this reason, we choose to only use volume-weighted profiles
and  remove gas cells with temperature smaller than 1 keV. We
discuss the consequences of this choice in more detail in Appendix~\ref{AppendixA}.

\subsection{The effect of gas dynamics}
\label{adiabatic}
The ADIArun simulation does not include any type of radiative cooling, star formation, 
or external heating process (i.e. no stellar or AGN feedback, and no UV background to model
reionization), and thus serves as a reference for the
more complex simulations. It tests the hydrodynamical evolution of the cluster under gravity.

Entropy profiles in fig.~\ref{allin} (first column, first row) show
the well-known features of any adiabatic simulation performed with
Godunov-type solvers: a self-similar power-law profile at large radii with a
flat core component.  The dotted line on the entropy plot
corresponds to the $r^{1.1}$ power-law inferred at large radii from
both SPH and AMR simulations of galaxy clusters by~\cite{voitetal05}.
The high entropy material following a power-law distribution is believed to 
originate from a large-scale expanding shock at a few ($\sim$ 2-3) virial radii
while the flat core component develops from the lower entropy material,
produced by small-scale turbulence and weak shocks in the depths of the gravitational
potential well \citep{tozzi&norman01}. Our results for the entropy profile
in ADIArun are in 
very good agreement with the numerical study performed
by~\cite{mitchelletal09}.  As pointed out by these authors, rather than producing
a flat entropy core, adiabatic cluster simulations with standard
SPH codes produce near power-law profiles down to the
resolution limit (smoothing-length) due to their inability
to capture Kelvin-Helmoltz instabilities \citep{agertzetal07}.

The entropy profiles in the late time evolution spanning $\sim$ 6 Gyrs from $z=0.58$ to $z=0$
for ADIArun vary only by $\sim 20 \%$ due to minor mergers and gas turbulence. 
Indeed, the amount of entropy in the adiabatic simulation of
this cluster is mainly determined by the major merger event lasting from $z \sim 1$ to $z \sim 0.6$ which
doubles the entropy level of the pre-merger phase.  
Meanwhile, the gas temperature increases as kinetic energy from the merger encounter is thermalised by shocks.
The heating from this major merger could potentially prevent the formation of a massive cooling flow
\citep{zuhoneetal10}.  However, even though radiative losses are not considered
in the ADIArun, we can compute the theoretical cooling time based on primordial atomic rates
(fig.~\ref{allin}, first column, fourth row).  This cooling time increases from 400 Myr in the pre-merger phase at 
$z=1.2$ to 700 Myr at the end of the merger $z=0.6$. Thus the major merger would only slightly delay the occurrence of a cooling catastrophe.
The fact that the cooling time is still very small ($t<1$ Gyr) after the major merger
in comparison to the $6$ Gyr or so which remain to reach $z=0$, means that several major
mergers of this type would be required to balance the gas radiative losses.  In
the following sub-sections, we will see that this ``sloshing'' mechanism
is not enough to prevent the cooling catastrophe.

Like the entropy profiles, density profiles in the ADIArun (fig.~\ref{allin}, first column,
second row) vary very little with time. Gas
concentrates in a flat core with low density ($\sim 0.04\, \rm cm^{-3}$).  In fact, the
maximum density reached in the cluster core is lower than the gas
density threshold for star formation ($\rho_0=0.1\, \rm H.cm^{-3}$).
Thus to trigger and maintain star
formation, the simulation needs to include a mechanism, like radiative
cooling,  which allows gas to efficiently concentrate
in the gravitational potential well (see section~\ref{cooling}).

The temperature profiles observed in fig.~\ref{allin} (first column,
third row) are consistent with gas in hydrostatic equilibrium and are therefore the direct consequence of the gas and total mass
distribution in the cluster (see \citealp{hansenetal11}).  For gas in pure hydrostatic equilibrium, the
pressure obeys:
\begin{equation} {d P_{\rm hydro}\over dr}= - \rho G {M_{\rm tot}(<r)
\over r}\, ,
\end{equation} where $\rho$ is the gas density, $G$ the gravitational
constant, $r$ the spherical radius, and $M_{\rm tot}(<r)$ the total
mass (gas, stars, DM and BHs) within $r$.  Fig.~\ref{allin} (first
column, fifth row) shows that the gas pressure is 90$\%$ of 
the hydrostatic pressure.  Some additional support for the cluster gas
may be provided by the global rotation of the cluster and its internal
turbulence \citep{dolagetal05, nagaietal07}.  To assess contributions
from the latter, we measure the gas velocity dispersion (fig.~\ref{allin}, first column, sixth row) in the inner
parts of the cluster. It is 100$-$200 km/s, which is 5$-$10 $\%$ of the
sound speed in the relaxed (post major merger) phase and reaches 15\% of the sound speed
after the merger (at $z=0.58$), in good agreement with recent
simulations from \cite{vazzaetal10}.  These values are extremely close
to what is needed to make up for the lack of internal energy support inferred from the pressure
profiles.

We do not plot the radial velocity profiles for any of the simulations, as pressure
profiles describe the dynamical state of the gas. However we comment that
for the ADIArun, the radial velocity is negligible as it represents
less than $1 \%$ of the sound speed in the cluster core.  This
demonstrates that, in the absence of any radiative cooling process,
the gas is in (almost) perfect equilibrium in the core and is unable
to concentrate in a dense galactic disc.

\subsection{The effect of atomic gas cooling}
\label{cooling}

\begin{figure*}
  \centering{\resizebox*{!}{4.5cm}{\includegraphics{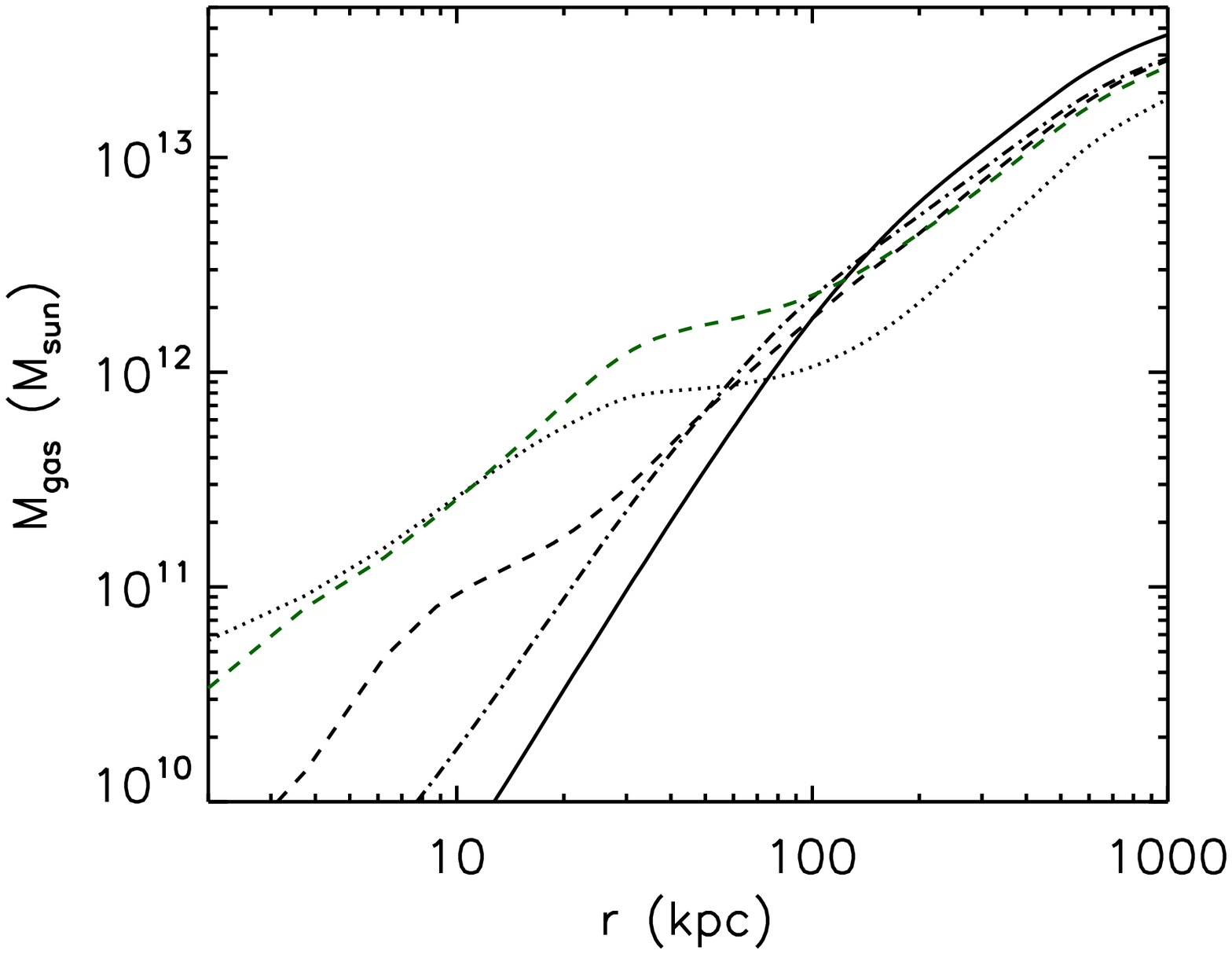}}}
  \centering{\resizebox*{!}{4.5cm}{\includegraphics{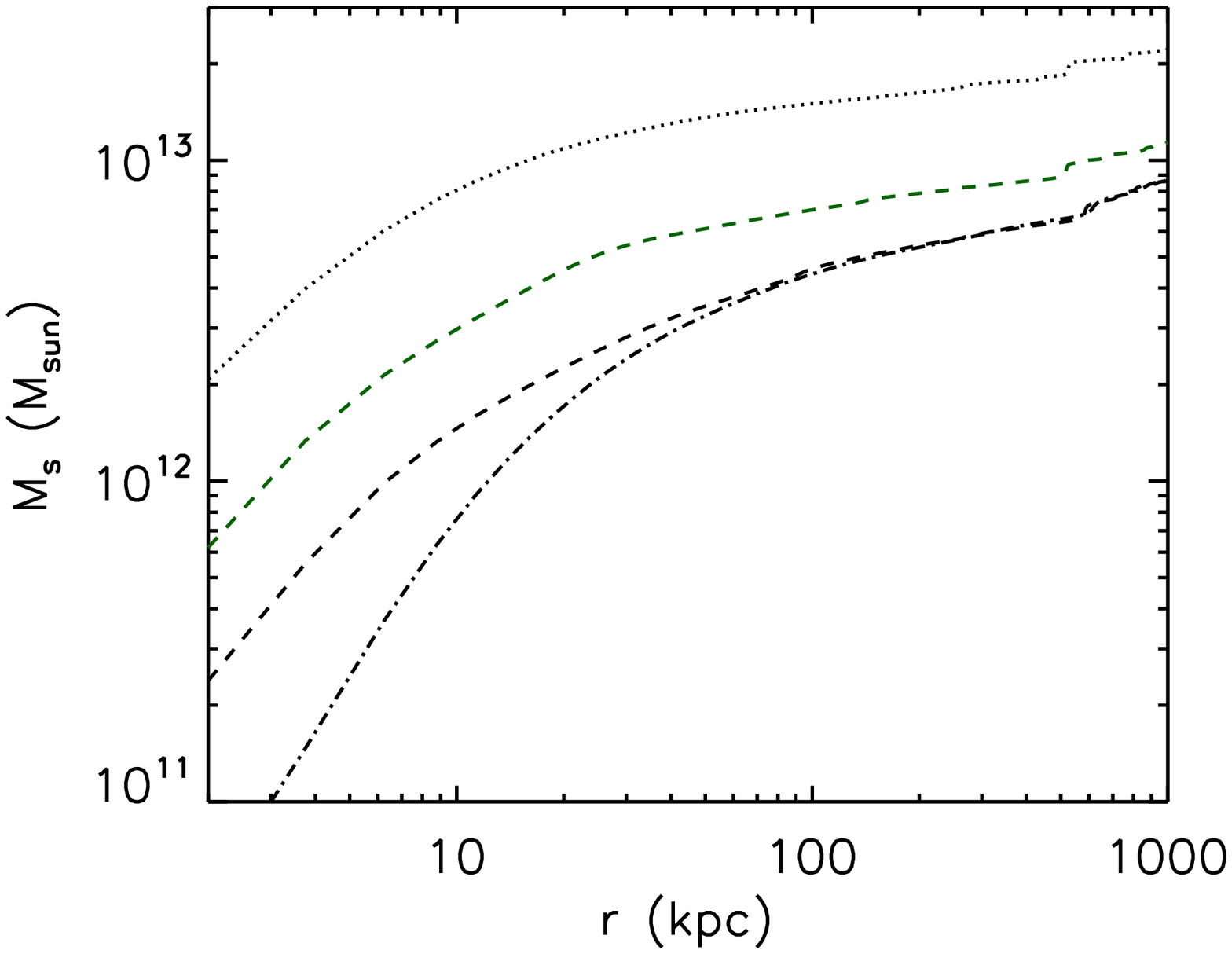}}}
  \centering{\resizebox*{!}{4.5cm}{\includegraphics{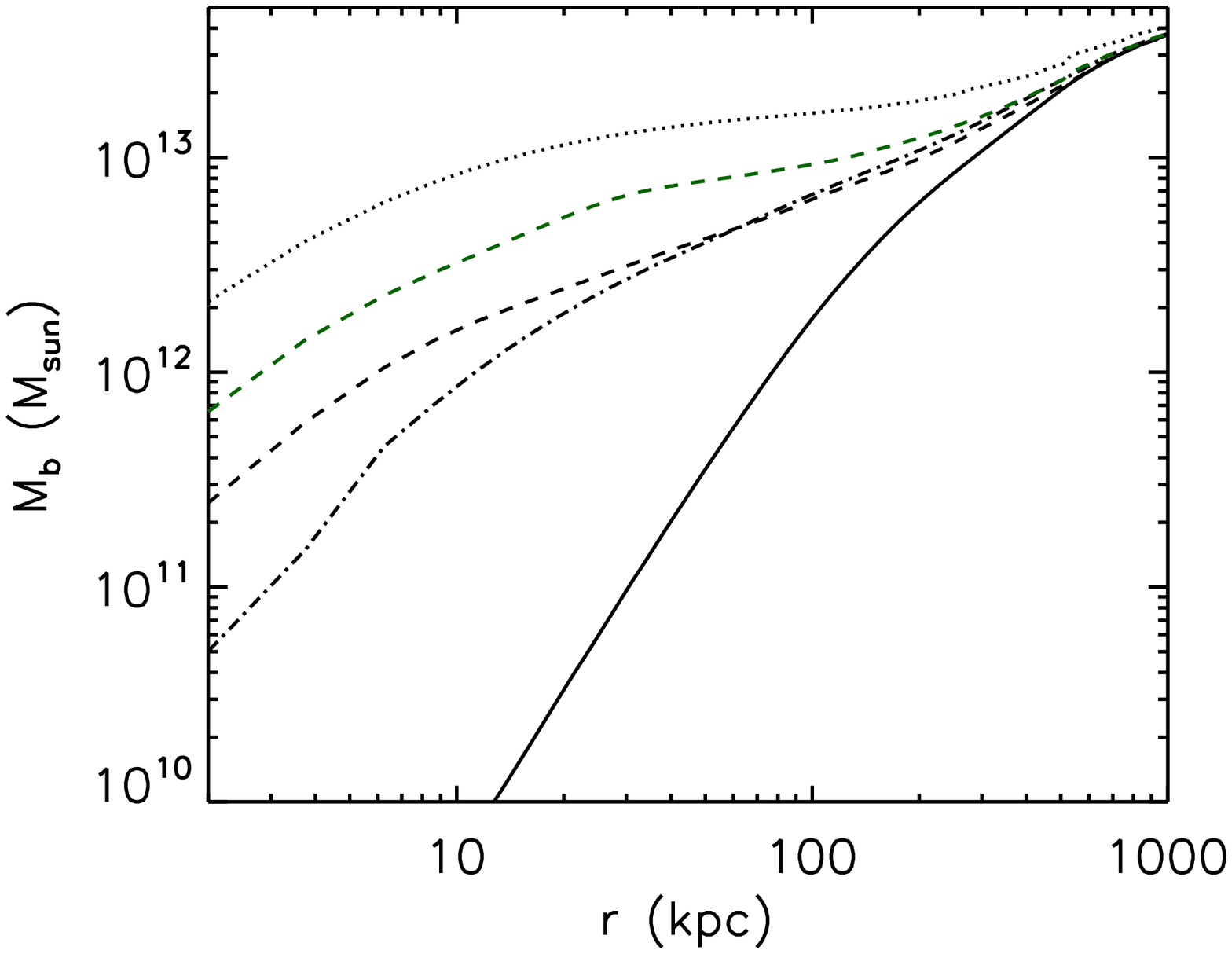}}}\\
  \centering{\resizebox*{!}{4.5cm}{\includegraphics{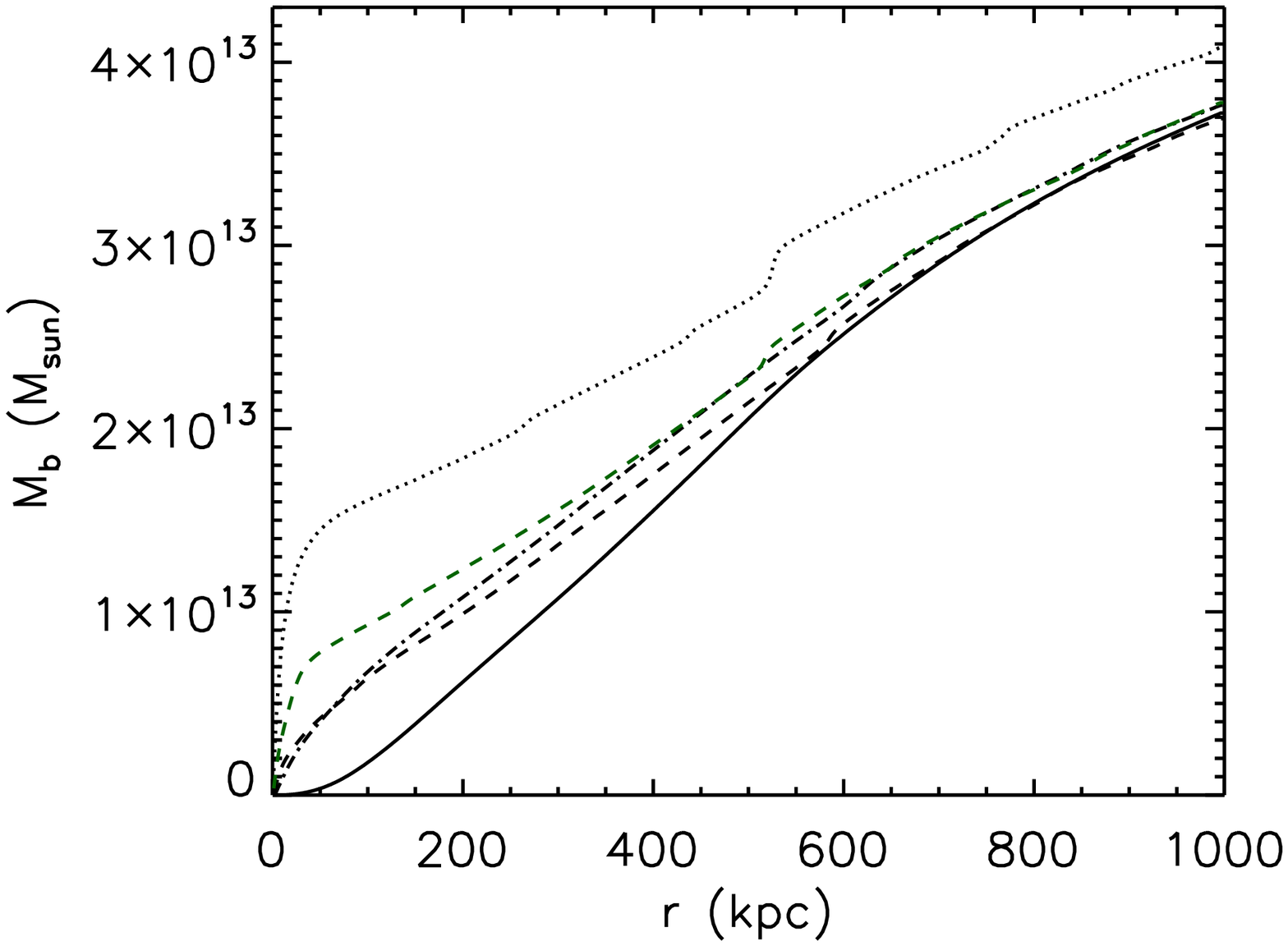}}}
  \centering{\resizebox*{!}{4.5cm}{\includegraphics{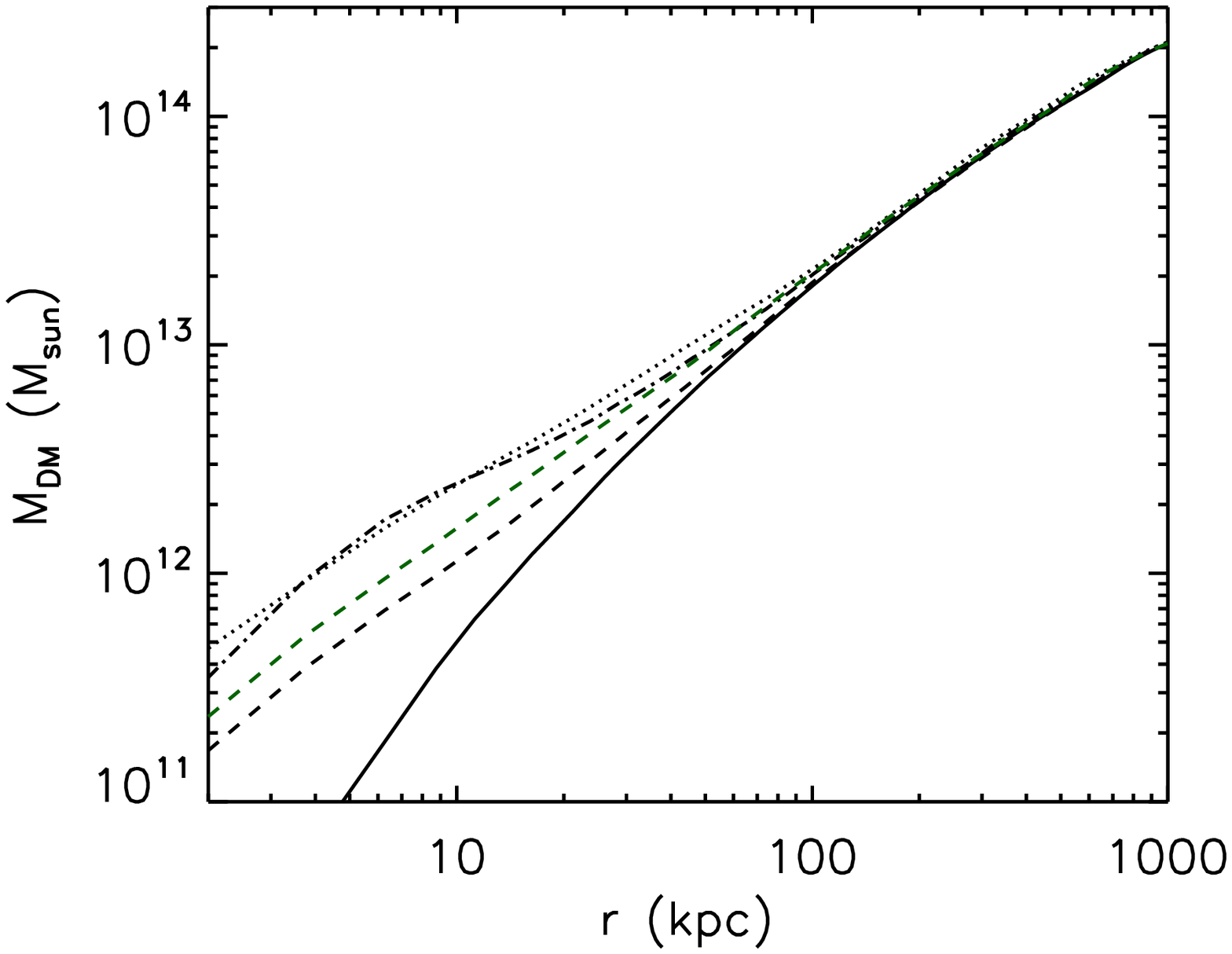}}}
  \centering{\resizebox*{!}{4.5cm}{\includegraphics{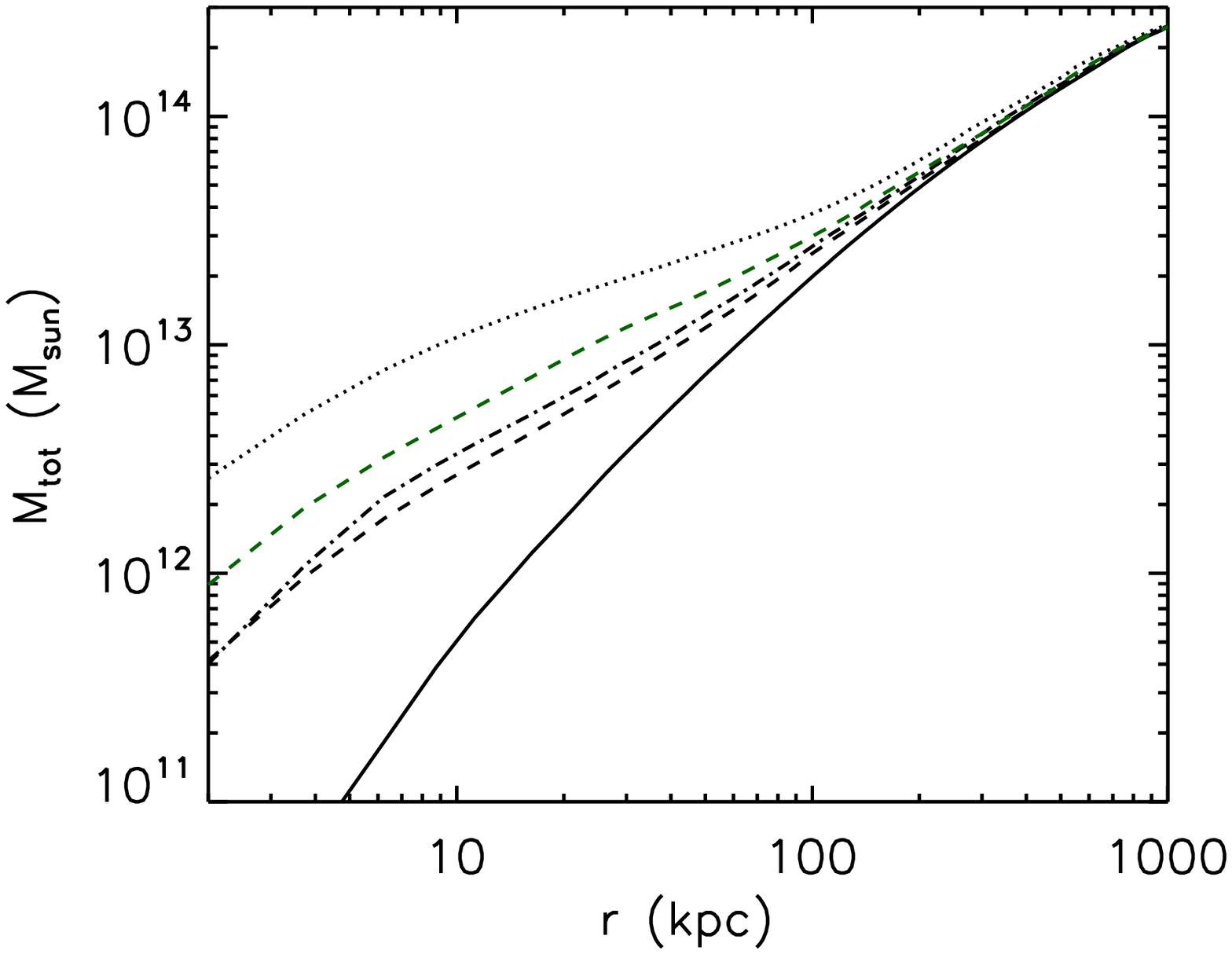}}}
  \caption{Cumulative gas (upper left), star (upper middle), baryon (logarithmic in upper right and linear in lower left), dark matter (lower middle) and total (lower right)
mass profiles of the cluster at $z=0$ for the ADIArun
(black solid line), the NOAGNrun (black dotted line), the AGNJETrun
(black dashed line), the AGNHEATrun (black dot-dashed line), and the
AGNOFFrun (green dashed line). The cumulative baryon mass is plotted on a linear scale in order to better display what is happening at large radii.}
    \label{mass_comp}
\end{figure*}

When atomic cooling and star formation are allowed (NOAGNrun), gas can cool down to very
low temperatures ($10^4$ K) compared to typical ICM temperatures
($T\sim 10^8$ K).  Cold gas condenses in the centres of 
gravitational potential wells to form galactic discs.  In turn, the
gravitational potential deepens, pulling more and more DM into
the centre.  This well-known mechanism called 'adiabatic contraction`
is described in detail in~\cite{blumenthaletal86} (see also
\citealp{gnedinetal04}).  
Displaying integrated mass profiles for the gas, stars, baryons and DM, Fig.~\ref{mass_comp} illustrates this effect:
as more baryons concentrate in the core of the cluster, more DM
particles are pulled in, forming a stronger DM core component.

Fig.~\ref{allin} (second column, first row) shows that at all times, the entropy
in the core of the NOAGNrun is one order of magnitude higher
than the entropy in the core of the ADIArun.  Furthermore,
the entropy in the NOAGNrun continuously increases with time in the cluster,
both inside and outside the core.  This entropy rise has been interpreted as
arising from the removal of low entropy gas as it cools and turns into stars
(note that the cooling time of the ICM is shorter than 6 Gyr, the time span between $z=0.58$ and $z=0$)
and its replacement with high-entropy material
coming from the outer parts of the cluster \citep{bryan00,
voit&bryan01}. Because of the Eulerian nature of grid codes, we cannot directly follow
the history of gas elements in our simulations. However using the Lagrangian SPH technique, or tracer particles within grid codes, \citet{mccarthyetal11} and \citet{vazza11} have validated the above picture by showing that when low-entropy material condenses
into the central disc, high-entropy material flows into the central parts of the halos replacing the depleted low-entropy gas.

Support for the above picture also comes from the gas density profiles (fig.~\ref{allin}, second row of second column) and the
cumulative mass profiles (fig.~\ref{mass_comp}).
The  gas density in the NOAGNrun strongly diminishes with time in the ICM
(fig.~\ref{allin}, second column, second row) as cold gas gets incorporated into
the galaxy.  
The small feature observed between $30$-$40$ kpc in the gas density 
and gas entropy profiles is characteristic of the flaring tail of the disc,
which is dense, warm and low entropy, and thus difficult to separate
from the `true' ICM.  Indeed, in Fig.~\ref{mass_comp} we plot the gas mass profile (dotted line)
for all the gas (ICM and gas in the galaxy) in the cluster. Since we include the galaxy's gas,  the cumulative gas mass
profile for the NOAGNrun (dotted line) shows a strong contribution
from the centre of the cluster. The outer
regions ($r>50$ kpc) in the cumulative gas mass profile are depleted of gas corroborating the picture that
gas from the outer regions moves inward, cools, gets incorporated into the galaxy component and forms stars.
Note that the total
baryon mass at $r_{500}\simeq 1$ Mpc is hardly affected (on the level of 10 \%) by
atomic gas cooling. Since the cooling time in these regions is larger than a Hubble time,  the accretion of material on
large scales is mainly driven by cosmological accretion, and not by
radiative processes.

More evidence for the above picture comes from pressure profiles (fig~\ref{allin}, second column, fifth row)
They reveal that the gas in the NOAGNrun is far from hydrostatic equilibrium within $r<50$ kpc
resulting in a massive
cooling flow ($v_{\rm r} > 1000$ km/s) towards the centre of the
cluster.  Since low-entropy material cools faster than high
entropy material (fig.~\ref{allin}, second column, fourth row), it
continuously flows onto the galaxy and is converted into stars. This removes
pressure support at small radii, causing
high-entropy gas at large radii to smoothly flow towards the centre as it is sucked into the cluster core.
This explains the difference
in the cumulative baryon mass seen at $r_{500}$ (table~\ref{tabnames} and
first panel in second row of fig.~\ref{mass_comp}) for the NOAGNrun (black dotted) compared to the ADIArun (black solid line).

Thus, we conclude that even though gas loses internal energy by
radiative losses, the net effect of cooling is to replace low-entropy
gas depleted onto the galaxy by high-entropy gas from large radii, and
to fill the ICM with gas at higher temperature (see fig.~\ref{allin},
third column, second row).  The rise in total mass (dotted curve in $M_{\rm tot}$ plot of fig.~\ref{mass_comp})
of the cluster core drives a gas internal energy increase to counterbalance
the extra gravity.  Consequently, the massive
reservoir of internal energy at large radii supplies the
cluster core with what is missing to support its own collapse: gas at
higher entropy and higher temperature.

The presence of this strong cooling flow is also characterized by a
vigorous, turbulent radial velocity dispersion (fig.~\ref{allin}, second column,
sixth row), but this cannot explain the increase in temperature as the
radial velocity dispersion remains very low compared to the gas sound
speed ($\sim 10-20 \%$).  In a nutshell, we see all the features of
a \emph{cooling catastrophe} in the NOAGNrun: gas density is irreversibly
depleted in the ICM, which leads to an unrealistic amount of gas flowing
towards the central galaxy, and larger values of entropy and
temperature in the cluster (see also \citealp{nagaietal07}).

\subsection{The effect of AGN feedback}
\label{AGNfbk}

\begin{figure}
  \centering{\resizebox*{!}{5.5cm}{\includegraphics{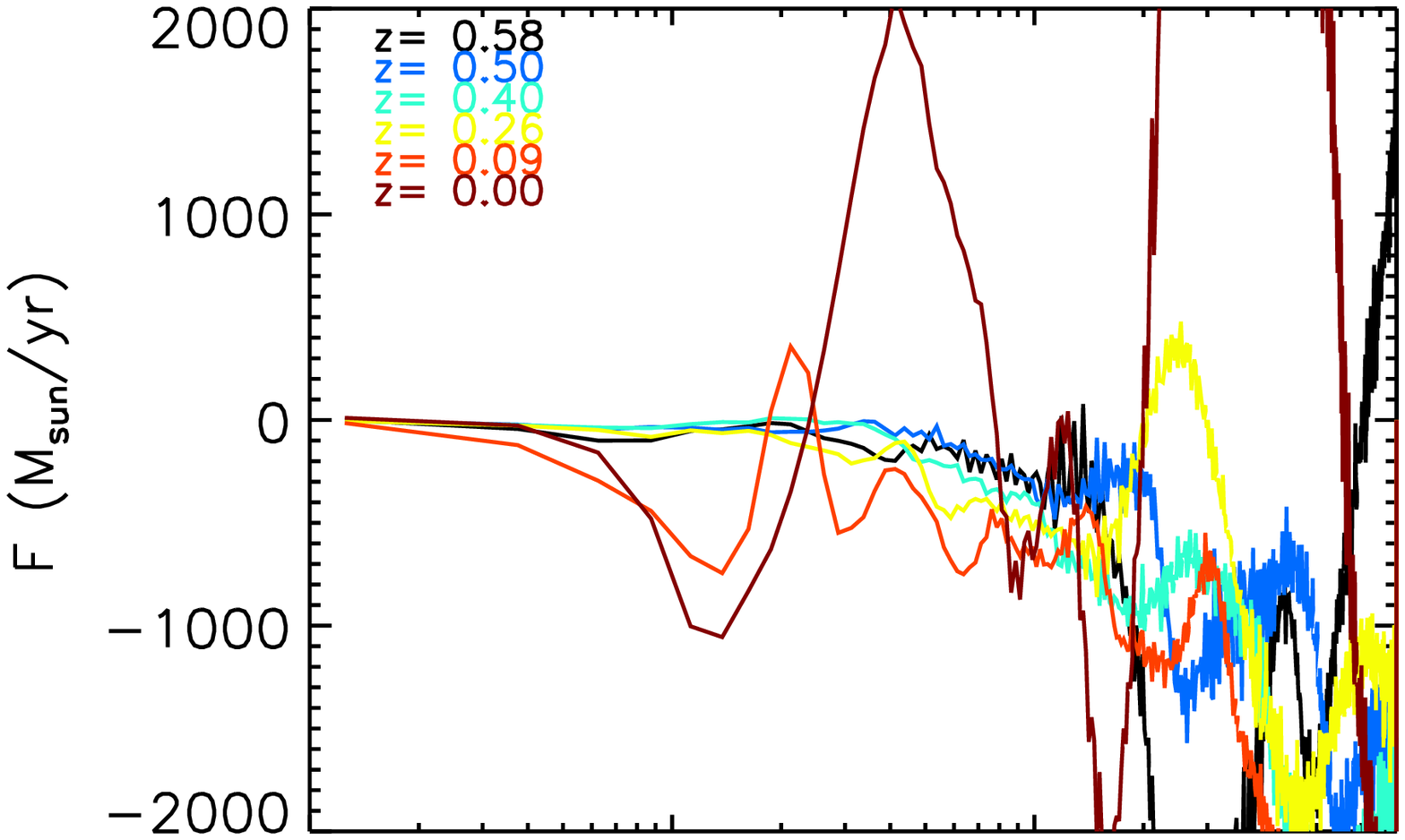}}\vspace{-1.6cm}}
  \centering{\resizebox*{!}{5.5cm}{\includegraphics{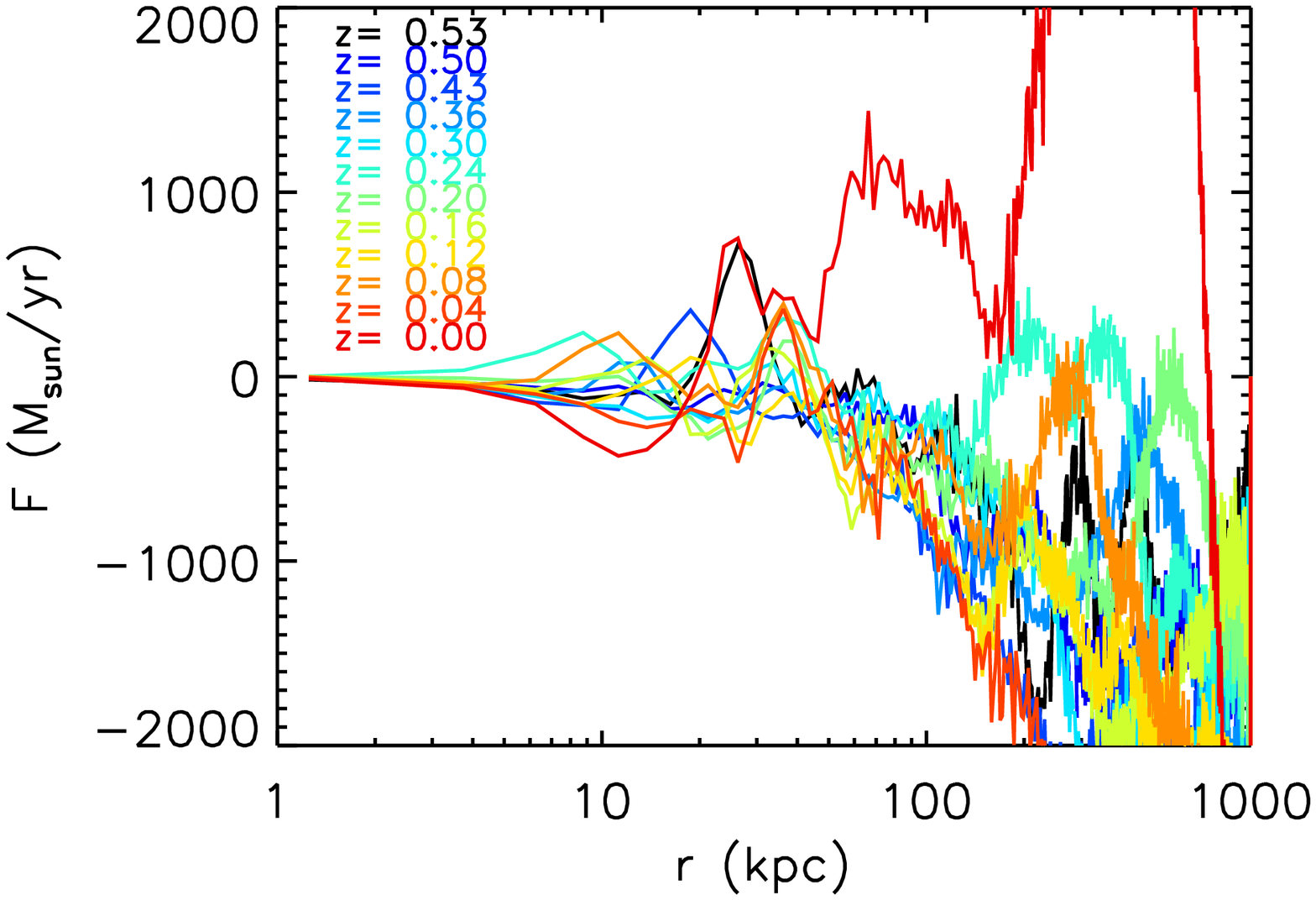}}}
  \caption{Time evolution for the net radial mass flux as a function of radius for the AGNJETrun (upper panel) and the AGNHEATrun
(lower panel). Colors correspond to fluxes measured at different redshifts as listed in their respective panels.}
    \label{Fluxnet_AGN}
\end{figure}

As was shown in~\cite{duboisetal10}, jet-induced AGN
feedback solves many problems associated with the cooling catastrophe:
galaxies are less massive, they form less stars especially at
late-times, and the cooling-flow is moderate and self-regulated.
In an effort to understand how this works and what leads to the entropy profile presented
in ~\cite{duboisetal10}, we re-examine the AGNJETrun simulation.

Following the cluster major merger at $z=0.58$, the
entropy in AGNJETrun shows a plateau in the core of the cluster at 100 keV.cm$^2$
(fig.~\ref{allin}, third column, first row), which slightly decreases
at late times as a moderate cooling flow develops and gas flows into
the central galaxy.  Nevertheless despite the cooling flow, high-entropy material does
not replace the low-entropy material. The AGN regulates the
amount of cold, low entropy  material that gets incorporated into the galaxy by ejecting some of it
back into the ICM,  preventing 
high-entropy gas at large radii from flowing in in the process.
Thus even though the cluster develops a small cooling flow, the amount of
low entropy gas which flows into the centre is limited.

Contrary to the supersonic cooling flow in the NOAGNrun, that in the AGNJETrun is quiescent and largely sub-sonic ($v_r\ll c_s$)
as can be seen from the radial velocity dispersion profiles (fig.~\ref{allin}, third column, sixth row).  According to
the cooling times plotted in figure~\ref{allin} (third column, fourth row), the cluster core should
experience a cooling catastrophe in less than 2-3 Gyrs (a few 100 Myr
in the very central $<10$ kpc region) if no source of feedback is
active.  It actually takes twice the time for the cluster to replenish
the cold gas component in the galaxy and trigger a strong AGN activity
at $z=0$.

Understanding the mechanism of jet-induced AGN feedback is complicated by
the variety of consequences it can have depending on the jet strength.
In general, jet-induced AGN feedback stirs the cluster gas.
Because the gas is compressible, kinetic energy imparted to the
ICM can eventually turn into thermal energy.  There are two ways
of doing this: weak jets can gently stir the ICM and increase its
turbulence, or strong jets can violently shock the
surrounding gas and reheat it.  While weak shocks increase the thermal energy 
indirectly by producing a high level of turbulence in the core that cascades into more
weak shocks that further increase the entropy, strong shocks
directly act on the entropy by increasing the temperature.
Unfortunately, disentangling these two modes is a difficult task that
is beyond the scope of this paper.

However, it appears that the effect of these different AGN jet modes on the ICM is to
produce various entropy profiles in the cluster at different times.
This could explain the large variety of entropy profiles in real X-ray
clusters independent of their mass range \citep{sandersonetal09,
prattetal10} as these would mainly dependent on the AGN activity.  Indeed with the AGNJETrun
simulation, we are able to produce a cluster that exhibits different
entropy profiles at different times: from a flat entropy core at
$z=0.58$ to a power-law entropy profile at $z=0.0$.

The gas density profiles (fig.~\ref{allin}, third column, second row) lend support to the picture 
that AGN jet feedback stems the gas flow to the cluster core.
Above an intermediate radius (100 kpc), the gas density
is extremely close to the gas density in the ADIArun simulation,
whereas in the NOAGNrun, gas is strongly depleted (gas density is four
times lower at 100 kpc at $z=0$ in the NOAGNrun compared to the
AGNJETrun).  These conditions at intermediate radii and above
hold over large time-scales, except at $z=0$ where a small cooling flow develops.
The core's gas density grows with time due to the cooling flow, but because
of the AGN jet feedback all this gas cannot cool and form stars as efficiently as in the NOAGNrun.
As a result the ICM density also grows in the core of the cluster.

The temperature profile (fig.~\ref{allin}, third column, third row) in the AGNJETrun simulation tends to an isothermal state at $z=0$. After 
gas is shocked by the major merger ($z=0.58$), 
the temperature is larger in the
central 10 kpc of the cluster. AGN activity is negligible during
this phase, (see \citealp{duboisetal10}) so it cannot explain the high temperature. Thereafter, gas cools 
diminishing the amount of thermal energy in the centre,
and creating a small cooling flow around $z=0$.

Interestingly, a thermal AGN feedback mode
(AGNHEATrun) rather than a kinetic mode (AGNJETrun) creates a high temperature core component.
We interpret this as a consequence of direct thermal energy
injection from the central AGN source which translates into a larger injection of entropy at small radii.  By not directly acting on the internal
energy, the kinetic (jet) mode  yields 
lower entropy profiles in the centre.

Like the jet AGN feedback, thermal AGN feedback prevents the cooling catastrophe.
Indeed, the cumulative stellar mass profile in fig.~\ref{mass_comp}, shows that the amount of stars
formed at $z=0$ is very similar for the AGNHEATrun and the AGNJETrun.  Gas
mass and total baryon mass distributions are also very similar at
intermediate and large ($r>25$ kpc) cluster radii, therefore
leading to similar ICM gas properties on these scales
(see fig.~\ref{allin}).  The only difference in the mass distributions
appear in the core of the cluster: the total mass distributions
(baryons plus dark matter) in the case of thermal energy and
kinetic energy inputs are almost superimposed, but differences 
of a factor of a few are present in the separate components in the
inner 20 kpc.  

In fig.~\ref{Fluxnet_AGN}, we have represented
the mass flux of gas at different radii and different times within
spherical shells centred on the cluster for the AGNJETrun and the
AGNHEATrun respectively. Negative and positive fluxes compensate to give a very faint net
flux in the central 10 kpc of the cluster (fig.~\ref{Fluxnet_AGN}) for both the AGNJETrun and the AGNHEATrun.
However at intermediate distances ($r > 10 \, \rm kpc$), larger negative or positive net
fluxes are measured, as a result of wave propagation from the central
AGN source.  The net flux in the cluster outskirts is largely
negative as the AGN feedback does not significantly impact the gas at
such large distances.

\begin{figure}
  \centering{\resizebox*{!}{6.5cm}{\includegraphics{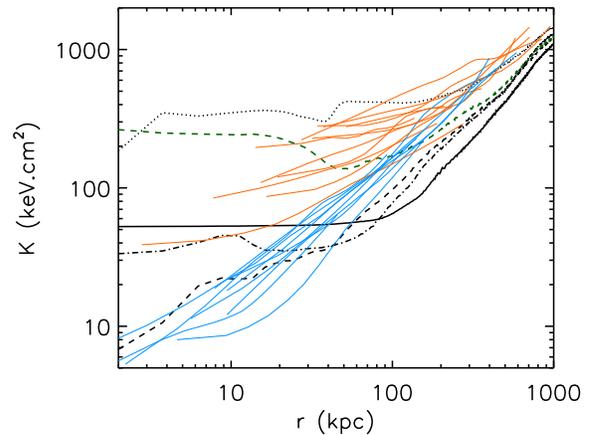}}}
  \caption{Entropy profiles of the cluster at $z=0$ for the ADIArun
(black solid line), the NOAGNrun (black dotted line), the AGNJETrun
(black dashed line), the AGNHEATrun (black dot-dashed line), and the
AGNOFFrun (green dashed line). Entropy profiles from real clusters
from \citet{sandersonetal09} with cool core clusters in light blue,
and non-cool core cluster in light orange are overplotted on top of
our simulation results.}
    \label{entro_comp}
\end{figure}

Fig.~\ref{entro_comp} shows that the 
simulation that includes AGN jet feedback (AGNJETrun) all the way down to $z=0$ reaches the lowest
entropy, with values below $10$ keV.cm$^2$ a few kpc away from the centre.  In contrast, the
simulation that has no feedback at anytime (NOAGNrun) has the largest entropy
floor in the core ($K\simeq 400$ keV.cm$^2$).  In fig.~\ref{entro_comp},
we also give the observational entropy profiles of 20
X-ray selected clusters with cool cores (light blue lines) or non-cool
cores (light orange lines) from \cite{sandersonetal09} in the 
temperature range 1-10 keV. The non-cool core clusters
show an entropy floor in the centre of the cluster (light orange lines),
whereas cool clusters possess entropy profiles declining towards
the centre (light blue lines).
Thus, the presence of jet AGN feedback allows the formation of a cool-core cluster
 whereas with the AGN heating mode the cluster is always maintained in a non-cool core configuration.
It is worth to note that without AGN feedback the entropy floor is kept very high in the core with an amplitude higher than those 
observed for the most massive of the non-cool core clusters~\citep{sandersonetal09}.

Finally, one might wonder how robust our profiles are to the choice of averaging method 
(spectroscopic-like/X-Ray or volume weighted) for the physical quantities. As discussed in detail in Appendix A, 
large differences between averages appear in the core of the cluster (inner 50 kpc or so) when a massive 
disc of cold gas has developed in the central galaxy, fed by an unchecked cooling catastrophe.  Otherwise, the agreement between all 
estimates is excellent, as demonstrated by the  $z > 0.40$ profiles  (black and dark blue curves in all panels) on Fig.~\ref{AGNOFF_comp}. 
Therefore, at least all the inner profiles we measure in the (more realistic) runs where the cooling catastrophe is prevented (AGNJET and AGNHEAT) 
are reliable.

\subsection{The effect of the pre-heating}
\label{agnoff_section}

\begin{figure}
  \centering{\resizebox*{!}{6.5cm}{\includegraphics{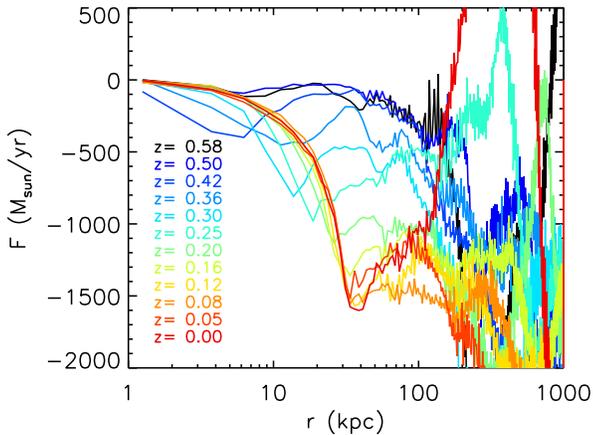}}}
  \caption{Time evolution as a function of radius of the net radial mass flux for the AGNOFFrun. Colors correspond to fluxes measured at different redshifts as listed in the panel.}
    \label{Flux_AGNOFF}
\end{figure}

In the interest of understanding whether a strong initial pre-heating
of the gas is alone able to prevent a cooling catastrophe for several Gyrs,
we run the following simple experiment: we restart the AGNJETrun simulation
after the major merger, i.e. from $z=0.58$, deactivating BH accretion and AGN feedback.
We call this simulation the AGNOFFrun.
In the literature, the term `pre-heating'  is generally
employed in a generic sense to describe a rise in the level of entropy of the proto-cluster gas (e.g. \citealp{kaiser91, evrard&henry91}) which 
can be caused by various physical processes. In the current work, we refer to pre-heating as the rise in entropy which is 
 self-consistently provided by the feedback processes modelled in the simulations before the final assembly of the cluster.
This explains why the end of the pre-heating phase coincides with the beginning of the last major merger undergone by the cluster at $z\sim1$.

Between $z=0.58$ and $z=0.42$, 2 Gyrs of evolution elapse. Fig.~\ref{allin}
(fifth column, fourth row) shows that this coincides with the cooling time for AGNOFFrun's ICM
gas at $z=0.58$ and $r=20$ kpc.  Therefore as expected, inside this radius, the entropy profile
strongly declines during this time interval (fig.~\ref{allin}, fifth column, first row), signaling a cooling flow into the centre.
More specifically the entropy in the cluster core ($r<10$ kpc) decreases from its values at $z=0.58$ ($K \sim 20\, \rm keV.cm^{2}$) to
very low values ($K<10 \, \rm keV.cm^2$) at $z=0.42$. Concomitantly the gas density in the centre 
of the cluster increases (fig.~\ref{allin}, fifth
column, second row) as internal energy is radiated away by gas cooling, leading to a loss of pressure support
(fig.~\ref{allin}, fifth column, fifth row).   
Nevertheless by redshift $z=0.42$, the
cluster has not yet reached the cooling catastrophe stage; the cooling
flow is still moderate and subsonic.
  
Gradually after $z=0.42$, the
cooling catastrophe develops: gas flows supersonically into the
gravitational potential well (see radial velocity dispersion profiles in fig.~\ref{allin}, fifth column, sixth row), concentrates in the cluster centre (fig.~\ref{mass_comp}, left panel in upper row) and
forms a cold disc component (see fig.~\ref{temp_map_nice}).  As the gas condenses in the centre, strong star formation occurs in
the central galaxy (see cumulative stellar mass profiles in upper middle panel of fig.~\ref{mass_comp}) resulting in a more massive
galaxy at $z=0$ in the AGNOFFrun than in the AGNJETrun. DM also concentrates
more in the cluster centre through adiabatic contraction (fig.~\ref{mass_comp}, middle panel in bottom row).
Bizarrely enough, it is also clear from this figure, that the AGNOFFrun cluster features a DM halo less
concentrated than in the AGNHEATrun. This, in turn, demonstrates that 
a different implementation of AGN feedback can have a significant and counterintuitive
impact on the $z=0$ DM concentration of clusters: the mode that
reduces the baryonic mass the most in the cluster core (AGNHEATrun) at
$z=0$ is also the mode which has the largest DM  halo concentration.
This is because the stellar component should be taken into account when calculating
adiabatic contraction as it responds in the same collisionless way as the
dark matter. In other words, when one compares the total mass of dark
matter {\em plus} stars (fig.~\ref{mass_comp}, right panel in bottom row, at
least in the center where the gas mass in negligible), the AGNOFFrun
becomes more concentrated  than its AGNHEATrun counterpart, as naively
expected.
  
We can monitor the effect of preheating on the progression of the cooling catastrophe in fig.~\ref{Flux_AGNOFF}.
It shows that as soon as the AGN feedback is turned off, the
low-entropy material quickly flows into the centre and becomes part of the central galaxy ISM. 
Gas located at larger distances is then subsequently accreted.  In other words, higher entropy material 
replaces low-entropy gas thereby increasing the  ICM entropy in the core of the
cluster (fig.~\ref{allin}, fifth column, first row).  In the meantime, fig.~\ref{allin} (fifth column, first row) shows that
the entropy continues to grow at larger
radii ($r>50$ kpc) because of the loss of pressure support at increasingly larger radii
as low entropy gas is removed from the cluster centre to form stars.
Unlike the supersonic flow at small radii, at large radii ($r>100$ kpc) the gas flows quiescently 
into the core (see gas radial velocity dispersion in fig.~\ref{allin}, fifth column, sixth row).
We also note that the entropy profile at $z=0$ is larger in the core
($r<50$ kpc) than at moderate distance ($100$ kpc) from the centre (see fig.~\ref{allin}, fifth column, first row).  
This cannot be the consequence of an adiabatic process, because 
the entropy profile would be monotonic. Therefore, there are two possible explanations
for the behavior of the $z=0$ entropy profile: (i) either it is the result of spurious entropy
produced by the grid code which is not a strictly entropy-conserving
scheme, or (ii) a strong shock developed somewhere in the cluster core causing
the gas to endure a non-reversible process.  Even though we cannot
entirely rule out the former possibility, there is plenty of evidence in favor of the latter.
As the central ICM rapidly flows onto a cold and dense galactic disc
component, there is a strong discontinuity both in gas density and
temperature at the disc interface. This is evidence of a strong
shock.  As a result, upwind of the shock the ICM is heated,
and both temperature and entropy increase.  
Figs.~\ref{entro_map_nice} and~\ref{temp_map_nice}, show this
high-entropy, hot region perpendicular to the
galactic disc.

Finally, an interesting feature of the AGNOFFrun simulation is that compared to the AGNJETrun
which includes kinetic AGN feedback all the way down to $z=0$, the gas
properties at large radii are hardly modified by the absence of
late-time AGN activity. This is because the cooling flow does not
extend to radii greater than $100$ kpc and the late AGN jet activity hardly perturbs
the gas beyond these distances.  In light of our previous analysis, we conclude that
pre-heating from AGN feedback at high redshift ($z>1$) cannot single-handedly
prevent a cooling catastrophe, but that it plays an important role in
(i) setting up the gas properties at
large distances from the centre, and (ii) to some extent softens
the consequences of the catastrophic cooling in the core (lower entropy,
temperature of the gas and even mass of dark matter and baryons). 

\subsection{The effect of metal cooling}
\label{metals}

\begin{figure*}
  \centering{\resizebox*{!}{3.4cm}{\includegraphics{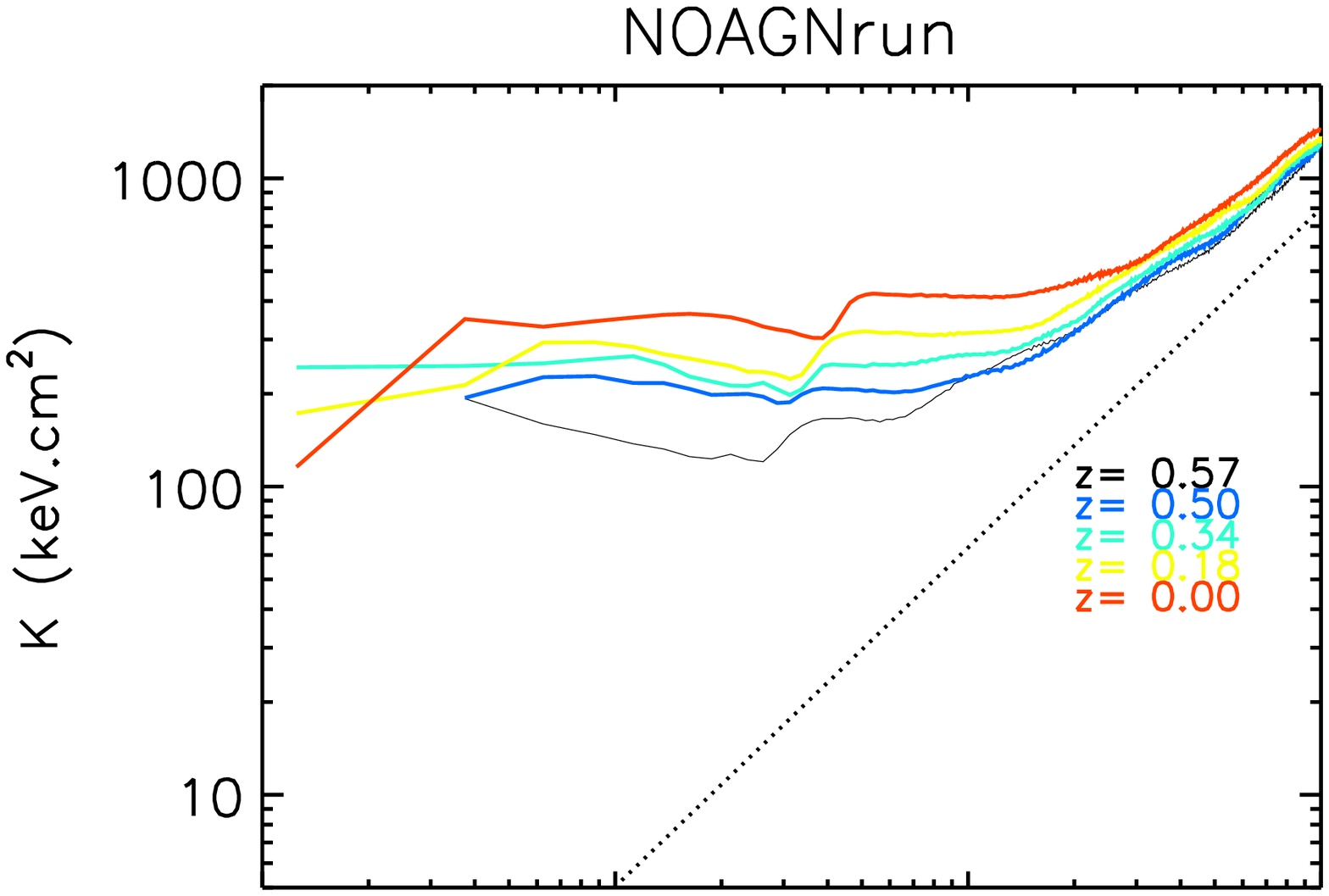}}\hspace{-1.40cm}}
  \centering{\resizebox*{!}{3.4cm}{\includegraphics{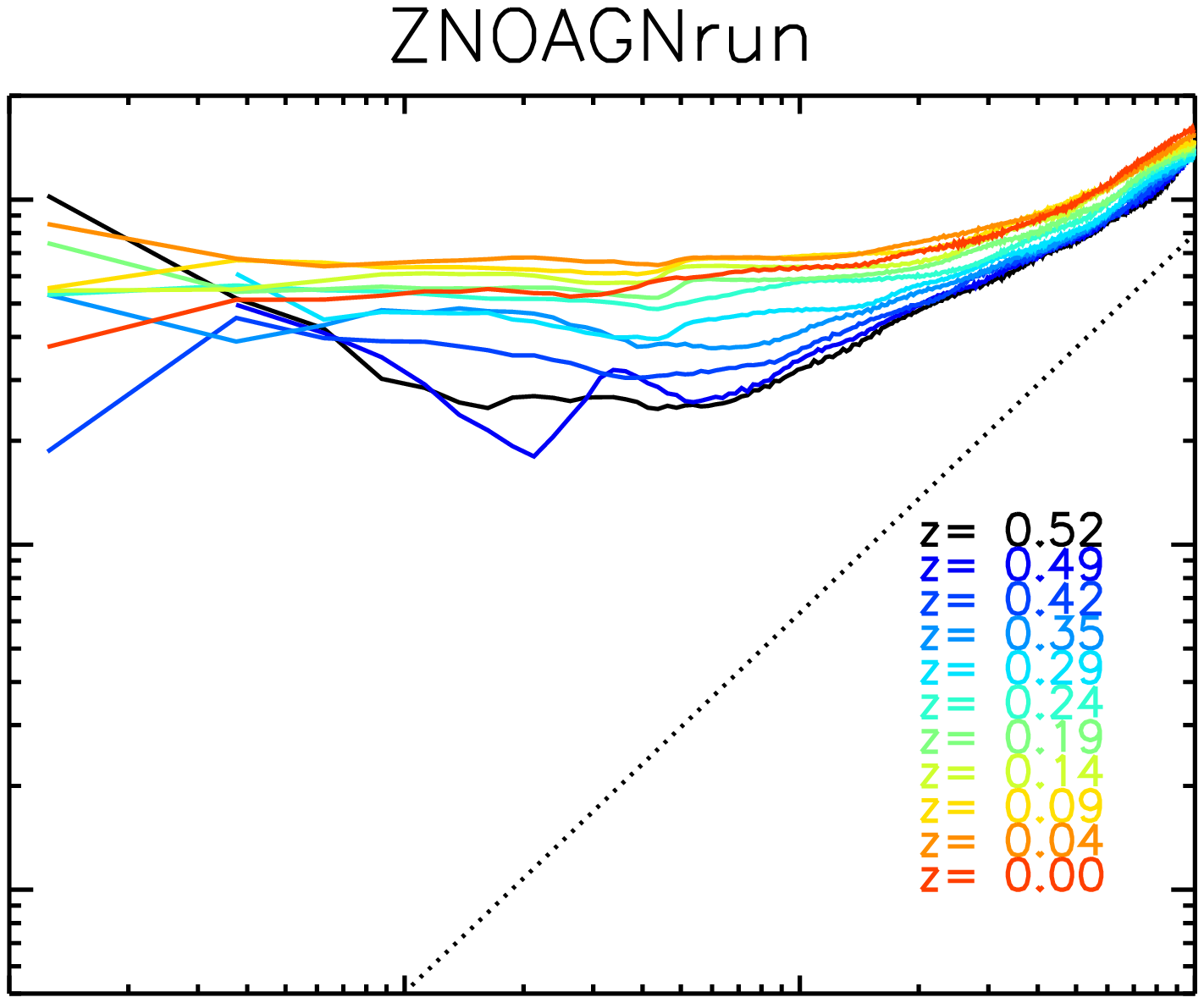}}\hspace{-1.40cm}}
  \centering{\resizebox*{!}{3.4cm}{\includegraphics{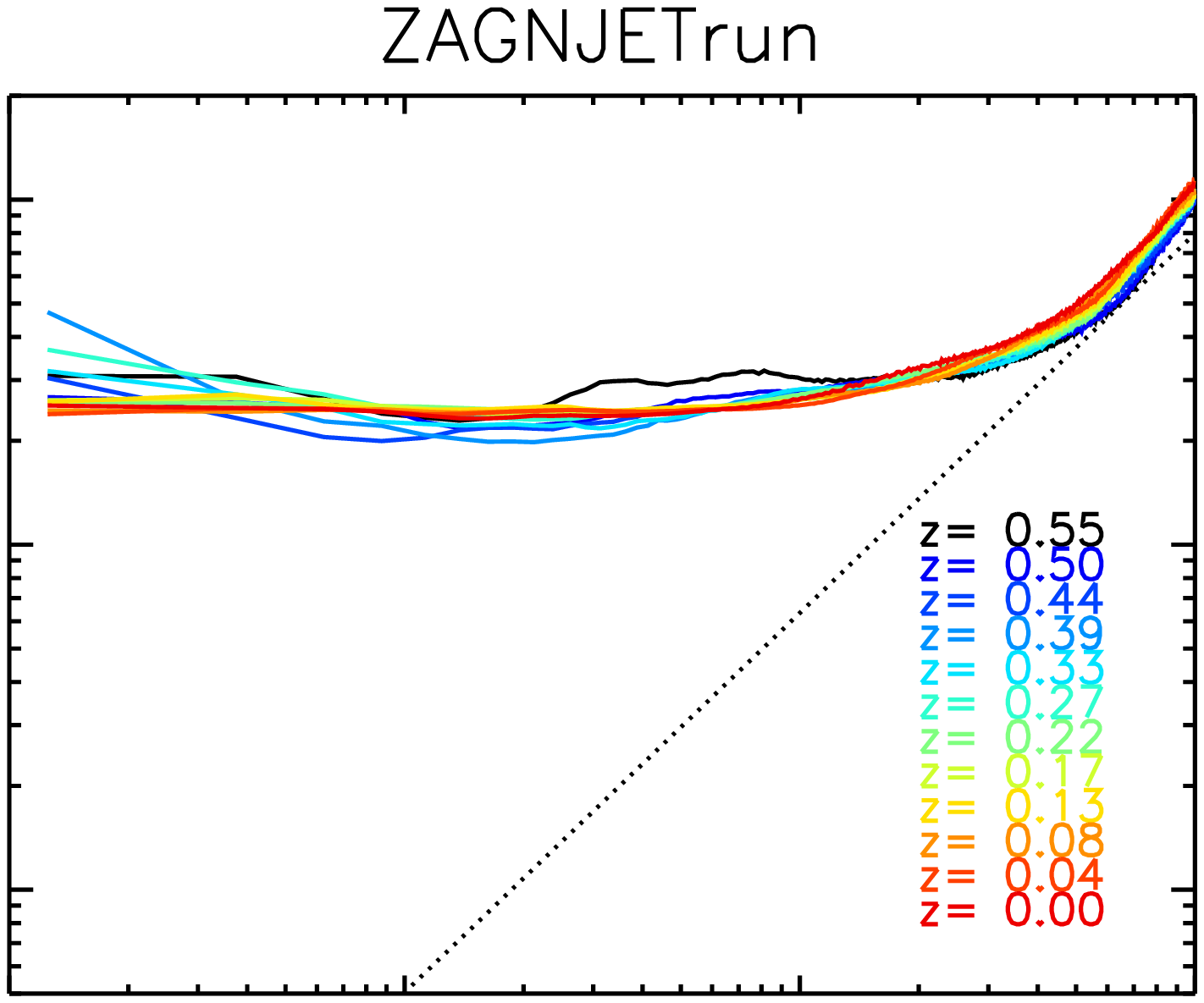}}\hspace{-1.40cm}}
  \centering{\resizebox*{!}{3.4cm}{\includegraphics{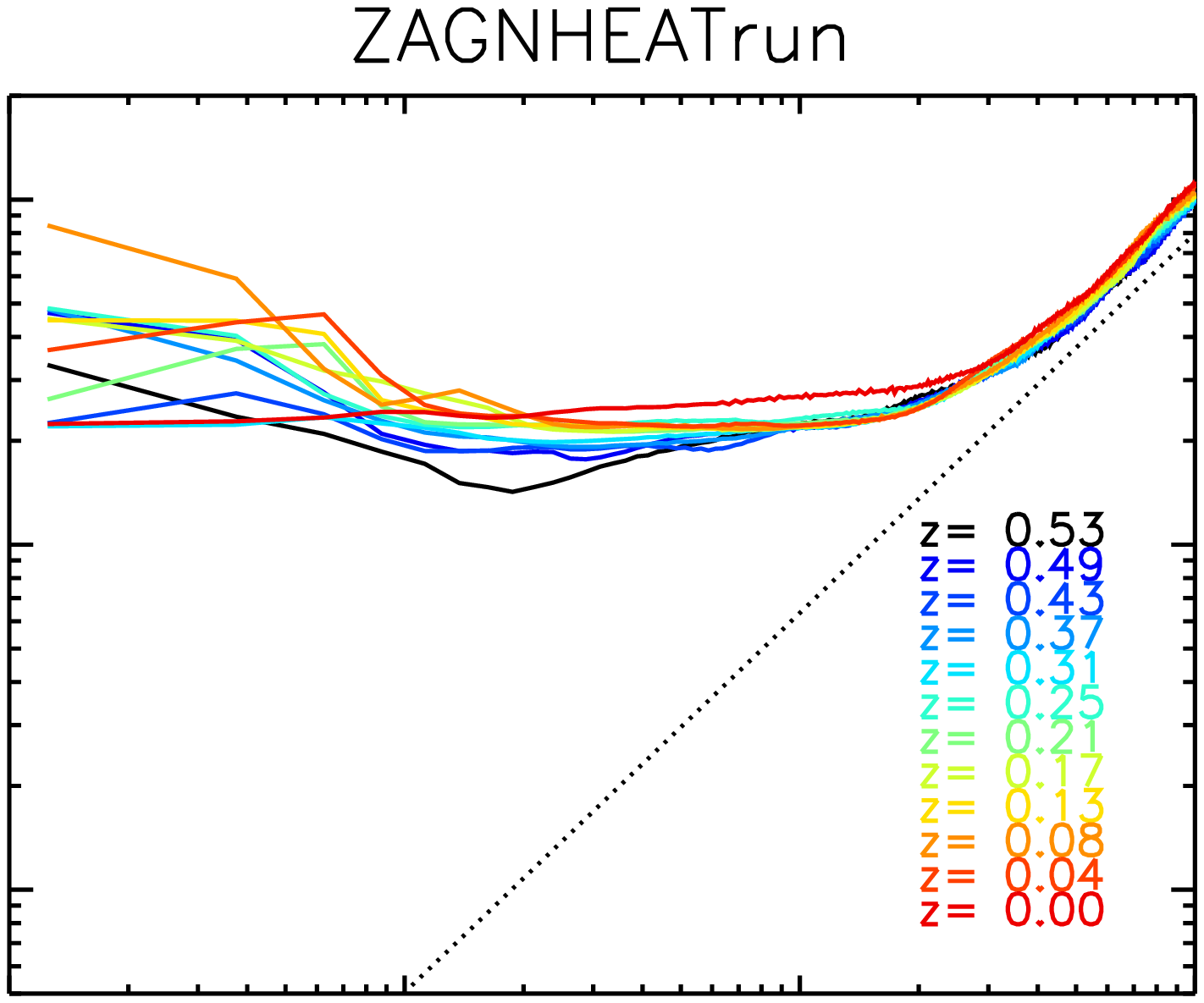}}\hspace{-1.40cm}}
  \centering{\resizebox*{!}{3.4cm}{\includegraphics{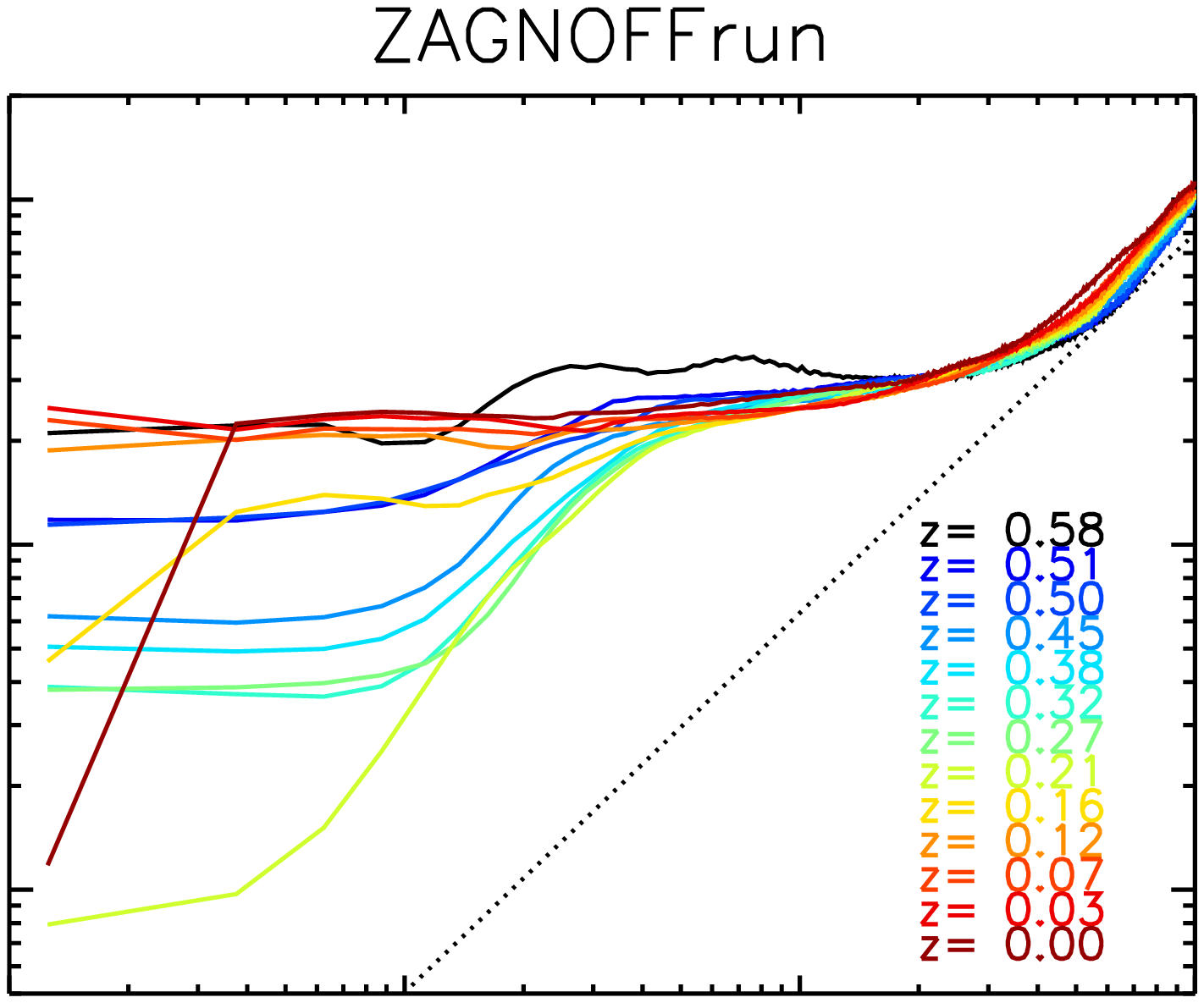}}\vspace{-0.99cm}}\\
  \centering{\resizebox*{!}{3.4cm}{\includegraphics{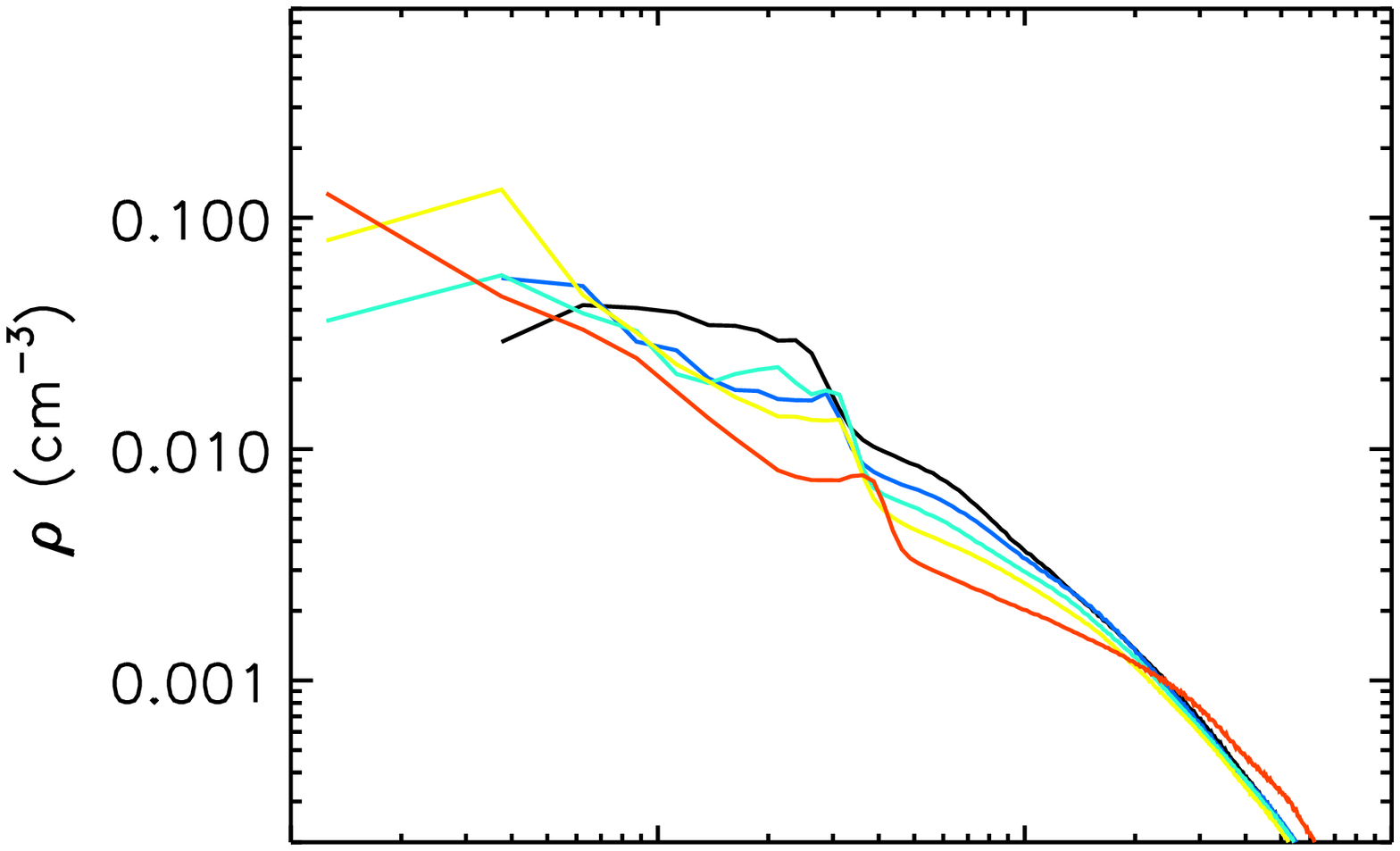}}\hspace{-1.40cm}}
  \centering{\resizebox*{!}{3.4cm}{\includegraphics{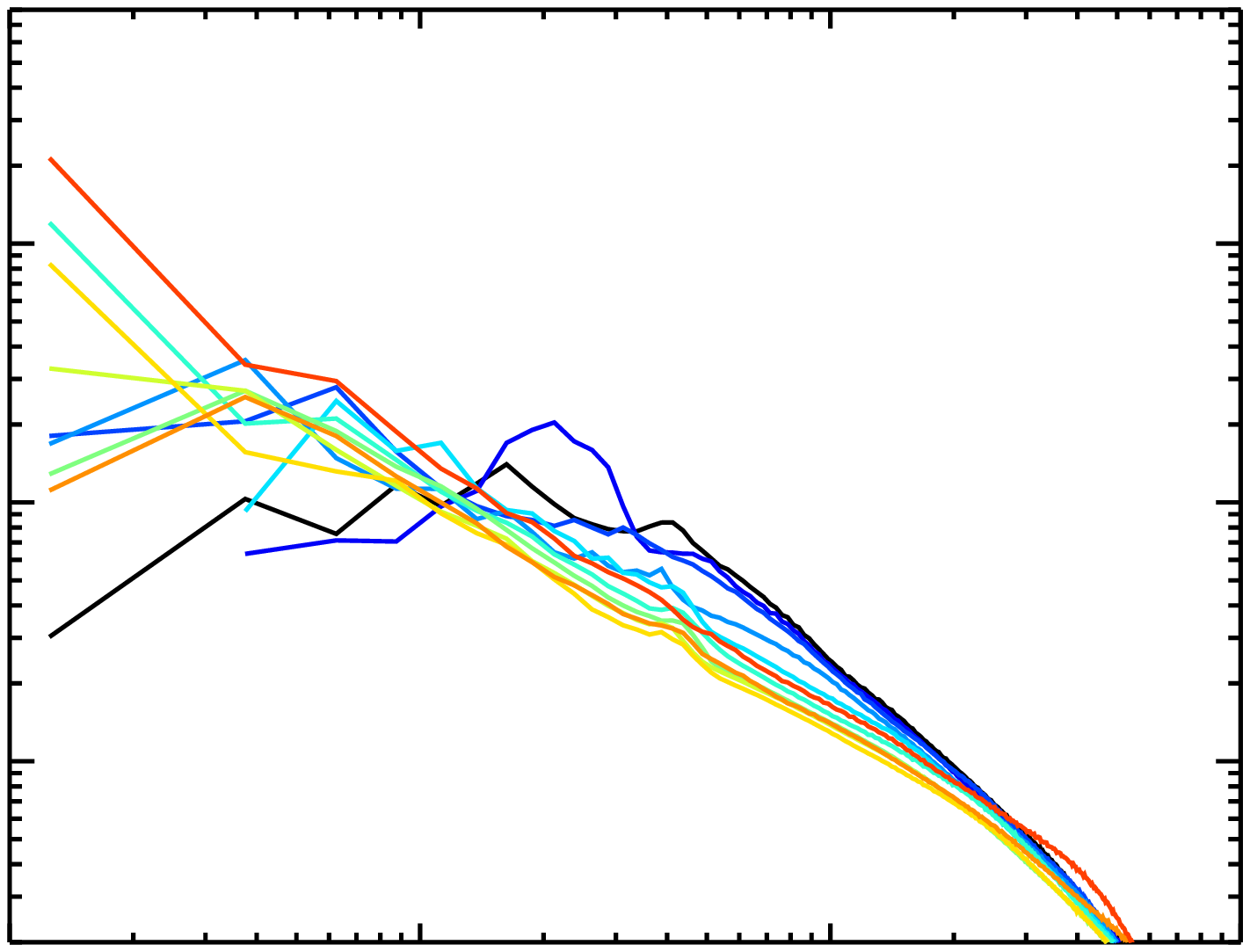}}\hspace{-1.40cm}}
  \centering{\resizebox*{!}{3.4cm}{\includegraphics{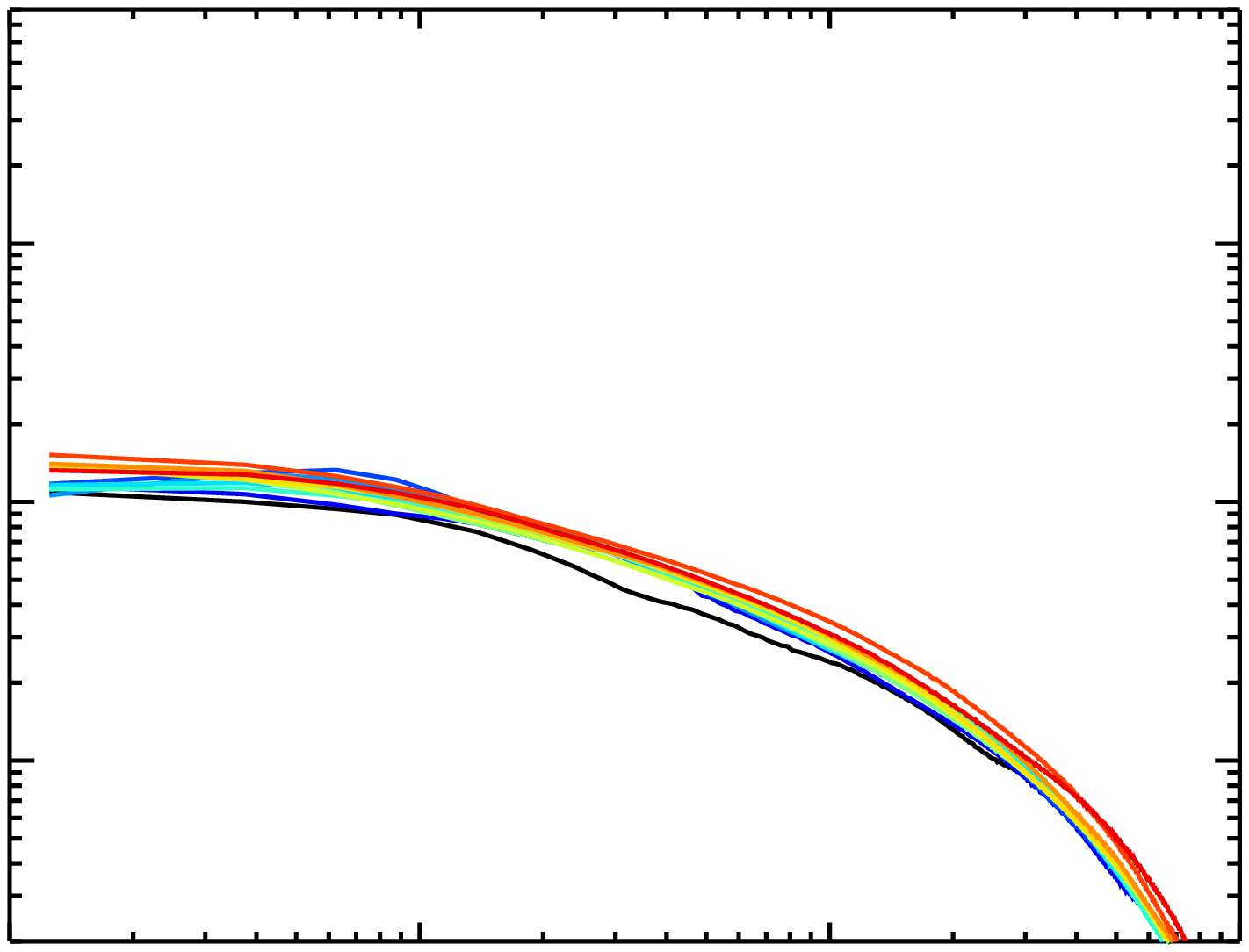}}\hspace{-1.40cm}}
  \centering{\resizebox*{!}{3.4cm}{\includegraphics{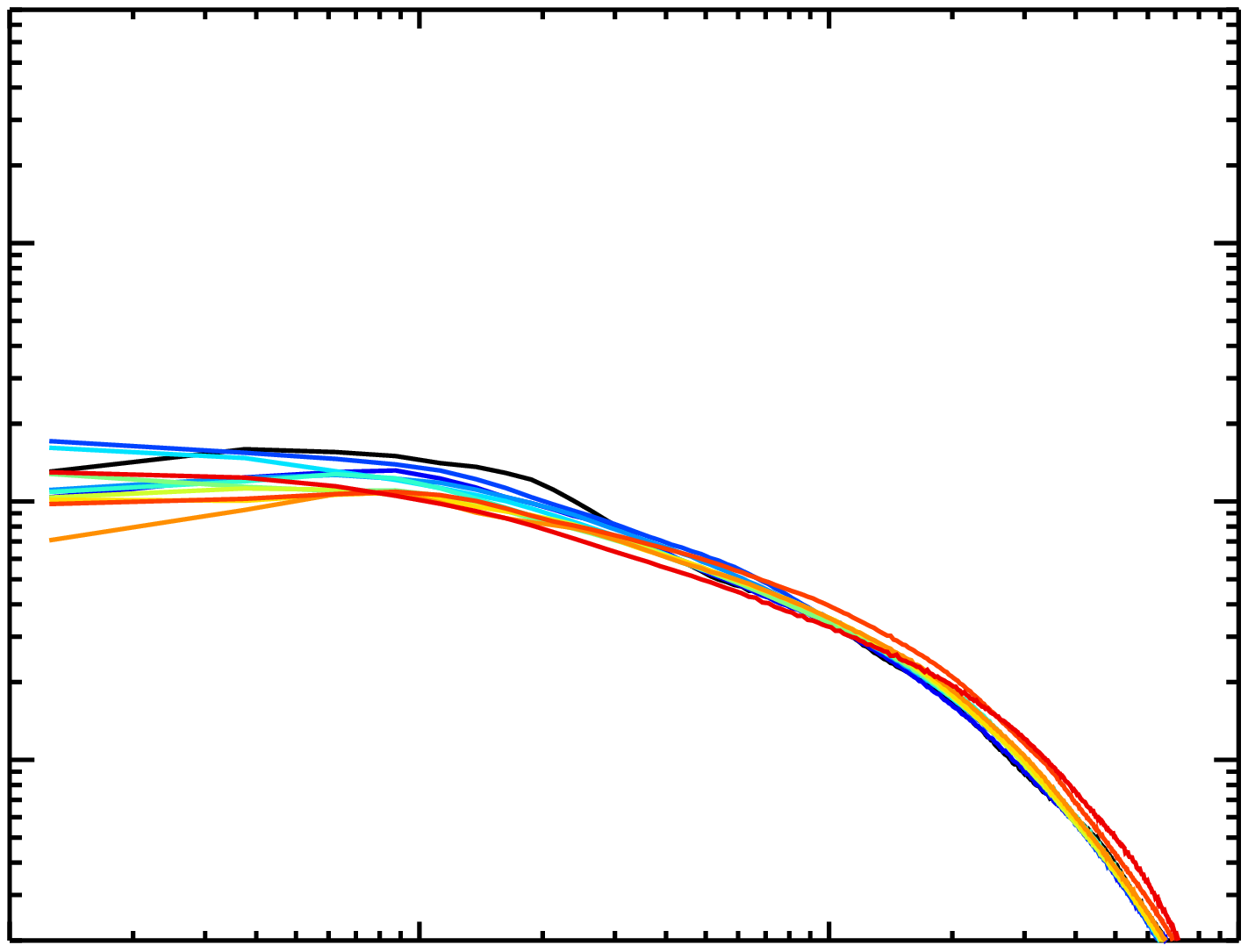}}\hspace{-1.40cm}}
  \centering{\resizebox*{!}{3.4cm}{\includegraphics{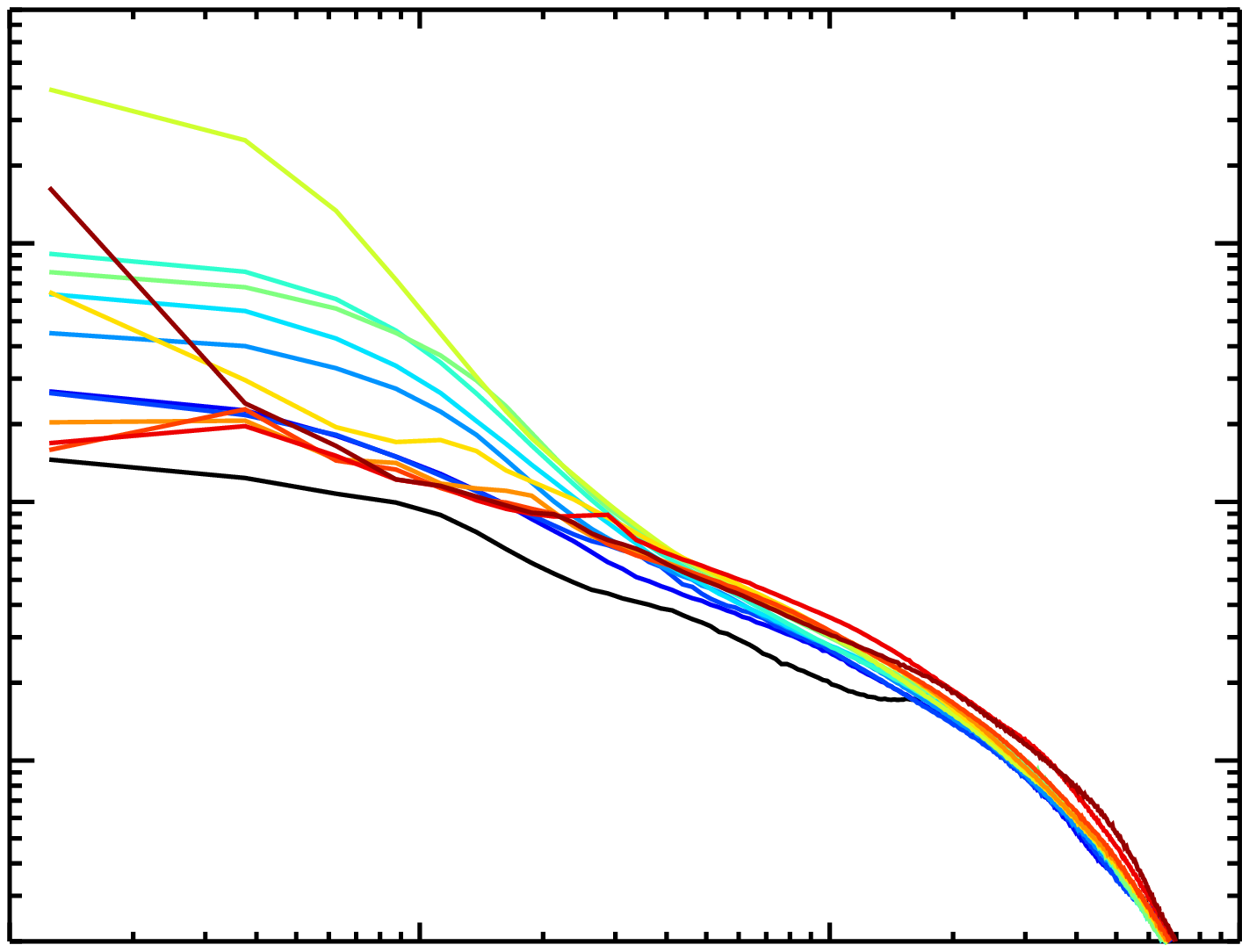}}\vspace{-0.99cm}}\\
  \centering{\resizebox*{!}{3.4cm}{\includegraphics{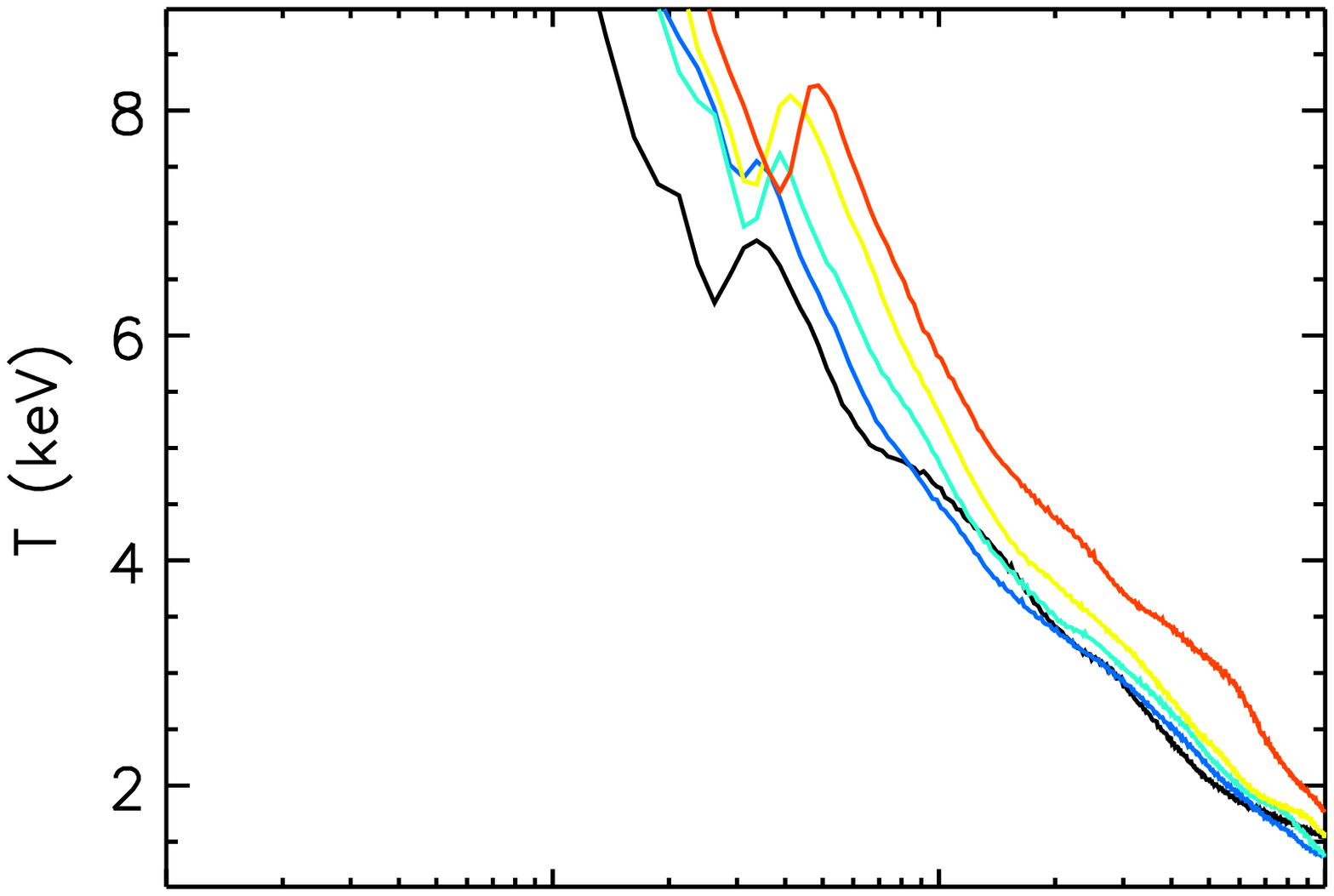}}\hspace{-1.40cm}}
  \centering{\resizebox*{!}{3.4cm}{\includegraphics{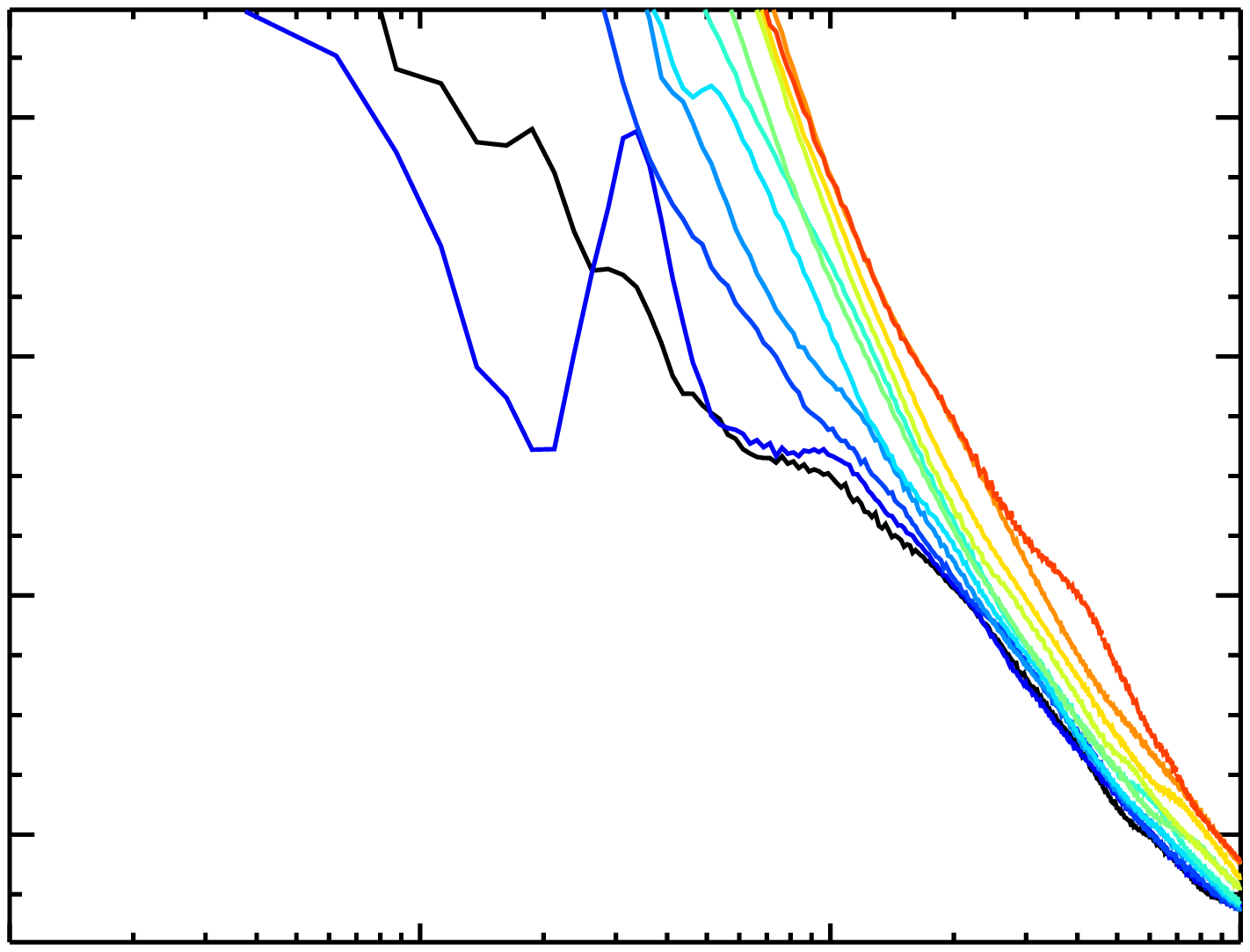}}\hspace{-1.40cm}}
  \centering{\resizebox*{!}{3.4cm}{\includegraphics{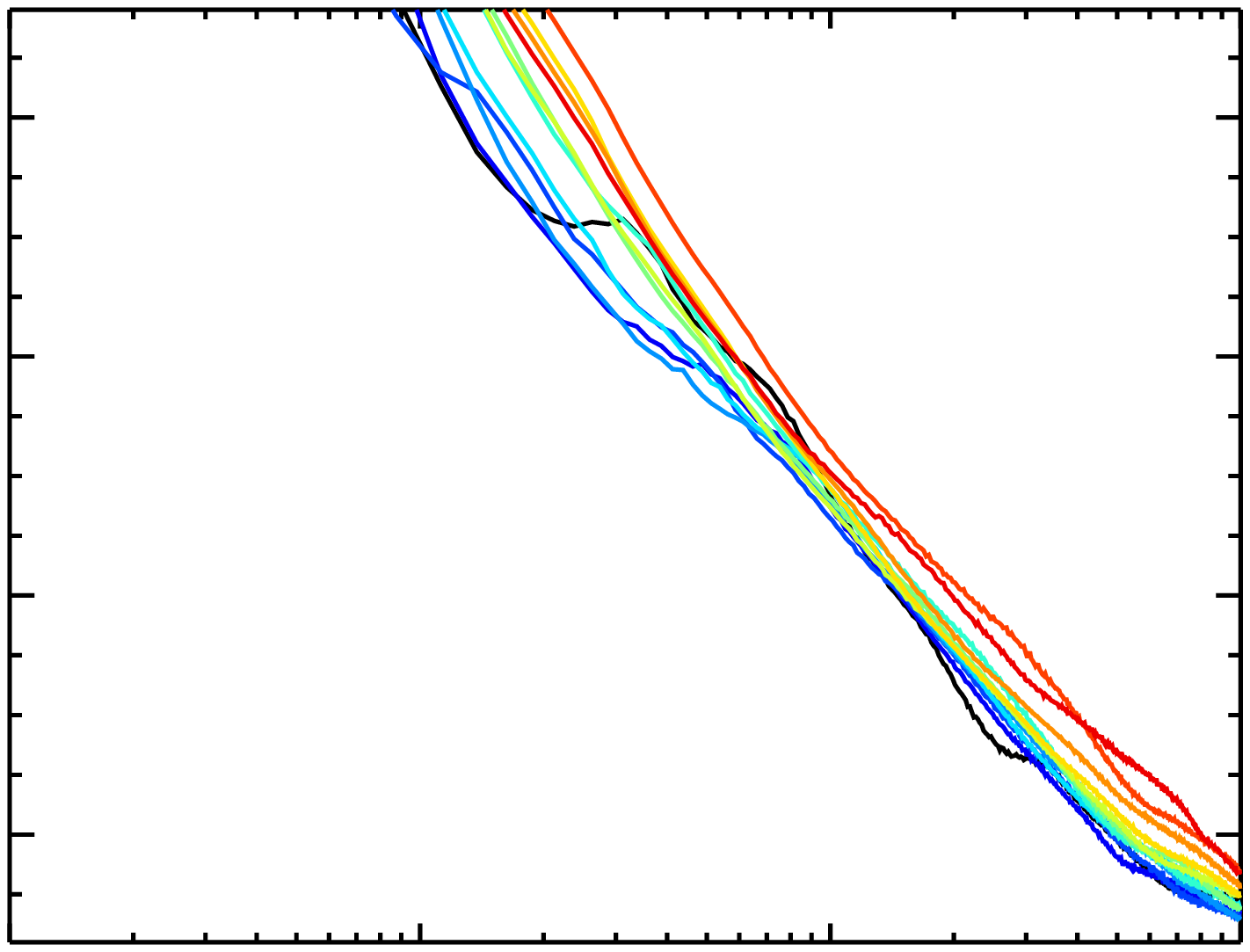}}\hspace{-1.40cm}}
  \centering{\resizebox*{!}{3.4cm}{\includegraphics{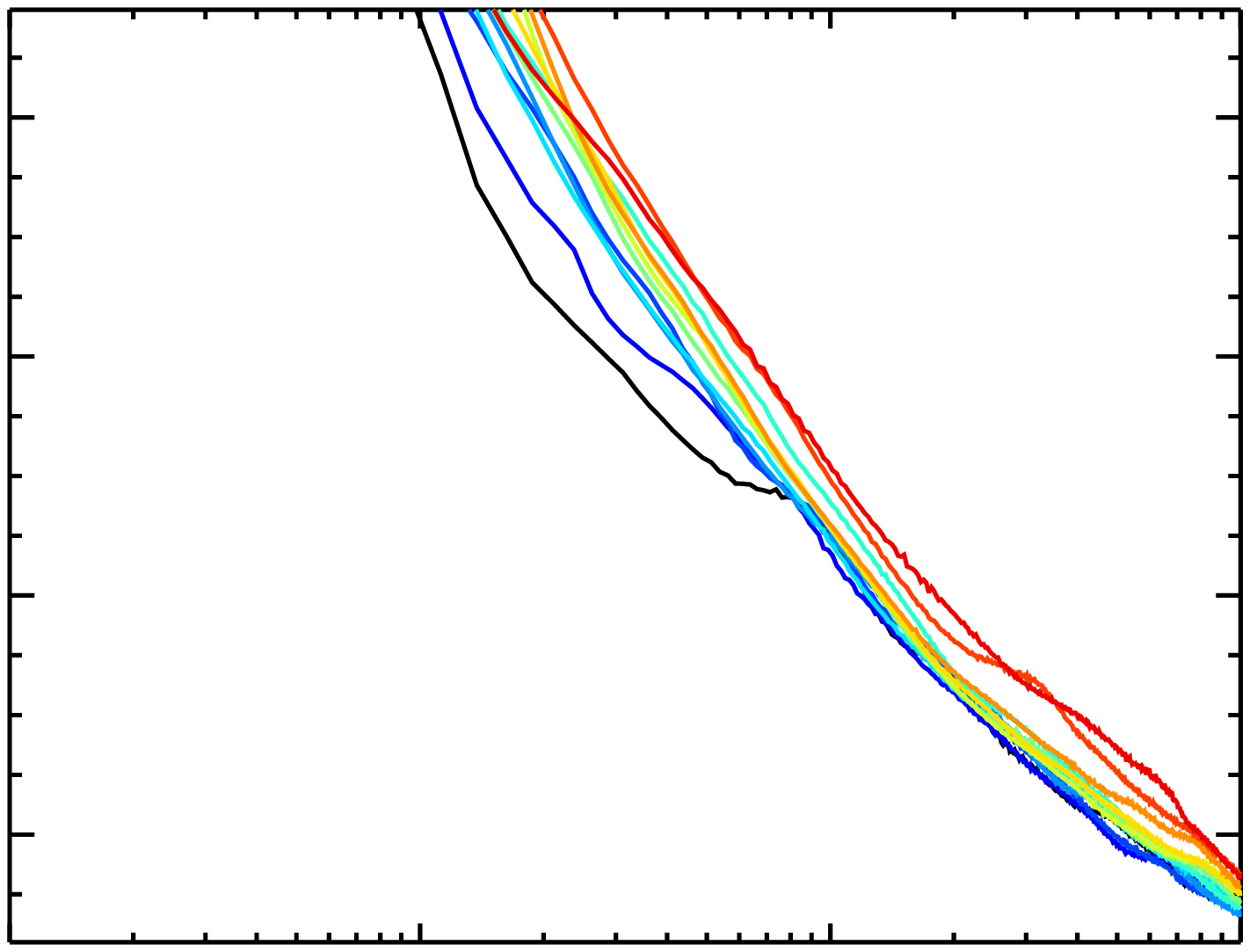}}\hspace{-1.40cm}}
  \centering{\resizebox*{!}{3.4cm}{\includegraphics{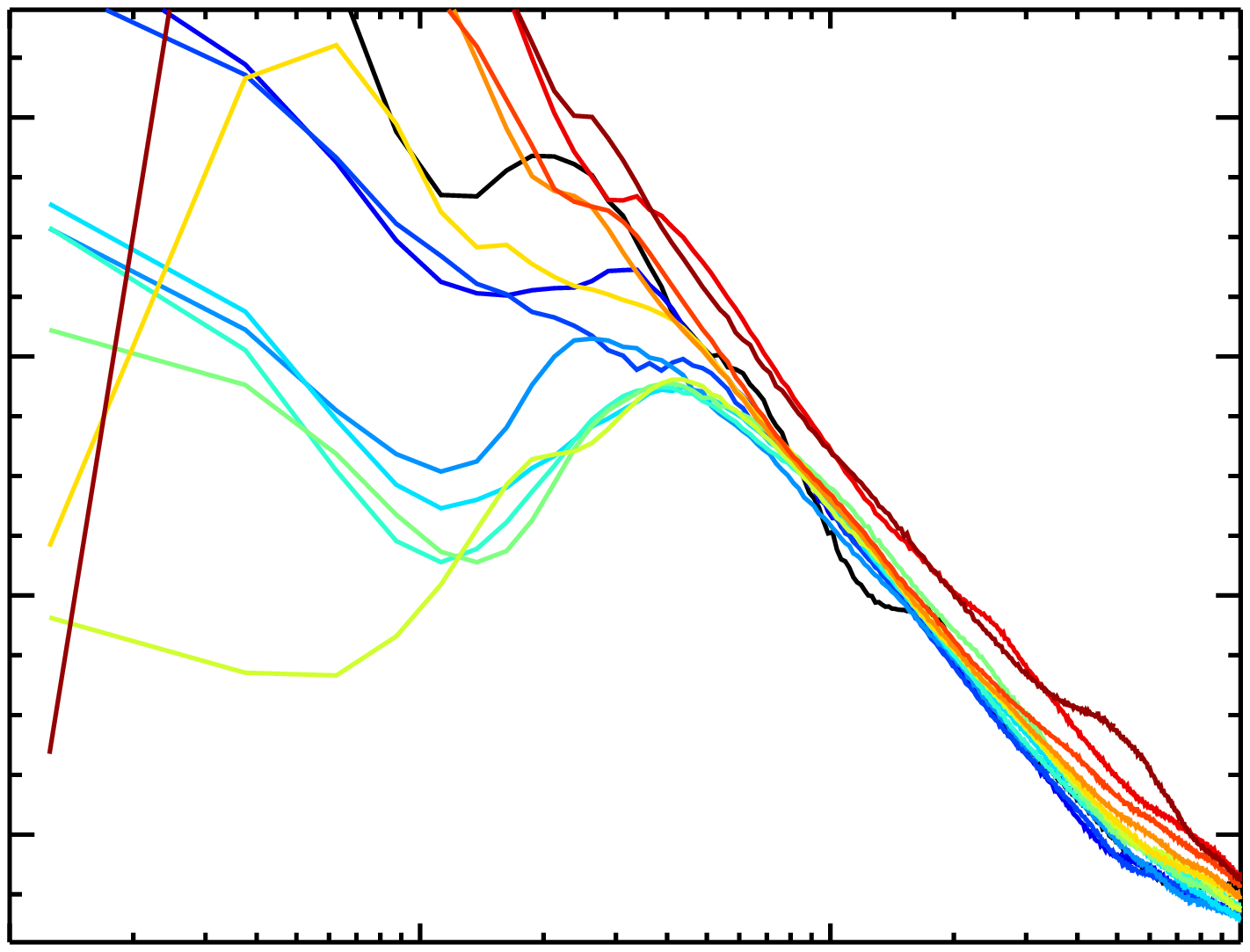}}\vspace{-0.99cm}}\\
  \centering{\resizebox*{!}{3.4cm}{\includegraphics{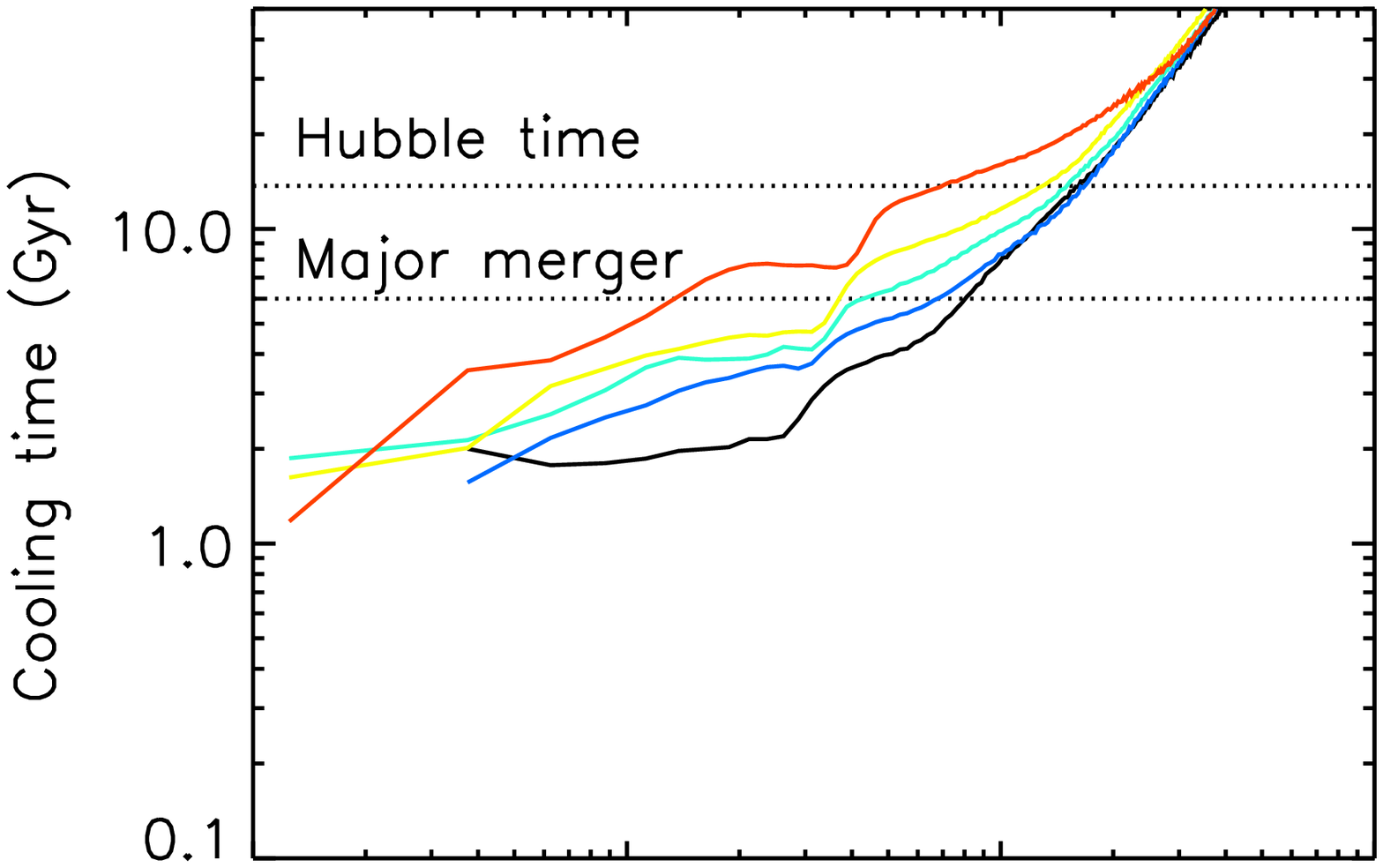}}\hspace{-1.40cm}}
  \centering{\resizebox*{!}{3.4cm}{\includegraphics{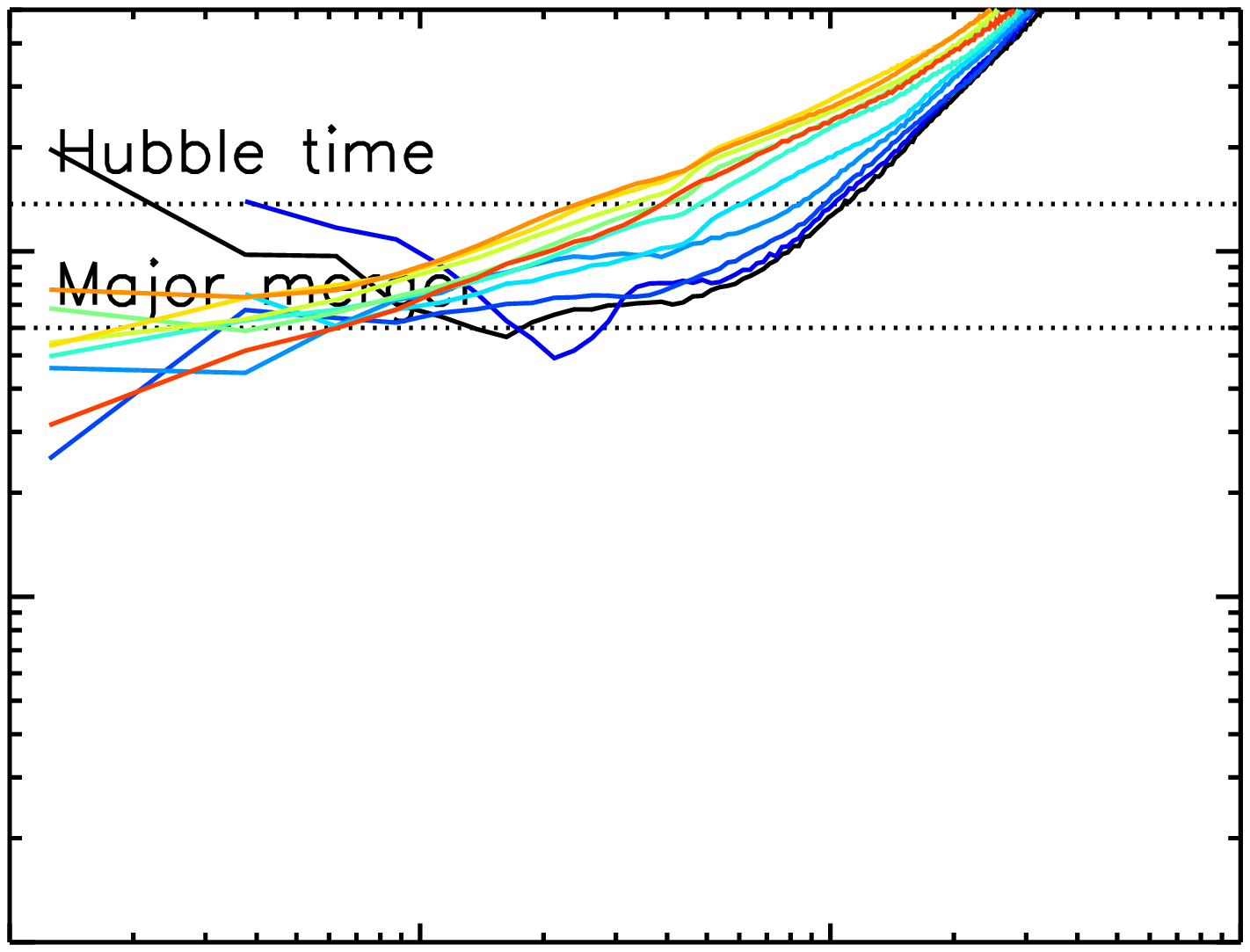}}\hspace{-1.40cm}}
  \centering{\resizebox*{!}{3.4cm}{\includegraphics{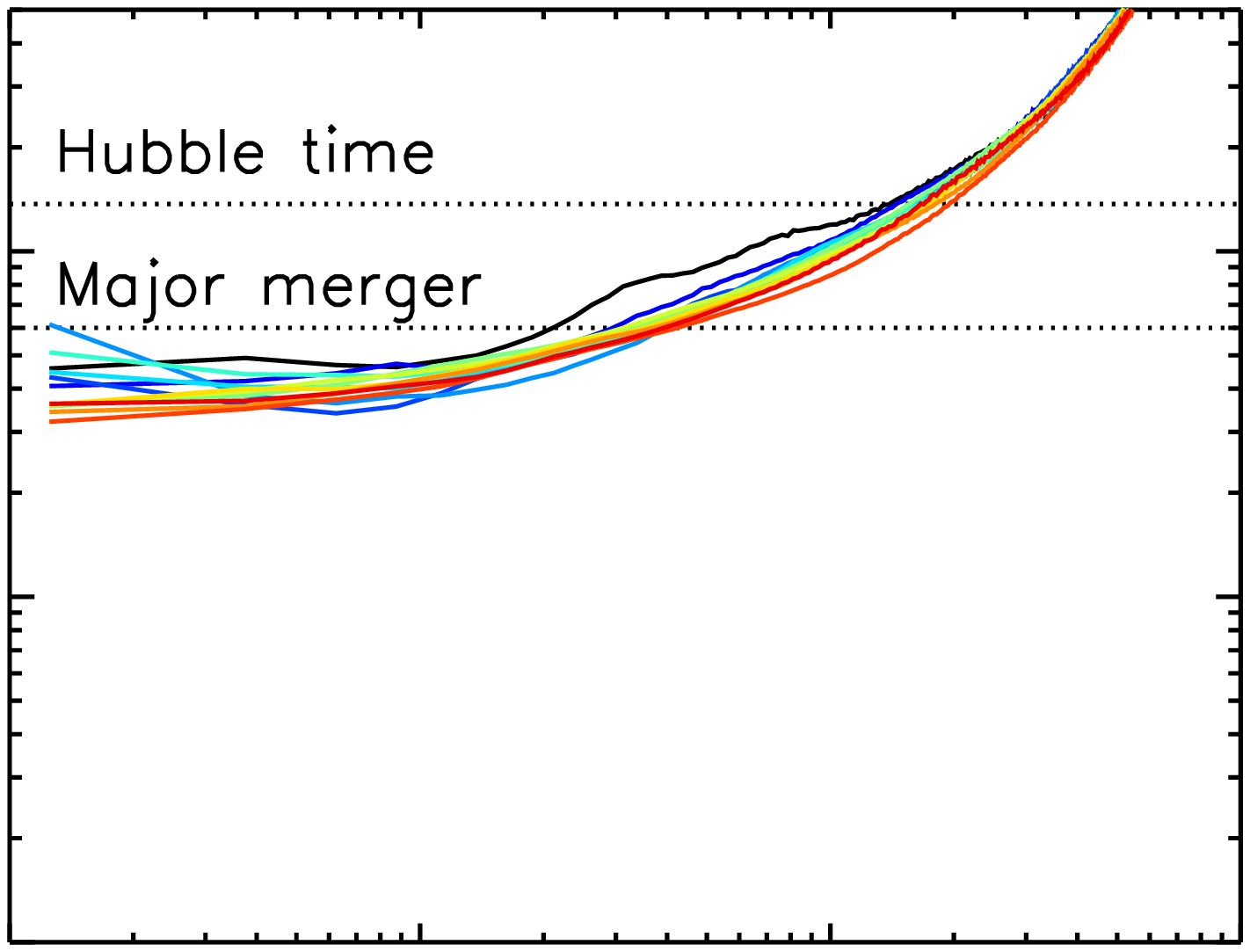}}\hspace{-1.40cm}}
  \centering{\resizebox*{!}{3.4cm}{\includegraphics{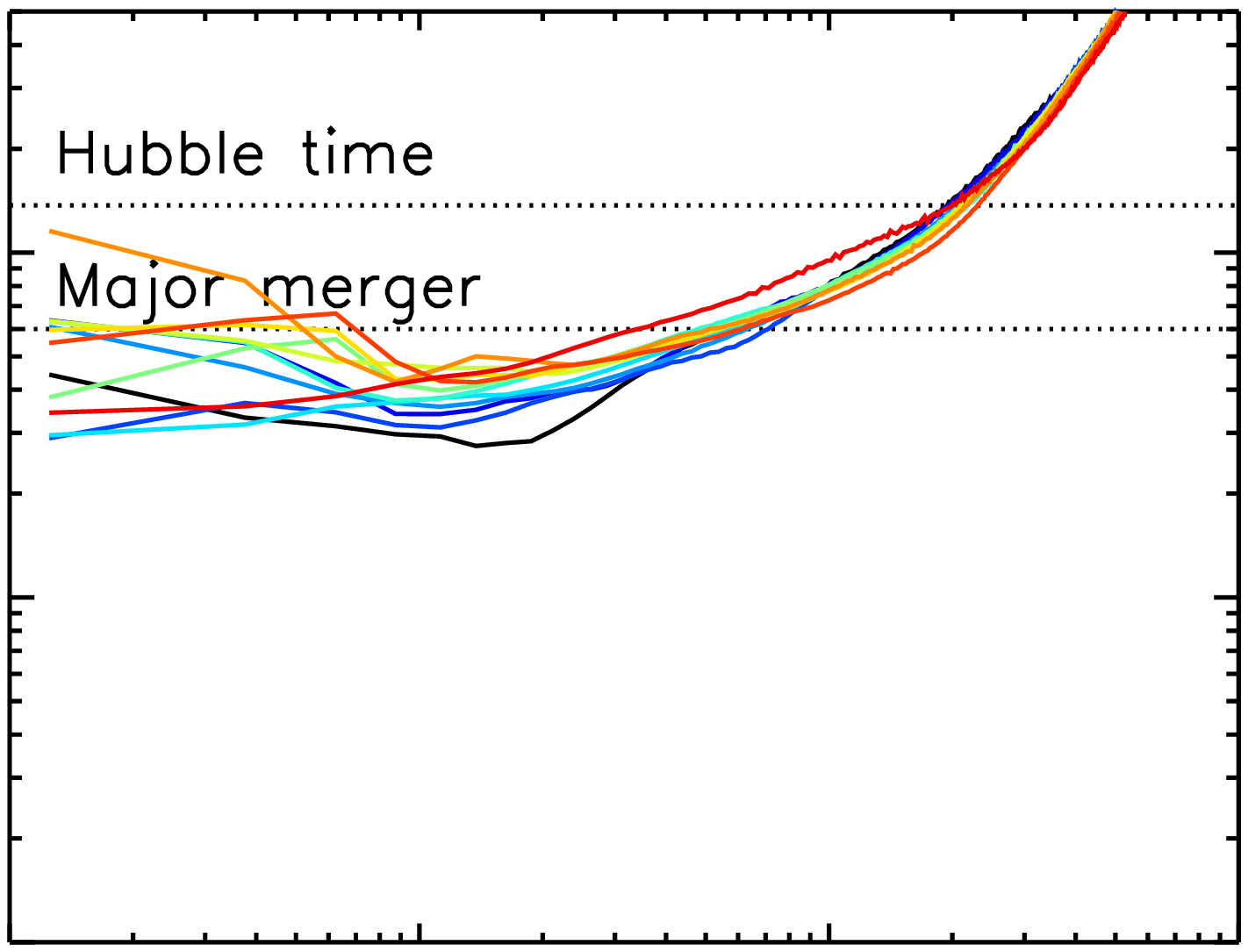}}\hspace{-1.40cm}}
  \centering{\resizebox*{!}{3.4cm}{\includegraphics{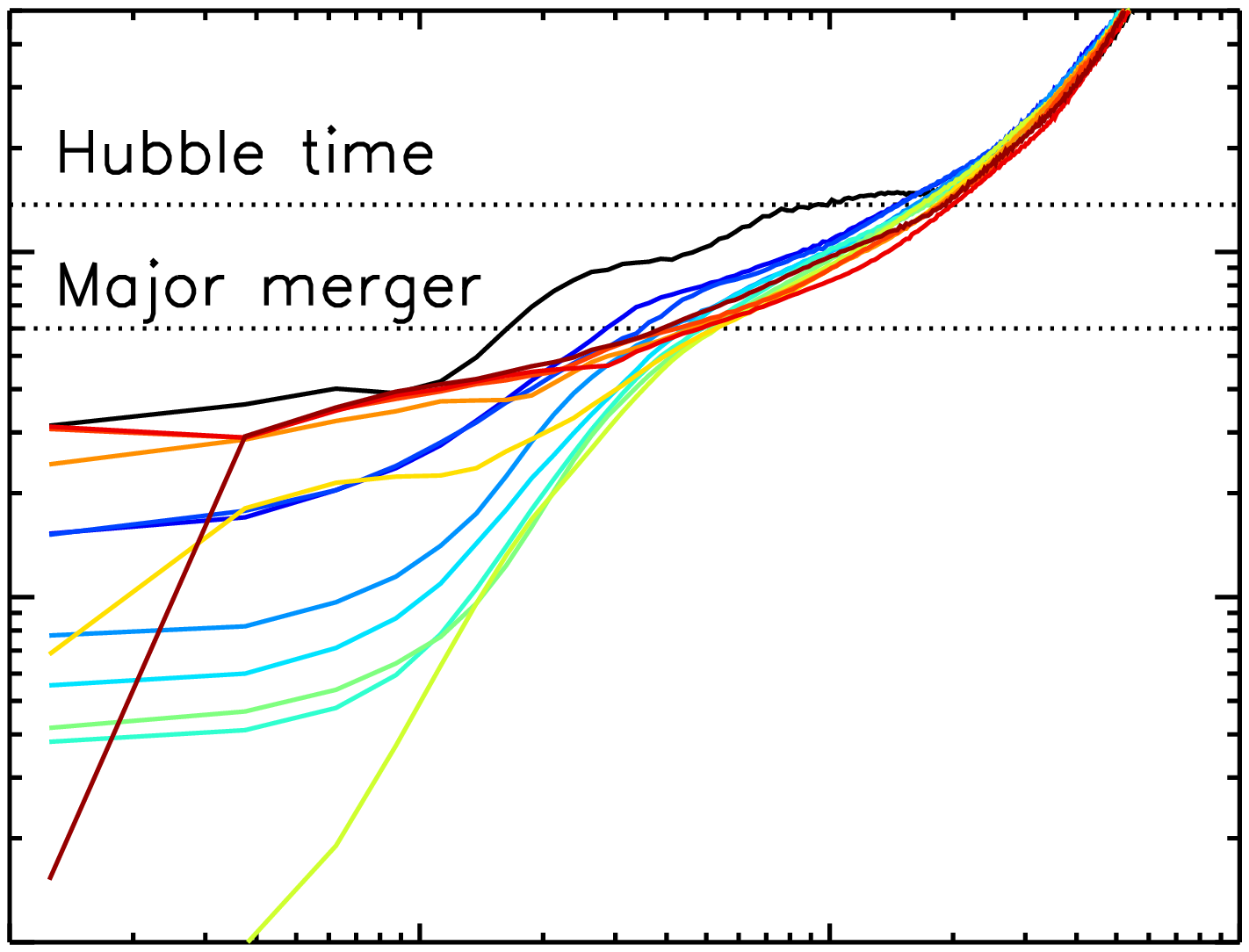}}\vspace{-0.99cm}}\\
  \centering{\resizebox*{!}{3.4cm}{\includegraphics{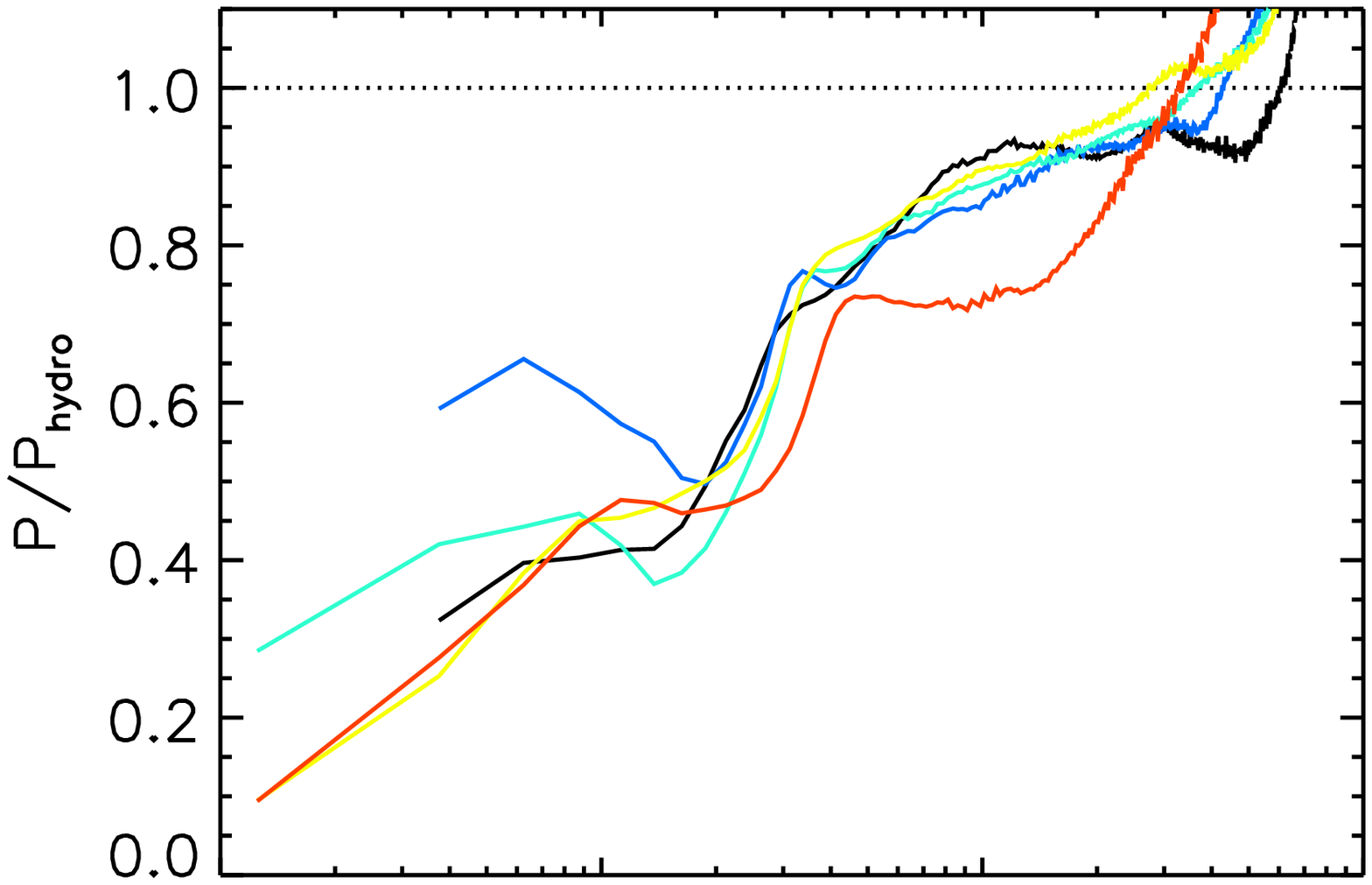}}\hspace{-1.40cm}}
  \centering{\resizebox*{!}{3.4cm}{\includegraphics{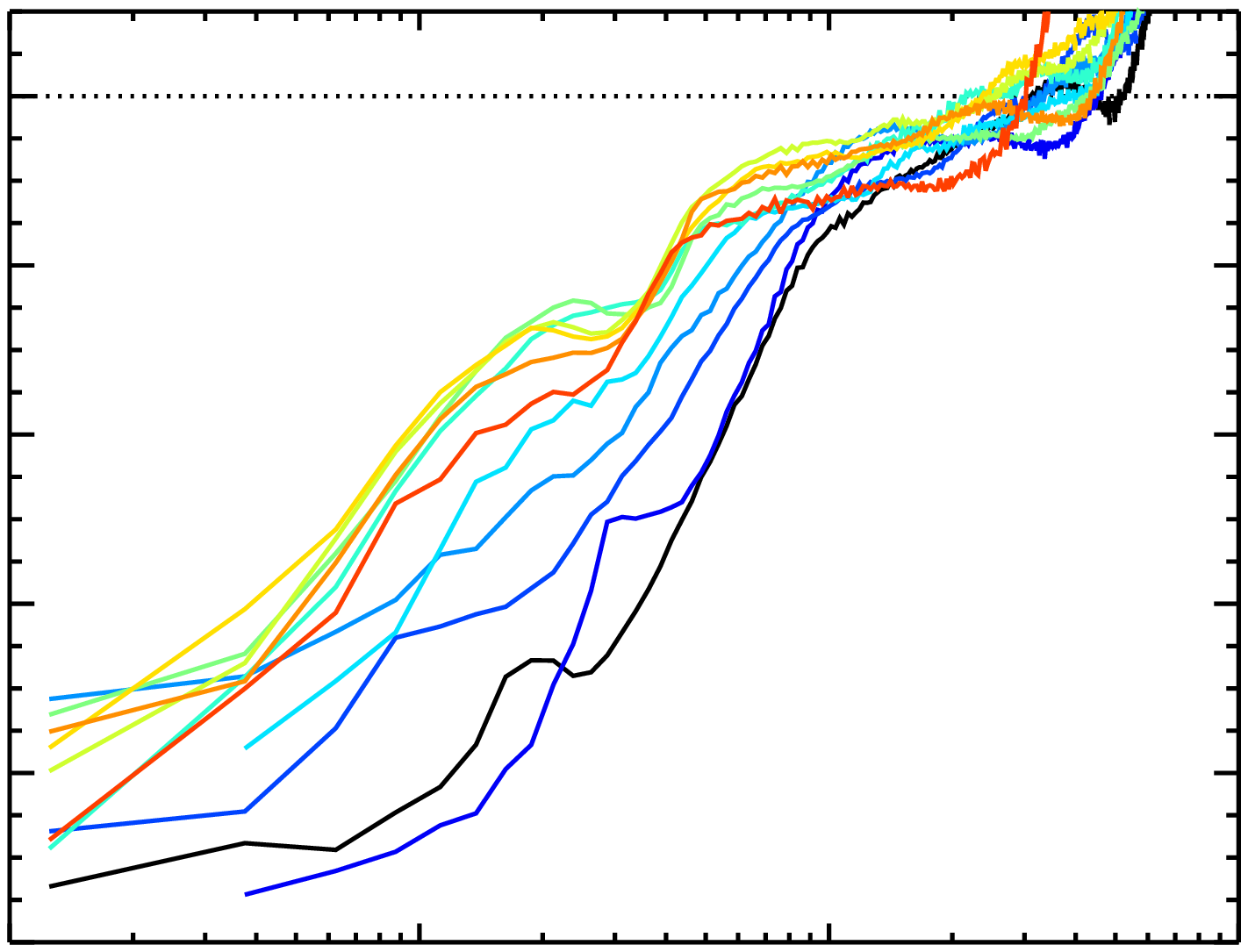}}\hspace{-1.40cm}}
  \centering{\resizebox*{!}{3.4cm}{\includegraphics{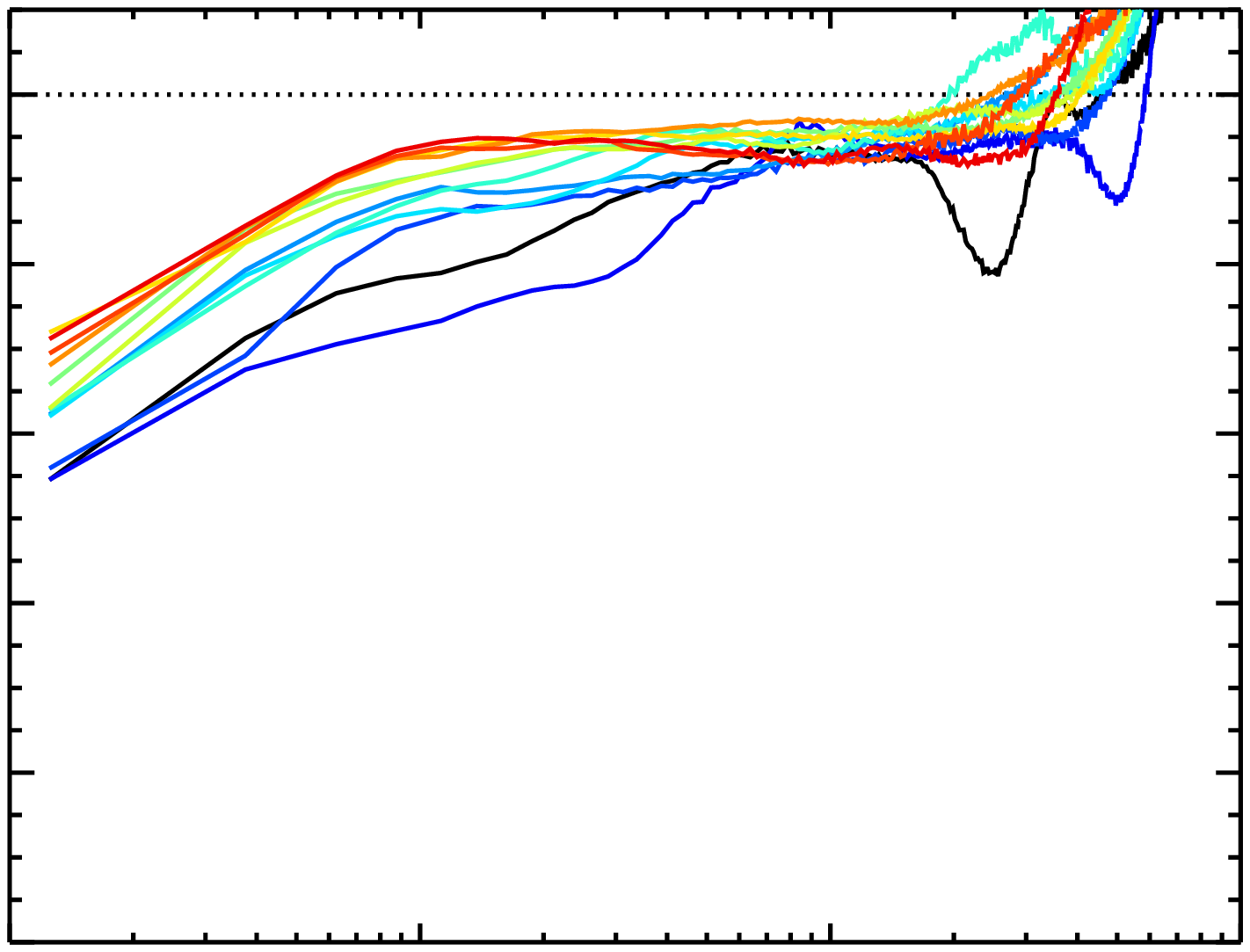}}\hspace{-1.40cm}}
  \centering{\resizebox*{!}{3.4cm}{\includegraphics{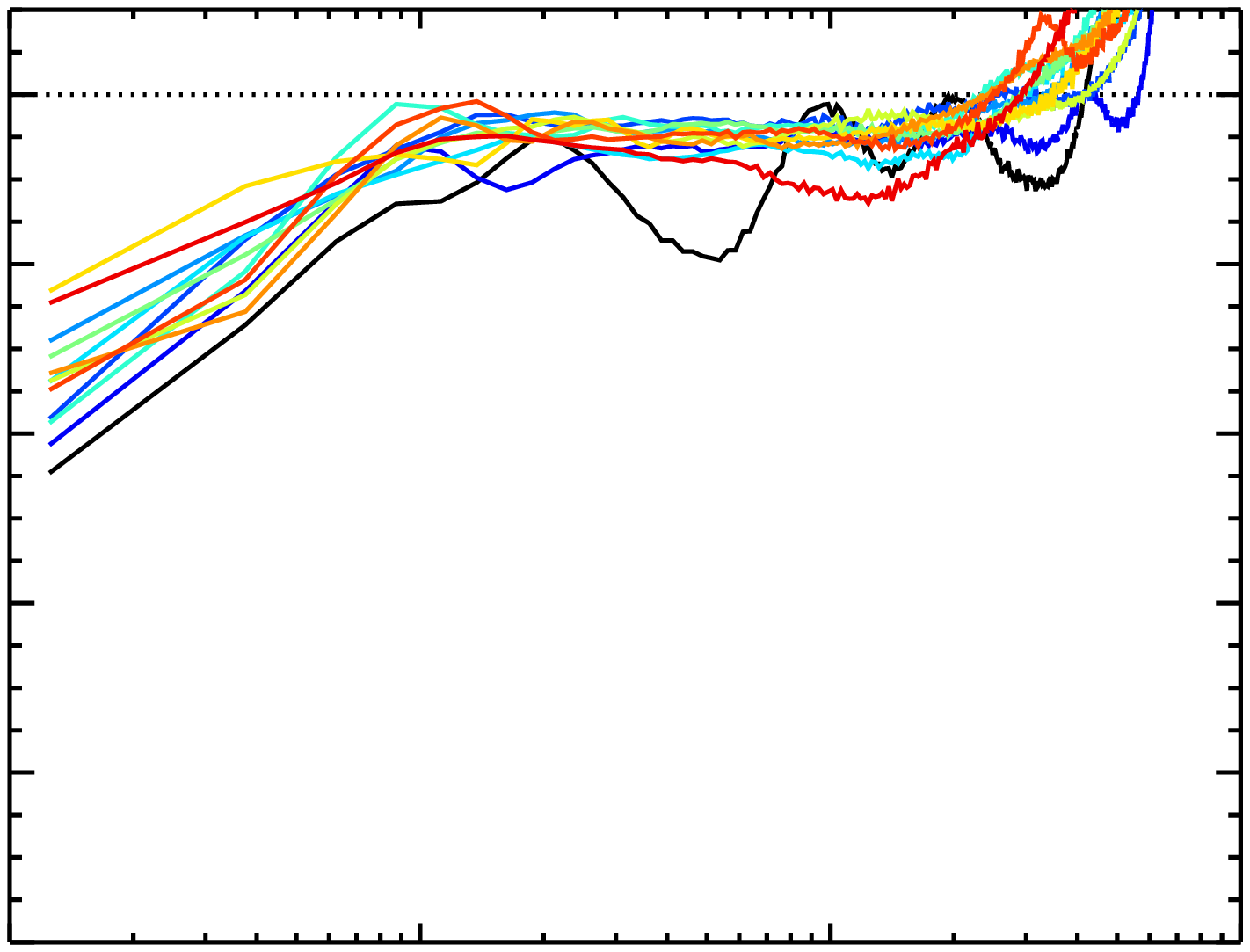}}\hspace{-1.40cm}}
  \centering{\resizebox*{!}{3.4cm}{\includegraphics{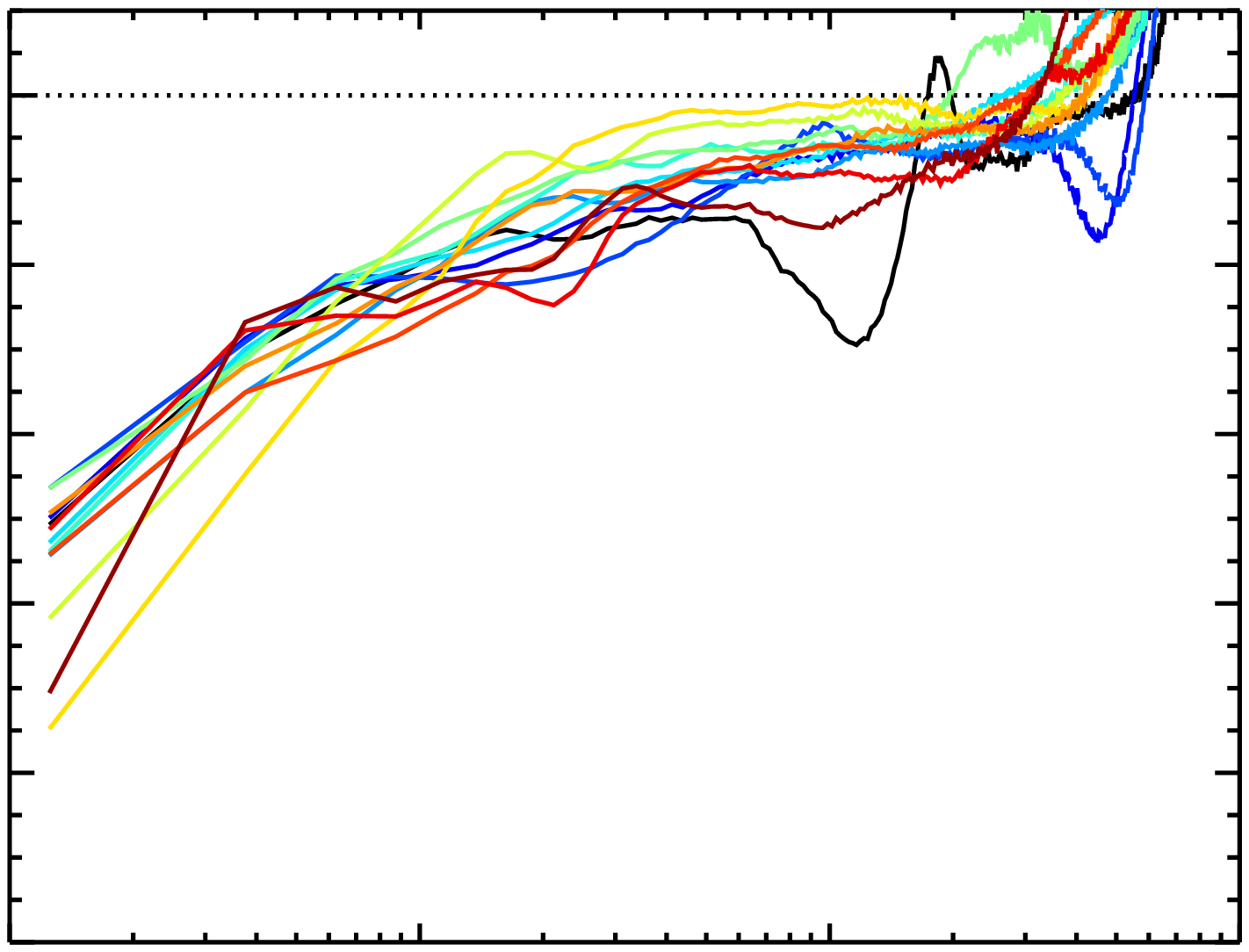}}\vspace{-0.99cm}}\\
  \centering{\resizebox*{!}{3.4cm}{\includegraphics{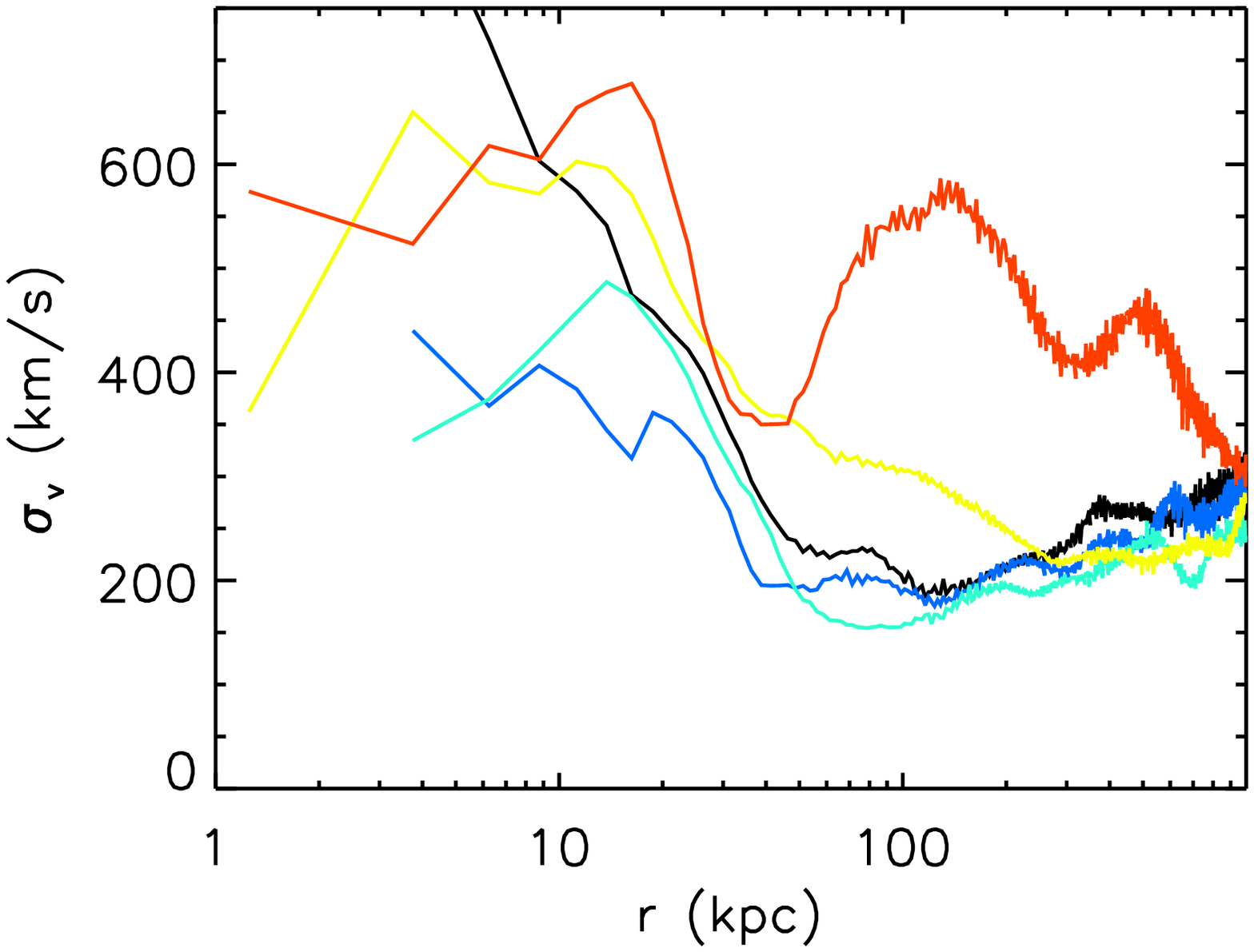}}\hspace{-1.40cm}}
  \centering{\resizebox*{!}{3.4cm}{\includegraphics{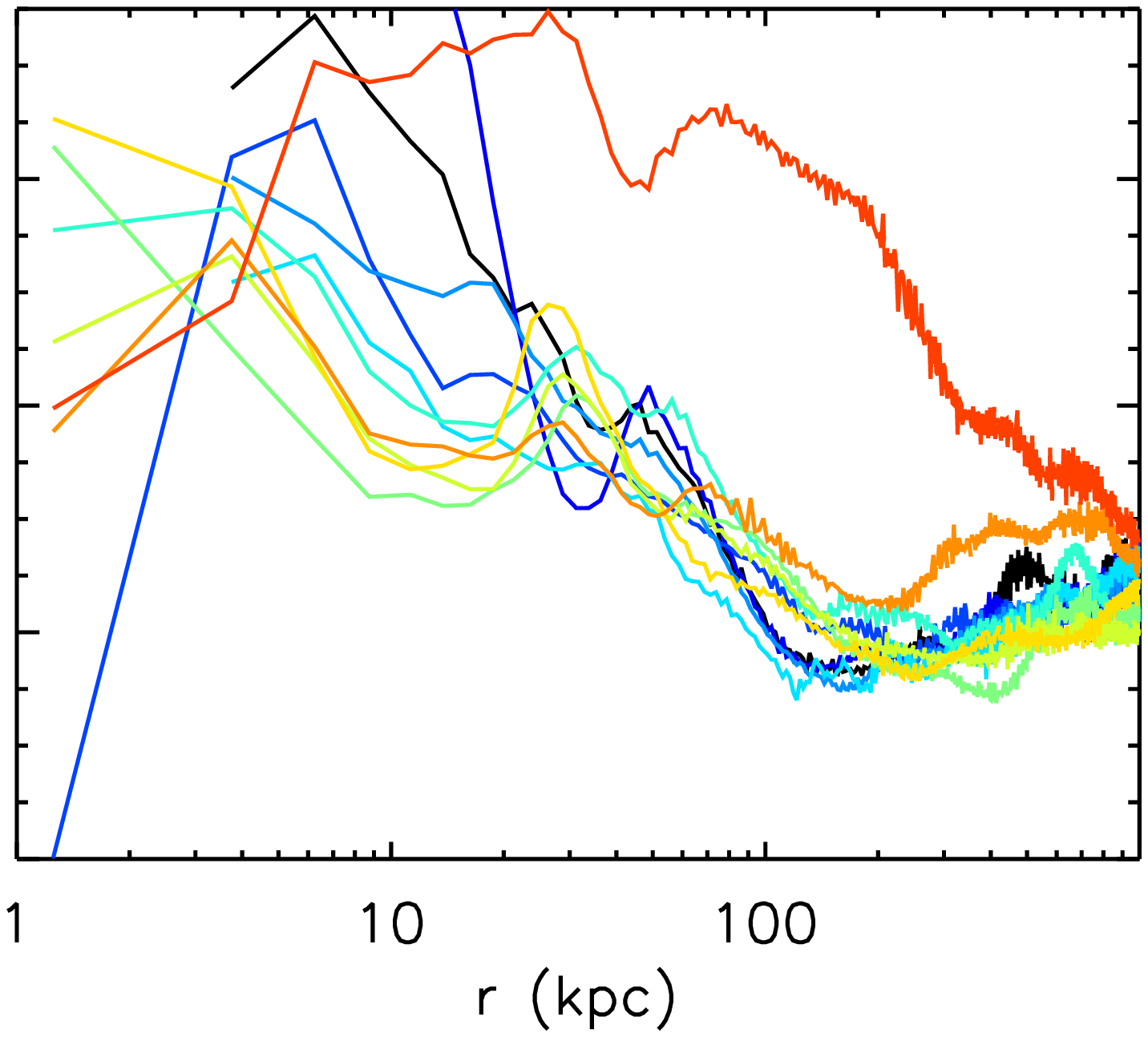}}\hspace{-1.40cm}}
  \centering{\resizebox*{!}{3.4cm}{\includegraphics{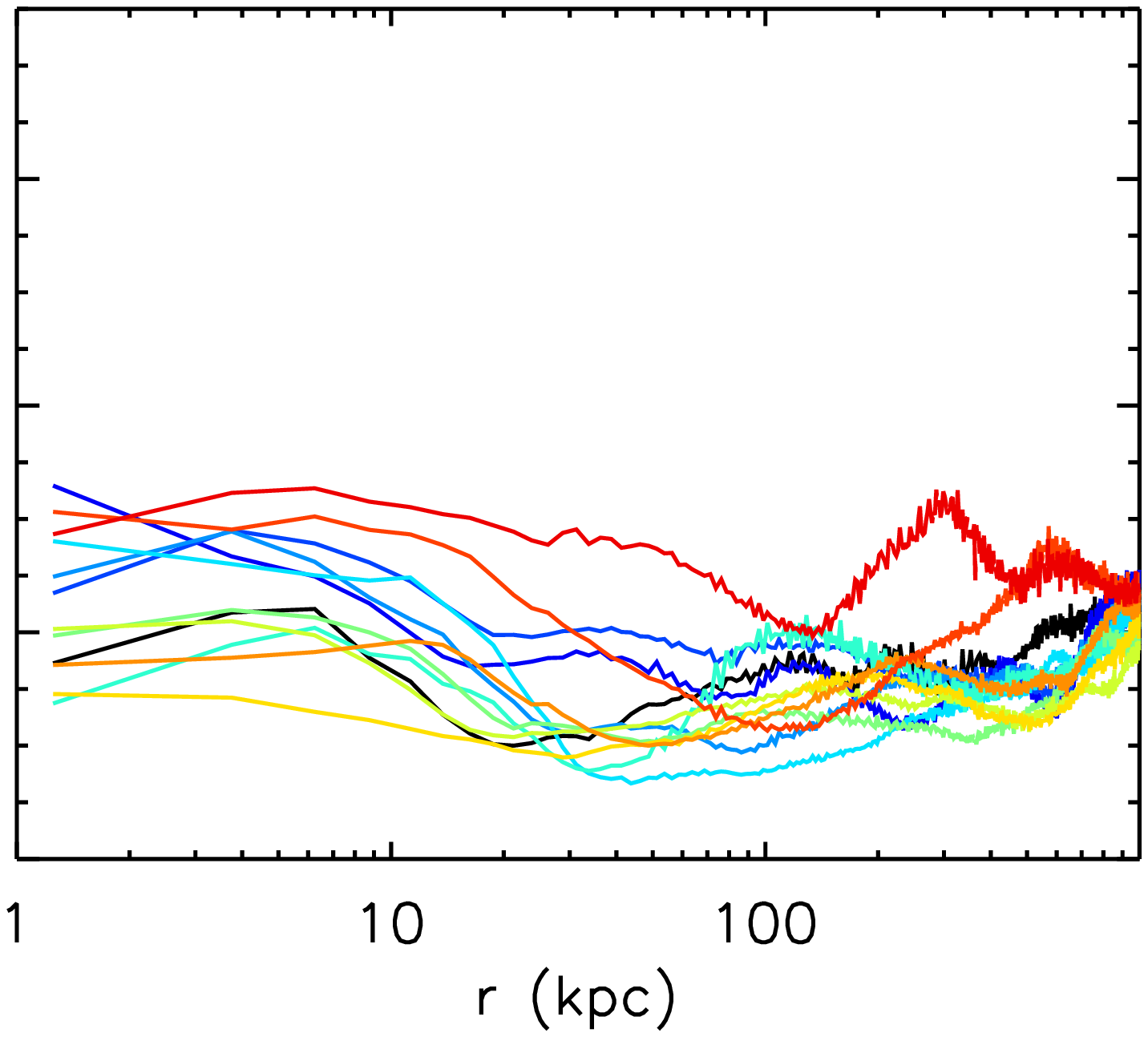}}\hspace{-1.40cm}}
  \centering{\resizebox*{!}{3.4cm}{\includegraphics{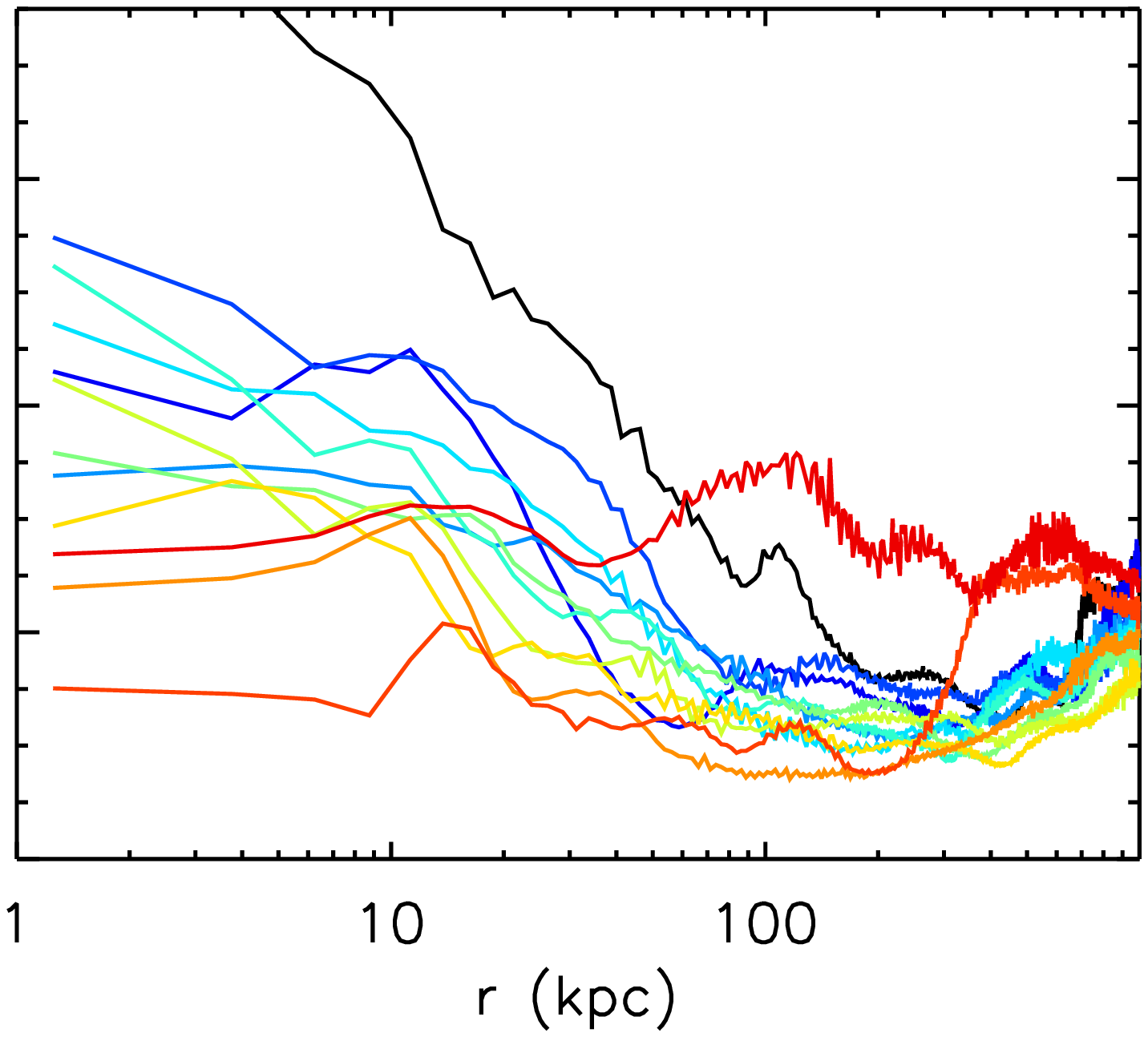}}\hspace{-1.40cm}}
  \centering{\resizebox*{!}{3.4cm}{\includegraphics{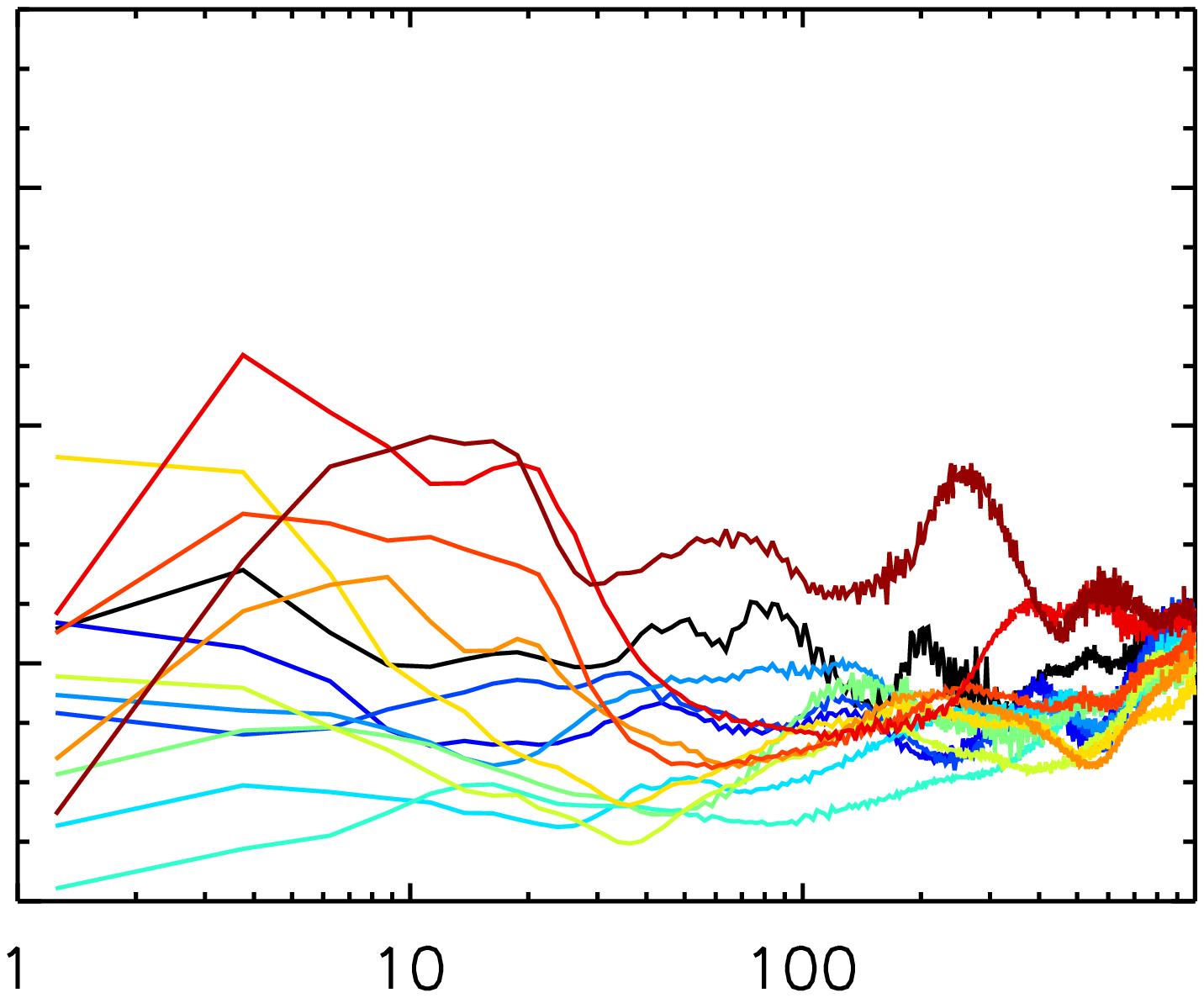}}}
  \caption{A comparison of the time evolution post major merger ($z=0.58$) of the ICM volume-weighted, angular averaged entropy (first row), density (second row), temperature (third row), gas cooling time (fourth row), ratio of gas pressure to hydrostatic pressure (fifth row) and radial gas velocity dispersion (sixth row) profiles for the NOAGNrun (first column), ZNOAGNrun (second column), ZAGNJETrun (third column), ZAGNHEATrun (fourth column), and ZAGNOFFrun (fifth column). Colors correspond to profiles at different redshifts listed on the entropy plots (first row). The dotted line in the entropy profiles corresponds to the $r^{1.1}$ power-law inferred at large radii from both SPH and AMR simulations of galaxy clusters by Voit et al. (2005).}
    \label{allin_Z}
\end{figure*}

\begin{figure*}
  \centering{\resizebox*{!}{4.5cm}{\includegraphics{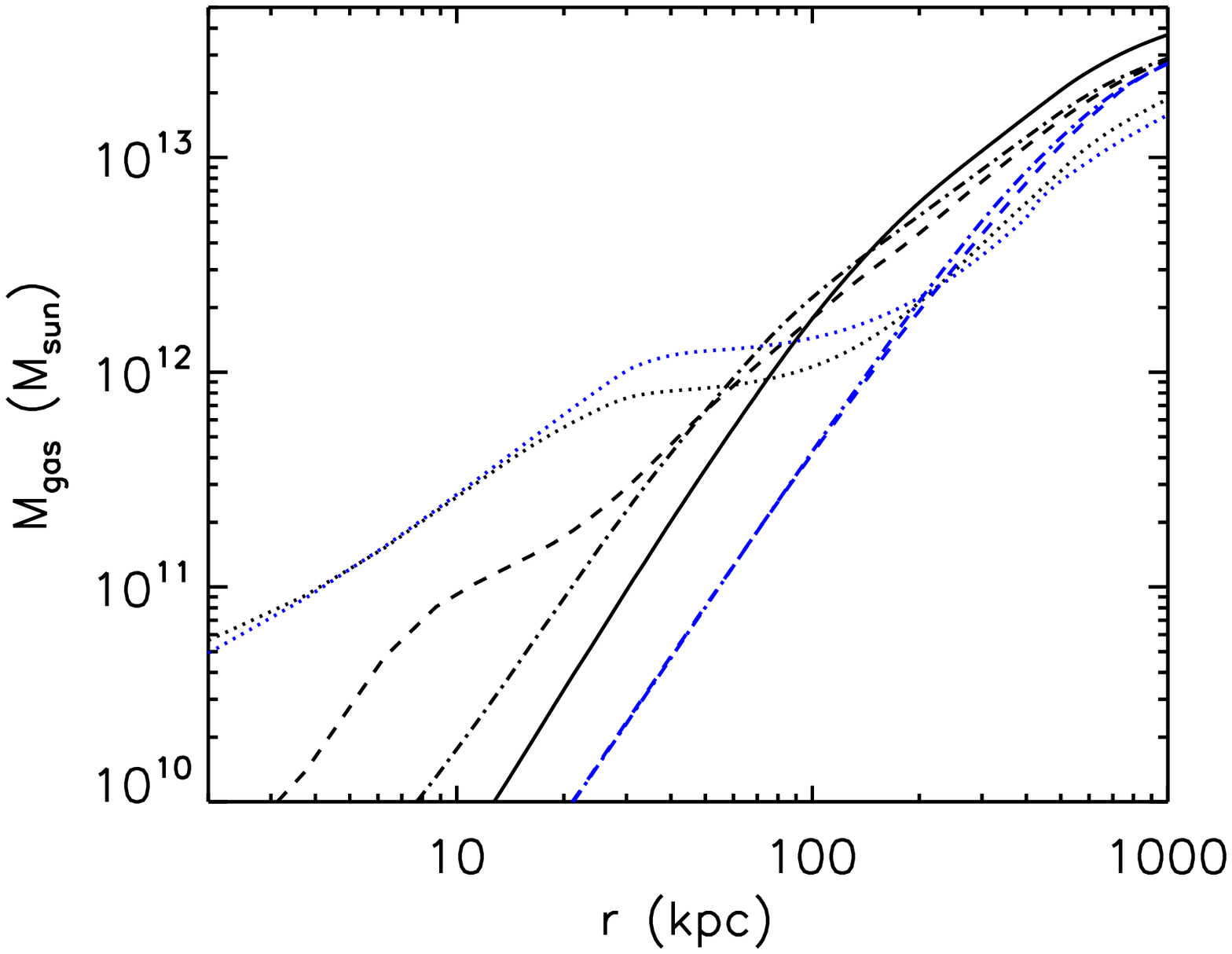}}}
  \centering{\resizebox*{!}{4.5cm}{\includegraphics{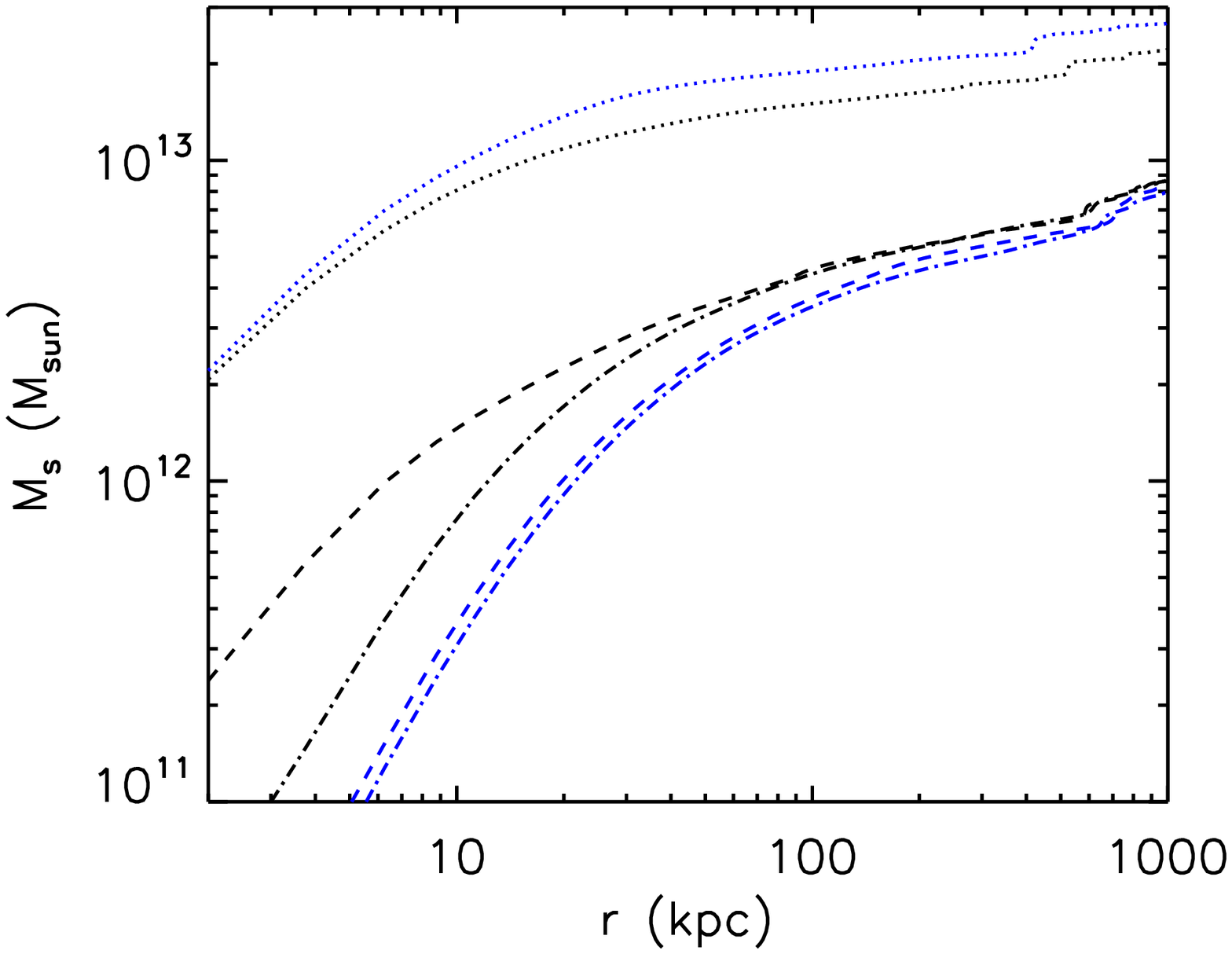}}}
  \centering{\resizebox*{!}{4.5cm}{\includegraphics{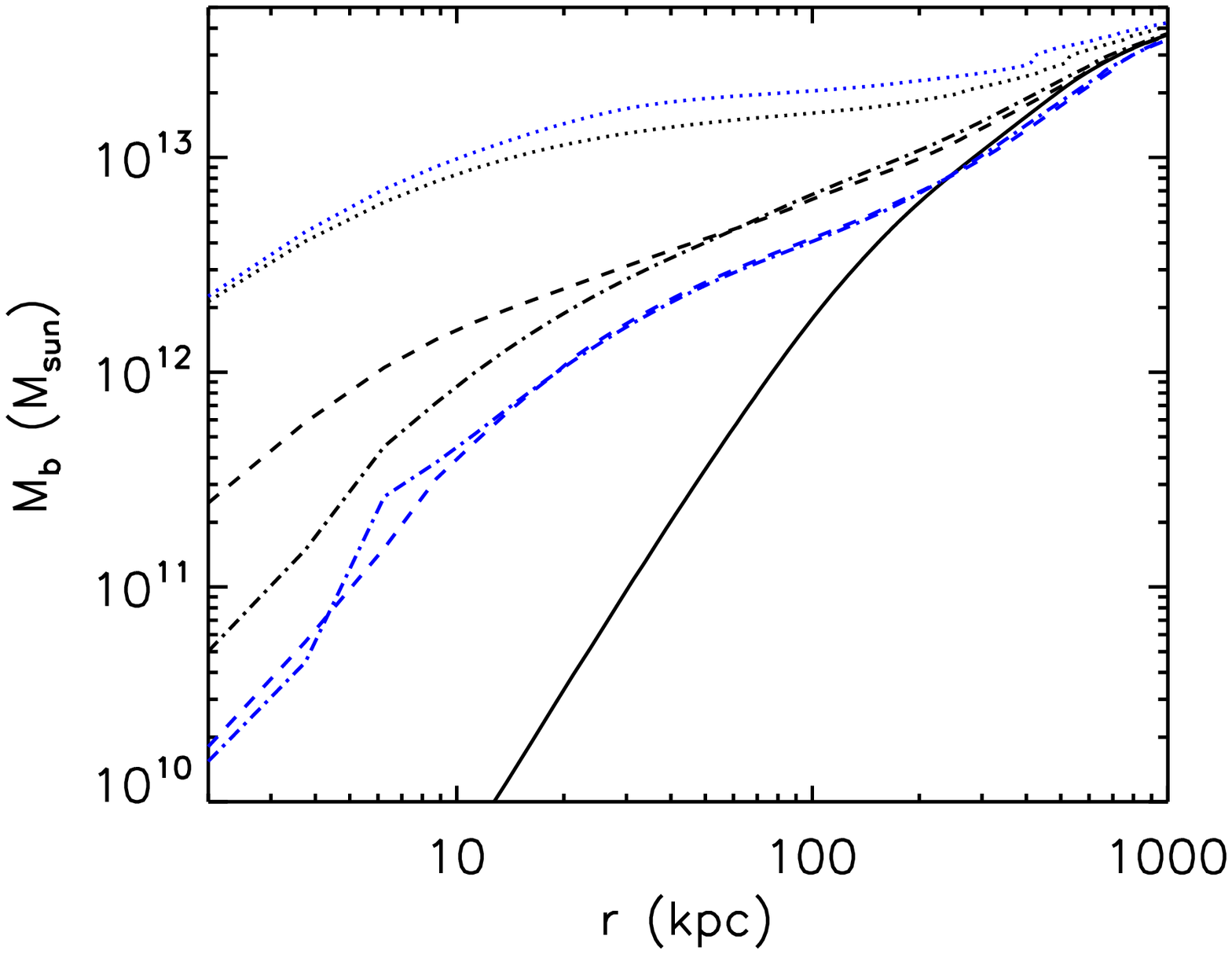}}}\\
  \centering{\resizebox*{!}{4.5cm}{\includegraphics{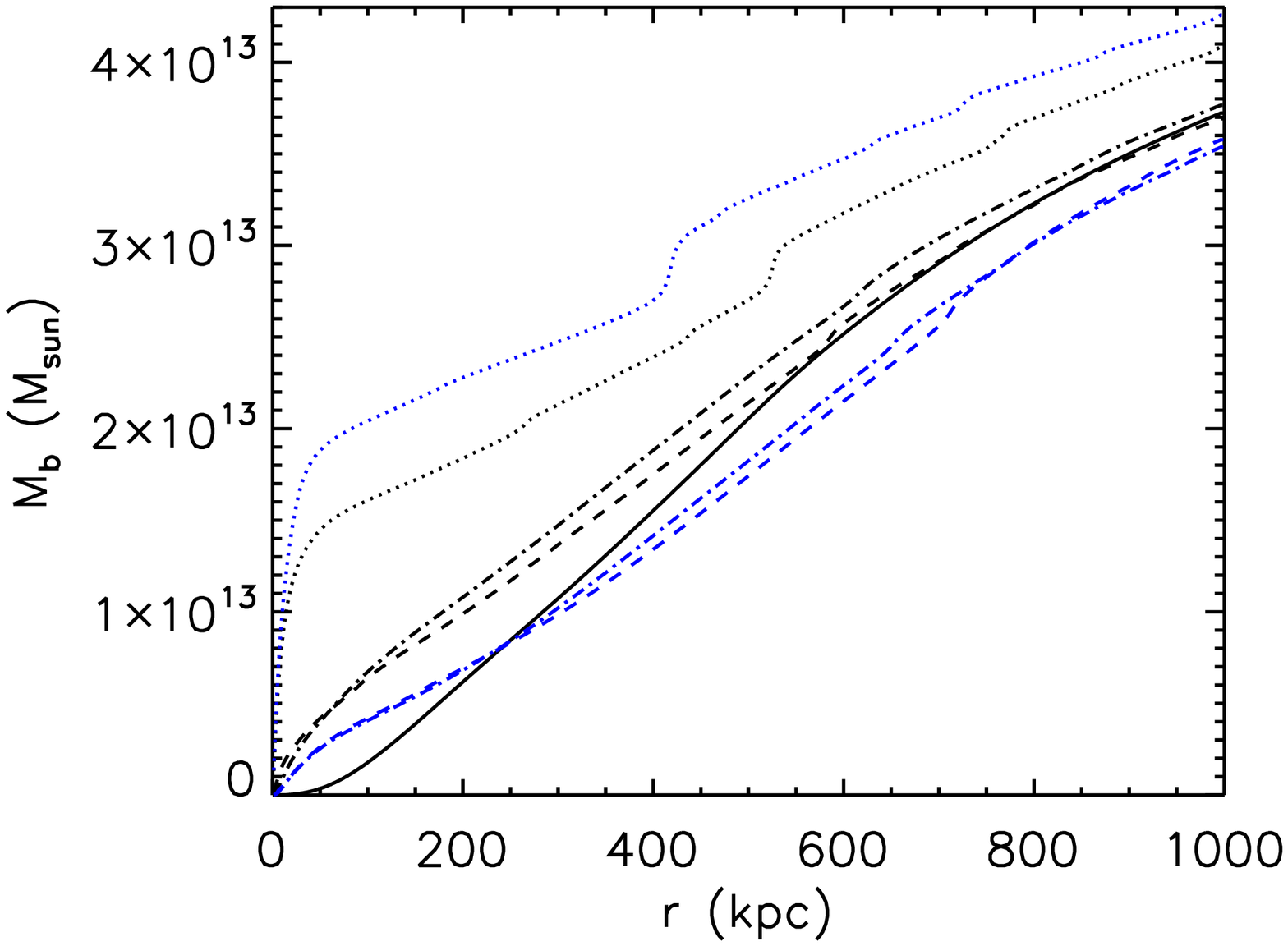}}}
  \centering{\resizebox*{!}{4.5cm}{\includegraphics{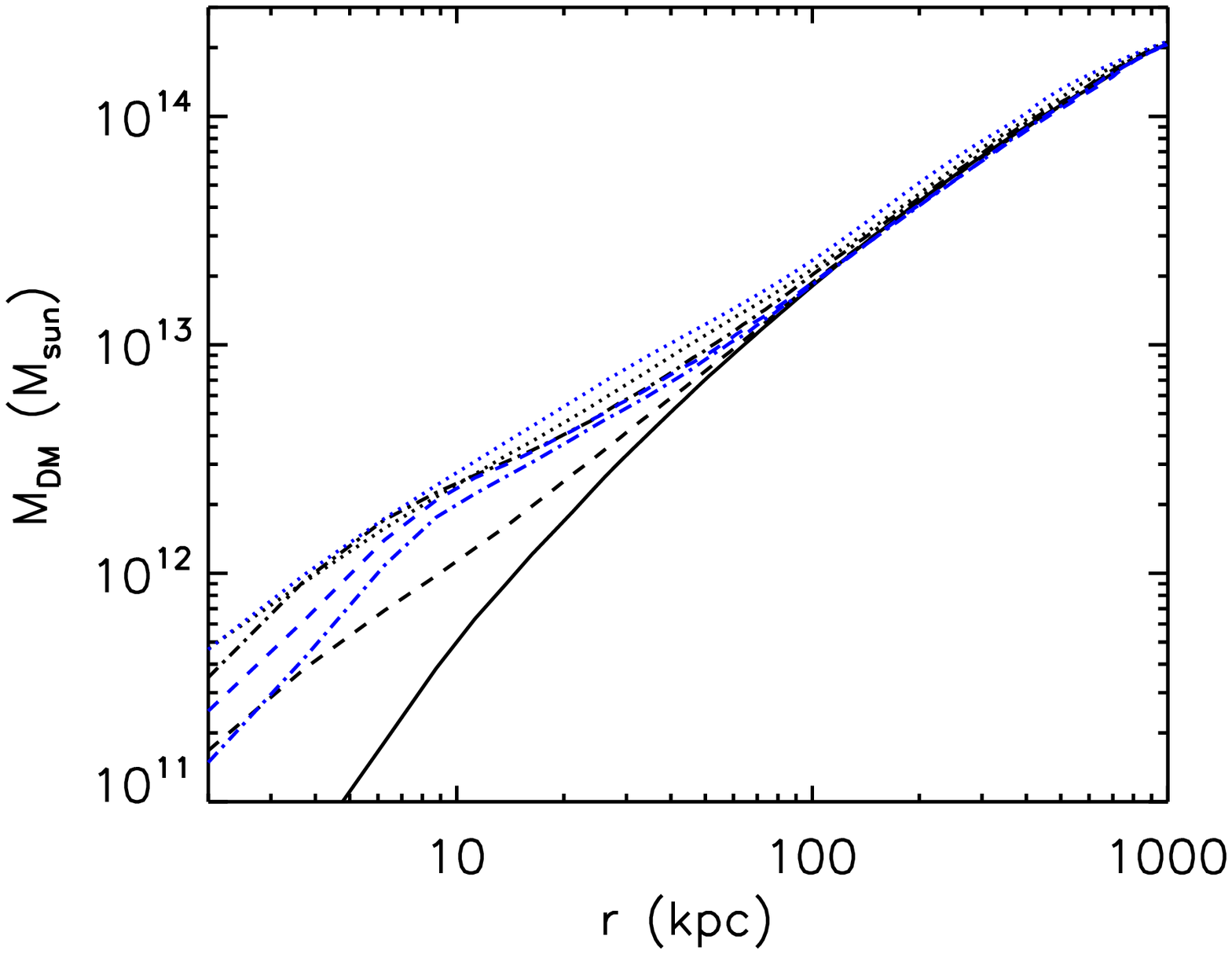}}}
  \centering{\resizebox*{!}{4.5cm}{\includegraphics{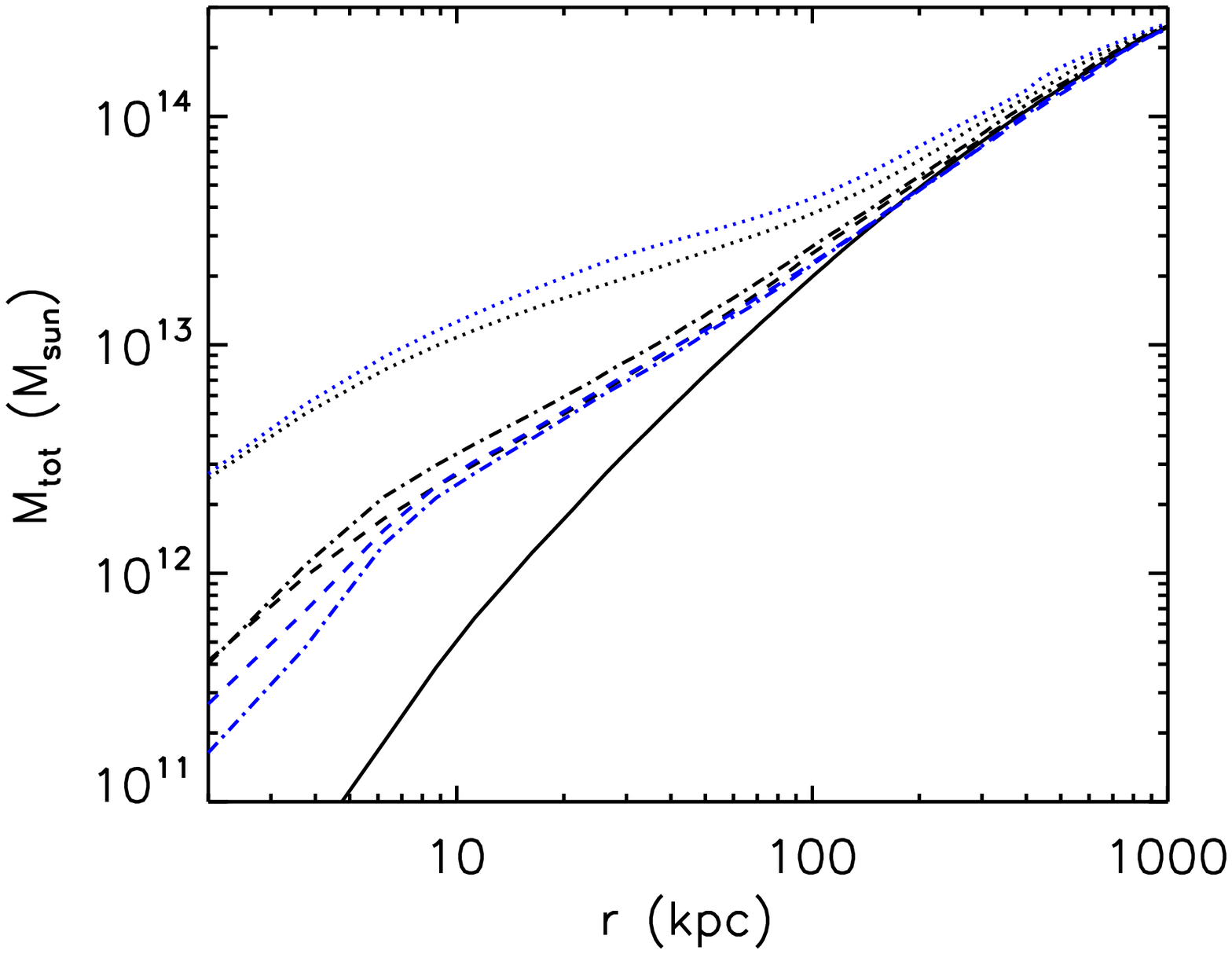}}}
  \caption{Cumulative gas (upper left), star (upper middle), baryon (logarithmic in upper right and linear in lower left), dark matter (lower middle) and total (lower right)
mass profiles of the cluster at $z=0$ for the ADIArun
(black solid line), the NOAGNrun (black dotted line), the AGNJETrun
(black dashed line), the AGNHEATrun (black dot-dashed line), the
ZNOAGNrun (blue dotted line), the ZAGNJETrun (blue dashed line), and the
ZAGNHEATrun (blue dot-dashed line). The cumulative baryon mass is plotted on a linear scale in order to better display what is happening at large radii.}
    \label{mass_comp_all}
\end{figure*}

Until now, this paper has discussed the results from the cluster resimulations without
supernovae and the accompanying stronger cooling due to metals
released in the explosions. In this section, we analyze versions of the cluster simulations with supernovae
(ZNOAGNrun, ZAGNJETrun, ZAGNHEATrun, ZAGNOFFrun) and focus on the effect of metal cooling on the
gas properties of the ICM.

Like Fig.~\ref{allin} for the no metal case, Fig.~\ref{allin_Z} shows the time evolution of different thermodynamical gas profiles for
the ICM when SN and hence metal enrichment is included in the simulations.  For ease of comparison,
we display the results from the simulation with primordial cooling (NOAGNrun) in the same panel.
In the absence of AGN feedback (ZNOAGNrun), the galaxy cluster
endures a strong cooling catastrophe as evidenced clearly by the gas pressure profiles (Fig.~\ref{allin_Z}, second
column, fifth row), even though energy from SN explosions is deposited into the
surrounding gas.  This is not a surprising result as it has been known for some time
that the energy from SN feedback
is not enough to suppress the cooling catastrophe in massive
structures \citep{nagaietal07}.  Cosmological re-simulations 
of the formation and evolution of a proto-galactic
structure in a Milky-Way like halo performed at sub-parsec resolution \citep{powelletal11} show that
the mass-loading factor of large-scale, supernova driven outflows is low and therefore they cannot
remove large amounts of gas from high-redshift galaxies, as was also shown in
an idealized context by \cite{dubois&teyssier08winds}.

Therefore, instead of alleviating the cooling catastrophe, SNe actually
produce a stronger one, with the total mass being more concentrated in the centre of the cluster
than in the absence of SN feedback (NOAGNrun,
fig.~\ref{mass_comp_all}).  This is because the metals released by SNe reduce the
cooling timescales at high redshift.
Indeed, even though this effect is very pronounced at lower
temperatures (T$< 10^7$K), and more
moderate for the ICM (20 \% variation at $T=3.5$ keV), the high redshift 
progenitors of the cluster are filled by material
at much lower temperature than a few keV, and so metals
lead to an increase of the gas cooling efficiency by about an
order of magnitude or more in these structures \citep{sutherland&dopita93}.
It results in larger cooling times at low redshift  (see cooling timescale profiles in
Fig.~\ref{allin_Z}, fourth row, first and second columns) because the cluster has suffered 
more from the cooling catastrophe and the gas that stays in the core of the ICM has a larger entropy. 

In other words, in the simulations with metals, a larger amount of gas
collapses early on into galaxy and group size dark matter halos and
forms stars. Halos thereby endure a more marked depletion of their gas
content than in the comparable simulation with no metals (NOAGNrun), leading to lower gas density profiles for ZNOAGNrun
(see Fig.~\ref{allin_Z}, second row, first and second columns).  As
explained in previous sections, as central regions get depleted, gas from the outer regions can flow into
them because of the lack of pressure support. Because this loss of pressure support associated with the cooling catastrophe in the ZNOAGNrun not only takes place earlier but also  
at larger radii than in the NOAGNrun, gas flows into the cluster
centre from larger distances in the ZNOAGNrun. Since the gas density
is lower at large radius but the temperature of the infalling shocked
IGM gas is the same as at smaller radii, the amount of entropy carried
by gas coming in from larger radii is larger.  
As a result, higher entropy gas than that in the NOAGNrun replaces the
low entropy material at small radii in the ZNOAGNrun, leading to entropy profiles that 
rise with time to a higher level in the run with metals (Fig.~\ref{allin_Z}, first and second column, first row).

The simulations ZAGNJETrun and ZAGNHEATrun, which include AGN
feedback with the kinetic and thermal modes respectively, show very
similar gas properties (Fig.~\ref{allin_Z}, third and fourth
columns), and mass distributions (Fig.~\ref{mass_comp_all}). Even the
concentration of the cluster DM profile which was noticeably
different in the AGNJETrun  and AGNHEATrun simulations is now very similar.   Thus, we
will not consider hereafter the different effects of these two modes on the gas, which are already described for their equivalent
simulations with no SN (AGNJETrun and AGNHEATrun) in section~\ref{AGNfbk}. Instead we focus
on how metals alter the effect of AGN feedback. First we compare the three simulations
with metals, where one has no AGN feedback (ZNOAGNrun) and the others do (ZAGNJETrun, ZAGNHEATrun). As
was the case without metals, for the simulations with metals (ZAGNJETrun and ZAGNHEATrun) AGN feedback
diminishes the amount of entropy in the ICM compared to
the simulation without AGN (ZNOAGNrun).  AGN feedback, again, suppresses the cooling
catastrophe, even though more efficient metal cooling at early times makes this more challenging than in the no metal case.
We emphasize that, as in the no metal case and despite a much larger amount of preheating from SN and AGNs 
at high redshift, only the runs where AGN feedback is present during the whole evolution of the cluster are able to prevent the 
cooling catastrophe. In the ZAGNOFFrun simulation (Fig.~\ref{allin_Z}, last column) where AGN feedback is switched off after the last major merger, the central galaxy 
ends up accreting cold gas at a rate of $\approx$ 500 $\rm M_\odot$/yr  by $z=0$.
All mass components  (gas, stars, DM) at $z=0$ in ZAGNJETrun and ZAGNHEATrun
are less concentrated in the centre of the cluster compared to the
ZNOAGNrun (Fig.~\ref{mass_comp_all}), and unlike in ZNOAGNrun, the gas supports itself against its own
collapse in ZAGNJETrun and ZAGNHEATrun (Fig.~\ref{allin_Z}, third and fourth columns, fifth row).

\begin{figure}
  \centering{\resizebox*{!}{6.5cm}{\includegraphics{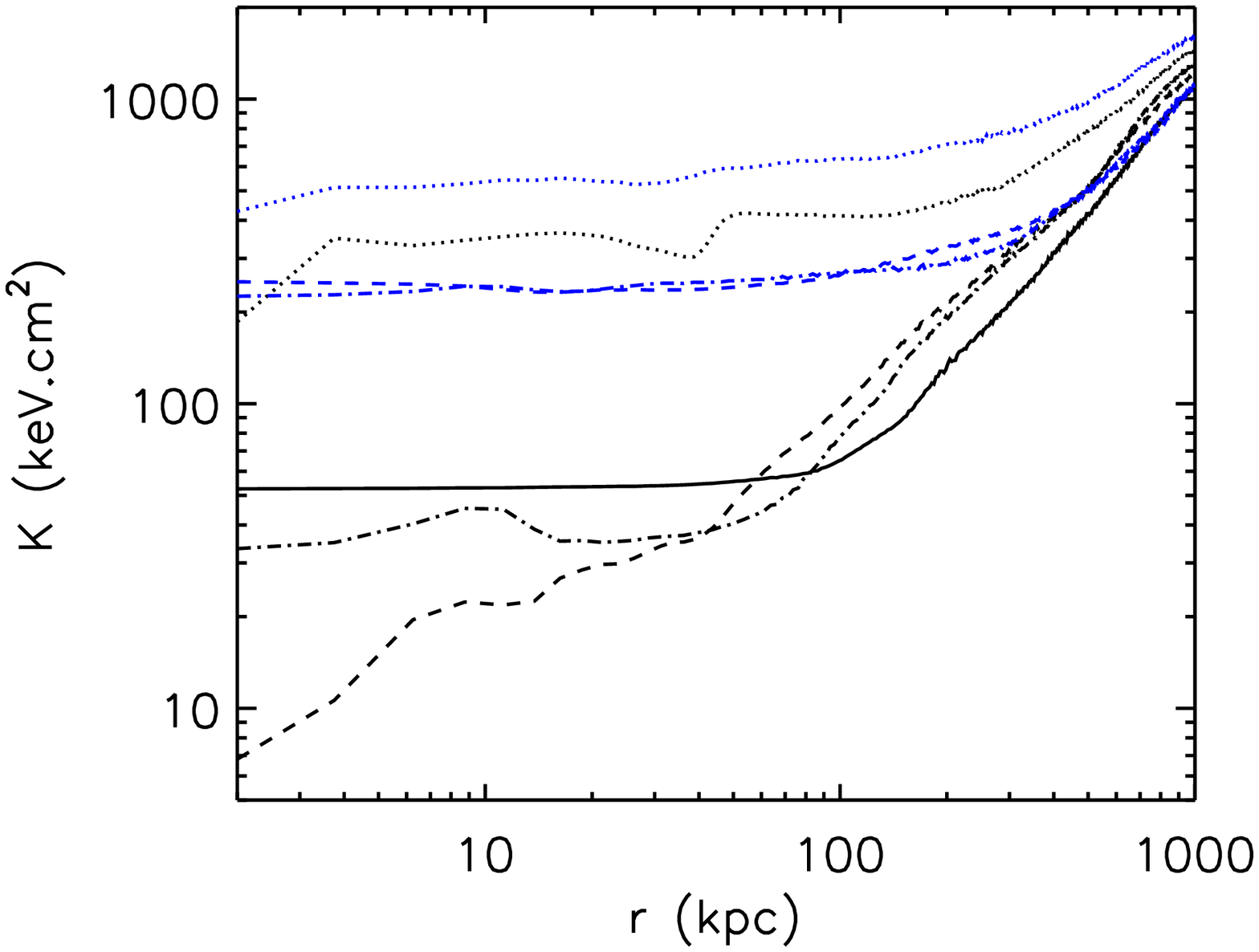}}}
  \centering{\resizebox*{!}{6.5cm}{\includegraphics{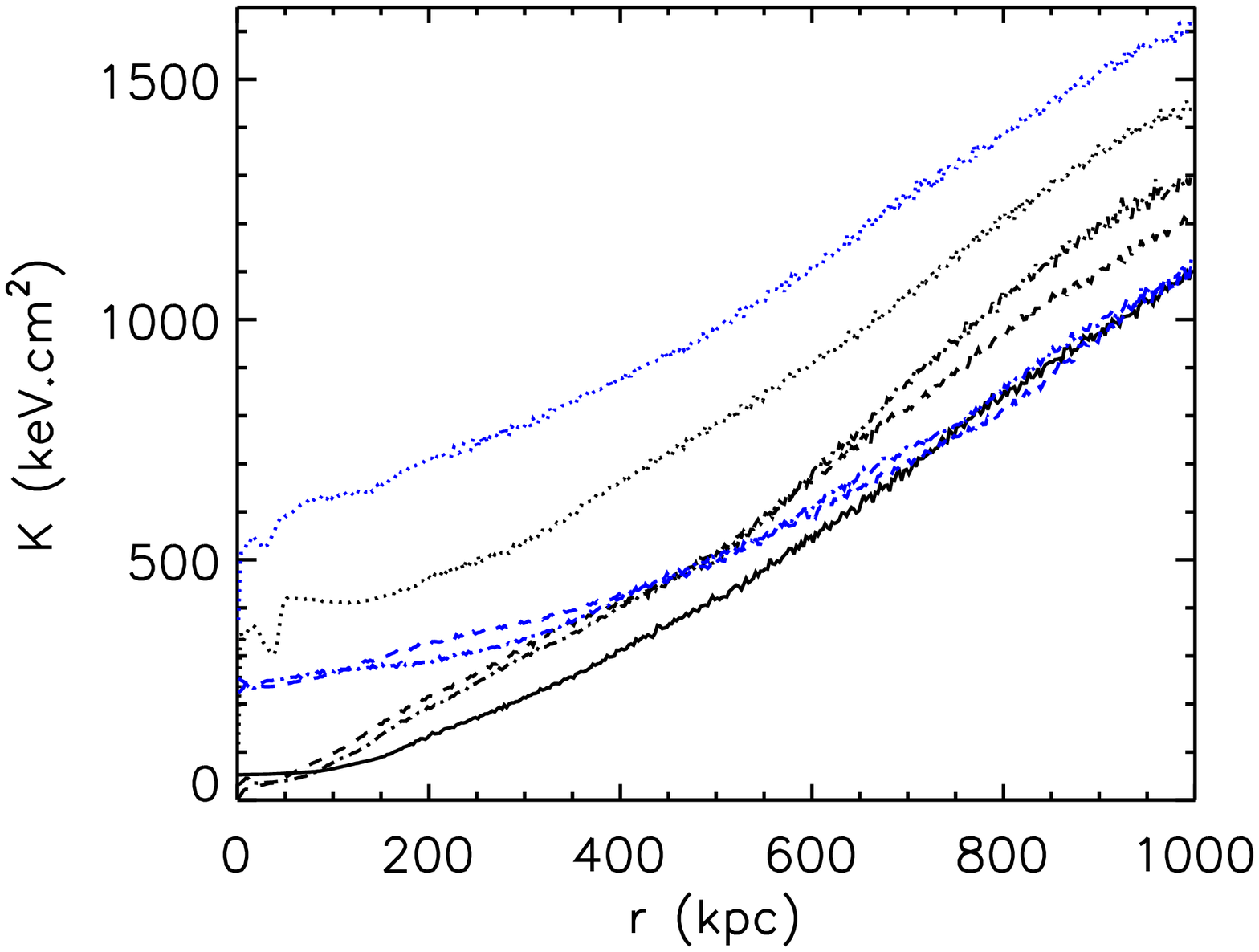}}}
  \caption{Entropy profiles of the cluster at $z=0$ for the ADIArun
(black solid line), the NOAGNrun (black dotted line), the AGNJETrun
(black dashed line), the AGNHEATrun (black dot-dashed line), the
ZNOAGNrun (blue dotted line), the ZAGNJETrun (blue dashed line), and
the ZAGNHEATrun (blue dot-dashed line).}
    \label{entro_comp_all}
\end{figure}

Next we examine the differences between the AGN feedback runs with and without metals.
The presence of metals plays a crucial role in how the AGN impacts the gas.
Comparing entropy profiles at $z=0$ for simulations with AGN feedback and
metals (ZAGNJETrun and ZAGNHEATrun), to the equivalent simulations without metals
(AGNJETrun and AGNHEATrun) in Fig.~\ref{entro_comp_all}, we see
that, in the simulations with metals, entropy levels in the cluster core are
larger by almost one order of magnitude.  The cause of this effect can be traced back
to the fact that metals increase the cooling efficiency of the gas in the
early phases of structure formation.  This leads to more fuel for AGN activity
at early times which pushes more gas to larger distances from the
centre of the halos. Note however, that this is only an integrated time
effect. As a consequence, very early on, before the AGN has
reached a mass when its feedback allows it to significantly impact the
fate of the intra-halo gas, baryons are more concentrated in the
ZAGNJETrun than in the AGNJETrun. This is reflected in the DM density
profile which is more concentrated in the ZAGNJETrun than in the
AGNJETrun (fig.~\ref{mass_comp_all} middle panel of bottom row), even
though the amount of baryons in the ZAGNJETrun at $z=0$ is much
smaller than in the AGNJETrun.

Nevertheless, in general, because of these early time effects,  
the integrated mass of baryons at $z=0$ at all radii
(Fig.~\ref{mass_comp_all}) is smaller when metals and AGN are present 
(ZAGNJETrun and ZAGNHEATrun) than for the equivalent AGN simulations without metals (AGNJETrun and AGNHEATrun).  
The total stellar mass at $z=0$ in the
cluster is also reduced by the combined effect of AGN activity and
metal release by SNe.  This underlines the fact that AGN feedback has a
stronger effect on the cold baryon content, because of the enhanced
cooling rates which allow the BH central engine to grow faster at
high redshift and thus quench star formation earlier on.
This is also reflected by the fact that the central BH is more massive when metal enhanced cooling is enabled (see table~\ref{tabnames}) and implies larger amounts of energy deposited into the ICM.

 We note that the main results we obtain when metal cooling and AGN feedback are included in the simulations (i.e. increased amount of pre-heating at high redshift, reduced stellar mass
for the central galaxies, too high level of entropy in the cluster core, central temperature 
profiles which are too steep) are broadly consistent with those obtained by \cite{fabjanetal10} for different clusters using a different numerical technique and a slightly different model of AGN thermal feedback.
We also remark that we still have an adiabatic contraction of the DM halo, contrary to what~\cite{teyssieretal11} obtained in their Virgo-like cluster with thermal AGN feedback. In this paper, we inject the energy of the thermal feedback on the same scale as the jet mode, close to the resolution limit, to facilitate a comparison
between the two runs. Note that this is different from the scale adopted by~\cite{teyssieretal11}, who injected the thermal energy into large bubbles, and explains why we are not able to halt adiabatic contraction and they are.
In short, injecting thermal energy on larger scales increases the effective efficiency of AGN feedback. A detailed study of parameter/resolution impact on our results 
is deferred to a companion paper (Dubois et al, in prep).

At late times $z<0.58$, there is no longer a cold gas component in the
centre of the cluster of ZAGNJETrun and ZAGNHEATrun (Fig.~\ref{allin_Z}, third and fourth columns, third row) and the gas remains in hydrostatic equilibrium
(Fig.~\ref{allin_Z}, third and fourth columns, fifth row) because cooling times are comparable to the last major merger times 
(Fig.~\ref{allin_Z}, third and fourth columns, fourth row), and
therefore cold gas only triggers some faint AGN feedback activity.  
Note however that such a reduced level of AGN feedback activity at low redshift is key to prevent a cooling 
catastrophe, as demonstrated by the gas profiles measured in the ZAGNOFFrun simulation (Fig.~\ref{allin_Z}, fifth column) which lacks 
such a low redshift AGN feedback activity.
The strong enhanced early phase of AGN activity in ZAGNJETrun and ZAGNHEATrun leads to lower entropy  compared to the no metal simulations AGNJETrun and AGNHEATrun
at large distances from the centre, because of the redistribution of the
gas from the centre of the cluster to the outskirts
(Fig.~\ref{entro_comp_all}, bottom panel).
However, the temperature profiles for ZAGNJETrun and ZAGNHEATrun seem rather unrealistic, when
compared to observations: they continuously increase with decreasing
radius with extremely high values (10--20 keV) in the central 10 kpc (Fig.~\ref{allin_Z}, third and fourth columnds, third row).  
This is a direct consequence of the mass distribution in the
cluster and the hydrostatic equilibrium.  It is possible that a more complete
AGN feedback prescription such as one that includes photo-heating
could, at least partially, alleviate this problem.  As suggested by
\cite{cantalupo10}, soft X-ray photons emitted by stars during strong
starburst episodes or by AGN, can offset radiative losses by photo-heating
the main ion coolants in proto-cluster and massive galaxy halos.  If this kind of feedback takes
place at high redshift, it could halt the cooling catastrophe 
without removing much mass from the centre.  This would increase
the central gas density, decrease the `apparent' cooling time (as
opposed to the effective cooling time, which is increased by the
photo-ionization effect), and allow for lower temperatures in the
cores of the clusters.  We defer a numerical investigation of this 
scenario to  future studies.

\begin{figure}
  \centering{\resizebox*{!}{3.6cm}{\includegraphics{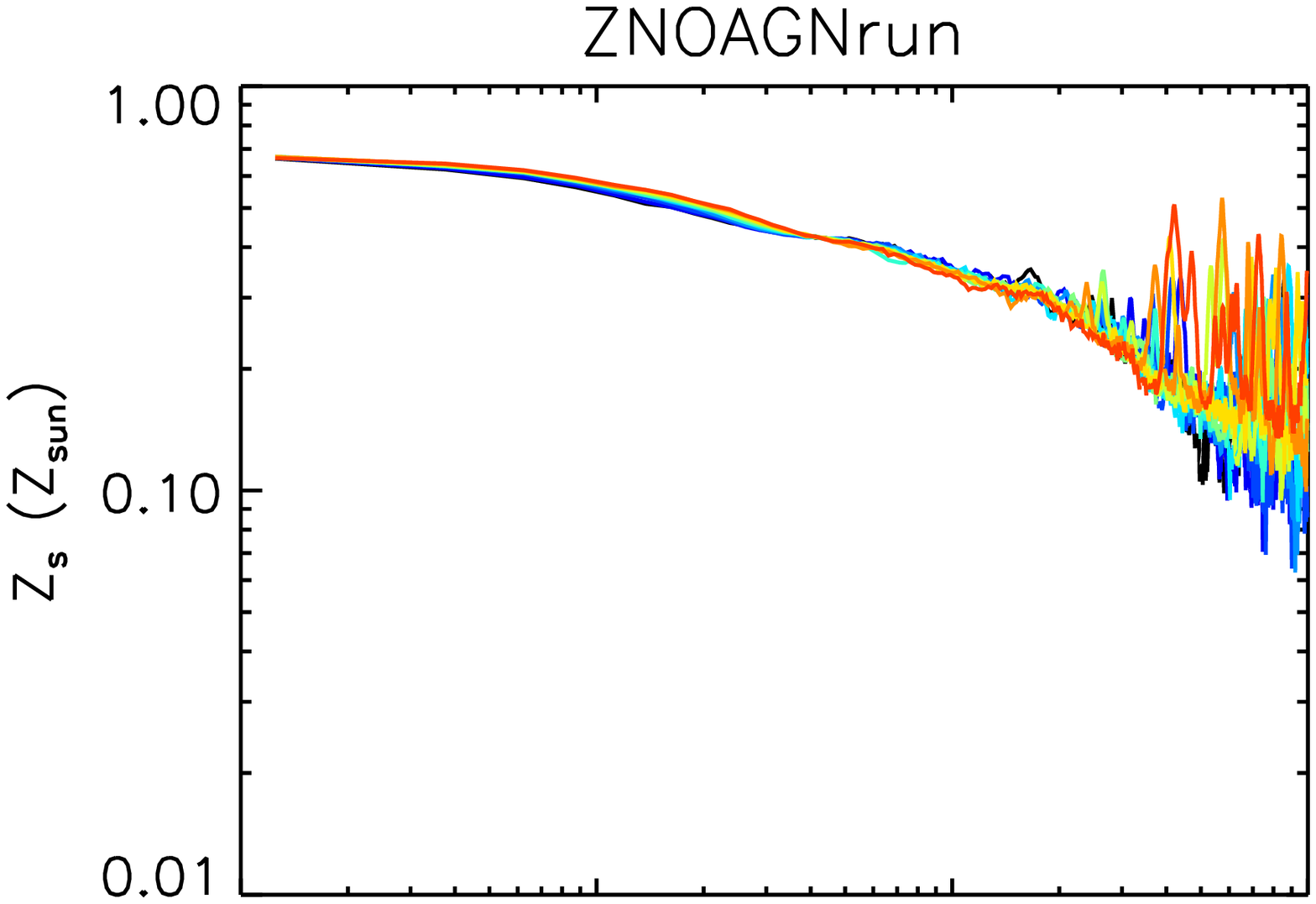}}\hspace{-1.47cm}}
  \centering{\resizebox*{!}{3.6cm}{\includegraphics{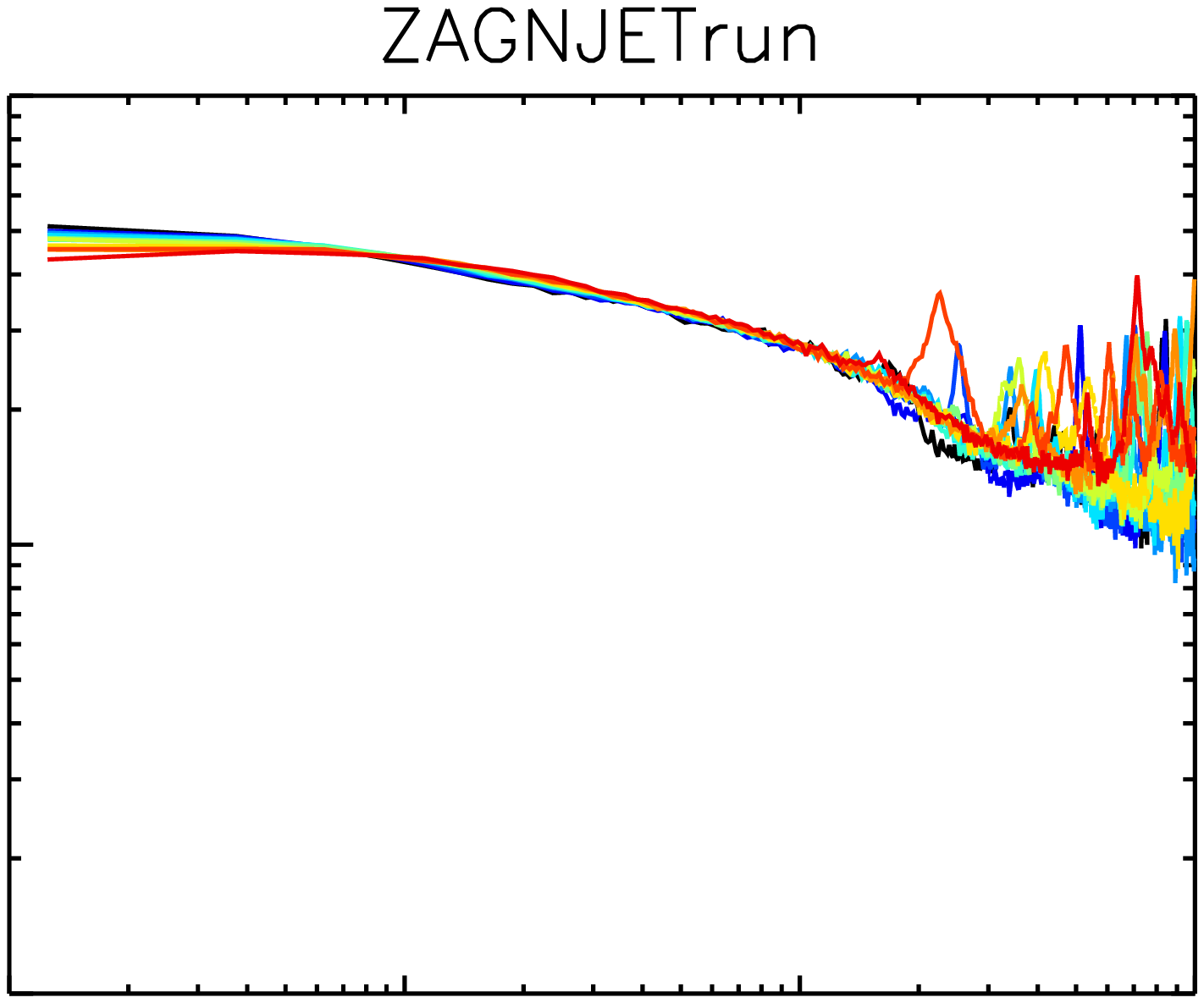}}\vspace{-1.04cm}}
  \\
  \centering{\resizebox*{!}{3.6cm}{\includegraphics{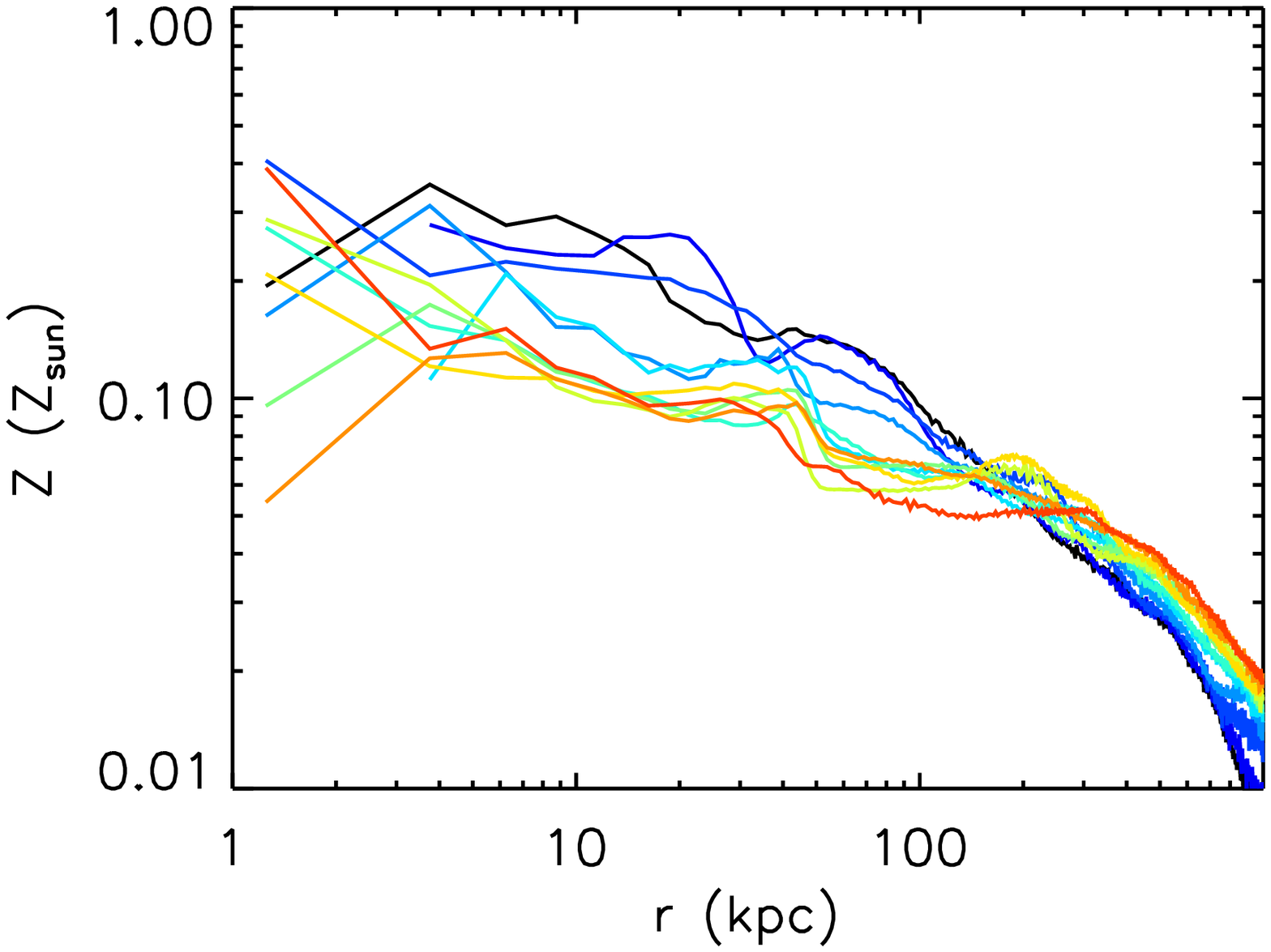}}\hspace{-1.47cm}}
  \centering{\resizebox*{!}{3.6cm}{\includegraphics{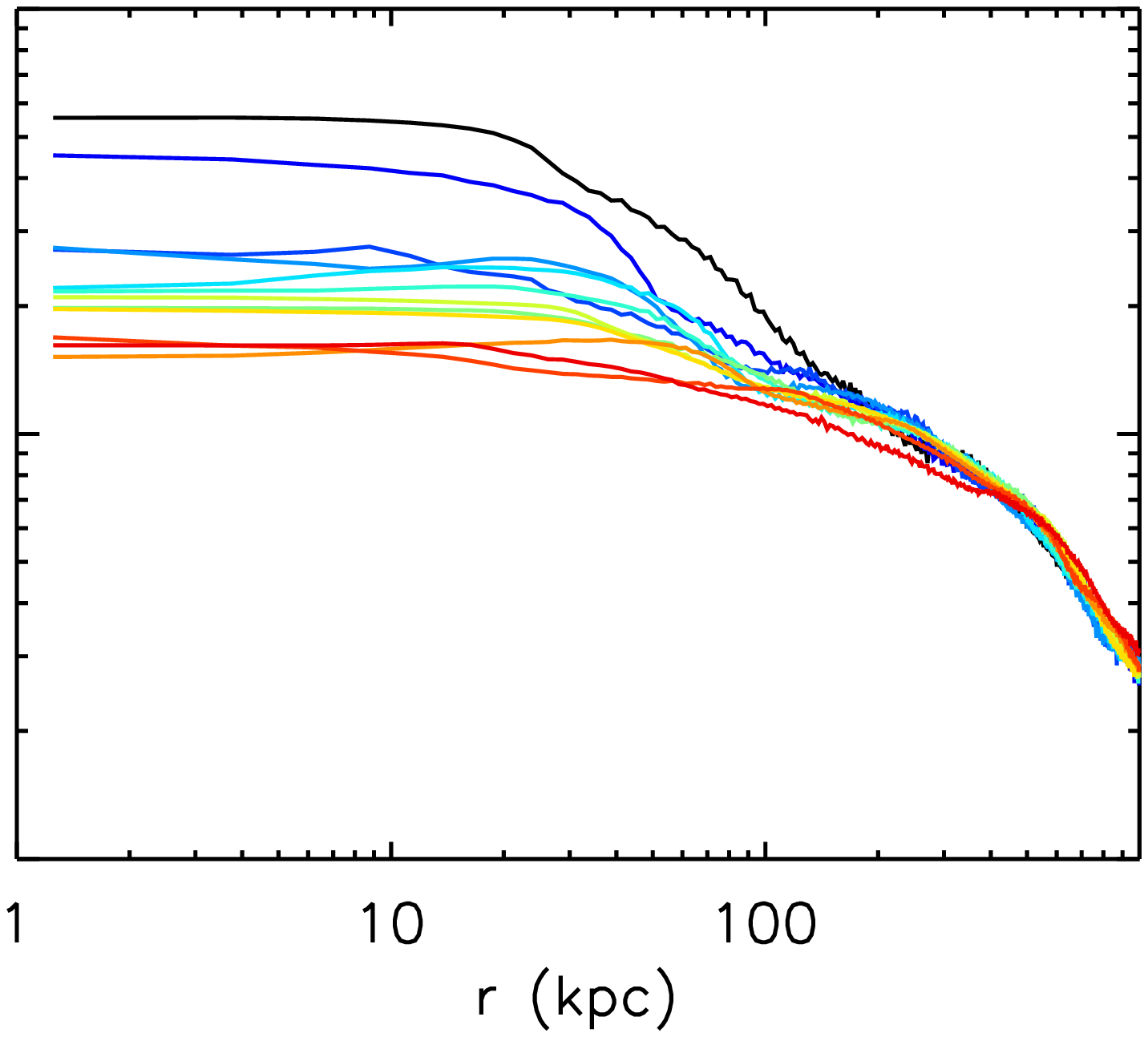}}}
  \caption{Stellar metallicity profiles (first row) and gas metallicity profiles (second row) for the ZNOAGNrun (left column) and the ZAGNJETrun (right column).}
    \label{Metal_Z}
\end{figure}

Finally, we must also point out that we measure a lower gas metallicity, $Z\simeq
0.15 \, Z_{\odot}$ at $z=0$, plotted in fig.~\ref{Metal_Z} 
and a steeper spatial gradient that those observed ($Z\simeq0.3$-$1
\, Z_{\odot}$ from \citealp{sandersonetal09}).  This is even worse for
the stellar metallicity, which is observed to be larger than
solar for giant ellipticals (e.g. \cite{casusoetal96}).  Aside from
uncertainties in the stellar yields and the stellar initial mass
function, there are (at least) two other reasons why we are not able
to achieve more realistic metal abundances. The first one is the
inability at a given DM mass resolution to resolve well enough the
star formation of small satellites galaxies that are responsible for the enrichment of the IGM with 
large-scale galactic winds \citep{dubois&teyssier08winds}.  The second
is the lack of modeling of both stellar winds and Type Ia SN that return a
large fraction of their stellar mass content back to the gas. We plan
to address both issues in a near future but the comparison of the no
metal and (low) metal runs that we have performed strongly suggest
that a higher amount of metals in the gaseous phase will simply lead
to more intense early AGN feedback.

\section{Conclusions}
\label{conclusion}

We have performed high-resolution resimulations of a cosmological galaxy cluster,
using standard implementations for galactic physics (e.g. radiative
cooling, star formation, supernova feedback, 
uniform UV background heating to model reionization)
and different prescriptions for AGN
feedback \citep{duboisetal10, teyssieretal11}. 

Whereas observations indicate that non-cool cores are highly correlated with the level of disturbance of the ICM (e.g. \citealp{sandersonetal09locuss, rossetti&molendi10}), 
suggesting that major mergers are responsible for the formation of non-cool core objects,  our set of simulations leads 
us to conclude that merger activity alone (i.e. not backed by AGN feedback) cannot sustain core profiles over 
timescales comparable to the Hubble time. It therefore seems difficult to argue that this mechanism can account for a 
significant fraction of the non-cool core cluster population. These simulations also show that 
pre-heating by SN and AGN feedback at high redshift alone, before the cluster has fully
assembled is not sufficient to prevent the occurrence of a cooling
catastrophe. Instead, they lend support to the view that AGN must
provide some amount of extra pressure support to the gas in the cluster core after the
galaxy cluster has formed. This late feedback is also key to shape the
thermodynamical properties of the central ICM. 

Indeed, we established that for a gas with pristine composition, the natural interaction between an AGN jet and the ICM 
which regulates the growth of the AGN's BH,  can produce entropy profiles
with different shapes: sometimes the profile has an
entropy floor in the core (i.e. the cluster will be classified as
non-cool core), while at other times, if cooling is sufficiently rapid
there will be a power-law entropy profile extending from
large radii down to the cluster centre (i.e. the cluster will belong
to the cool core cluster category). Hence, AGN jet feedback appears to be a
good candidate to explain the variety of entropy profiles observed in galaxy clusters
\citep{sandersonetal09}, provided metal cooling is neglected.

Without significant AGN feedback at
intermediate and low redshift, clusters rapidly transfer their
low-entropy ICM material to a galactic disc and replace it with
the high-entropy gas stocked in their outskirts. As a result, a  
high-entropy floor and large amounts of baryons are systematic
features of cluster cores.  Moreover, as the cooling catastrophe 
ensues, gigantic starbursts erupt within the massive galactic discs
anchored at the bottom of the gravitational potential well of these clusters. 
Taken at face value, these results seem to contradict the recent
findings of \cite{mccarthyetal11}'s simulations of galaxy
groups. In these lighter structures, they conclude that, on average, turning off the AGN feedback at
intermediate redshift ($0.375<z<1.50$) lowers the entropy of the gas
in the core compared to simulations where AGN feedback is permitted to
proceed down to $z=0$.  Assuming that the differences between our results cannot be attributed to
numerics (SPH versus Eulerian grids), we interpret them as an 
indication of a fundamental difference in the way feedback mechanisms
can operate in galaxy groups and clusters. For the former,
the effects of high redshift AGN feedback are still felt at $z=0$,
whereas for the latter they are wiped out by the stronger gravitational
forces on time scales comparable to the assembly time and therefore 
need to be sustained at later epochs. 

Finally, we assessed the effect of metal cooling on the gas properties
of the ICM.  Metals ejected by SNe
accelerate the gravitational collapse of baryons by sapping their
internal energy. As a result, in the absence of an
AGN feedback mechanism, the cooling catastrophe is exacerbated. However, when AGN feedback 
is turned on, AGN activity is fueled by a more rapid growth of the
central BH engine, and thus has a more dramatic impact on the ICM and the baryon
content of galaxies. Indeed as the gas concentrates in structures earlier, 
larger BH accreting at a higher rate can push the gas further away
from the centers of less massive halos, resulting in less concentrated
gas and stellar profiles in the final galaxy cluster at low redshift.  Since the gas
density decreases, and entropy scales as
$K\propto n_e^{-2/3}$, larger entropy floors are logically present in the core
of metal rich galaxy clusters. These effects systematically lead to the formation 
of non-cool core clusters regardless of how AGN feedback was implemented in the simulations, 
which is in obvious contradiction with the observed cool core / non-cool core dichotomy.

There are several outstanding issues which need to be addressed in
more detail with follow up work. First we need to extend our study to both lower-mass galaxy
groups and the most-massive superclusters to verify that our conjecture regarding 
the different impact of early AGN pre-heating on groups and clusters
holds.  At the same time, large number statistics are mandatory to
support our view that the two types of entropy profiles (cool and non-cool
cores) are simply linked to different evolutionary stages of cluster
assembly histories. It also remains to be shown that we can
retrieve this result when metals are included in the simulations. 
Finally, the metallicities of both the simulated ICM and central massive
galaxy under-estimate observations so far. It will therefore be important to quantify 
how resolution, stellar winds and type Ia SNe alter metallicity profiles, cooling times,
and, thus, the gas properties of the ICM. The effect of X-ray photo-ionisation
of ion species which act as coolants of the proto-cluster intra-halo gas must also be explored
as it is possible that it will offset the large amount of extra metal
cooling. In summary, we find that the cooling core structure is extremely sensitive both to 
the physics and the AGN feedback modelling, and consequently requires a more careful investigation.

\section*{Acknowledgment}

We would like to thank the anonymous referee for his very constructive report that helped us to improve the quality of this paper. We thank Andrei Kravtsov and Arif Babul for useful
discussions. We thank Francesco Vazza and Steen Hansen for their comments. We also thank Alastair Sanderson for providing his observational data points. YD is supported by an STFC Postdoctoral Fellowship. The
simulations presented here were run on the TITANE cluster at the
Centre de Calcul Recherche et Technologie of CEA Saclay on allocated
resources from the GENCI grant c2009046197, and on the DiRAC facility jointly funded
by STFC, the Large Facilities Capital Fund of BIS and the University
of Oxford. This research is part of the Horizon-UK project. JD and AS' research is supported by Adrian Beecroft, the Oxford Martin
School and STFC.

\begin{appendix}

\section{Volume-weighted versus X-ray-weighted profiles}
\label{AppendixA}

\begin{figure*}
  \centering{\resizebox*{!}{3.4cm}{\includegraphics{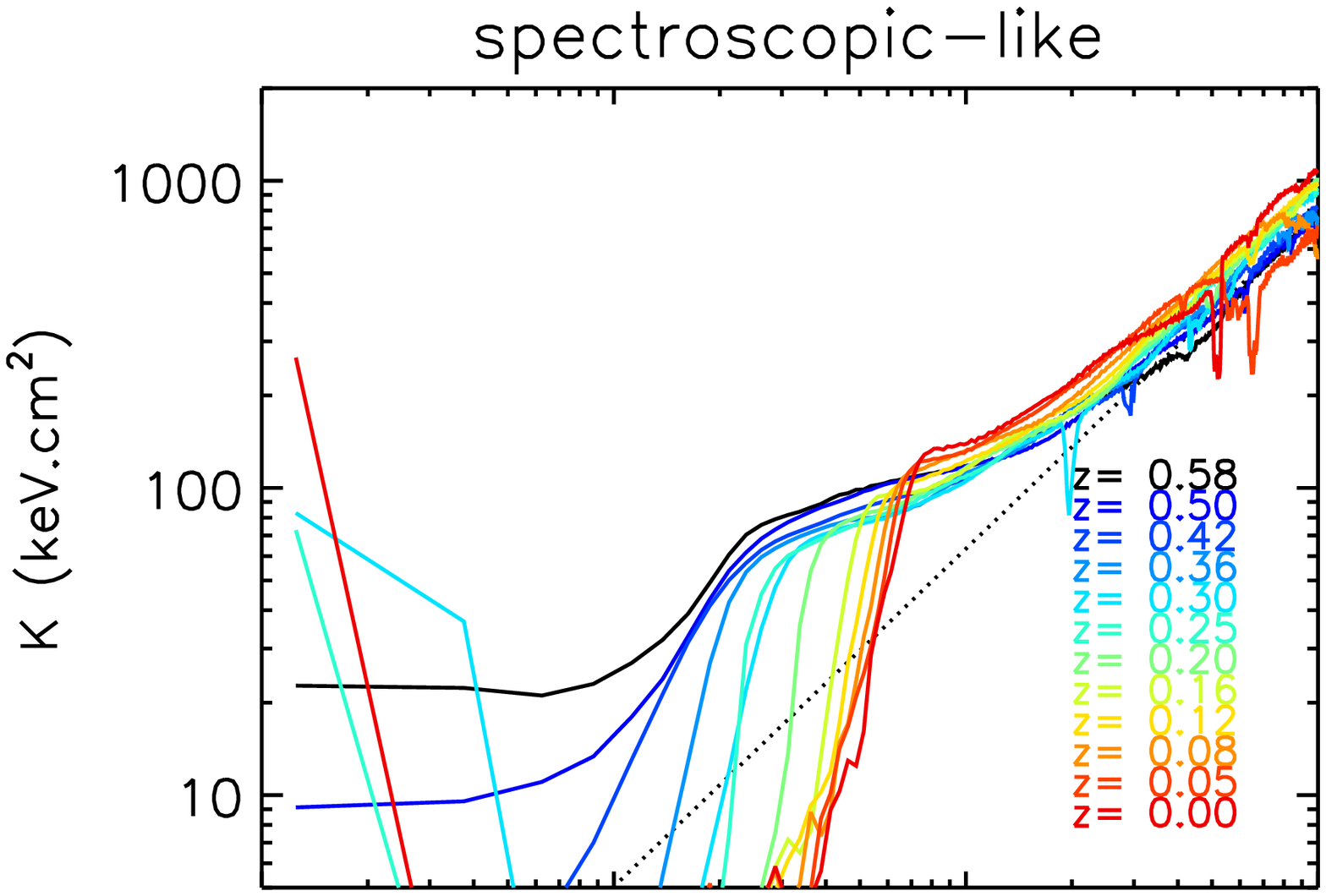}}\hspace{-1.40cm}}
  \centering{\resizebox*{!}{3.4cm}{\includegraphics{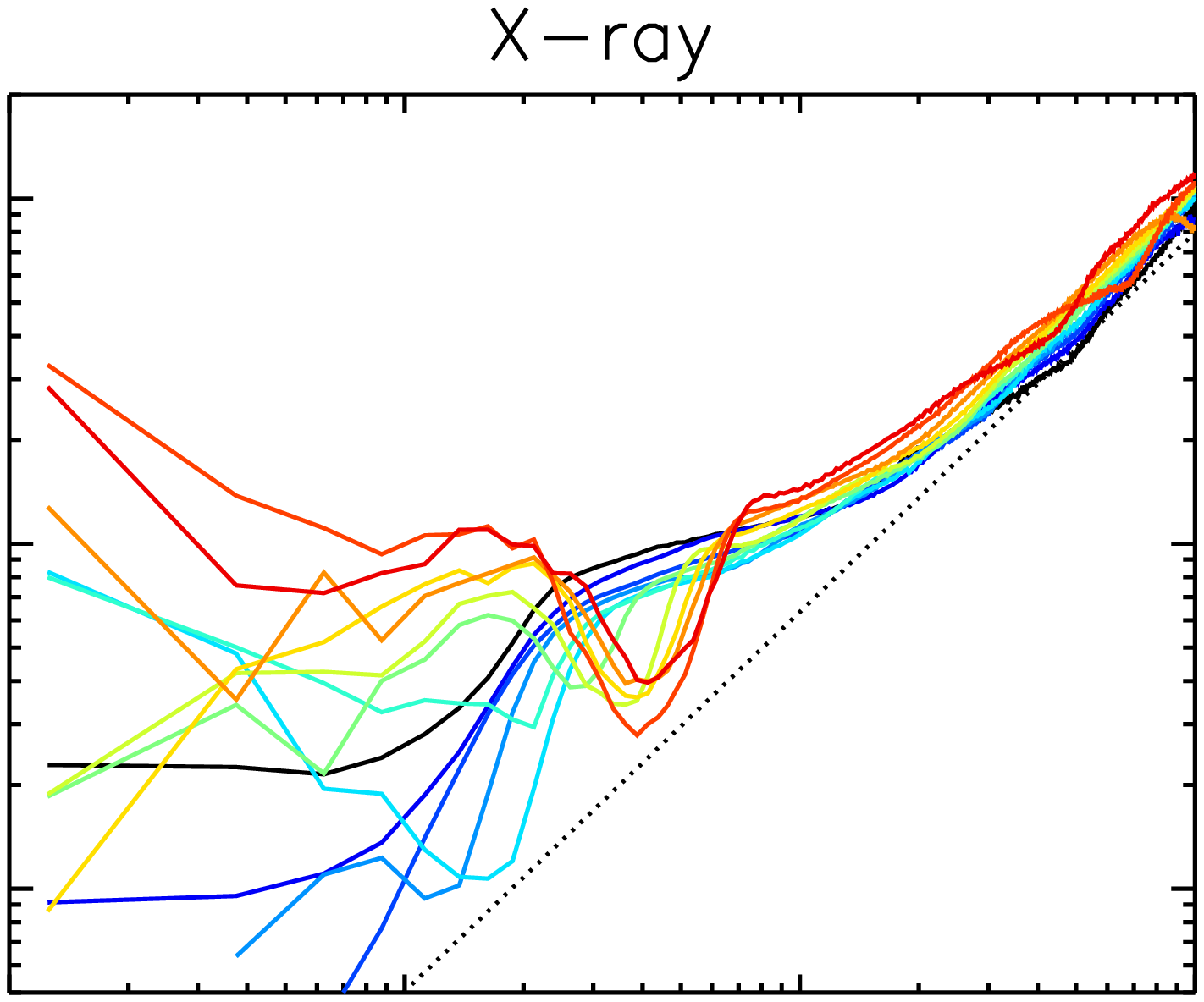}}\hspace{-1.40cm}}
  \centering{\resizebox*{!}{3.4cm}{\includegraphics{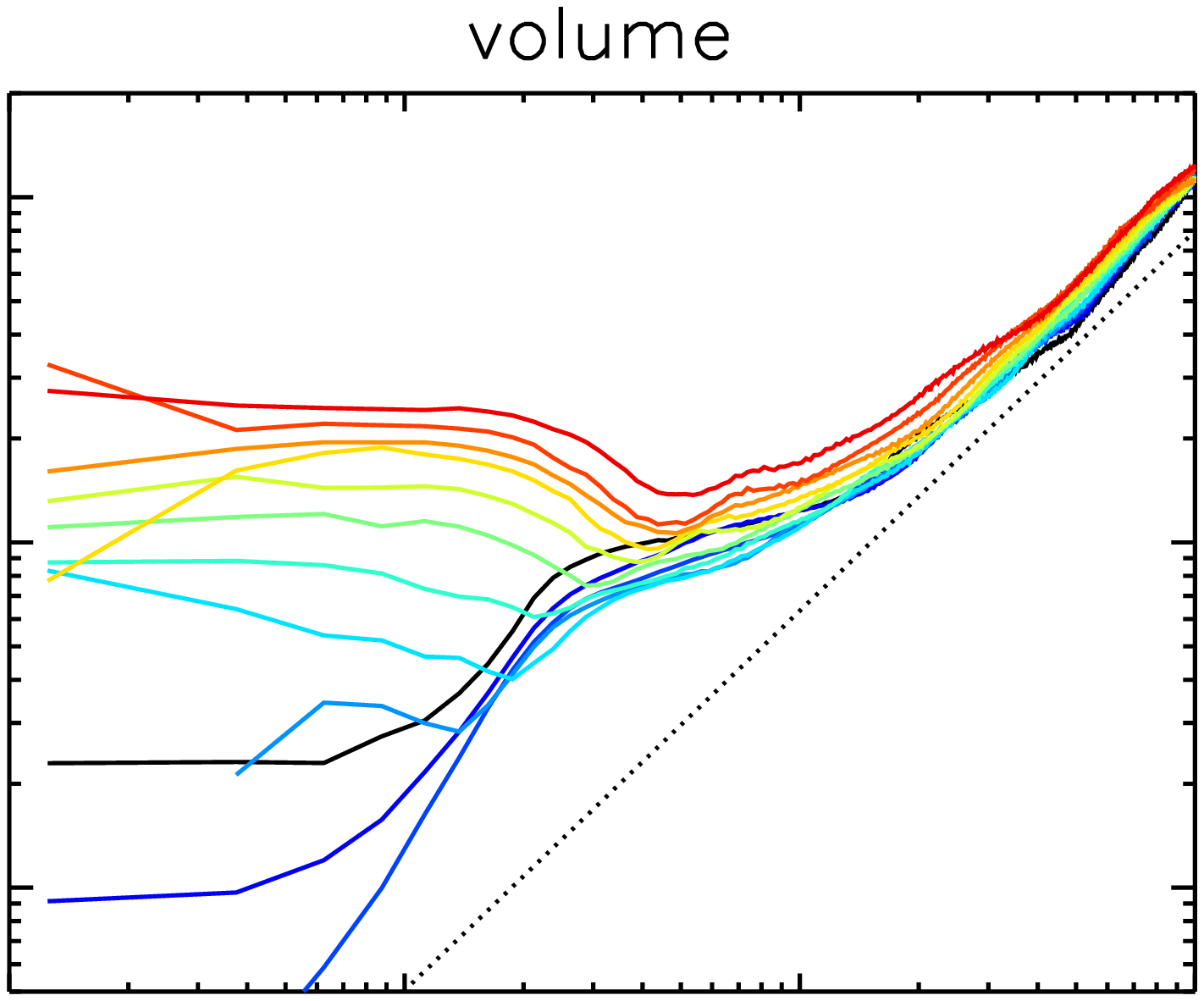}}\vspace{-0.99cm}}
\\
  \centering{\resizebox*{!}{3.4cm}{\includegraphics{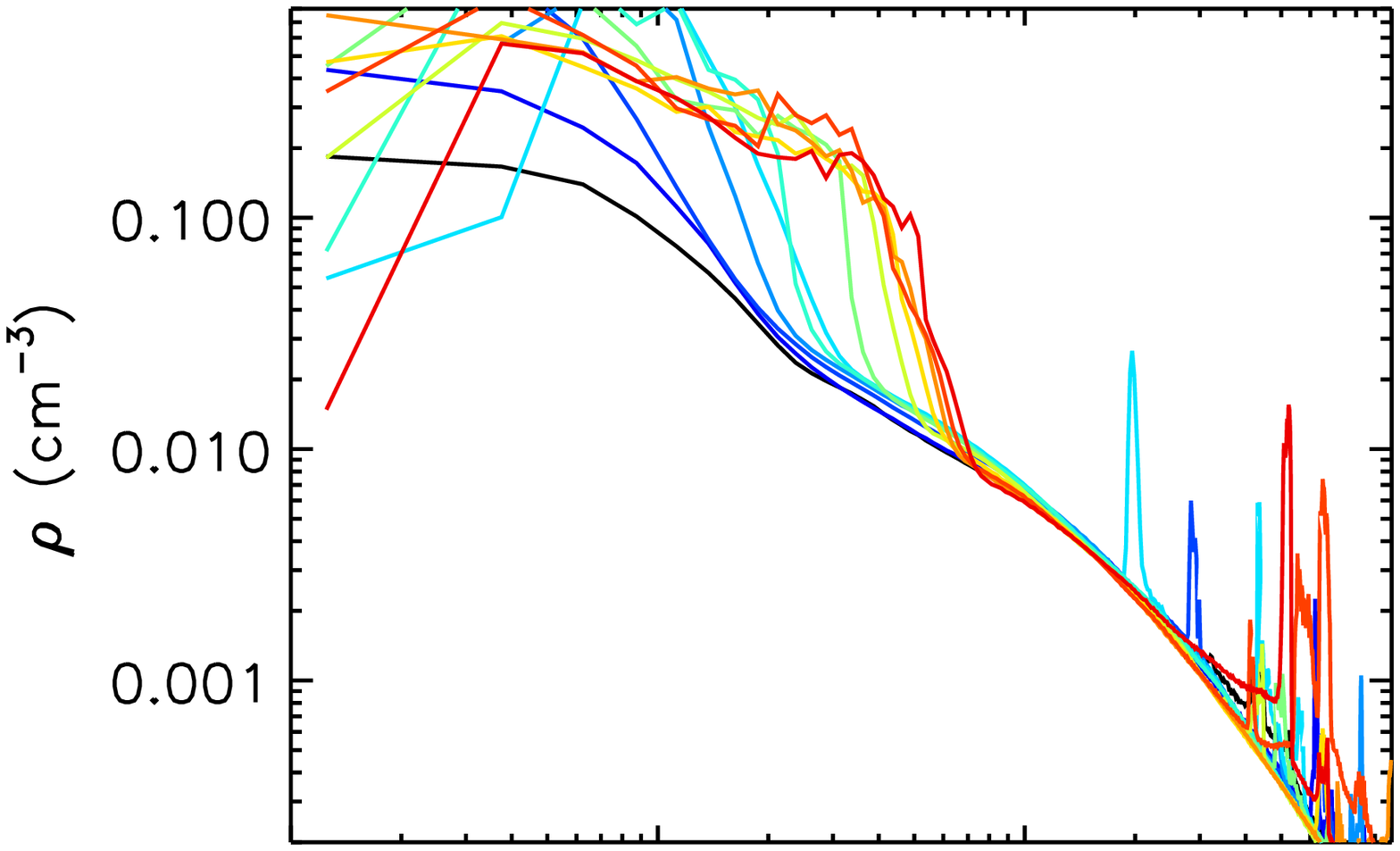}}\hspace{-1.40cm}}
  \centering{\resizebox*{!}{3.4cm}{\includegraphics{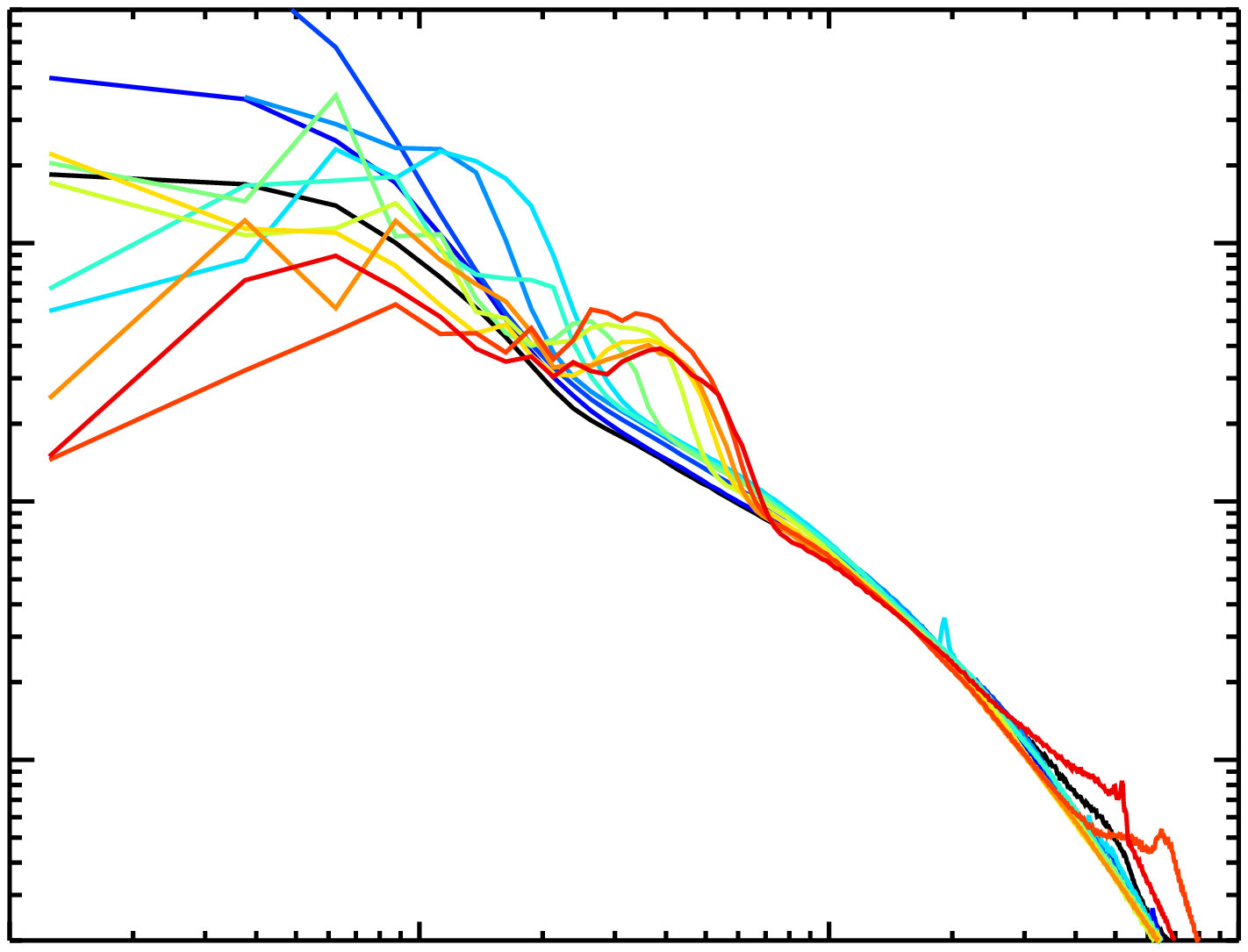}}\hspace{-1.40cm}}
  \centering{\resizebox*{!}{3.4cm}{\includegraphics{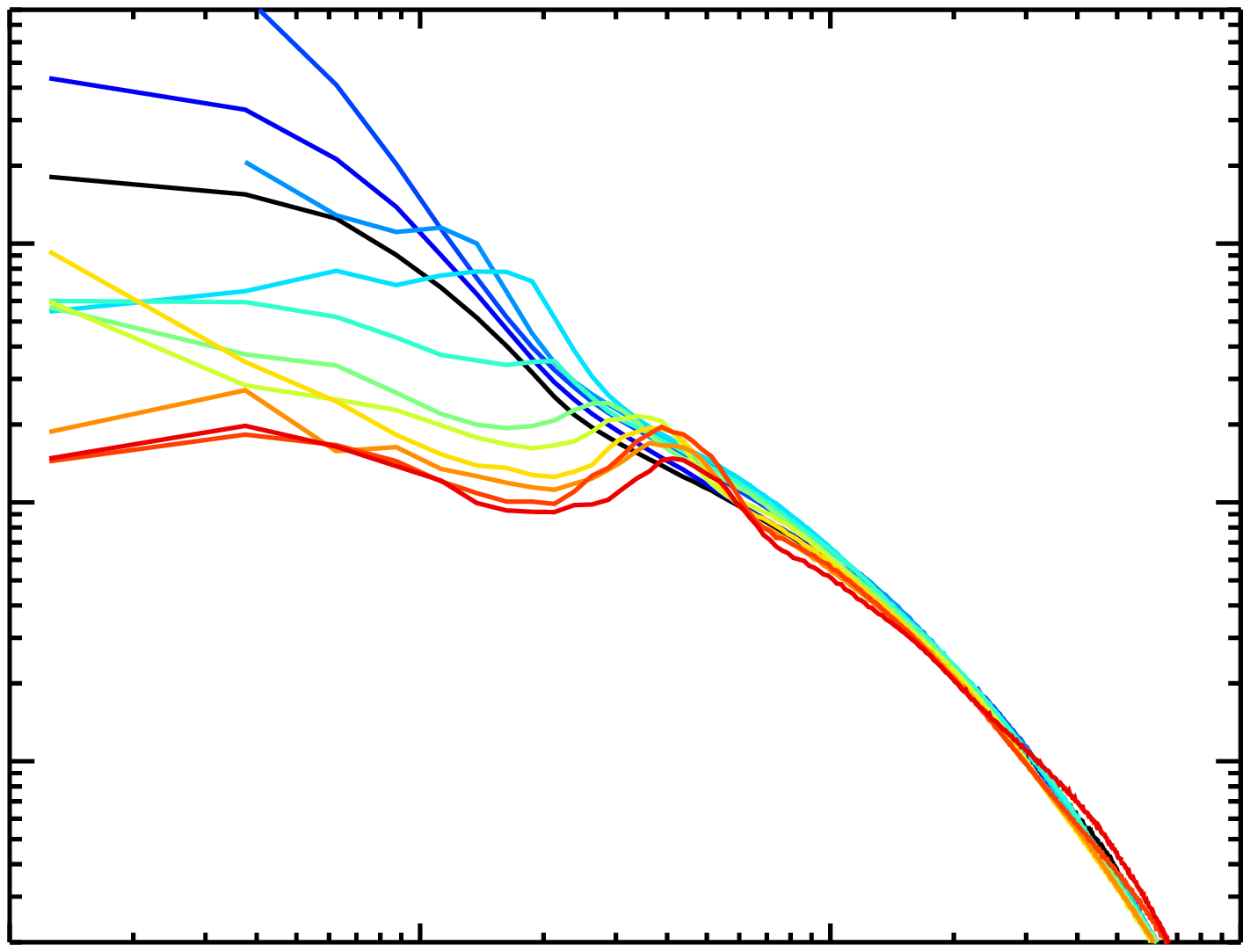}}\vspace{-0.99cm}}
\\
  \centering{\resizebox*{!}{3.4cm}{\includegraphics{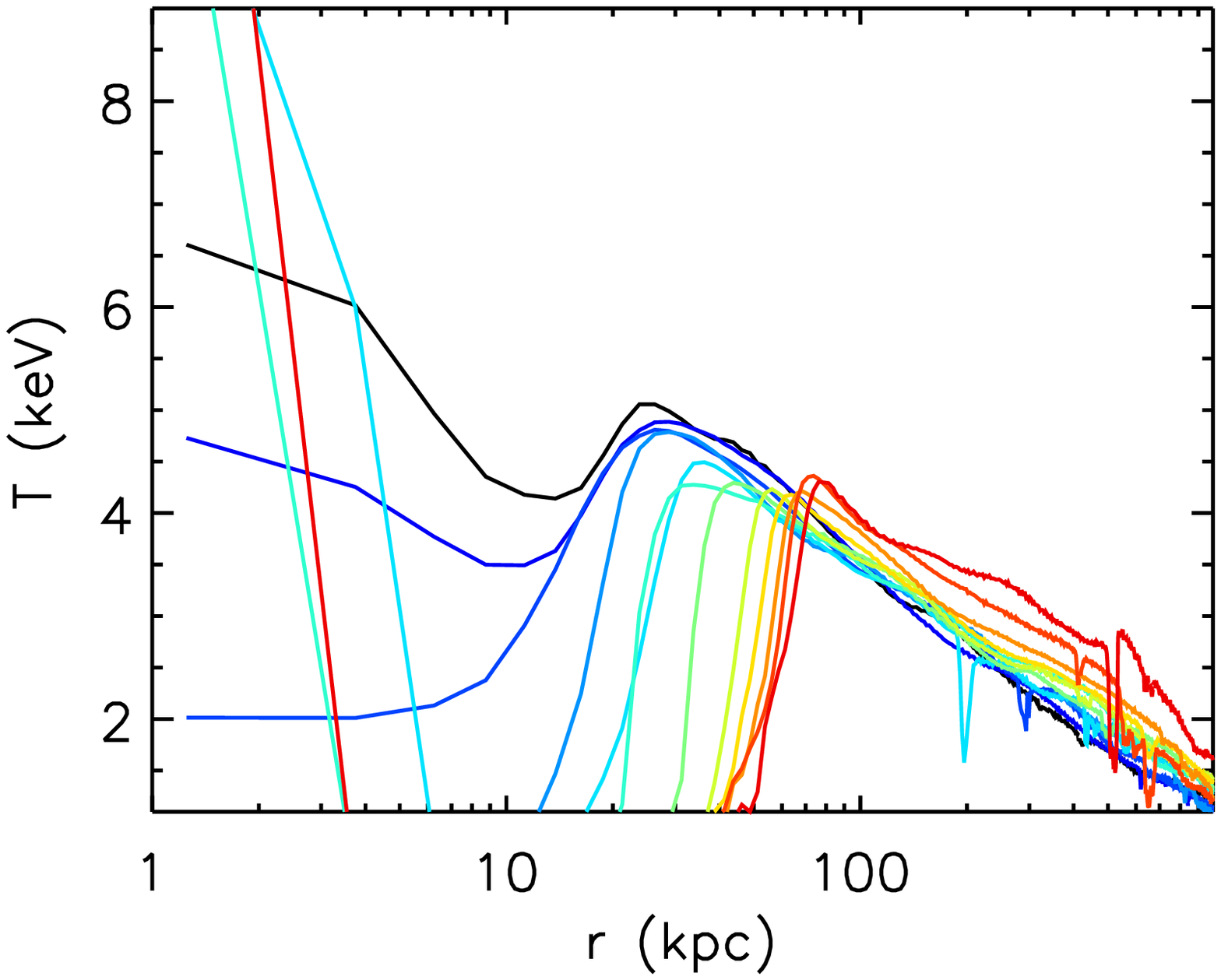}}\hspace{-1.40cm}}
  \centering{\resizebox*{!}{3.4cm}{\includegraphics{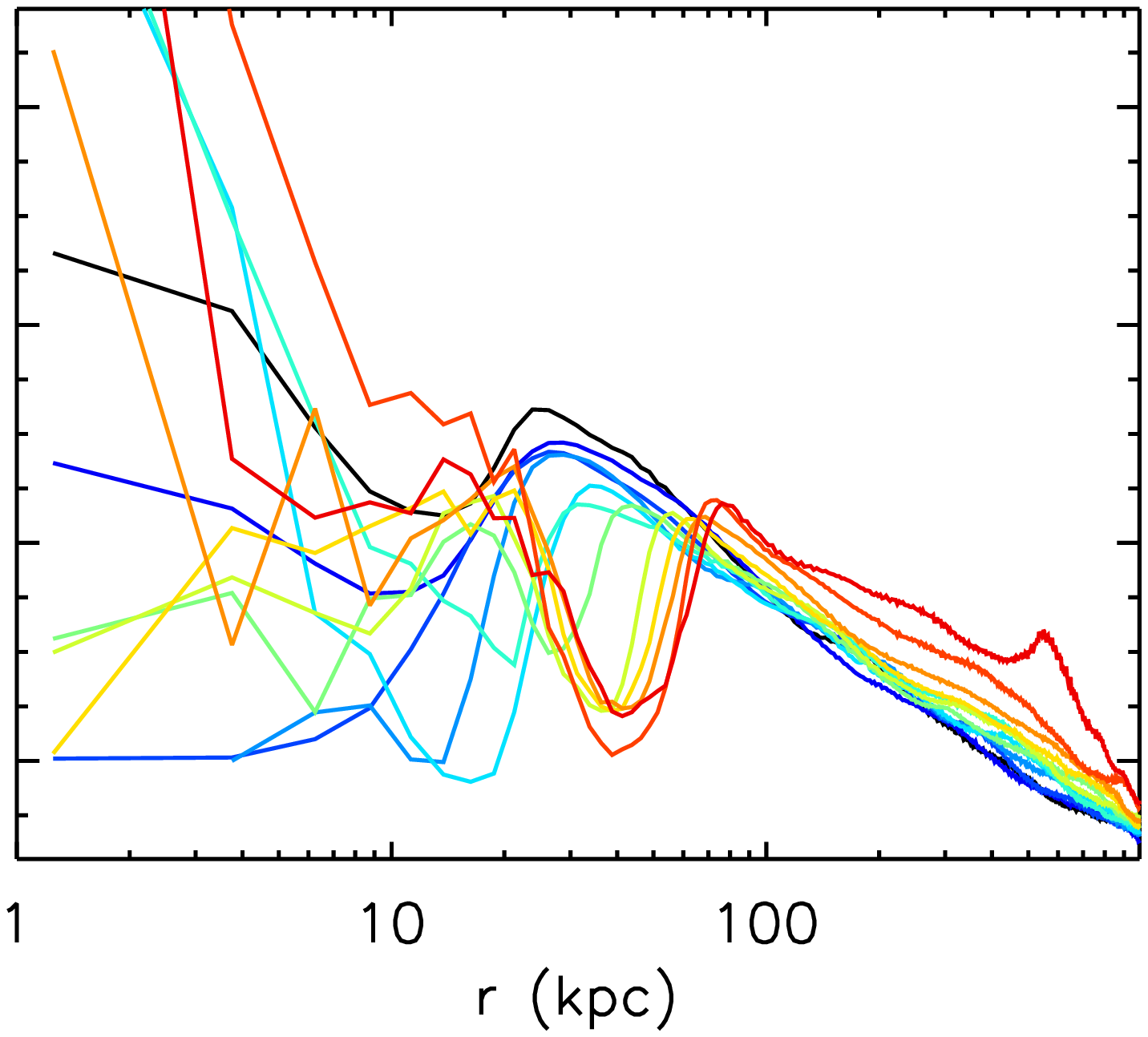}}\hspace{-1.40cm}}
  \centering{\resizebox*{!}{3.4cm}{\includegraphics{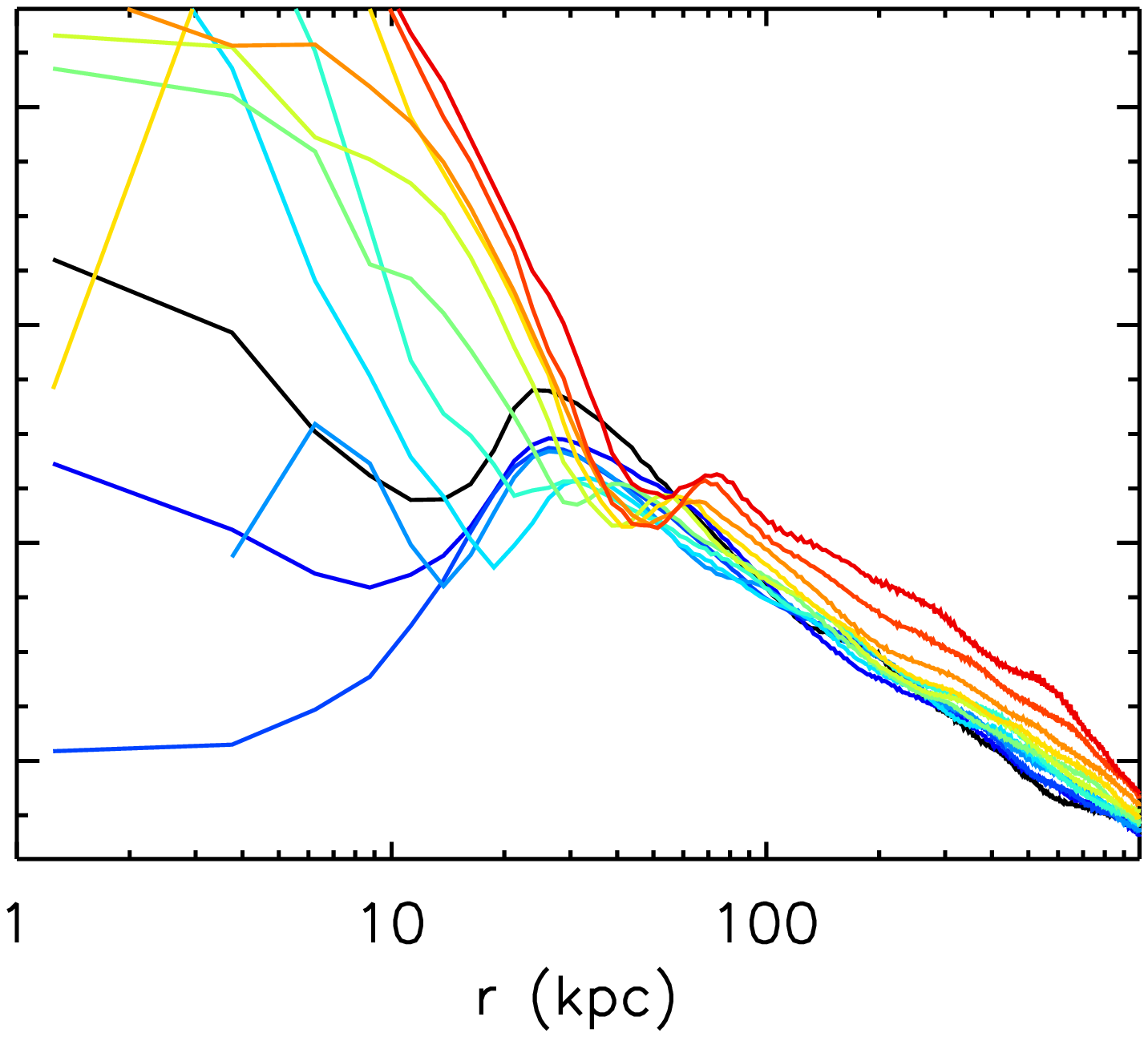}}}
  \caption{A comparison of spectroscopic-like weighting (first column), X-ray weighting (second column) and
volume weighting (third column) of entropy, gas density, and temperature profiles
for the AGNOFFrun. Cells with temperature below $T=0.3$ keV are excised in the profiles using spectroscopic-like weighting, while
those with temperature below $T=1$ keV are excised for both the X-ray and volume weighted profiles. Colors correspond to
profiles measured at different redshifts listed in the panel of the entropy profiles for the spectroscopic-like weighting case. }
\label{AGNOFF_comp}
\end{figure*}

A commonly adopted way to measure ICM gas properties is to
weight them by their X-ray emission ($\propto n_{\rm
e}^2\sqrt{T}$).  Indeed, as observations of ICM gas are often done
in X-ray (although the Sunyaev-Zeldovich (SZ) effect can also be a good probe of the gas
temperature, density, etc., \citealp{kayetal08, prokhorovetal10}),
it seems natural for one to mass-weight or emission-weight 
one-dimensional gas profiles when comparing simulations with real data.
However, the problem with such weightings is that strong density contrasts
will mask the contribution of diffuse gas. This typically happens
when a cold, gaseous galactic disc is present in the centre of a galaxy
cluster.  The central parts of the profiles are dominated by the ISM
emission because of its large density, and so the temperature will
shift towards the peak of radiative cooling around $10^4$-$10^5$ K. 
One solution often used  is to excise this ISM dominated emission by 
removing all gas elements colder than some temperature ($\sim0.1$
keV),  which corresponds to the lower temperature limit of X-ray instruments (\emph{Rosat,
Chandra, XMM-Newton}). Another form of weighting which might provide
a better comparison to observations is spectroscopic-like weighting
\citep{mazzottaetal04}. In this appendix, we discuss why none of these
prescriptions is entirely satisfactory.

In fig.~\ref{AGNOFF_comp}, we compare spectroscopic-like weighting (first column), X-ray weighting (second column) and
volume weighting (third column) of entropy, gas density, and temperature profiles
for the AGNOFFrun. 
The cluster in this simulation does not harbour any cold
ISM component at $z=0.58$ (recall it is restarted from the $z=0.58$ output of AGNJETrun), and then
very quickly develops a cold and massive galactic disc in the centre
(see section \ref{agnoff_section} for details).  
The first thing to note is that the slopes and the levels of entropy,
density, and temperature, are very insensitive to the weighting method
sufficiently far away from the centre ($r>80$ kpc), independent of redshift.
Furthermore, when no cold gas disc is present (e.g. at $z=0.58$, $z=0.50$, and $z=0.42$)
there exist no significant difference between the different weighting
methods all the way to the centre of the cluster.

It is only when the cold ISM has built up significantly
that the profiles start to differ strongly.  Even when 
cold gas cells (defined as cells with temperature $T<0.3$ or $T<1$
keV) are excluded from the analysis,  X-ray and spectroscopic-like
profiles are dominated by the disc component.  The reason for this is that
even though most of the cold gas in galaxies is at very low
temperatures ($T\ll0.1$ keV), the gas at the interface between the disc
and the ICM lies close to a very strong discontinuity, and as such is
extremely sensitive to any kind (numerical or physical) of diffusion
process that mixes the two phases together, especially because it is extremely difficult to resolve the cooling length ($\lambda_{\rm cool}=c_s t_{\rm cool}$) .  This disc
`skin' is extremely difficult to peel from the profiles  except by
applying a prohibitive density/temperature cut-off. 
In addition to these 'skins', large galactic discs also sport a flaring component at large
radii, which is dense ($\rho\sim0.1\, \rm cm^{-3}$), hot ($T\sim1$
keV ), and low entropy ($K \ll 100$ keV.cm$^2$).  As X-ray emission is
proportional to $\rho^2$, this means that even when suppressing cold gas
cells ($T< 1$ keV), profiles will still be dominated by the disc tail
component. The characteristic signature of such a disk tail is apparent in the X-ray
weighted profiles of fig.~\ref{AGNOFF_comp} (middle column) with cells below a temperature of $0.3$ keV 
removed. At $\sim$ 40 kpc from the centre, a dip is visible 
in both the entropy and temperature profiles, and
a bump shows up in the density profiles.  In contrast, volume-weighting the
profiles, allows us to mostly probe the volume filling ICM
without being strongly affected by the disc component of the central
galaxy. For this reason, we adopted volume-weighted
quantities throughout this paper.

\end {appendix}

\bibliographystyle{mn2e}
\bibliography{author}

\label{lastpage}

\end{document}